\documentclass[12pt,a4paper]{article}

\usepackage{epsfig}
\usepackage{axodraw}
\usepackage{cite} 
\usepackage{url}
\usepackage{enumerate}
\usepackage{multirow}
\usepackage{longtable}

\usepackage{bold-extra}

\newlength{\abstwidth}
\newlength{\captivewidth}
\begin{document}
 
\renewcommand{\topfraction}{0.9}    
\renewcommand{\bottomfraction}{0.9} 
\renewcommand{\textfraction}{0.1}   
\renewcommand{\floatpagefraction}{0.8}
 
\newcommand{\mrm}[1]{\mathrm{#1}}
\newcommand{\mbf}[1]{\mathbf{#1}}
\newcommand{\mtt}[1]{\mathtt{#1}}
\newcommand{\tsc}[1]{\textsc{#1}}
\newcommand{\tbf}[1]{\textbf{#1}}
\newcommand{\ttt}[1]{\texttt{#1}}
\newcommand{\br}[1]{\overline{#1}}
\newlength{\tmplen}
\newcommand{\clab}[1]{\tiny\settowidth{\tmplen}{\scriptsize#1}%
\colorbox{white}{\textcolor{white}{#1}}\hspace*{-1.27\tmplen}\scriptsize#1}

\def\lsim{\mathrel{\rlap{\lower4pt\hbox{\hskip1pt$\sim$}}
    \raise1pt\hbox{$<$}}}                
\def\gsim{\mathrel{\rlap{\lower4pt\hbox{\hskip1pt$\sim$}}
    \raise1pt\hbox{$>$}}}                
\newcommand{\half}{$\frac{1}{2}$}        
 
\newcommand{\alphas}{\alpha_{\mathrm{s}}}
\newcommand{\alphaem}{\alpha_{\mathrm{em}}}
\newcommand{\pT}{\ensuremath{p_{\perp}}}
\newcommand{\pTs}{\ensuremath{p^2_{\perp}}}
\newcommand{\kT}{\ensuremath{k_{\perp}}}
\newcommand{\pTmin}{p_{\perp\mathrm{min}}}
\newcommand{\pTmax}{p_{\perp\mathrm{max}}} 
\newcommand{\pTp}{{p'}_{\perp}} 
\newcommand{\pTo}{p_{\perp 0}}
\newcommand{\pTevol}{p_{\perp\mathrm{evol}}}
\newcommand{\pTPOW}{p_{\perp\mathrm{POWHEG}}}
\newcommand{\ECM}{E_{\mathrm{CM}}}
\newcommand{\mmin}{\mathrm{min}}
\newcommand{\mmax}{\mathrm{max}}
\newcommand{\MeV}{\ensuremath{\!\ \mathrm{MeV}}}
\newcommand{\GeV}{\ensuremath{\!\ \mathrm{GeV}}}
\newcommand{\TeV}{\ensuremath{\!\ \mathrm{TeV}}}
\newcommand{\mb}{\ensuremath{\!\ \mathrm{mb}}}
\newcommand{\rem}{\ensuremath{\mathrm{rem}}}
 
\renewcommand{\b}{\mathrm{b}}
\renewcommand{\c}{\mathrm{c}}
\renewcommand{\d}{\mathrm{d}}
\newcommand{\e}{\mathrm{e}}
\newcommand{\f}{\mathrm{f}}
\newcommand{\g}{\mathrm{g}}
\renewcommand{\j}{\mathrm{j}}
\newcommand{\J}{\mathrm{J}}
\newcommand{\hrm}{\mathrm{h}}
\newcommand{\n}{\mathrm{n}}
\newcommand{\p}{\mathrm{p}}
\newcommand{\q}{\mathrm{q}}
\newcommand{\s}{\mathrm{s}}
\renewcommand{\t}{\mathrm{t}}
\renewcommand{\u}{\mathrm{u}}
\newcommand{\A}{\mathrm{A}}
\newcommand{\D}{\mathrm{D}}
\renewcommand{\H}{\mathrm{H}}
\newcommand{\K}{\mathrm{K}}
\newcommand{\Q}{\mathrm{Q}}
\newcommand{\W}{\mathrm{W}}
\newcommand{\Z}{\mathrm{Z}}
\newcommand{\bbar}{\overline{\mathrm{b}}}
\newcommand{\cbar}{\overline{\mathrm{c}}}
\newcommand{\dbar}{\overline{\mathrm{d}}}
\newcommand{\fbar}{\overline{\mathrm{f}}}
\newcommand{\nbar}{\overline{\mathrm{n}}}
\newcommand{\pbar}{\overline{\mathrm{p}}}
\newcommand{\qbar}{\overline{\mathrm{q}}}
\newcommand{\sbar}{\overline{\mathrm{s}}}
\newcommand{\tbar}{\overline{\mathrm{t}}}
\newcommand{\ubar}{\overline{\mathrm{u}}}
\newcommand{\Bbar}{\overline{\mathrm{B}}}
\newcommand{\Dbar}{\overline{\mathrm{D}}}
\newcommand{\Qbar}{\overline{\mathrm{Q}}}
\newcommand{\qval}{\ensuremath{\q_{\mrm{v}}}}
\newcommand{\qsea}{\ensuremath{\q_{\mrm{s}}}}
\newcommand{\qcmp}{\ensuremath{\q_{\mrm{c}}}}
\newcommand{\val}{\ensuremath{{\mrm{v}}}}
\newcommand{\sea}{\ensuremath{{\mrm{s}}}}
\newcommand{\cmp}{\ensuremath{{\mrm{c}}}}
 
\newcommand{\sg}{\tilde{\mathrm{g}}}
\newcommand{\sq}{\tilde{\mathrm{q}}}
\newcommand{\sqd}{\tilde{\mathrm{d}}}
\newcommand{\squ}{\tilde{\mathrm{u}}}
\newcommand{\sqc}{\tilde{\mathrm{c}}}
\newcommand{\sqs}{\tilde{\mathrm{s}}}
\newcommand{\st}{\tilde{\mathrm{t}}}
\newcommand{\schi}{\tilde{\chi}}
 
\newenvironment{Itemize}{\begin{list}{$\bullet$}%
{\setlength{\topsep}{0.2mm}\setlength{\partopsep}{0.2mm}%
\setlength{\itemsep}{0.2mm}\setlength{\parsep}{0.2mm}}}%
{\end{list}}
\newcounter{enumct}
\newenvironment{Enumerate}{\begin{list}{\arabic{enumct}.}%
{\usecounter{enumct}\setlength{\topsep}{0.2mm}%
\setlength{\partopsep}{0.2mm}\setlength{\itemsep}{0.2mm}%
\setlength{\parsep}{0.2mm}}}{\end{list}}
 
\sloppy
 
\pagestyle{empty}
 
\begin{flushright}
LU TP 10-26\\
MCnet/10/20\\
November 2010
\end{flushright}
 
\vspace{\fill}
 
\begin{center}
{\LARGE\bf Interleaved Parton Showers\\[4mm] and Tuning Prospects}\\[10mm]
{\Large R.~Corke\footnote{richard.corke@thep.lu.se} and %
T.~Sj\"ostrand\footnote{torbjorn@thep.lu.se}} \\[3mm]
{\it Theoretical High Energy Physics,}\\[1mm]
{\it Department of Astronomy and Theoretical Physics,}\\[1mm]
{\it Lund University,}\\[1mm]
{\it S\"olvegatan 14A,}\\[1mm]
{\it S-223 62 Lund, Sweden}
\end{center}
 
\vspace{\fill}
 
\begin{center}
{\bf Abstract}\\[2ex]
\begin{minipage}{\abstwidth}
General-purpose Monte Carlo event generators have become important
tools in particle physics, allowing the simulation of exclusive
hadronic final states. In this article we examine the 
\tsc{Pythia 8} generator, in particular focusing on its parton-shower
algorithms. Some relevant new additions to the code are introduced, that
should allow for a better description of data. We also implement and
compare with $2 \to 3$ real-emission QCD matrix elements, to check how well
the shower algorithm fills the phase space away from the soft and collinear
regions. A tuning of the generator to Tevatron data is performed for two
PDF sets and the impact of first new LHC data is examined.
\end{minipage}
\end{center}
 
\vspace{\fill}
 
\clearpage
\pagestyle{plain}
\setcounter{page}{1}

\section{Introduction}
\label{sec:intro}
The production of exclusive hadronic final states involves many
aspects, ranging from calculable perturbative to incalculable 
nonperturbative physics. General-purpose Monte Carlo (MC) event 
generators remain vital tools in the simulation of such events.
Here, typically, leading-order (LO) matrix elements are combined with 
parton showers and hadronisation models. Additionally, when the 
incoming beams are comprised of hadrons, parton distributions functions 
(PDFs) and multiple parton interactions (MPIs) are required to account 
for the (multi)parton content of the hadrons. Alternatively, a large 
number of specialized generators concentrate on providing a more accurate 
perturbative description, e.g. by higher-order calculations. There 
need not be a contradiction between the two approaches; indeed a large 
amount of work has been done to combine them, in order to get the best 
possible overall description.

Two of the most widely used general-purpose MC tools have been 
\tsc{Pythia 6} \cite{Sjostrand:2006za} and HERWIG \cite{Marchesini:1983bm,
Marchesini:1987cf}. These two programs both strive to describe the same
physics, but differ in the models used to do so. One key difference between
the two programs is in their shower algorithms; both are based on
DGLAP evolution \cite{Gribov:1972ri,Altarelli:1977zs,Dokshitzer:1977sg},
supplemented with Sudakov form factors \cite{Sudakov:1954sw}, but while
HERWIG uses angular ordering of emissions, \tsc{Pythia 6} instead offers 
showers with either virtuality ($Q^2$) or transverse-momentum ($\pT$)
ordering. Even with all the differences between the two programs, it is
remarkable the range of physics which both are able to describe.

Despite the successes, both continue to evolve. They have been ported from 
Fortran to C++, but neither of these new versions is a simple rewrite.
New features continue to be added, including changes to their shower 
algorithms. For example, Herwig++ \cite{Bahr:2008pv} has 
a modified shower-ordering variable to be better suited for heavy 
particles. Other new shower developments include the release of 
the Sherpa event generator \cite{Gleisberg:2008ta}, featuring a 
Catani-Seymour dipole shower \cite{Catani:1996vz,Nagy:2006kb,Schumann:2007mg},
and Vincia, based on the dipole-antenna picture \cite{Giele:2007di}.

The original \tsc{Pythia 6} shower was virtuality-ordered. Combined with 
the MPI model \cite{Sjostrand:1987su}, a large amount  of Tevatron data 
can successfully be described. Later, a transverse-momentum-ordered shower
was added in conjunction with an updated MPI model
\cite{Sjostrand:2004ef,Sjostrand:2004pf}. An advantage to this $\pT$
ordering is the implicit inclusion of coherence effects
\cite{Gustafson:1986db}, also present in an angular-ordered shower, but
only crudely implemented in the $Q^2$-ordered shower. Another feature is 
a dipole-style approach to recoils, as pioneered by the \tsc{Ariadne} 
program \cite{Gustafson:1986db,Gustafson:1987rq,Lonnblad:1992tz}. 
The use of a $\pT$ evolution variable for both the shower and MPI 
frameworks also allows for interleaving, i.e.\ for a single downwards 
evolution in $\pT$ where, at any stage, either a shower emission or 
an MPI can take place. In \tsc{Pythia 6}, initial-state radiation (ISR) 
and MPIs are interleaved, while final-state radiation (FSR) is only
considered afterwards. The argument is that it is primarily MPI and ISR
that compete for beam momentum. Also new is the inclusion of showers off 
all MPI subsystems, limited to just the hard scattering in the
$Q^2$-ordered showers.

\begin{figure}
\begin{minipage}{0.5\linewidth}
\includegraphics[scale=0.65]{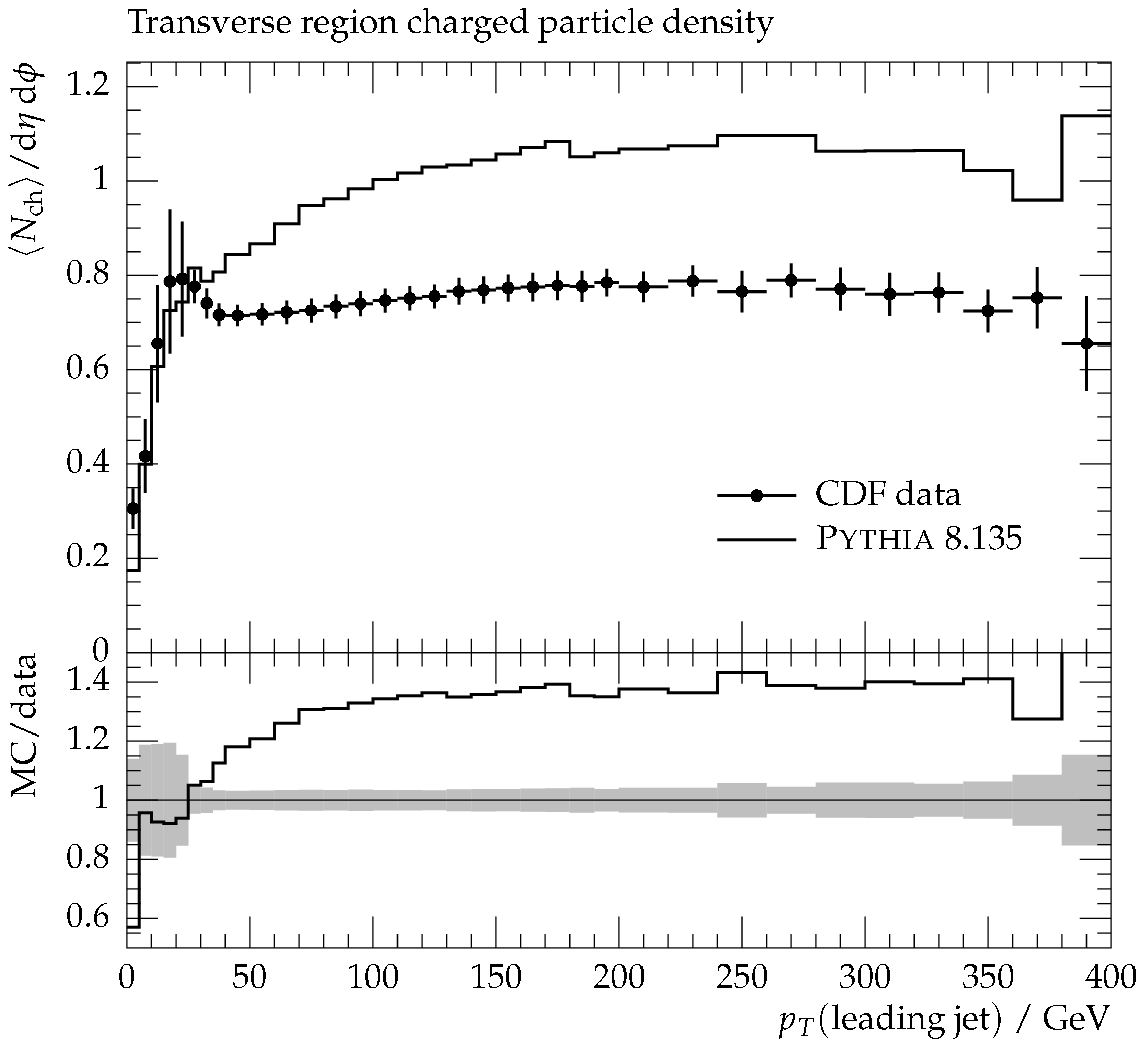}
\vspace{-3mm}\begin{center}{\scriptsize(a)}\end{center}\centering
\end{minipage}
\begin{minipage}{0.5\linewidth}
\includegraphics[scale=0.65]{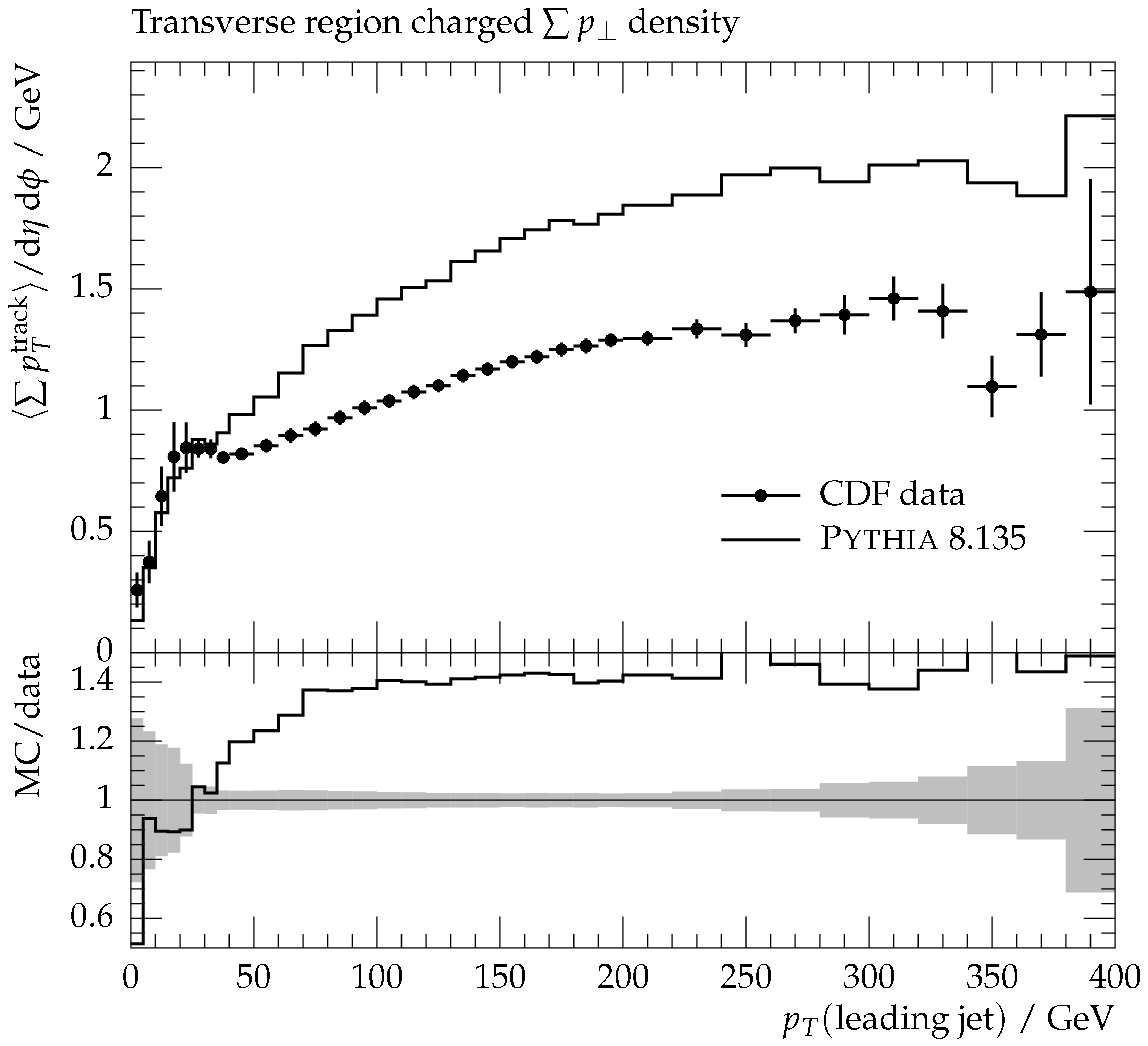}
\vspace{-3mm}\begin{center}{\scriptsize(b)}\end{center}\centering
\end{minipage}
\caption{Rick Field's leading-jet underlying-event analysis implemented in
Rivet. Results for \tsc{Pythia 8.135}, default tune, for transverse region
charged (a) number and (b) $\sum \pT$ density 
\label{fig:intro-rivet}}
\end{figure}

In \tsc{Pythia 8}, the $Q^2$-ordered showers are no longer available,
while the $\pT$-ordered showers are updated \cite{Sjostrand:2007gs}. 
The main change is the additional interleaving of FSR along with
ISR and MPI. It has been possible to obtain a good agreement with much
data also in this new model, but some time ago it was noted that problems exist 
in the description of underlying-event (UE) physics when the model has
been tuned to minimum-bias (MB) data \cite{Hoeth:2009zz}. This is exemplified in
Fig.~\ref{fig:intro-rivet}, which shows the transverse region charged
(a) number and (b) $\sum \pT$ density for \tsc{Pythia 8.135} in 
Rick Field's analysis of the UE activity as a function of the 
leading-jet $\pT$ \cite{Field:2002vt}. Although the default tune
is able to describe minimum-bias multiplicities, 
the activity in the underlying event clearly rises too quickly with jet 
$\pT$. With this in mind, we study the new shower framework of 
interleaved FSR and ISR, and introduce modest changes to both that
significantly improve agreement.

Although the shower description is an approximation of QCD primarily 
intended to be accurate in the soft and collinear limits, it can also be used 
to fill the phase space away from these regions. One direct way to examine 
how well the shower does away from these limits is through a comparison 
to $2 \to 3$ real-emission matrix elements. We will perform such a 
study in this article, for the improved shower framework we present. 

This should be put in context. In preparing for LHC phenomenology, 
there has been a large focus on techniques to combine higher-order 
matrix elements with parton showers. One goal is to describe many hard 
and widely separated jets, achieved by algorithms such as MLM or 
CKKW(-L) \cite{Alwall:2007fs,Catani:2001cc,Lonnblad:2001iq}, while another is 
to combine full NLO calculations with showers, with algorithms such as 
MC@NLO and POWHEG \cite{Frixione:2002ik,Frixione:2003ei,Frixione:2007vw}.
Attempts to combine both NLO calculations and higher-order Born-level
matrix elements have also been made
\cite{Giele:2007di,Bauer:2008qh,Lavesson:2008ah,Hamilton:2010wh}.
\tsc{Pythia 8} comes with matching built in for a few processes
\cite{Miu:1998ju,Norrbin:2000uu}, and can be interfaced to external
matching programs \cite{Corke:2010zj}, but QCD multijets has not been 
high on the list. An eventual goal would be to have at least 
a complete built-in matching of the shower to the QCD $2 \to 3$ matrix 
elements for the first shower emission, e.g.\ in the spirit of 
\cite{Norrbin:2000uu}. 

There are at least two good reasons why the default showers should 
still be fixed up to give as good a description of the three-jet 
(and higher-jet) region as possible. One is that the CKKW-L-matching 
approach is based on obtaining Sudakov form factors from trial showers, 
so the more accurate the shower, the more precise the matching. The 
other is that one can foresee that time-consuming matching approaches 
will primarily be applied to the hardest interaction of an event, 
whereas further MPIs will be handled to the accuracy provided by 
the default showers.

The formulation of sound physics models is central to the development of
a successful generator, but it is not enough. By the nonperturbative,
and therefore incalculable, nature of much of the physics, and by the
uncertainty that comes from shower approximations and unknown higher
orders in the perturbative regime, generators inevitably come to
contain a number of free parameters. Tuning of these parameters thus 
becomes a key part of obtaining an agreement with data. Assuming jet
universality, LEP data can be used to tune FSR and hadronisation.
The $\pT$ spectrum of $\Z^0$ provides a handle on ISR (and, at low $\pT$,
on primordial $\kT$). PDFs are a topic unto themselves, addressed further
in later sections, still leaving the MPI model, the crosstalk between
ISR and FSR, beam remnants, colour flow issues, diffractive physics and
more. The underlying-event analyses here serve as vital ingredients
\cite{Field:2002vt,Kar:2009kc,Field:2009zz}. Many of the tunes for the
$Q^2$-ordered showers of \tsc{Pythia 6} are based on Rick Field's original
Tune A \cite{Field:2005sa}.

The tuning process has now been systematised to
a large extent, and Rivet \cite{Buckley:2010ar} and Professor
\cite{Buckley:2009bj} have become the tools of choice for many MC
programs. For \tsc{Pythia 6}, the Professor team have released both a
$Q^2$-shower tune, Pro-Q20, and a $\pT$-shower tune, Pro-pT0
\cite{Buckley:2009bj,Buckley:2009vk}. For the $\pT$-ordered showers, Peter
Skands has also released the ``Perugia'' tunes, where a central parameter
set is supplemented by 8 related variations that attempt to systematically
explore different changes in the parameter sets \cite{Skands:2010ak}.
The current default tune for \textsc{Pythia 8} is based on a tuning to LEP
data by Hendrik Hoeth, using the Rivet + Professor framework,
supplemented by hadron collider data comparisons by Peter Skands.
Other tunes are often developed within the experimental collaborations
\cite{Moraes:2007rq,AtlasMC09} and tunes that incorporate (preliminary) LHC
data have also started to appear \cite{AtlasTune}. While the new energies
probed by the LHC offer an exciting opportunity to test and constrain models
and model parameters inside generators, data from lower energy runs
($\sqrt{s} = 900\GeV$ and $2.36\TeV$) currently appears slightly
incompatible with earlier Tevatron results \cite{AtlasUE}. It remains to be
seen if a ``global tune'' that can encompass both LHC and Tevatron data is
possible.

In Section 2, the shower algorithms of \tsc{Pythia 8} are reviewed, 
and some developments made to allow an improved description of
underlying-event and minimum-bias data. Other parts of the event generation
framework are also outlined, especially those not previously documented
and relevant for tuning exercises. In Section 3, a comparison of the shower 
is made to $2 \to 3$ QCD real-emission matrix elements. In Section 4, 
some initial tunes of the modified framework are made to Tevatron data, 
while Section 5 contains a first look at the new LHC data. Finally, 
in Section 6, a summary and outlook is given.

\section{Event-generation framework}
\label{sec:evtgen}

\subsection{Interleaved evolution}
\label{sec:interleaved}

The \tsc{Pythia}~8 showers are ordered in transverse momentum 
\cite{Sjostrand:2004ef}, both for ISR and for FSR. Also, multiparton 
interactions (MPI) are ordered in $\pT$ \cite{Sjostrand:1987su}.
This allows a picture where MPI, ISR and FSR are interleaved in one 
common sequence of decreasing $\pT$ values \cite{Sjostrand:2007gs}. 
This is most important for MPI and ISR, since they are in direct 
competition for momentum from the beams, while FSR (mainly) 
redistributes momenta between already kicked-out partons.
The interleaving implies that there is one combined evolution equation
\begin{eqnarray}
\frac{\d \mathcal{P}}{\d \pT}&=& 
\left( \frac{\vphantom{\left(\right)} \d\mathcal{P}_{\mrm{MPI}}}{\d \pT}  + 
\sum   \frac{\vphantom{\left(\right)} \d\mathcal{P}_{\mrm{ISR}}}{\d \pT}  +
\sum   \frac{\vphantom{\left(\right)} \d\mathcal{P}_{\mrm{FSR}}}{\d \pT} \right)
\nonumber \\ 
 & \times & \exp \left( - \int_{\pT}^{p_{\perp\mrm{max}}} 
\left( \frac{\vphantom{\left(\right)} \d\mathcal{P}_{\mrm{MPI}}}{\d \pT'}  + 
\sum   \frac{\vphantom{\left(\right)} \d\mathcal{P}_{\mrm{ISR}}}{\d \pT'}  +
\sum   \frac{\vphantom{\left(\right)} \d\mathcal{P}_{\mrm{FSR}}}{\d \pT'} 
\right) \d \pT' \right) ~,
\label{eq:combinedevol}
\end{eqnarray}
that probabilistically determines what the next step will be.
Here the ISR sum runs over all incoming partons, two per
already produced MPI, the FSR sum runs over all outgoing partons, 
and $p_{\perp\mrm{max}}$ is the $\pT$ of the previous step.
ISR is described by backwards evolution \cite{Sjostrand:1985xi}, wherein
branchings are constructed from the hard process, back to the shower
initiators. Starting from a single hard interaction,
eq.~(\ref{eq:combinedevol}) can be used repeatedly to construct a complete
parton-level event of arbitrary complexity.

The decreasing $\pT$ scale can be viewed as an evolution towards
increasing resolution; given that the event has a particular structure
when activity above some $\pT$ scale is resolved, how might that 
picture change when the resolution cutoff is reduced by some infinitesimal
$\d \pT$? That is, let the ``harder'' features of the event 
set the pattern to which ``softer'' features have to adapt.
It does not have a simple interpretation in absolute time;
all the MPIs occur essentially simultaneously (in a simpleminded
picture where the protons have been Lorentz contracted to pancakes),
while ISR stretches backwards in time and FSR forwards in time. 
The closest would then be to view eq.~(\ref{eq:combinedevol}) as an 
evolution towards increasing formation time.

\subsection{Dipoles vs. Feynman graphs}
\label{sec:dipvsfeyn}

\begin{figure}
\centering
\begin{picture}(160,160)(-80,-90)
\SetWidth{2}
\ArrowLine(-60,60)(0,30)
\Gluon(-60,-60)(0,-30){5}{5}
\Gluon(0,-30)(0,30){5}{4}
\ArrowLine(0,30)(60,60)
\Gluon(0,-30)(60,-60){5}{5}
\SetColor{OliveGreen}
\DashArrowArcn(-60,2)(45,60,-60){2}
\SetColor{Blue}
\DashArrowArcn(0,-120)(71,130,50){5}
\SetColor{Red}
\DashArrowArcn(60,2)(45,240,120){8}
\Text(0,-90)[]{(a)}
\end{picture}
\begin{picture}(200,160)(-100,-90)
\SetWidth{2}
\LongArrow(0,0)(80,00)
\Gluon(0,0)(-80,0){5}{6}
\LongArrow(0,0)(56,57)
\Gluon(0,0)(30,-74){5}{6}
\SetColor{OliveGreen}
\DashLine(80,8)(42,8){2}
\DashArrowLine(40,8)(-80,8){2}
\SetColor{Blue}
\DashArrowLine(-80,-2)(28,-74){5}
\SetColor{Red}
\DashArrowLine(32,-74)(58,57){8}
\Text(5,-90)[]{(b)}
\end{picture}\caption{(a) One possible Feynman graph and colour flow for the 
$\q \g \to \q \g$ process, and (b) the resulting colour dipoles
between the scattered partons and the beam remnants
\label{fig:qgflow}}
\end{figure}
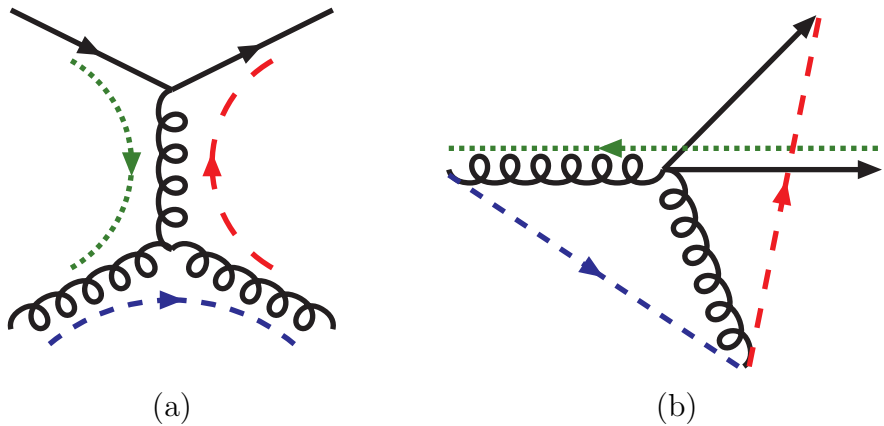

Before the detailed presentation of the initial- and final-state showers,
it is useful to put them in the context of the overall philosophy.
As a simple illustration, consider the process $\q \g \to \q \g$.
One of the contributing graphs is a $t$-channel exchange of a gluon,
with one color-anticolour being annihilated in the process, another
created, and one colour flowing through from the initial to final state,
Fig.~\ref{fig:qgflow}a. This can be transformed to Fig.~\ref{fig:qgflow}b,
illustrating how the partons move out from the interaction vertex. In this
picture, the incoming particles are represented by fictitious outgoing
antiparticles; from a colour point of view, these ``holes'' also represent
the beam remnants left behind by the hard interactions. The three colour
lines of the Feynman graph here transforms into the presence of three
colour dipoles. By the nature of the endpoints, they can be classified as
final-final (FF), final-initial (FI) or initial-initial (II). We begin by
considering the first kind, but with many of the statements relevant for
all three.

In planar QCD, i.e.\ in the limit of infinitely many colours,
$N_C \to \infty$ \cite{'tHooft:1973jz}, the radiation pattern of soft gluons
becomes a simple sum of the radiation from each dipole separately.
Interference between the dipoles are suppressed by a factor
$\mathcal{O}(1/N_C^2)$ and so can often be neglected. When emitted gluons
are not soft, however, recoil effects come into play; the endpoint partons
of a radiating dipole obtain shifted momenta, affecting all dipoles they belong
to. It is therefore not possible to consider emission from the different
dipoles independently. Instead, one should impose a well-defined common
order in which possible emissions are considered, such as $\pT$ 
\cite{Gustafson:1986db,Gustafson:1987rq}.  Each new gluon emission also 
means that the radiating dipole is split into two, so the number of 
dipoles will increase as the shower evolves.
    
While radiation should be viewed as emanating from the dipole-as-a-whole,
it is necessary to impose a strategy for how the recoil of an
emission should be shared between the two endpoints of a dipole.
There is no unique prescription, but obviously the original dipole
direction should not change more than necessary. One possible choice 
is to view the emission as being associated with one of the two 
endpoint partons, the radiator, while the other parton acts as a recoiler. 
If viewed in the rest frame of the dipole, the recoiler does not 
change direction by the emission, but it obtains a reduced absolute 
momentum such that the overall four-momentum of the dipole is preserved 
during the emission. The radiator and the emitted parton obtain opposite 
$\pT$ kicks by the branching, but the distribution in azimuthal angle 
is isotropic, at least to first approximation. All partons, both before 
and after the emission, are put on mass shell.

By the radiator-recoiler strategy, the radiation of a dipole is 
split into that of its two endpoints. This can be done in a smooth 
manner, such that radiation close to either endpoint is associated
with that parton radiating, whereas large-angle radiation is  
split between the two in reasonable proportions. In this way, it becomes 
possible to describe gluon radiation from a quark and gluon end by the
splitting kernels    
\begin{eqnarray}
P_{\q \to \q \g} & = & C_F \frac{1 + z^2}{1 - z} ~, \\
\frac{1}{2} P_{\g \to \g \g} & = & \frac{N_C}{2} \frac{1 + z^3}{1 - z} ~, 
\end{eqnarray}
respectively, with a continuous transition between the two for a 
$\q\g$ dipole. 

The dipole picture can also be extended to apply to initial-state partons.
Here the effect of radiation is that the incoming parton has to have
a larger momentum than previously considered, in the spirit of 
backwards evolution. There are then four kinds of dipole ends,
FF, FI, IF and II, classified by whether the radiator and recoiler 
are in the final or initial state.  

\begin{figure}
\centering
\begin{picture}(160,160)(-80,-90)
\SetWidth{2}
\ArrowLine(-60,60)(0,20)
\ArrowLine(0,20)(0,-20)
\ArrowLine(0,-20)(-60,-60)
\Gluon(0,20)(60,40){5}{5}
\Photon(0,-20)(60,-40){5}{5}\Text(70,-40)[]{$\Z^0$}
\SetColor{OliveGreen}
\DashArrowLine(-60,65)(60,45){2}
\SetColor{Red}
\DashArrowLine(60,32)(-60,-55){5}
\Text(0,-80)[]{(a)}
\end{picture}
\begin{picture}(200,160)(-80,-90)
\SetWidth{2}
\ArrowLine(-60,60)(0,20)
\ArrowLine(0,20)(0,-20)
\ArrowLine(0,-20)(-60,-60)
\Gluon(-40,47)(30,60){5}{5}
\Gluon(0,20)(60,40){5}{5}
\Photon(0,-20)(60,-40){5}{5}\Text(70,-40)[]{$\Z^0$}
\DashCArc(45,0)(55,0,360){5}
\Text(20,-80)[]{(b)}
\end{picture}\caption{(a) Colour flow (dipole) topology for 
$\q\qbar \to \Z^0 \g$, and (b) in \textsc{Pythia} the emission of 
a second gluon would give a recoil to the whole existing $\Z^0 \g$ 
system rather than only to the dipole-connected first gluon
\label{fig:Zgflow}}
\end{figure}
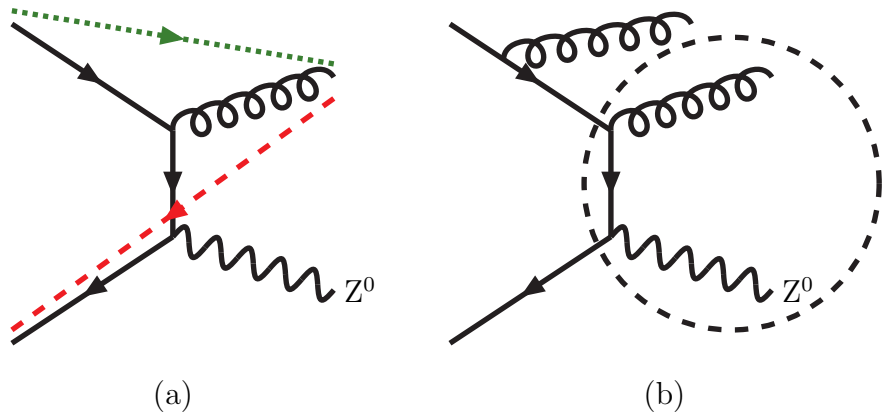

For ISR, however, the dipole picture is known to have shortcomings.
Specifically, it is in direct conflict with results from traditional
Feynman-diagram calculations. As an example, consider the production
of a $\Z^0$. For the emissions of a first gluon the two approaches
would agree, but this emission would lead to a picture with two 
IF dipoles, stretched from the incoming $\q$ and $\qbar$ to the gluon,
respectively, Fig.~\ref{fig:Zgflow}a. Any further radiation would
therefore occur from these two dipoles, and their daughters, but never
again have any impact on the kinematics of the $\Z^0$. With Feynman
diagrams, on the other hand, the $\Z^0$ takes a recoil that is 
modified as further gluon emissions are considered, and typically
resummation techniques \cite{Dokshitzer:1978hw} are used to sum up 
the effects of infinitely many gluon emissions on the $\pT$ spectrum 
of the $\Z^0$. This is a well-known problem e.g.\ with Catani-Seymour (CS)
dipoles \cite{Catani:1996vz}. Therefore CS-based dipole showers do not
necessarily apply the CS recoil strategy to the letter. Furthermore, when
matching procedures are used, hard emissions are described by matrix
elements and only softer radiation would be affected by the dipole
picture.

Given these shortcomings, the ISR in \textsc{Pythia} is not based on the 
CS-style dipole picture. Instead, an ISR emission is allowed to give 
a recoil to all partons ``downstream'' of it, i.e. that have been 
created by the hard process or by previously considered emissions 
at a larger $\pT$ scale, Fig.~\ref{fig:Zgflow}b. This applies both 
to the II and IF dipole ends. In the latter case one would still 
expect a memory of the colour flow topology to show up as an azimuthal 
anisotropy of radiation, studied further in later sections.
For the FF and FI types, a more standard dipole picture is used, but we 
note and address an issue with the allowed emission region for the 
FI case. As it turns out, these moderate changes to the IF and FI dipole
handling do help address the problem with tuning to Tevatron data. 

\subsection{Transverse-momentum-ordered showers}

The current \tsc{Pythia} parton shower orders FSR emissions in terms of 
a $\pT^2$ evolution variable, with an additional energy-sharing variable
$z$ in the branching.  For QCD emissions the DGLAP evolution 
equations lead to the probability for the splitting of parton $a \to bc$
\begin{equation}
\d \mathcal{P}_a = \frac{\d \pT^2}{\pT^2}  
\sum_{b, c} \frac{\alphas(\pT^2)}{2\pi}
P_{a \to bc}(z) \, \d z
~,
\end{equation}
where $P_{a \to bc}$ are the DGLAP splitting kernels. This
inclusive quantity can be turned into an exclusive one by requiring that,
for the first (``hardest'') emission below the starting scale $\pTmax^2$, 
no emissions can have occurred at a scale larger than $\pT^2$.
The no-emission probability is the Sudakov form factor
\begin{equation}
S_a(\pTmax^2, \pT^2) =  \exp \left( 
- \int_{\pT^2}^{\pTmax^2} \frac{\d \pTp^2}{\pTp^2} \sum_{b,c} 
\int^{z_{\mmax}(\pTp^2)}_{z_{\mmin}(\pTp^2)} \d z \,
\frac{\alphas(\pTp^2)}{2 \pi} \, P_{a \to bc}(z) \right) ~.
\end{equation}
Its introduction turns the unnormalised distribution into a normalised 
one, 
\begin{equation}
\d \mathcal{P}_a = \frac{\d \pT^2}{\pT^2}  
\sum_{b, c} \frac{\alphas(\pT^2)}{2\pi}
P_{a \to bc}(z) \, \d z \,\, S_a(\pTmax^2, \pT^2)  ~,
\label{eq:py_FSR_evol}
\end{equation}
i.e.\ with unit integral over the full phase space. In practice, 
a lower cutoff, $\pTo^2$, is introduced to keep the shower away from 
soft and collinear regions, which leads to a fraction of events with 
no emissions inside the allowed region.

For initial-state radiation (ISR), using backwards evolution, 
a given parton $b$ entering a hard scattering is unresolved into a parton
$a$ which preceded it. Here, the parton distribution functions, reflecting
the contents of the incoming hadron, must be taken into account. This leads
to a Sudakov with the form
\begin{equation}
S_b(x, \pTmax^2, \pT^2) = 
\exp \left(- \int^{\pTmax^2}_{\pT^2} \d\pTp^2 \, \sum_{a,c} 
\int^{z_{\mmax}(\pTp^2)}_{z_{\mmin}(\pTp^2)} \d z \,
\frac{\alphas \left( \pTp^2 \right)}{2\pi} \, P_{a \to bc} (z) \,
\frac{x'f_a(x',\pTp^2)}{xf_b(x,\pTp^2)} \right)
\label{eq:ISRsudakov}
~,
\end{equation}
where $z = x/x' = x_b/x_a$, and a corresponding normalised distribution
\begin{equation}
\d \mathcal{P}_b = \frac{\d \pT^2}{\pT^2}  
\sum_{a, c} \frac{\alphas(\pT^2)}{2\pi}
P_{a \to bc}(z) \frac{x'f_a(x',\pT^2)}{xf_b(x,\pT^2)}
\, \d z \,\, S_b( x, \pTmax^2, \pT^2)  ~.
\label{eq:py_ISR_evol}
\end{equation}
Currently both the running renormalisation and factorisation 
shower scales, i.e.\ the scales at which $\alphas$ and the PDFs are 
evaluated, are chosen to be $\pT^2$.

For the following, a relevant aspect is that, while $\pT$ ordering is 
used both for MPI, ISR and FSR in \tsc{Pythia}, the $\pT$ definition
is slightly different in the three components. For an MPI the $\pT$ 
is the expected one; the transverse momentum of the two scattered partons 
in a $2 \to 2$ process, defined in a Lorentz frame where the two 
incoming beams are back-to-back. To understand why the same choice is
not made for ISR or FSR, consider a $\q \to \q\g$ branching, 
where the $\pT$ of the emitted gluon is defined with respect to the 
direction of the initial quark. The $\pT$ as a function of the 
gluon emission angle $\theta$ increases up till $90^{\circ}$, but then 
decreases again, $\pT \to 0$ for $\theta \to 180^{\circ}$. Thus, an 
ordering in such a $\pT$ would classify a $\sim 180^{\circ}$ emission 
as collinear and occurring late in the evolution, although it would 
involve a more off-shell propagator than an emission at $90^{\circ}$. 
It could also erroneously associate a $1/\pT^2$ divergence with the 
$\theta \to 180^{\circ}$ limit. Therefore it is natural to choose an 
evolution variable that does not turn over at $90^{\circ}$. 

To this end, consider a branching $a \to b c$ (e.g.\ $\q \to \q \g$), 
where $z$ is defined as the lightcone momentum along the axis of $a$
that $b$ takes. Then
\begin{equation}
p_{\perp\mrm{LC}}^2 = z(1-z)m_a^2 - (1-z)m_b^2 - z m_c^2 ~,
\end{equation}
and this equation can be used as inspiration to define evolution variables
\begin{eqnarray}
&&\mrm{ISR}: \pTevol^2 = (1-z) Q^2 \label{eq:pTISR}\\
&&\textnormal{with $m_b^2 = -Q^2$ and $m_a^2 = m_c^2 = 0$,} \nonumber \\
&&\mrm{FSR}: \pTevol^2 = z (1-z) Q^2 \label{eq:pTFSR}\\
&&\textnormal{with $m_a^2 = \phantom{-}Q^2$ and $m_b^2 = m_c^2 = 0$,} \nonumber
\end{eqnarray}
which are monotonous functions of the virtuality $Q^2$. In the actual 
branching kinematics, $z$ in replaced by a Lorentz invariant definition
while the $Q^2$ interpretation is retained.

\subsubsection{Initial-state showers}
\label{sec:spaceshower}
For the initial-state shower, the radiating dipole is always chosen such
that the recoiler is the incoming parton from the other side of the
subcollision. Thus the whole enclosed systems share the recoil of a new
emission, unlike e.g. a Catani-Seymour dipole scheme, as discussed
previously.

The $z$ definition in eq.~(\ref{eq:pTISR}) is chosen to be
$z = m^2_{br}/m^2_{ar}$, where $r$ is the recoiler, which allows
for a straightforward bookkeeping of the $x$ values to be used in PDFs. 
The actual 
$\pT$ of $b$ and $c$ then becomes 
\begin{equation}
p_{\perp b,c}^2 = (1-z)Q^2 - \frac{Q^4}{m_{ar}^2} = 
\pTevol^2 - \frac{\pTevol^4}{p_{\perp\mrm{evol,max}}^2} ~,
\label{eq:pTbc}
\end{equation}
where $p_{\perp\mrm{evol,max}}^2 = (1-z) Q_{\mrm{max}}^2 = 
(1-z)^2 m_{ar}^2$ is the kinematically possible upper limit,
not to be confused with the starting scale for the downwards
evolution in $\pTevol^2$. Usually $\pTevol^2 \ll p_{\perp\mrm{evol,max}}^2$,
so that $p_{\perp b,c}^2 \approx \pTevol^2$. 

To understand how the two dipole ends combine to produce the radiation
pattern, again consider $\q(1) \qbar(2) \to \g(3) \Z^0(4)$, 
cf.\ Fig.~\ref{fig:Zgflow}a. Defining $\hat{\theta}$ as the emission
angle of the gluon in the rest frame of the process, and with the
the Mandelstam variables 
$\hat{t}, \hat{u} = -\hat{s}(1 \mp \cos\hat{\theta})/2$, for the gluon
being emitted by the incoming $\q$, one has
\begin{equation}
\frac{\d\pTevol^2}{\pTevol^2} = \frac{\d Q^2}{Q^2} = 
\frac{\d \hat{t}}{\hat{t}} = 
\frac{\d (\cos\hat{\theta})}{1 - \cos\hat{\theta}} = 
(1 + \cos\hat{\theta}) \frac{\d (\cos\hat{\theta})}{1 - \cos^2\hat{\theta}}
~,
\label{eq:thetaISR}
\end{equation}
while an emission by the incoming $\qbar$ is exactly the same,
except for a sign flip from $\hat{t} \to \hat{u}$. It then follows
that the quark is responsible for a fraction $(1 + \cos\hat{\theta})/2$
of the total amount of radiation at an angle $\hat{\theta}$, i.e.\ 
there is a smooth transition in the middle of the event. One simplifying
factor is that $z = M_{\Z}^2 / \hat{s}$ is independent of the radiator side. 
Another is that we have considered the process for fixed incoming
partons, so that PDF weights do not enter. If viewed instead as
backwards evolution from $\q \qbar \to \Z^0$, a different 
PDF ratio would enter for emissions on the two sides, but the rest
would be the same. The simple relations between the evolution and the
Mandelstam variables makes this shower convenient for some matching tasks
\cite{Miu:1998ju}.  Further details of the kinematics, also for massive
quarks, can be found in \cite{Sjostrand:2004ef}.

The direct relation between $\pTevol$ and emission angle 
(for fixed $z$) implies that typically central emissions will be 
considered first, and then successively emissions closer in angle or, 
equivalently, rapidity to the two beam directions. This ordering need 
not be strict however, since $z$ will also vary from one emission to 
the next. This implies that the resulting dipoles can zigzag back and
forth in rapidity. We note that in CCFM-based 
\cite{Ciafaloni:1987ur,Catani:1989sg} approaches
\cite{Andersson:1995ju} (as well as in angular-ordered showers, 
of course) emissions are ordered in rapidity,
but then need not be ordered in $\pT$. Hard emissions tend to be 
ordered in both, however, and it is only in the soft region that 
a random walk in $\pT$ sets in. While the amount of emitted partons 
need not be incorrect in our $\pT$-ordered approach, there is a worry 
that the zigzagging colour connections will give too large dipoles,
which then gives more FSR and hadronisation activity than if colours 
were angularly ordered. Ultimately this is an issue that could be 
studied by the rate of forward jets and forward particle production
in general. Different modifications to the existing algorithm could
be envisioned. For now, however, we will content ourselves with an 
option where rapidity-unordered ISR emissions are vetoed, subsequent
to the first emission off each dipole end. That is, while $\pT$ is 
always ordered, the rapidity may or may not be. 

Perturbation theory breaks down in the $\pT \to 0$ limit, and thus
some cutoff procedure is necessary for showers. The simplest is a
fixed lower $\pTmin$ cutoff. However, a more attractive possibility 
is a smooth damping. Actually, the damping introduced within the 
MPI framework, see below in Sec.~\ref{sec:MI}, could be viewed as
a reduced ``resolved'' partonic content of the two incoming hadrons,
and thus be the same for the ISR description. If so, the shower rate 
should be reduced roughly like
\begin{equation}
\frac{\d\mathcal{P}}{\d\pT^2} \propto \frac{\alphas(\pT^2)}{\pT^2}
\to \frac{\alphas(\pTo^2 + \pT^2)}{\pTo^2 + \pT^2} ~. 
\label{eq:turnoffISR}  
\end{equation}
We will allow for either possibility in the following.

\subsubsection{Azimuthal asymmetries}
\label{sec:aziasym}

For an ISR emission, as described previously, the dipole is always taken as
stretching between the two incoming partons. In the colour-dipole picture,
however, it will involve a colour line that is either stretched
along the beam (collision) axis, or out to a final-state parton. This can
lead to azimuthal asymmetries, as follows. In the first emission from a
$\q\qbar \to \Z^0$ event, the colour line stretched between the two beams means
there is no preferred azimuthal direction, and emissions are isotropic in
$\varphi$. For a second emission, the colour line is instead now stretched
from the radiating incoming parton to the final state. This colour dipole
configuration leads to an enhancement of radiation near the azimuthal angle
of this final-state parton. For the emission of a soft gluon from this
dipole, and with a convenient separation of radiation between the two
dipole ends, it is possible to derive what this distribution should look
like \cite{Webber:1986mc}.  To be specific, if the radiator is moving along
the $+z$ direction and its colour partner along a (reasonably small) polar angle
$\theta_{\mrm{par}}$ and azimuthal angle $\varphi_{\mrm{par}}$, then soft
emissions should be distributed roughly like
\begin{equation}
\frac{\d\mathcal{P}}{\d\varphi} \propto 
\frac{1 - r \cos(\Delta\varphi)}{1 + r^2 - 2 r \cos(\Delta\varphi)}
~~~~\mathrm{with}~\Delta\varphi = \varphi - \varphi_{\mrm{par}} ~,
\end{equation} 
where $r = \theta / \theta_{\mrm{par}}$. For small $r$ the bias is tiny, 
but then increases as $r \to 1$. It turns partly negative for 
$r > 1$, and there integrates to 0 --- the angular-ordering condition.
It must be stressed, however, that the behaviour near $r \to 1$ 
depends on how the dipole is split into two ends, and also obtains 
corrections when the energy of the radiated parton is not vanishingly 
small. As already discussed, we also optionally allow for cases where
emissions are not rapidity-ordered, meaning $r > 1$, where azimuthal
anisotropies again should fall off in the limit of a large $r$. 
Instead, we define an alternative
\begin{equation}
r = N \frac{2 \theta \theta_{\mrm{par}}}{\theta^2 + \theta_{\mrm{par}}^2}
~,
\label{eq:azistr}
\end{equation}
where $N$ is a free parameter to be tuned later. It has a natural value 
of $1/2$ by the above small-angle behaviour, but we allow for somewhat
different values, though never getting close to unity.

Another reason for anisotropies is the plane polarization of the gluon.
That is, the planes defined by the production and by the decay of a gluon,
respectively, tend to be correlated \cite{Webber:1986mc}. If the gluon 
is produced in a $\g \to \g\g$ branching then a $\g \to \g\g$ decay 
leads to a small tendency for the two planes to line up. By contrast,
a $\g \to \q\qbar$ branching gives a strong tendency for the two planes
to be at $90^{\circ}$ to each other. The branching $\g \to \q\qbar$
is less frequent than $\g \to \g\g$, so to first approximation it is 
expected that the net effect should be vanishingly small 
\cite{Bengtsson:1988qg}. This is also what we observe, both when 
applied for ISR and for FSR. 

\subsubsection{Final-state showers}
\label{sec:timeshower}

For FSR, the dipole picture is again used but, unlike ISR, the recoil
of emissions is here assumed to be taken entirely by the other end of 
the colour dipole, as already discussed in Sec.~\ref{sec:dipvsfeyn}.
For the case of FF dipoles all relevant kinematics equations 
are already published \cite{Sjostrand:2004ef}, and we only sketch some 
relevant points here. 

In the rest frame of a dipole $ar$, the branching $a \to b c$ at an
evolution scale $\pTevol$ implies that parton $a$ acquires a virtuality 
$Q^2 = \pTevol^2/(z(1-z))$. Its energy is thereby increased from the 
original $m_{ar}/2$ to $(m_{ar}^2 + Q^2)/(2 m_{ar})$, and the recoiler 
energy reduced accordingly. Parton $b$ then takes a fraction $z$ of the
increased radiator energy, and $c$ takes $1-z$. This choice gives an 
exact match e.g.\ to the singularity structure of the well-known
$\q\qbar \to \q(1)\qbar(2)\g(3)$ branching in $\e^+ \e^-$ annihilation:
\begin{equation}
\frac{\d p_{\perp\mrm{evol},\q}^2}{p_{\perp\mrm{evol},\q}^2} \,
\frac{\d z_{\q}}{1 - z_{\q}} + 
\frac{\d p_{\perp\mrm{evol},\qbar}^2}{p_{\perp\mrm{evol},\qbar}^2} \,
\frac{\d z_{\qbar}}{1 - z_{\qbar}} = 
\frac{\d x_1 \, \d x_2}{(1 - x_2) x_3} + 
\frac{\d x_1 \, \d x_2}{(1 - x_1) x_3} = 
\frac{\d x_1 \, \d x_2}{(1 - x_1) (1 - x_2)} 
\end{equation}
with $x_i = 2 E_i/\ECM$. In the soft-gluon limit, $x_3 \to 0$, the
kinematics simplifies to 
$1 - x_2 \propto m_{\q\g}^2 \propto 1 - \cos\hat{\theta}$ and
$1 - x_1 \propto m_{\qbar\g}^2 \propto 1 + \cos\hat{\theta}$,
where $\hat{\theta}$ is the emission angle of the gluon with
respect to the original $\q\qbar$ axis. Thus one obtains exactly 
the same continuous transition from $\q$- to $\qbar$-emitted gluons 
as for ISR, eq.~(\ref{eq:thetaISR}). For nonvanishing gluon energy,
the angular dependence becomes more complicated, but retains the
same qualitative features. 

The relation between the formal $\pTevol$ and the 
$p_{\perp b} = p_{\perp c}$ of the daughters with respect to the 
mother direction is somewhat lengthy. It shares the same physical 
properties as eq.~(\ref{eq:pTbc}) for ISR; at small values $\pTevol^2$ 
and $p_{\perp b,c}^2$ follow suit, and so correspond to identical 
$1/\pT^2$ singular behaviours, but the latter then turns around
and vanishes when $\pTevol^2$ approaches the kinematical limit.
 
While FF dipoles are familiar from and constrained by
$\e^+ \e^-$ physics, the FI ones are less well studied.
In the \tsc{Pythia 6} shower, FSR was deferred until after all ISR,
such that the FI dipoles at the end of the ISR evolution were typically
characterised by a low $\pT$ scale, meaning they did not radiate
much. Indeed, it was shown that disregarding emission from them 
altogether had only a tiny influence on event properties. Instead,
with ISR and FSR interleaved, typically each event contains one or a 
few FI dipole ends that should be evolved all the way from the 
$\pT$ scale of the hard interaction. 

The kinematics of an FI branching gives some differences relative 
to an FF one. In the dipole rest frame a fraction $Q^2 / m_{ar}^2$
of the recoiler energy is given from the recoiler to the emitter, 
as above. That is, the emitter four-momentum is increased to 
\begin{equation} 
p'_a = p_a + \frac{Q^2}{m_{ar}^2} p_r ~,
\end{equation}
which is valid in any Lorentz frame. But the recoiler is not a
final-state particle, so the increase of $a$ momentum is not 
compensated anywhere in the final state. Instead the incoming parton 
that the recoiler represents must have its momentum increased, 
not decreased, by the same amount as the emitter. That is, its $x$ 
value needs to be scaled up as
\begin{equation} 
x'_r = \left( 1 + \frac{Q^2}{m_{ar}^2} \right) x_r ~.
\end{equation}
The dipole mass $m_{ar}$ and the squared subcollision mass
$\hat{s}$ are increased in the process, the latter by the same 
factor as $x_r$.
As with ISR, the increased $x$ value leads to an extra PDF weight
\begin{equation}
\frac{x'_r f_r(x'_r,\pT^2)}{x_r f_r(x_r,\pT^2)}
\end{equation}
in the emission probability and Sudakov form factor. This ensures 
a proper damping of radiation in the $x'_r \to 1$ limit.

The above equations are valid in the massless limit. Incoming 
particles are always bookkept as exactly massless, so that $p'_r$ 
remains parallel with the beam axis if $p_r$ was. If the recoiler
has an on-shell mass $m_r^2$, then its energy is already from the
onset bigger than the recoiler one in the dipole rest frame,
and the momentum and $x$ are not rescaled by as much
\begin{equation} 
x'_r = \left( 1 + \frac{Q^2 - m_a^2}{m_{ar}^2 - m_a^2} \right) x_r ~.
\end{equation}
  
The issue that remains is how the emission off the FI dipole end 
should be damped as a function of location along the dipole.
It has already been shown that there is a natural fall-off at around 
$90^{\circ}$ in the rest frame of the dipole. For an FF and II 
dipole this offers a convenient subdivision, with the two ends
together providing a smooth coverage. What must now be asked
is how to subdivide FSR and ISR in an FI dipole. The natural choice 
is to view the dipole in a boosted+rotated frame where the radiator 
is along the $+z$ axis, with an energy equal to its original $\pT$, 
and the recoiler along $-z$ with energy $m_{ar}^2 / (4\pT)$. This is 
close to the Breit frame definition in Deeply Inelastic Scattering, 
with $\pT$ as the hard scale instead of (the DIS variable) $Q/2$ and 
$m_{ar}^2$ instead of $W^2$. The Breit frame offers a convenient 
separation between asymmetric current and target hemispheres. 

A more precise motivation is offered by the fractal picture of 
dipole emission. While originally formulated for FSR 
\cite{Gustafson:1990qi}, the same kinematics argument can be applied 
here, as follows. Consider a dipole stretched between two partons 
$i$ and $j$ along the $\pm z$ axis. At some arbitrary (e.g.\ cutoff) 
scale $m_0$ the rapidity range open for emissions is then 
$\Delta y = \ln(m_{ij}^2/m_0^2)$. Assume a gluon $k$ is emitted at 
an angle $\theta$, sufficiently soft that recoil effects can be 
neglected. Then the new range(s) allowed subsequently is
\begin{eqnarray}
\Delta y' & = &  \ln(m_{ik}^2/m_0^2) + \ln(m_{kj}^2/m_0^2)
= \ln \left( \frac{2 E_i E_k (1 - \cos\theta)}{m_0^2} \,
\frac{2 E_j E_k (1 + \cos\theta)}{m_0^2} \right) \nonumber \\
& = & \ln \left( \frac{4 E_i E_j}{m_0^2} \, 
\frac{E_k^2 \sin^2\theta)}{m_0^2} \right)
= \Delta y + 2 \ln(p_{\perp k}/m_0) ~.
\end{eqnarray}
Here the unchanged $\Delta y$ term can be associated with radiation
essentially along the $z$ axis. In our case this would correspond not 
only to II dipoles but also to the IF ends of FI dipoles, which are
both considered as a continuous chain along the whole allowed rapidity 
range. The last term is the additional radiation to be associated with 
the two new FI dipole ends.

In summary, what we see is that the $\pT$ scale of the hard emission
serves a double purpose. On the one hand, it sets the maximum scale 
for subsequent emissions, on the other it delimits the phase space
that emissions from an FI dipole end can populate without 
doublecounting.

This clarification especially concerns cases when the two
relevant scales, the dipole mass $m_{ar}$ and the radiator scale
$\pT$ are well separated. Actually, this is precisely what you 
expect to happen by $t$-channel gluon exchange; a (relatively)
small change of momentum but a complete swap of colours. So
in a process such as $\u\d \to \u\d$ the outgoing $\u$ will be 
hooked up with the beam remnant of the incoming $\d$, which 
most of the time will be in the other hemisphere of the event,
with $m_{ar} \gg \pT$. Such a dipole would have a large rapidity
range to radiate inside, if hemispheres are separated in the 
dipole rest frame, and produce a doublecounting of FSR and ISR
effects. As we will show, indeed it does appear to make a difference
in event properties.

Finally, we need to specify a convenient expression for the damping
at the border between the two uneven hemispheres. Again consider the
``Breit''-frame arrangement, with parton $a$ having a momentum $\pT$
along the $+z$ axis. In the branching $a \to b c$ parton $b$ takes 
the fraction $z$, where we assume $z < 1/2$, so that $b$ is the 
softer one. Both have transverse momentum $p_{\perp b,c}$. 
If $z$ is interpreted as lightcone momentum then rapidity 0 for $b$
is seen to correspond to $p_{\perp b,c} = p_{+b} = z p_{+a} = z 2 \pT$. 
The complete kinematics is more messy, but qualitatively it holds 
that $p_{\perp b,c} = z (1-z) \pT$ is a reasonable measure
of rapidity 0, give or take factors of 2, and with $1-z$ inserted 
for symmetry. We therefore introduce a suppression weight
\begin{equation}
\mathcal{W} = \frac{z (1-z) \pT}{z (1-z) \pT + p_{\perp b,c}} 
= \frac{\pT p_{\perp b,c}}{\pT p_{\perp b,c} + p_{\perp b,c}/(z(1-z)) } 
= \frac{\pT p_{\perp b,c}}{\pT p_{\perp b,c} + Q^2 } 
\end{equation}
to smoothly kill the unwanted radiation in the backward hemisphere.
Note that this damping comes on top of the ``natural'' subdivision
of radiation in the dipole rest frame. Whichever one is more restrictive 
will dominate.

\subsection{Other model details}

In later sections, parameters relating to other parts of the
event generation framework enter when tuning to data. In
the rest of this section, then, these parts of the generator are outlined.

\subsubsection{Total cross sections}
\label{sec:sigmatot}

The total cross section for $\p\p$ collisions (excluding the Coulomb
contribution) is subdivided into elastic and inelastic ones, with the
latter further subdivided into diffractive and non-diffractive 
contributions. Many diffractive topologies are allowed, but the 
most common and the only ones currently implemented in \textsc{Pythia}
are single and double diffraction. We therefore write
\begin{equation}
\sigma_{\mrm{tot}}(s) = \sigma_{\mrm{el}}(s) + \sigma_{\mrm{sd}}(s) +
    \sigma_{\mrm{dd}}(s) + \sigma_{\mrm{nd}}(s) ~,
\end{equation}
with $s$ the squared center-of-mass energy.
While the class of elastic events is unambiguously defined, the other 
three are less easily distinguished. Single and double diffraction should 
be characterised by the presence of a rapidity gap, with either one or
both of the incoming protons excited to a higher-mass system, whereas 
non-diffractive events should have no rapidity gaps. In practice, however,
generated diffractive systems can fluctuate in mass up to the kinematical 
limit, in which case there is no gap, while generated non-diffractive 
events can still display gaps, through fluctuations in the hadronisation
process. The experimental definition of diffractive events therefore does
not have a one-to-one correspondence with the underlying picture in a
generator. This opens up the possibility for alternative experimental
definitions, but it is important that the exact one is clearly specified.

The ``minimum bias'' subset is also poorly defined as all events
that are triggered within the context of some specific detector setup.
In practice, elastic and low-mass diffractive events are excluded,
leaving inelastic non-diffractive events with a modest high-mass
diffractive contamination. Within the \textsc{Pythia} generator 
``minbias'' is used as a convenient shorthand for the more clumsy
and nondescriptive ``inelastic non-diffractive'' classification.

The different cross sections can be calculated from Regge theory, once
a few free parameters have been extracted from data itself. This works 
fine at lower, fixed-target energies. It also appears to work for
the total cross section up to higher energies, as e.g. in the DL 
parameterisation used by us \cite{Donnachie:1992ny}. In this approach the 
diffractive (and elastic) cross sections increase too fast with energy, 
however, and ultimately exceed the parameterised total cross section. 
This signals the need for higher-order corrections to dampen the growth. 
The parameterisations developed for \textsc{Pythia}
\cite{Schuler:1993wr,Schuler:1996en}
introduce such corrections in an empirical fashion. In a program
like \textsc{Phojet} a more sophisticated eikonalisation approach 
is chosen instead, which leads to a slower growth of diffractive
cross sections than in \textsc{Pythia}. Other arguments have also been 
raised why the diffractive cross section should cease to increase
\cite{Goulianos:1995vn,Goulianos:2009zz}.
A recent ATLAS study of the diffractive cross 
section at $7\TeV$ (although not currently fully corrected for detector
effects) suggests that \textsc{Pythia} is on the high side, though not
overly so, while \textsc{Phojet} rather undershoots \cite{AtlasDiff}.
 
It seems likely that some tuning downwards of diffractive rates will
be required. While it has been possible for a user to set the respective 
cross sections by hand, this would have to be done separately for each 
energy. We have now introduced an alternative simple mechanism to 
dampen the growth relative to the parameterisation of 
\cite{Schuler:1993wr}, without (significantly) affecting fixed-target 
phenomenology. For each of the diffractive rates it is thus possible 
to specify a maximum value that will be approached asymptotically
\begin{equation} 
\sigma_i^{\mrm{mod}}(s) = 
\frac{\sigma_i^{\mrm{old}}(s) \, \sigma_i^{\mrm{max}}}%
{\sigma_i^{\mrm{old}}(s) + \sigma_i^{\mrm{max}}} ~.
\end{equation}
We do not propose that $\sigma_i^{\mrm{mod}}(s)$ should approach asymptotia 
at current energies, only that the increase should be slowed down. 
Since the total and elastic cross sections are not affected, 
a reduction in diffractive cross sections means an increase of the 
non-diffractive one.

\subsubsection{Multiple interactions}
\label{sec:MI}

Let us now study the inelastic non-diffractive event class, and focus on 
just the contribution from MPI. The probability for an
interaction is then given by
\begin{equation}
\frac{\d \mathcal{P}_{\mrm{MPI}}}{\d {\pT}} =
\frac{1}{\sigma_{\mrm{nd}}} \frac{\d \sigma}{\d \pT} \;
\exp \left( - \int_{\pT}^{p_{\perp i-1}}
\frac{1}{\sigma_{\mrm{nd}}} \frac{\d \sigma}{\d \pT'} \d \pT' \right)
~.
\label{eq:MPIevol}
\end{equation}
Here, interactions are supposed to be generated in a decreasing sequence
of $\pT$ values, $p_{\perp 1} > p_{\perp 2} > p_{\perp 3} > \ldots$,
starting from some maximum scale, like the $\pT$ scale of the hard 
interaction or the upper kinematical limit. The exponential is a 
standard ``Sudakov'' factor that restores unitarity. It encodes the 
probability that there is no interaction intermediate in $\pT$ between 
the previous ($i - 1$) and the current ($i$) one.
 
The $\d \sigma / \d \pT$ is given by the perturbative QCD $2\rightarrow2$ 
cross section 
\begin{equation}
\frac{\d\sigma}{\d\pTs} = \sum_{i,j} \int \d x_1 \int \d x_2 \,
f_i(x_1, \pT^2) \, f_j(x_2, \pT^2) \, \frac{\d \hat{\sigma}}{\d\pTs} ~,
\label{eq:dsigmadpt}
\end{equation}
representing the convolution of the hard partonic cross section, 
$\d \hat{\sigma} / {\d\pTs}$, with the two incoming parton densities,
$f_i$ and $f_j$. This cross section is dominated by $t$-channel gluon
exchange, and diverges roughly as $\d \pT^2 / \pT^4$ (further steepened by
the rise of the PDFs at small $x$). To avoid this
divergence, the idea of colour screening is introduced. The concept of a
perturbative cross section is based on the assumption of free incoming
states, which is not the case when partons are confined in colour-singlet
hadrons. One therefore expects a colour charge to be screened by the
presence of nearby anti-charges; that is, if the typical charge separation
is $d$, gluons with a transverse wavelength $\sim 1 / \pT > d$ are no
longer able to resolve charges individually, leading to a reduced effective
coupling. This is introduced by reweighting the interaction cross section
such that it is regularised according to
\begin{equation}
\frac{\d \hat{\sigma}}{\d \pT^2} \propto
\frac{\alphas^2(\pT^2)}{\pT^4} \rightarrow
\frac{\alphas^2({\pT^2}_0 + \pT^2)}{({\pT^2}_0 + \pT^2)^2}
,
\label{eq:pt0}
\end{equation}
where $\pTo$ (related to $1 / d$ above) is now a free parameter in the
model. To be more precise, it is the physical cross section $\d \sigma / \d
\pT^2$ that needs to be regularised, i.e.  the convolution of
$\d \hat{\sigma} / \d \pT^2$ with the two parton densities,
eq.~(\ref{eq:dsigmadpt}). One is thus at liberty to associate the screening
factor with the incoming hadrons, half for each of them, instead of with
the interaction. Such an association also gives a recipe to regularise the
ISR divergence, as already noted.

Not only $\pTo$ itself, as determined e.g.\ from Tevatron data, 
comes with a large uncertainty, but so does the energy scaling of this 
parameter. The ansatz for the energy dependence of $\pTo$ is that
it scales in a similar manner to the total cross section, 
i.e.\ driven by an effective power related to the Pomeron intercept
\cite{Donnachie:1992ny}, which in turn could be related to the 
small-$x$ behaviour of parton densities. This leads to a scaling
\begin{equation}
\pTo(E_{\mathrm{CM}}) = p_{\perp0}^{\mathrm{ref}} \times \left(
\frac{E_{\mathrm{CM}}}{E_{\mathrm{CM}}^{\mathrm{ref}}}\right)%
^{E_{\mathrm{CM}}^{\mathrm{pow}}} ~,
\label{eq:pT0scaling}
\end{equation}
where $E_{\mathrm{CM}}^{\mathrm{ref}}$ is some convenient reference energy 
and $p_{\perp0}^{\mathrm{ref}}$ and $E_{\mathrm{CM}}^{\mathrm{pow}}$ are
parameters to be tuned to data. This same scaling can also be adopted when
regularising the ISR divergence, with the possibility to set the parameters
used in the shower independently of those used in the MPI model.

The so-far Poissonian nature of the framework is then changed, first by
requiring that there is at least one interaction, such that there is a
physical event, and second by including an impact parameter, $b$. 
For a given matter distribution, $\rho(r)$, the time-integrated
overlap of the incoming hadrons during collision is given by
\begin{equation}
\mathcal{O}(b) =
\int \d t \int \d^3 x \;
\rho(x, y, z) \;
\rho(x + b, y, z + t)
,
\end{equation}
after a suitable scale transformation to compensate for the boosted nature
of the incoming hadrons. This matter profile, $\rho(r)$ may then provides new
model parameters within the MPI framework. In the original MPI model,
accompanying the $Q^2$-ordered showers of \tsc{Pythia~6}, the default
selection was a double Gaussian
\begin{equation}
\rho(r) \propto \frac{1 - \beta}{a_1^3}
                \exp \left( - \frac{r^2}{a_1^2} \right)
              + \frac{\beta}{a_2^3}
                \exp \left( - \frac{r^2}{a_2^2} \right) ~,
\end{equation}
such that a fraction $\beta$ of the matter is contained in a radius $a_2$,
which in turn is embedded in a radius $a_1$ containing the rest of the
matter. Peter Skands subsequently noted that the inclusion of radiation off
all scattering subsystems in an event provides a dynamical source of
fluctuations, mirroring the effect of this double Gaussian matter
distribution, and was able to achieve good agreement with data using only a
single Gaussian profile, introducing no free parameters 
\cite{Skands:2010ak}. Both of these options are available in 
\tsc{Pythia~8}. A further option, intermediate in the sense that it has 
one free parameter, $E^{\mrm{pow}}_{\exp}$, is to have an overlap function
for the convolution of the two incoming matter distributions of the form
\begin{equation}
  \exp \left( -b^{E^{\mrm{pow}}_{\exp}} \right) ~.
\label{eq:exppow}
\end{equation}

\subsubsection{Parton densities}
\label{sec:pdfs}

Both hard processes, showers and multiple interactions make use of 
parton densities. Since these are not specified from first principles,
different parameterisations are on the market. The CTEQ5L \cite{Lai:1999wy} 
PDF set, dating from 1999, is currently the default choice for both 
\tsc{Pythia~6} and \tsc{Pythia~8}. A large range of PDF sets has always 
been available by linking to the LHAPDF library \cite{Whalley:2005nh}.

Recently, a selection of newer PDF sets has been made available directly 
in \tsc{Pythia~8} \cite{Kasemets:2010sg}, to make the \tsc{Pythia}
program more easy to install and run standalone, with a reproducible
behaviour. One also saves some computer time by not having to go via 
interfaces. These new sets include members from the CTEQ6 and MSTW2008 
families \cite{Pumplin:2002vw,Martin:2009iq}, as well as specially 
modified sets designed for LO event generators, such as those from 
the CT09 and MRST* families \cite{Lai:2009ne,Sherstnev:2007nd}. 
These modified sets have been adopted by experimental collaborations 
\cite{AtlasMC09}.

The reason for the modified sets is the following. Precision tests of
QCD today normally involve comparisons with NLO matrix elements 
convoluted with NLO PDFs. NLO expressions are not guaranteed to be 
positive definite, and do not have a simple probabilistic interpretation.
For PDFs, specifically, it is well-known that the gluon has a tendency
to start out negative at small $x$ and $Q^2$, and only turn positive 
by the QCD evolution towards larger $Q^2$. In the MRST/MSTW
sets this negativity is explicit, while it is masked in the CTEQ sets
by picking an ansatz form that cannot be negative, but which still gives a
very small gluon contribution. The combination of LO matrix elements and
NLO PDFs thus gives an abnormally small rate of interactions at small $x$
and $\pT$, relative to either a pure LO or a pure NLO combination. This
region may not be important for New Physics searches, but it is crucial for
the modeling of MPI, i.e.\ minimum-bias and underlying-event physics, and
also plays a role for showers.

So if NLO PDFs are not convenient for use in LO generators, also 
LO PDFs have their problems. Usually NLO calculations give an enhanced 
rate relative to LO ones, the ``K factor'' is above unity, and offer 
a better description to data. This introduces a tension in LO PDF fits, 
where data sensitive to a specific $x$ range attempts to pull in more 
of the total momentum, at the expense of other $x$ ranges. The solution 
of the new modified PDFs is to allow the momentum sum rule to be broken 
(typically by 10 -- 15\%), so that the whole $x$ range can be enhanced. 
  
These new PDFs therefore offer the hope to allow improved descriptions 
of data within a LO framework for MPI and showers. They are not necessarily
a panacea, however \cite{Kasemets:2010sg}, and comparing results between the
modified sets and more traditional ones will therefore be useful.
\tsc{Pythia~8} also allows one PDF set to be used for the hard process and
another for MPI and showers, so that one could use NLO for the former
while still having a sensible behaviour in the $\pT \to 0$ limit with the
latter.

\subsubsection{Primordial $\kT$}
\label{sec:pkt}
When the combined MPI/ISR/FSR $\pT$ evolution has come to an end, 
the beam remnants will consist of the remaining valence content of the 
incoming hadrons, as well as any companion (anti)quarks to kicked-out
sea quarks. These remnants must carry the remaining fraction of
longitudinal momentum. \tsc{Pythia} will pick $x$ values for each component
of the beam remnants according to distributions such that the valence
content is ``harder'' and will carry away more momentum. In the rare case
that there is no remaining quark content in a beam, a gluon is assigned to
take all the remaining momentum.

The event is then modified to add primordial $\kT$. Partons are expected to
have a non-zero $\kT$ value just from Fermi motion within the incoming
hadrons. A rough estimate based on the size of the proton gives a value of
$\sim 0.3\GeV$, but when comparing to data, for instance the $\pT$
distribution of $\Z^0$ at CDF, a value of $\sim 2\GeV$ appears to
be needed. The current solution is to decide a \kT value for each
initiator parton taken from a hadron based on a Gaussian whose width is
generated according to an interpolation
\begin{equation}
\sigma(Q, \widehat{m}) = 
\frac{Q_{\frac{1}{2}} \, \sigma_{\mrm{soft}} + Q  \, \sigma_{\mrm{hard}}}%
{Q_{\frac{1}{2}} + Q } \, 
\frac{\widehat{m}}{\widehat{m}_{\frac{1}{2}} + \widehat{m}}  ~,
\end{equation}
where $Q$ is the hardness of a sub-collision ($\pT$ for a $2 \to 2$
QCD process) and $\widehat{m}$ its invariant mass, $\sigma_{\mrm{soft}}$ 
and $\sigma_{\mrm{hard}}$ is a minimal and maximal value, and 
$Q_{\frac{1}{2}}$ and $\widehat{m}_{\frac{1}{2}}$ the respective scale 
giving a value halfway between the two extremes. Beam remnants are assigned
a separate width $\sigma_{\mrm{remn}}$ comparable with $\sigma_{\mrm{soft}}$.
The independent random selection of primordial $\kT$ values gives a
net imbalance within each incoming beam, which is shared between 
all initiator and remnant partons, with a reduction factor 
$\widehat{m} / ( \widehat{m}_{\frac{1}{2}} + \widehat{m} )$ for initiators of 
low-mass systems. With the $\kT$'s of the two initiators of a system
known, all the outgoing partons of the system can be rotated and  
Lorentz boosted to the relevant frame. During this process, the
invariant mass and rapidity of all systems is maintained by appropriately
scaling the lightcone momenta of all initiator partons.

\subsubsection{Colour reconnection}
\label{sec:cr}
The final step at the parton level, before hadronisation, is colour
reconnection.  The idea of colour reconnection can be motivated by noting
that MPI leads to many colour strings that will overlap in physical space,
which makes the separate identity of these strings questionable.
Alternatively, moving from the limit of $N_C \rightarrow \infty$ to $N_C =
3$, it is not unreasonable to allow these strings to be connected
differently due to a coincidence of colour. Adapting either of these
approaches, dynamics is likely to favour reconnections that reduce the
total string length and thereby the potential energy.

In the old MPI framework,
good agreement to CDF data is obtained if 90\% of additional
interactions produces two gluons with ``nearest neighbour'' colour
connections \cite{Field:2002vt}. More recently, an annealing algorithm
has been used \cite{Sandhoff:2005jh,Skands:2007zg}, again requiring
a significant amount of reconnection to describe data.
In \tsc{Pythia 8}, colour reconnection is currently performed by giving 
each system a probability to reconnect with a harder system
\begin{equation}
\mathcal{P} = \frac{{\pT}_{\mrm{Rec}}^2}{({\pT}_{\mrm{Rec}}^2 + \pT^2)} ~,
~~~~~~~~~
{\pT}_{\mrm{Rec}} = R \times \pTo,
\label{eq:crec}
\end{equation}
where $R$ is a user-tunable parameter and $\pTo$ is the same
parameter as in eq.~(\ref{eq:pt0}). 

With the above probability for reconnection, it is 
easier to reconnect low-$\pT$ systems, which can be viewed as them having 
a larger spatial extent, such that they are more likely to overlap with 
other colour strings. When a reconnection is allowed,
the partons of the lower-$\pT$ systems are attached to the 
existing higher-$\pT$ colour dipoles in a way that minimizes the total
string length. 

Currently, however, all of this is only a convenient 
ansatz. More than that, given the lack of a firm theoretical basis, 
the need for colour reconnection has only been established within the 
context of specific models.

\subsubsection{Diffraction}
\label{sec:diffraction}

The diffractive treatment in \tsc{Pythia~8} has been extended from the
simple one implemented in \tsc{Pythia~6}, to share many more features with 
the sophisticated description of non-diffractive events \cite{Navin:2010kk}.
This is possible by using the Ingelman--Schlein \cite{Ingelman:1984ns} 
picture, wherein single diffraction is viewed
as the emission of a Pomeron pseudoparticle from one incoming proton,
leaving that proton intact but with reduced momentum, followed by the 
subsequent collision between this Pomeron and the other proton. The
Pomeron is to first approximation to be viewed as a glueball state with
the quantum numbers of the vacuum, but by QCD interactions it will also
have a quark content. The Pomeron--proton collision can then be handled 
as a normal hadron--hadron non-diffractive event, using the full machinery
of MPI, ISR, FSR and other aspects as already described above.

The simple factorisation into a Pomeron flux times a Pomeron-proton 
collision should be viewed as an effective picture, i.e.\ as a first
approximation to physics that is much more complicated than that. 
Indeed the HERA studies show deviations from perfect factorisation
\cite{Slominski:2009zz}, and diffractive $\W$ and jet production at 
the Tevatron is reduced by a significant factor relative to naive 
expectations \cite{Affolder:2000vb}. Nevertheless, this factorisation 
does allow what hopefully is a realistic picture for the bulk of the 
diffractive cross section.

In the generation, the first step is to use the parameterised diffractive
cross sections, Sec.~\ref{sec:sigmatot}. Once a single diffractive
topology has been decided on, the second step is to use the 
Pomeron flux $f_{\mrm{I\!P/p}}(x_{\mrm{I\!P}}, t)$ to pick the 
fraction $x_{\mrm{I\!P}}$ that the Pomeron takes out of the 
proton momentum, and the squared momentum transfer $t$ of the 
emitted Pomeron, which is related to the scattering angle of the proton.
To first approximation 
\begin{equation}
f_{\mrm{I\!P/p}}(x_{\mrm{I\!P}}, t) \approx \frac{1}{x_{\mrm{I\!P}}} \,
\exp(B t) ~,
\end{equation}
where $B$ is an energy-dependent slope parameter. The squared mass 
of the diffractive $\mrm{I\!P}\p$ system $X$ is given by
$M_X^2 = x_{\mrm{I\!P}} s$, so it follows that 
$\d x_{\mrm{I\!P}} / x_{\mrm{I\!P}} = \d M_X^2 / M_X^2 = \d \ln M_X^2$.
Since the rapidity range over which particle production can occur
is proportional to $\ln M_X^2$, it follows that all kinematically 
possible ranges are about equally probable, and by complementarity,
the same applies to the size of the rapidity gap. This is only a first
approximation, however; in the program four alternative 
$f_{\mrm{I\!P/p}}(x_{\mrm{I\!P}}, t)$ shapes can be used, and all of 
them are somewhat more peaked towards lower masses than the ansatz
above indicates. 

Note that only the relative shape of $f_{\mrm{I\!P/p}}(x_{\mrm{I\!P}}, t)$ 
is of interest to us. If the absolute normalization were taken at face
value it would lead to a too rapidly increasing $\sigma_{\mrm{sd}}(s)$,
as already indicated in Sec.~\ref{sec:sigmatot}. Instead, we implicitly 
assume that the screening corrections appear as an energy-dependent 
but $x_{\mrm{I\!P}}$- and $t$-independent reduction factor. 

The third step is to set up the partonic state of the diffractive system. 
Here we make a distinction between low-mass and high-mass diffraction,
not because physics is expected to be discontinuous, but because we 
need to consider such a wide range of diffractive masses $M_X$, where
perturbation theory may be applied for the higher masses but not for the 
lower ones. A pragmatic dividing line is set at $10\GeV$; below it 
everything is soft, above it the fraction of soft events smoothly 
drops towards zero.

The soft description is simulated as a mix of two components.
In one, the Pomeron is assumed to kick out one of the valence quarks
of the proton, so that the partonic final state consists of a single
string stretched between the kicked-out quark and the remnant diquark.
In the other, the Pomeron instead kicks out a gluon. This gives a 
hairpin topology, where the string is stretched from a quark in the 
remnant to the kicked-out gluon and then back to the diquark of the
remnant. At the lower end of the range, extending to $1.2\GeV$, i.e.\ 
around the $\Delta$ mass, the former is expected to dominate, and then
gradually the latter takes over. Above $10\GeV$, the perturbative
description starts to be used.

The Pomeron PDFs are significantly worse known than those of the proton.
Quite aside from the considerable measurement problems, the data usually 
probe a convolution of the Pomeron flux with its PDF. Therefore the two 
cannot be specified independently. It is not even 
guaranteed that the two factorise. Nevertheless a few alternatives
are chosen, both toy-model ones and four H1 sets. Default in \tsc{Pythia~8} 
is the H1 2006 Fit B LO distribution. Note that the H1 sets use an 
arbitrary normalisation of the flux, such that the PDFs do not come out 
normalised to unit momentum sum, but have a sum of roughly $1/2$.
 
Strictly speaking the Pomeron PDF (convoluted with its flux) 
contributes to the overall PDF of a proton. This implies that the 
diffractive component ought to be subtracted from the normal PDFs
used for non-diffractive events. In reality the diffractive contribution
comes out to be a tiny fraction of the event rate at medium-to-large
$\pT$ scales, and so this doublecounting correction has been neglected
for now. 

In the description of the MPI interaction probability, 
eq.~(\ref{eq:MPIevol}), $\sigma_{\mrm{nd}} = \sigma_{\mrm{nd}}^{\p\p}(s)$ 
plays an important role as a normalisation factor. For the handling 
of a diffractive system it should be replaced by 
$\sigma_{\mrm{tot}}^{\mrm{I\!P}\p}(s)$, which unfortunately is not 
directly measurable. As with PDFs one reason is that the separation
from the Pomeron flux factor is not unambiguous, and another is that
we have allowed for a damping of $\sigma_{\mrm{nd}}^{\p\p}(s)$ away from
the result of a naive convolution. Instead, an order-of-magnitude
estimate is used, as follows. At low energies 
$\sigma_{\mrm{nd}}^{\p\p}(s)$ is in the ballpark $25$--$30\mb$, and 
then slowly increases. By contrast $\sigma_{\mrm{nd}}^{\pi\p}(s)$
is slightly below $20\mb$, i.e.\ roughly $2/3$, as suggested by quark
counting rules. The same number is now assumed for 
$\sigma_{\mrm{tot}}^{\mrm{I\!P}\p}(s)$, by analogy between the two-gluon
$\mrm{I\!P}$ state and the two-quark $\pi$ state. The larger colour 
charge of gluons could imply that the Pomeron couples more strongly to 
protons than pions do. By the same token the Pomeron wave function 
could be smaller, which would act in the opposite direction, so we 
will assume that such effects cancel. As a first guess, we therefore 
take $20\mb$ to be a reasonable number to use in eq.~(\ref{eq:MPIevol}). 
This assumes, however, that the Pomeron PDFs are normalised to unit 
momentum sum. Since the H1 fits are normalised only to roughly half 
of this, $\sigma_{\mrm{nd}}$ needs to be reduced accordingly, in order 
for the ratio to balance out. The actual number used therefore is $10\mb$. 
As a sanity check we note that this gives an average charged multiplicity 
for diffractive systems at a given $M_X$ that is comparable with  
non-diffractive $\p\p$ events at the same energy. 

Double diffraction is handled in the same spirit. Specifically, the
final state will here consist of two Pomeron-proton collision, each 
with properties as already described. What needs some more consideration
is the form of the combined two-side Pomeron flux  
$f_{\mrm{I\!P/p}\times\mrm{I\!P/p}}(x_{\mrm{I\!P}1}, x_{\mrm{I\!P}2}, t)$,
that in particular determines the masses of the two diffractive systems.
The old machinery \cite{Schuler:1993wr} here gives a nontrivial correlation,
while some of the alternative forms have only been formulated for single 
diffraction and therefore here are used by minimal extension, which 
is likely to be too simpleminded.

\subsubsection{Hadronisation}
\label{sec:hadronisation}

Once the partonic configuration has been specified, including its colour
flow, the normal Lund string fragmentation machinery can be used to turn it
into a set of primary hadrons \cite{Andersson:1983ia}. Many of these
are unstable and subsequently decay further.

Here we assume that fragmentation parameters tuned to $\e^+ \e^-$ 
data, notably from LEP, can be carried over unchanged for application
to $\p\p$ collisions. Such a ``jet universality'' assumption is by no means 
obvious. In $\e^+ \e^- \to \gamma^*/\Z^0 \to \q \qbar$ there is only 
one original dipole that may radiate further, typically giving rise to 
one string that does not bend over so much that it can overlap with
itself. By contrast, the MPI structure of  $\p\p$ events ensures that 
many colour strings overlap in space and time during the hadronisation 
process. The assumption that they then can be treated as completely 
independent of each other (once colour reconnection has been taken 
into account) may be overly optimistic. All kinds of collective effects
could be imagined, bordering on those associated with a quark--gluon
plasma. This could affect e.g.\ the fraction of strangeness production,
which is known to be higher in heavy-ion collisions than in $\e^+ \e^-$
ones \cite{Abelev:2007xp}. For now, however, such potential issues are 
left aside. 

\section{Comparison to 3-jet matrix elements}

\subsection{$2 \to 3$ real emission matrix elements}

In this section, we compare the first shower emission against $2 \to 3$
real-emission matrix elements in QCD events. Eventually, we foresee a full
matching of the first emission to these matrix elements, as is done e.g.
for electroweak gauge boson production, but here we begin with some simple
kinematic comparisons of the default shower. Even with such a matching,
emissions after the first will still be handled by the standard shower
machinery, and it is questionable if it would be used in all MPI subsystems
rather than just the hardest process in an event. It is therefore important
to have the best understanding of the shower as possible. There is quite
some freedom in how to handle recoil effects, and when the shower has been
designed to give sensible tails away from soft and collinear regions, the manner
in which this is done grows in importance. Indeed, as we will show, in the
soft and collinear regions, the showers give good agreement with the ME
behaviour, while further out in phase space, the accuracy does degrade, but
without exceedingly large deviations. In the rest of this section, some
details of the $2 \to 3$ ME implementation are given, before comparisons
are made, firstly to examine the azimuthal asymmetries introduced in
Sec.~\ref{sec:aziasym}, and secondly to study the rates and kinematics of
shower emissions.

The formulae for the squared matrix elements in the massless QCD
approximation, summed and averaged over spin, are taken from
\cite{Berends:1981rb}, with crossing applied where necessary. These
expressions are compact, and so easy to input, but the tradeoff is the lack
of information for generating the colour structure of events; for the
kinematic distributions we consider here, this is not a serious limitation.
In what follows, the outgoing partons are labeled 3, 4 and 5, and ordered
such that ${\pT}_3 > {\pT}_4 > {\pT}_5$. The renormalisation scale is
expected to relate to each vertex of the process, and by default is taken
to be the geometric mean of the squared transverse masses of the three
outgoing particles, $\mu_r^2 = \sqrt[3]{ m_{\perp3}^2 * m_{\perp4}^2 *
m_{\perp5}^2 }$. The factorisation scale, instead, in a picture comparable
to backwards evolution in ISR, would be related to the smaller scales in
the event, and is taken to be the geometric mean of the two smallest
squared transverse masses of the three outgoing particles, $\mu_f^2 =
\sqrt{ m_{\perp4}^2 * m_{\perp5}^2 }$.

A phase space generator has been written that is specially adapted for
these processes, taking into account the important role of the soft and
collinear singularities.  The variables ${\pT}_3$, ${\pT}_5$, $y_3$, $y_4$
and $y_5$ are used to generate the phase space. ${\pT}_3$ is first picked
according to a $\d^2{\pT}_3 / {\pT}_3^4$ distribution and then ${\pT}_5$
according to $\d^2{\pT}_5 / {\pT}_5^2$. $\varphi_3$ and $\varphi_5$ are
picked flat, before ${\pT}_4$ is finally reconstructed. All three
rapidities are picked flat in their respective allowed range. Phase space
cuts are placed on the two ${\pT}$ variables, ${\pT}_3^{\mmin}$ and
${\pT}_5^{\mmin}$, and, additionally, there is an $R = \sqrt{(\Delta
\eta)^2 + (\Delta \varphi)^2}$ separation cut, $R_{\mrm{sep}}^{\mmin}$,
applied to all three possible pairs of partons. ${\pT}_4$ is not explicitly
constrained, but must lie in the range specified by ${\pT}_3$ and
${\pT}_5$.

\renewcommand{\arraystretch}{1.15}
\begin{table}
\begin{center}
\begin{tabular}{|c|p{20mm}|p{20mm}|p{20mm}|p{20mm}|}
\cline{2-5}
\multicolumn{1}{c}{} &
\multicolumn{4}{|c|}{\textbf{Cross section ($\mu\b$)}} \\
\cline{2-5}
\multicolumn{1}{c}{} &
\multicolumn{2}{|c|}{\textbf{Tevatron}} &
\multicolumn{2}{|c|}{\textbf{LHC}} \\
\hline
\textbf{Cuts} &
\multicolumn{1}{|c|}{\textbf{Ref}} &
\multicolumn{1}{|c|}{\textbf{\tsc{Pythia}}} &
\multicolumn{1}{|c|}{\textbf{Ref}} &
\multicolumn{1}{|c|}{\textbf{\tsc{Pythia}}} \\
\hline
1 & \hfill  1404 & \hfill   1409 & \hfill   23938 & \hfill  24164 \\
2 & \hfill 2.786 & \hfill  2.798 & \hfill  171.58 & \hfill 172.97 \\
3 & \hfill 0.191 & \hfill  0.191 & \hfill   17.87 & \hfill  17.92 \\
\hline
\end{tabular}
\end{center}
\caption{Overall cross sections for $2 \to 3$ partonic processes, for three
sets of cuts (see text) at the Tevatron ($\p\pbar$, $\sqrt{s} = 1.96\TeV$) and
LHC ($\p\p$, $\sqrt{s} = 14\TeV$). Results from \tsc{Pythia} are compared
to summed cross sections from AlpGen and MadEvent
\label{tab:2to3comp}}
\end{table}

To validate these processes, the cross sections and kinematics have been
compared to AlpGen \cite{Mangano:2002ea,Mangano:2001xp,Caravaglios:1998yr}
(excluding processes with no final state gluons) and MadEvent
\cite{Alwall:2007st} (for processes with no final state gluons).
For all numbers shown here, the CTEQ6L1 PDF set has been used, with a first
order running of $\alphas$, with $\alphas(M_\W) = 0.13$. In both AlpGen and
MadEvent, the factorisation and renormalisation scales have been set to
follow the \tsc{Pythia} defaults. In all programs, the allowed incoming
and outgoing flavours have been set to include the charm quark and below.
In Tab.~\ref{tab:2to3comp}, overall cross sections are given for three sets
of cuts at Tevatron ($\p\pbar$, $\sqrt{s} = 1.96\TeV$) and LHC
($\p\p$, $\sqrt{s} = 14\TeV$) energies:
\begin{itemize}
\item[1)] ${\pT}_3^\mmin = 5.0\GeV$, ${\pT}_5^\mmin = 5.0\GeV$,
          $R_{\mrm{sep}}^{\mmin} = 0.1$
\item[2)] ${\pT}_3^\mmin = 50.0\GeV$, ${\pT}_5^\mmin = 5.0\GeV$,
          $R_{\mrm{sep}}^{\mmin} = 0.1$
\item[3)] ${\pT}_3^\mmin = 50.0\GeV$, ${\pT}_5^\mmin = 25.0\GeV$,
          $R_{\mrm{sep}}^{\mmin} = 1.0$
\end{itemize}
The difference between the reference and \tsc{Pythia} cross sections is
less than 1\% for all three sets of cuts. In the PS, the amount of ISR is
roughly 70\%, 60\% and 90\% respectively, and so dominates, even
when using a rather small $R_{\mrm{sep}}^{\mmin}$. In
Tab.~\ref{tab:2to3class}, the different classes of $2 \to 3$ processes are
given, with indicative cross sections for each process for the first set of
cuts at Tevatron and LHC energies. Overall, the cross section is dominated
by $\g\g$ and $\q\g$ scattering, and the effect grows at LHC energies, in
particular driven by the large increase in small-$x$ gluons.

\renewcommand{\arraystretch}{1.15}
\begin{table}
\begin{center}
\begin{tabular}{|l|l|r@{.}l|r@{.}l|}
\cline{3-6}
\multicolumn{2}{c}{} &
\multicolumn{4}{|c|}{\textbf{Cross section ($\mu\b$)}} \\
\hline
\textbf{Class} & \textbf{Processes}   &
\multicolumn{2}{|c|}{\textbf{Tevatron}} &
\multicolumn{2}{|c|}{\textbf{LHC}} \\
\hline
\texttt{HardQCD:gg2ggg}     &       $\g\g \to \g\g\g$      &  \phantom{00}717&46 & 15236&92 \\
\texttt{HardQCD:qqbar2ggg}  &       $\q\qbar \to \g\g\g$   &    0&51 &     2&47 \\
\texttt{HardQCD:qg2qgg}     &       $\q\g \to \q \g \g$    &  525&65 &  6776&26 \\
\texttt{HardQCD:qq2qqgDiff} &       $\q\q' \to \q \q' \g$,
                             $\q \qbar' \to \q \qbar' \g$  &   45&47 &   404&31 \\
\texttt{HardQCD:qq2qqgSame} &       $\q\q \to \q \q \g$,
                       $\qbar \qbar \to \qbar \qbar \g$    &    8&92 &    94&51 \\
\texttt{HardQCD:qqbar2qqbargDiff} &
                               $\q\qbar \to \q' \qbar' \g$ &    0&55 &     2&68 \\
\texttt{HardQCD:qqbar2qqbargSame} &
                                 $\q\qbar \to \q \qbar \g$ &   16&08 &    81&93 \\
\texttt{HardQCD:gg2qqbarg}        & $\g\g \to \q \qbar \g$ &   71&22 &  1302&09 \\
\texttt{HardQCD:qg2qqqbarDiff} & $\q\g \to \q\q'\qbar'$,
                              $\qbar\g \to \qbar\qbar'\q'$ &   18&05 &   197&95 \\
\texttt{HardQCD:qg2qqqbarSame} & $\q \g \to \q \q \qbar$,
                             $\qbar \g \to \qbar \qbar \q$ &    6&02 &    66&01 \\
\hline
\end{tabular}
\end{center}
\caption{The different classes of event implemented and their corresponding
processes. Primes refer to different quark/anti-quark flavours and processes
include all possible permutations of the initial or final state. Cross
sections from \tsc{Pythia} are given for the first set of cuts as detailed in
the text
\label{tab:2to3class}}
\end{table}

\subsection{Parton shower comparison}
\label{sec:pscomp}

The comparison we wish to make is between the $2 \to 3$ matrix elements and
$2 \to 2$ events with one additional PS emission (hereafter
referred to as $2 \to 2 \otimes \mrm{PS}$). Overall, the generation
must be such that the comparison is as fair as possible, but still allowing
a reasonable efficiency, so the following settings have been applied:
\begin{itemize}
\item[1)] The CTEQ6L1 PDF set is used everywhere.
\item[2)] The coupling strength for the hard process, ISR and FSR is set
          constant, $\alphas = 0.130$.
\item[3)] QED radiation is disabled.
\item[4)] The number of incoming and outgoing flavours possible is set
          equal for both the ME and PS (up to and including the charm
          quark).
\item[5)] All other stages of the event generation framework are switched
          off.
\item[6)] To improve efficiency (without affecting the outcome), a cut
          $\hat{p}_{\perp \mmin} = 0.4 * {\pT}_3^\mmin$ is used in
          generating the underlying $2 \to 2$ events.
\end{itemize}
Even with these settings, a couple of issues remain. One is the different
scale at which PDF factors are evaluated in ISR,
eq.~(\ref{eq:py_ISR_evol}). Specifically, both the numerator and
denominator of this factor are evaluated at a scale $\pTs$, giving an
implicit running $\alphas$ when compared to the ME result. A quick check
using typical scales and $x$ values for the above cuts shows that this
effect is not large, and we choose not to correct for this.

More important is the presence of Sudakov form factors in the PS results.
To generate inclusive $2 \to 2 \otimes \mrm{PS}$ distributions, which can
be compared against the ME results, a $2 \to 2$ hard process is selected,
then a parton shower emission allowed to create a $2 \to 2 \otimes
\mrm{PS}$ event. This event is then analysed to determine if it meets the
required cuts, and if so, it is accepted and added to the statistics. The
emission is then vetoed; \textsc{Pythia} will revert the event to its
original $2 \to 2$ state, and then continue the shower evolution from the
scale of the now-vetoed branching. Thus, a single $2 \to 2$ event, in the
course of its evolution, can give rise to several $2 \to 3$ configurations.
In fact, for an evolution variable $t$, ranging between 0 and 1, the
distribution becomes a Poissonian with an average $\int^{1}_{0} p(t) \d t$.
By the veto algorithm, the Sudakov form factor is exactly compensated by
the Poissonian distribution
\begin{equation}
p(t) \exp \left( - \int_0^t ~ p(t') ~ \d t' \right) \cdot
\sum_{n = 0}^{\infty} ~  \frac{1}{n!}
\left( \int_0^t ~ p(t') ~ \d t' \right)^n = p(t) ~,
\end{equation}
where the sum runs over the $n$ $2 \to 3$ events that may have been
generated at $t' < t$.

\subsection{The $\alpha$ angle}
\label{sec:alphaME}

We begin by studying the azimuthal asymmetries introduced for ISR in
Sec.~\ref{sec:aziasym}. The strength of the asymmetry, tuned through the $N$
parameter of eq.~(\ref{eq:azistr}), enters in all following studies, and
one goal here is to fix its value. The question to be asked is, how
strongly does the plane of an initial-state emission correlate with its
colour partner in the final state, when present. The distribution can be
calculated in the soft limit, but obtains corrections away from this
region; hard emissions, especially in the case when strict rapidity
ordering is not enforced (as for the first emission and optionally for
subsequent ones) are expected to be constrained to a much lower degree. In
both the ME calculation and in data, there is no clean separation between
ISR and FSR, so we instead use the $\alpha$ angle observable, used by the
CDF collaboration in order to study colour coherence effects in hadron
colliders \cite{Abe:1994nj}.

At the hadron level, the idea is to search for a leading jet which is hard
enough such that ``soft'' radiation is still hard enough to form secondary
jets. The study is based on QCD events, where at least
three jets must be present (after jet clustering with the CDF fixed-cone
algorithm, $R_{\mrm{cone}} = 0.7$). Both the hardest and second hardest
jets must lie in the region $|\eta| < 0.7$ and the leading jet must have
$E_{\perp 1} > 110\GeV$. The two leading jets must be approximately
back-to-back in the transverse plane, $||\varphi_1 - \varphi_2| - \pi| <
\pi/9$, while the third jet must have $E_{\perp 3} > 10\GeV$. These
cuts should isolate a region in which the shower approximation is still
valid, and azimuthal asymmetries should appear as a correlation between the
directions of the second and third jets. 

Defining $\Delta\eta = \eta_3 - \eta_2$ and
$\Delta\varphi = \varphi_3 - \varphi_2$, an
$R_{23} = \sqrt{\Delta\eta^2 + \Delta\varphi^2}$ cut is introduced,
$1.1 < R_{23} < \pi$. The lower limit is placed due to clustering cone-size
effects, while the upper limit avoids the region where $|\Delta\varphi|$
approaches its maximal value of $\pi$. The $\alpha$ angle is then defined by
\begin{equation}
 \alpha = \mathrm{tan}^{-1}
 \left( \frac{\mathrm{sgn}(\eta_2) \Delta\eta}
             {| \Delta\varphi |} \right) ~.
\end{equation}
A full discussion of the details of the different $\alpha$-angle regions is
given in \cite{Abe:1994nj}. Here, we restrict ourselves to azimuthal
correlations only. The primary effect should be a shift of events towards
the endpoints of the distribution, $\alpha \rightarrow \pm \pi/2$, such
that ISR is biased to sit in the plane of its colour-connected partner,
$\Delta \varphi \to 0$. It should be noted that $\alpha < 0$ will be
slightly favoured due to the third jet having access to central rapidity
regions, where phase space is larger.

A similar study has also been done by the D0 collaboration, but instead using
$\beta = \mathrm{tan}^{-1} (\mathrm{sgn}(\eta_2) \Delta\varphi / \Delta\eta)$
as their observable \cite{Abbott:1997bk}. The overall conclusions of both
of these studies is that parton shower MCs that implement colour coherence,
either through the choice of evolution variable or through an angular veto
are able to describe data.

First, a comparison is made against the $2 \to 3$ ME, using the second set
of cuts from the previous section. In this kinematic region, the softest
jet is expected to come from radiation and should be relatively soft
compared to the underlying $2 \to 2$ process, where the parton shower
is expected to be most accurate. At the parton level, many of the cuts used in
the full experimental analysis are no longer relevant, and so the only
restriction placed is that $0.5 < R_{23} < 1.0$. The lower cut, although no
longer necessary for clustering reasons, is included to enhance the effect
of ISR by removing some contributions from FSR, which will be highest in
the region of low $R_{23}$. The upper cut helps to isolate the region
where the azimuthal weighting should have the biggest effect ($r \sim 1$,
Sec.~\ref{sec:aziasym}).

\begin{figure}
\begin{minipage}{\linewidth}
\centerline{\footnotesize(a)}\vspace{-3mm}
\includegraphics[scale=0.6,angle=270]{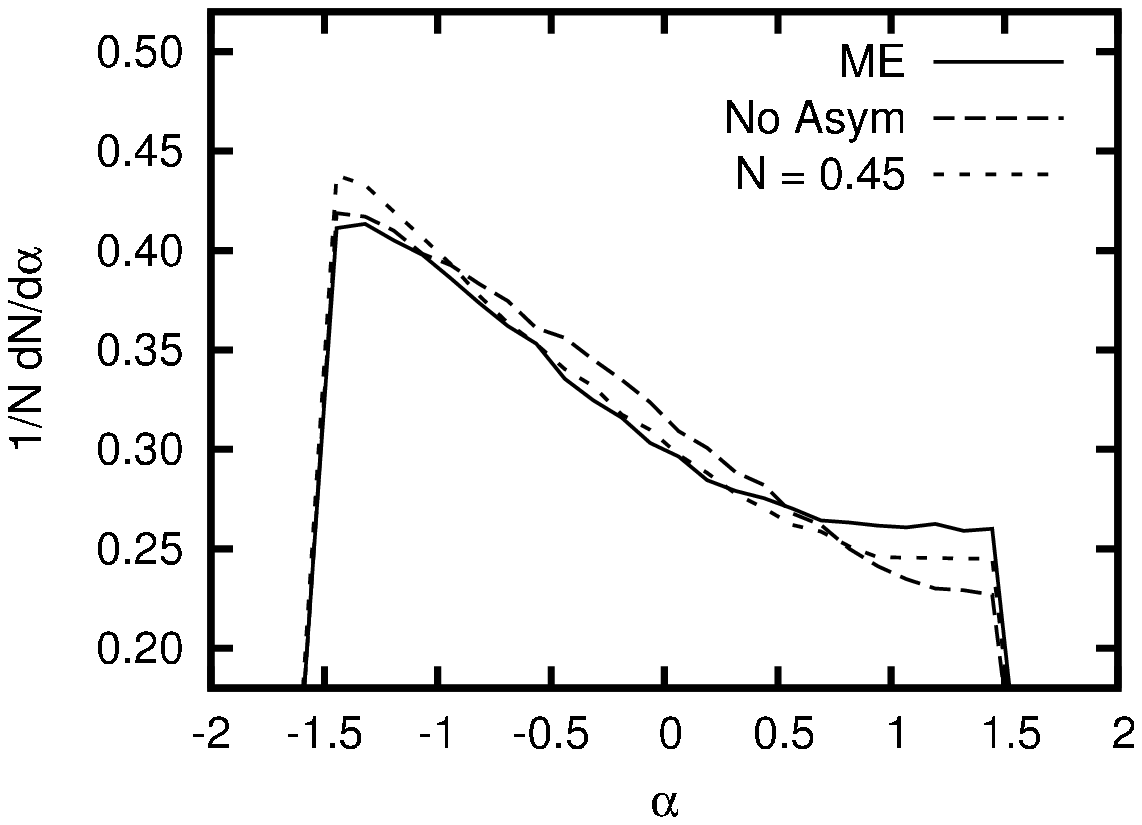}
\includegraphics[scale=0.6,angle=270]{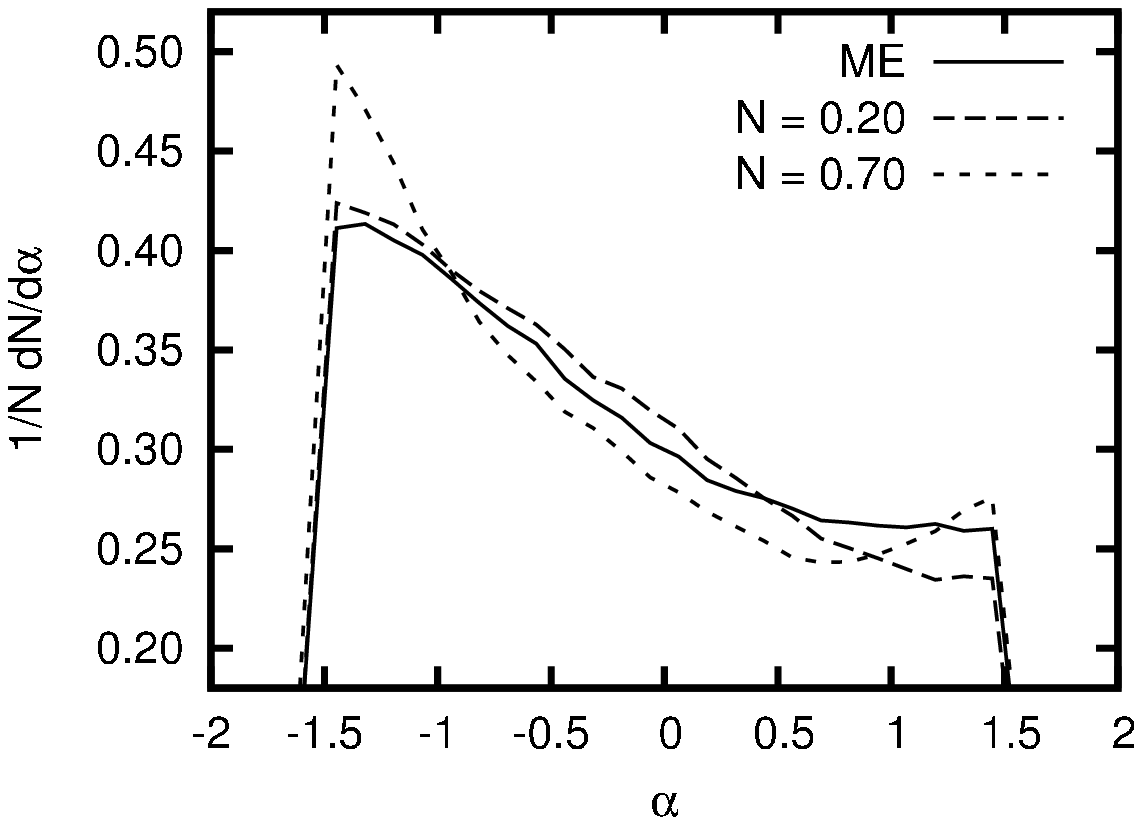}
\end{minipage}\vspace{2mm}
\begin{minipage}{\linewidth}
\centerline{\footnotesize(b)}\vspace{-3mm}
\includegraphics[scale=0.6,angle=270]{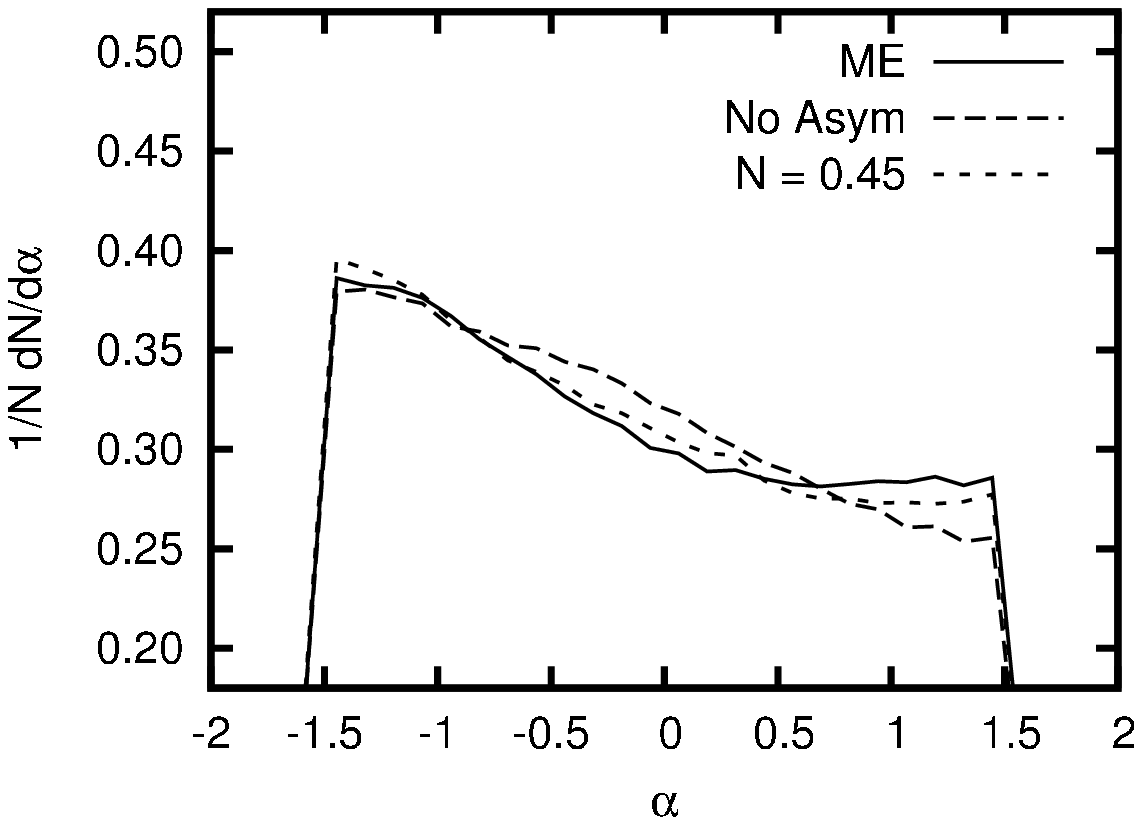}
\includegraphics[scale=0.6,angle=270]{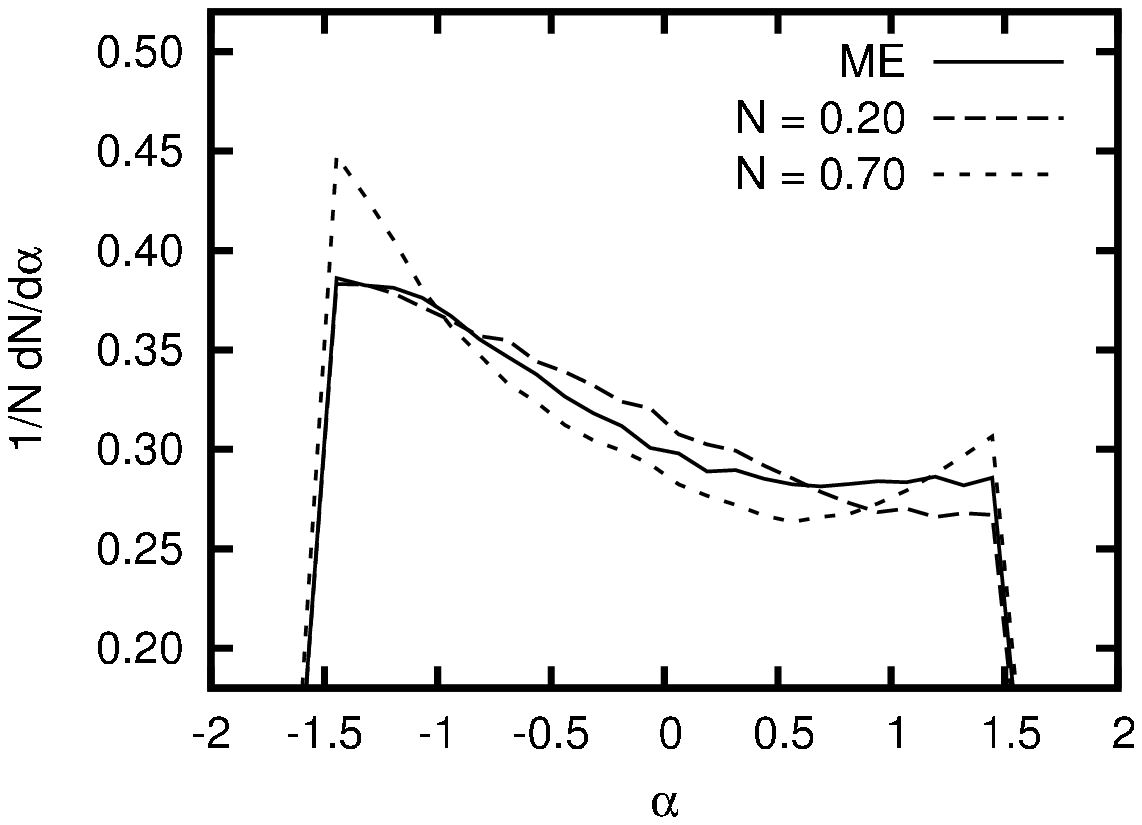}
\end{minipage}
\caption{$\alpha$-angle comparison for different strengths of asymmetry at
(a) Tevatron ($\p\pbar$, $\sqrt{s} = 1.96\TeV$) and (b) LHC ($\p\p$,
$\sqrt{s} = 14\TeV$) energies
\label{fig:2to3-alpha}}
\end{figure}

Results are shown in Fig.~\ref{fig:2to3-alpha} for different strengths of
asymmetry, $N$, at (a) Tevatron ($\p\pbar$, $\sqrt{s} = 1.96\TeV$) and (b)
LHC ($\p\p$, $\sqrt{s} = 14\TeV$) energies. For the PS, it is clear that
when all initial-state azimuthal weighting has been switched off, there is
an excess of events in the region of small $|\alpha|$.  When the weighting
is switched on, the results are as expected; events are shifted towards the
endpoints of the distribution, with the effect growing as $N$ is increased.
One unexpected feature is the agreement of the PS and ME at $\alpha =
-\pi/2$ when no azimuthal weighting is switched on. This suggests that ISR
emitted toward the central region already has some natural bias towards small
$\Delta \varphi$. The extra weighting, then, overshoots in this region, but
does have the desired effect at $\alpha = \pi/2$, where the slope of the
tail is brought up. In moving to higher energies, it should be noted that
the same cuts are used, meaning phase space effects will affect the exact
shape of the distribution and the balance between the positive and negative
$\alpha$ regions. Considering only the first emission, a value of $N =
0.45$ gives a reasonable agreement across the whole $\alpha$ range and for
both energies.

\begin{figure}
\begin{minipage}{\linewidth}
\centerline{\footnotesize(a)}\vspace{-1mm}
\centerline{
\includegraphics[scale=0.65]{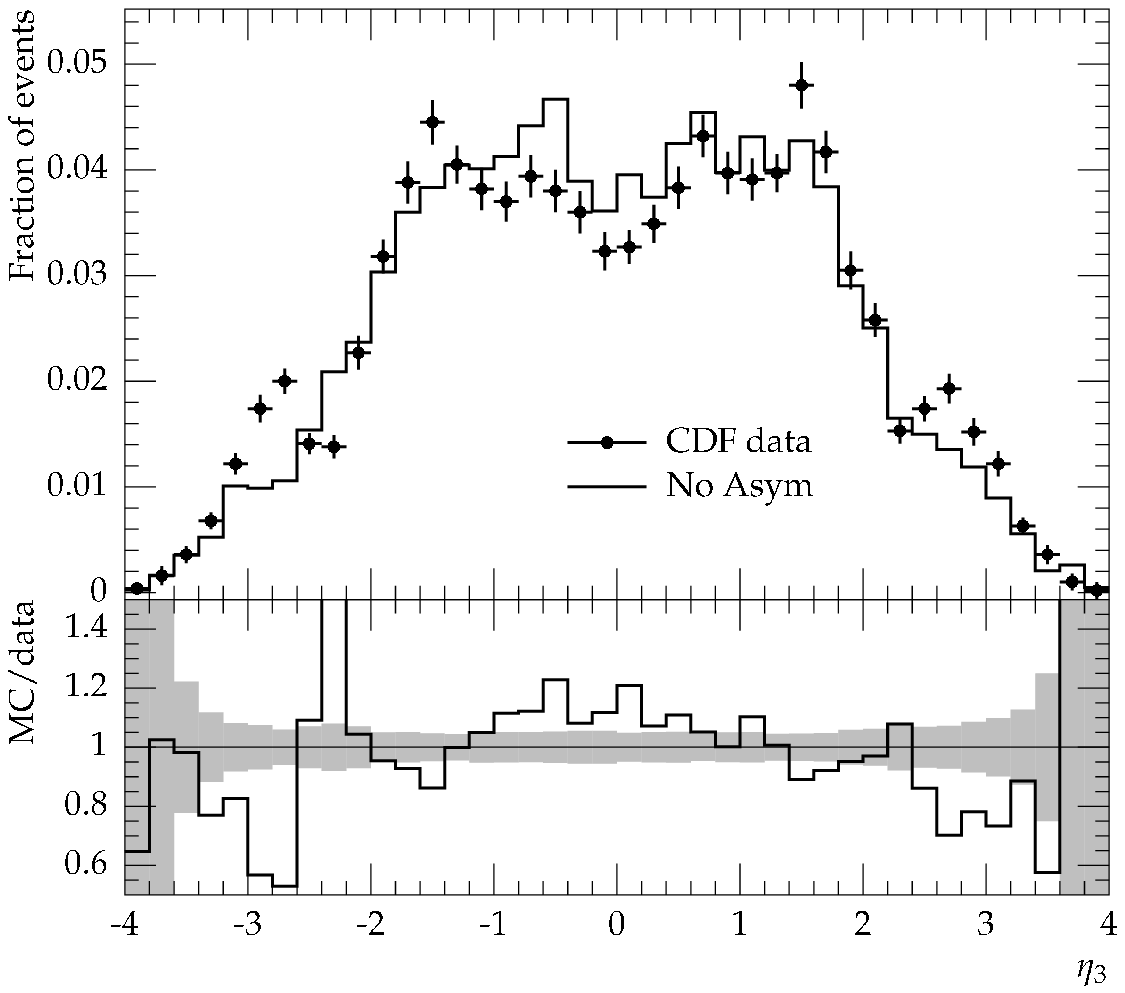}\hspace{4mm}
\includegraphics[scale=0.65]{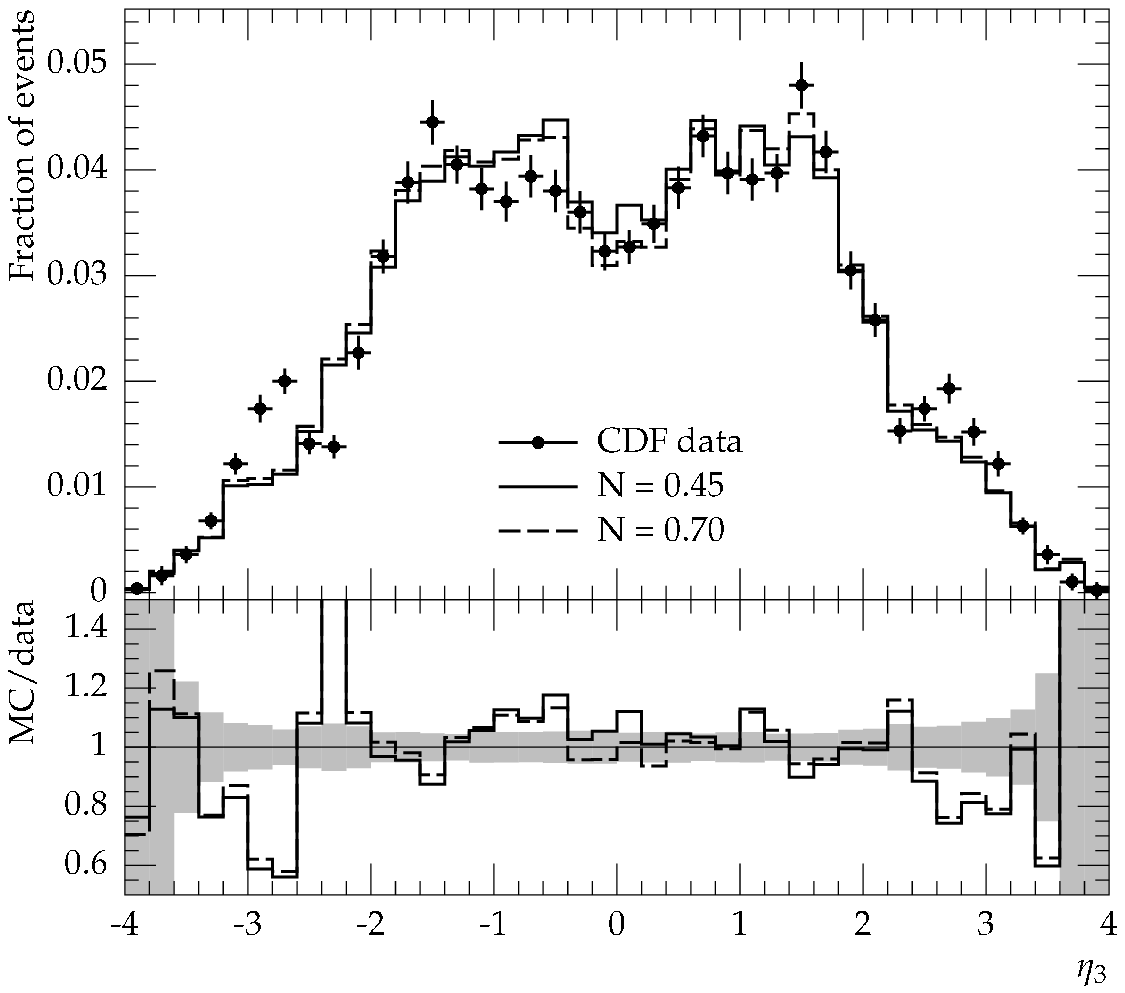}
}
\end{minipage}\vspace{4mm}
\begin{minipage}{\linewidth}
\centerline{\footnotesize(b)}\vspace{-1mm}
\centerline{
\includegraphics[scale=0.65]{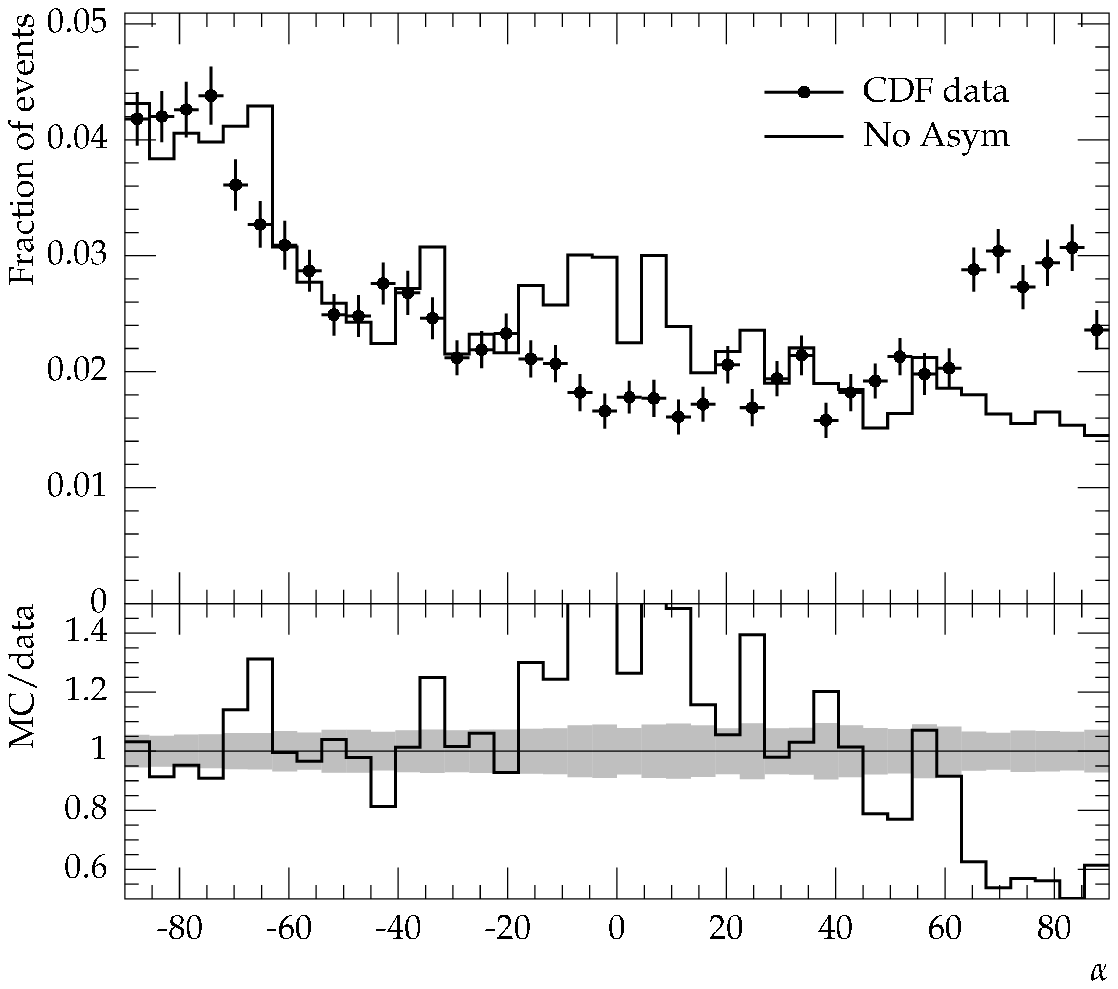}\hspace{4mm}
\includegraphics[scale=0.65]{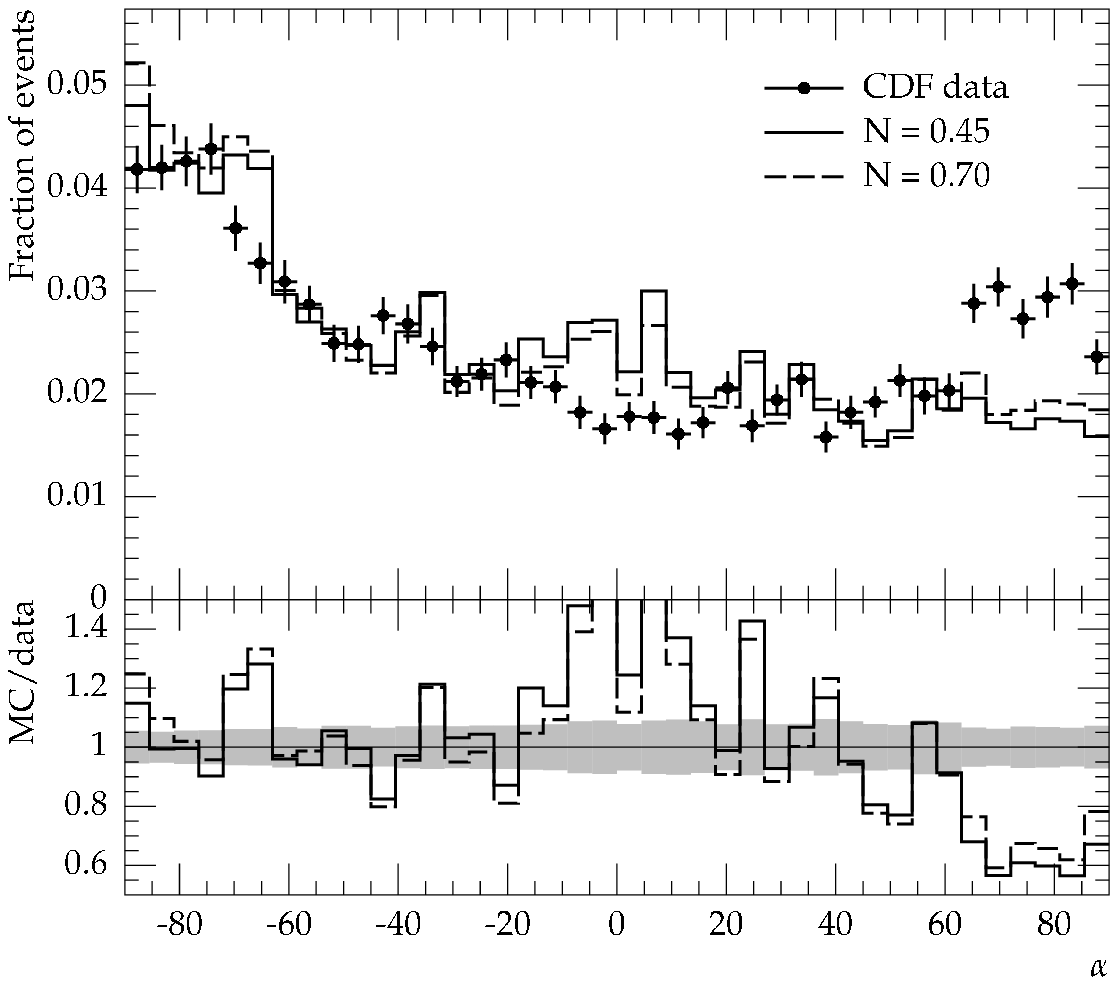}
}
\end{minipage}
\caption{Rivet CDF $\alpha$-angle analysis ($\p\pbar$, $\sqrt{s} =
1.80\TeV$)
\label{fig:rivet-alpha}}
\end{figure}

This is not the whole story however; before event generation is
complete, there is still the rest of the shower evolution and additional
non-perturbative effects to be added. Afterwards, some memory of
the parton-level asymmetries should remain, which, in the $\alpha$-angle
observable, should show up in the same way as in the ME comparison.
Additionally, at the hadron level we also study the the pseudorapidity
distribution of the third jet (no explicit $R_{23}$ separation cut is
applied to this observable). It has been shown that colour coherence
effects lead to a broadening in this distribution, as jets are more likely
to be emitted at higher rapidity; a good agreement is also a sign that
coherence effects are well modeled.

The default tune of \textsc{Pythia 8.142} is used, and only those settings
related to the azimuthal asymmetries in ISR changed. The output of the
generator is fed through the CDF $\alpha$-angle Rivet analysis and
bin-by-bin detector corrections added to the final results. The results
are given in Fig.~\ref{fig:rivet-alpha}, for (a) $\eta_3$ and (b) $\alpha$.
The width of the $\eta_3$ distribution is already well described, even
without the extra azimuthal weighting. Increasing the strength of the
asymmetry does lead to a very slight broadening of the distribution, but
the largest difference is in the dip at $\eta_3 = 0$. In general, the
high-$\pT$ jets will prefer to sit at central rapidities. Biasing $\Delta
\varphi$ to smaller values, in combination with the implicit $R$ separation
cut from the jet algorithm, will shift the third jet slightly further out
in rapidity.

The results for the $\alpha$-angle are similar to the ME
comparison. At the smallest values, there is a small overshoot, while at
higher values, the agreement is improved. At $\alpha = \pi/2$, the tail
does not come up to the levels of the data, but is no longer falling.
Indeed, with the weighting that has been introduced, bringing the tail up
to match the data is a hard task, indicating that additional correlations
in rapidity may be involved in later stages of the event-generation
process. Based on the CDF data, then, a stronger asymmetry is favoured, and
a value $N = 0.70$ is adopted as the default value for all the following
sections.

\subsection{Kinematic distributions}

The main freedom in adjusting the PS to better fill the phase space is
given by the starting scale of the initial- and final-state showers. For
those hard processes which do not contain any final-state particles which
may be shower produced, the shower can begin at the kinematical limit, so
as to populate the full phase space without risk of double counting. For
the QCD processes considered here, where the hard process already contains
particles which may be produced in the shower, double counting becomes an
issue and the shower is instead started at the factorisation scale of the
hard process. In the case of a $2 \to 2$ process, this defaults to the
smaller of the squared transverse masses of the two outgoing particles.
Neglecting parton masses, this implies that the $\pT$ of the shower jets
are below the $2 \to 2$ $\pT$ scale. This appears to be a reasonable
starting point, but note that the recoil of an emitted parton can boost one
of the original two jets to have a smaller $\pT$ than the emitted jet.
Effects can also go in the opposite direction, where a parton emitted at
the hard $\pT$ scale boosts both original partons to a higher $\pT$,
opening an unfilled region.

\begin{figure}
\centering
\includegraphics[scale=0.43,angle=270]{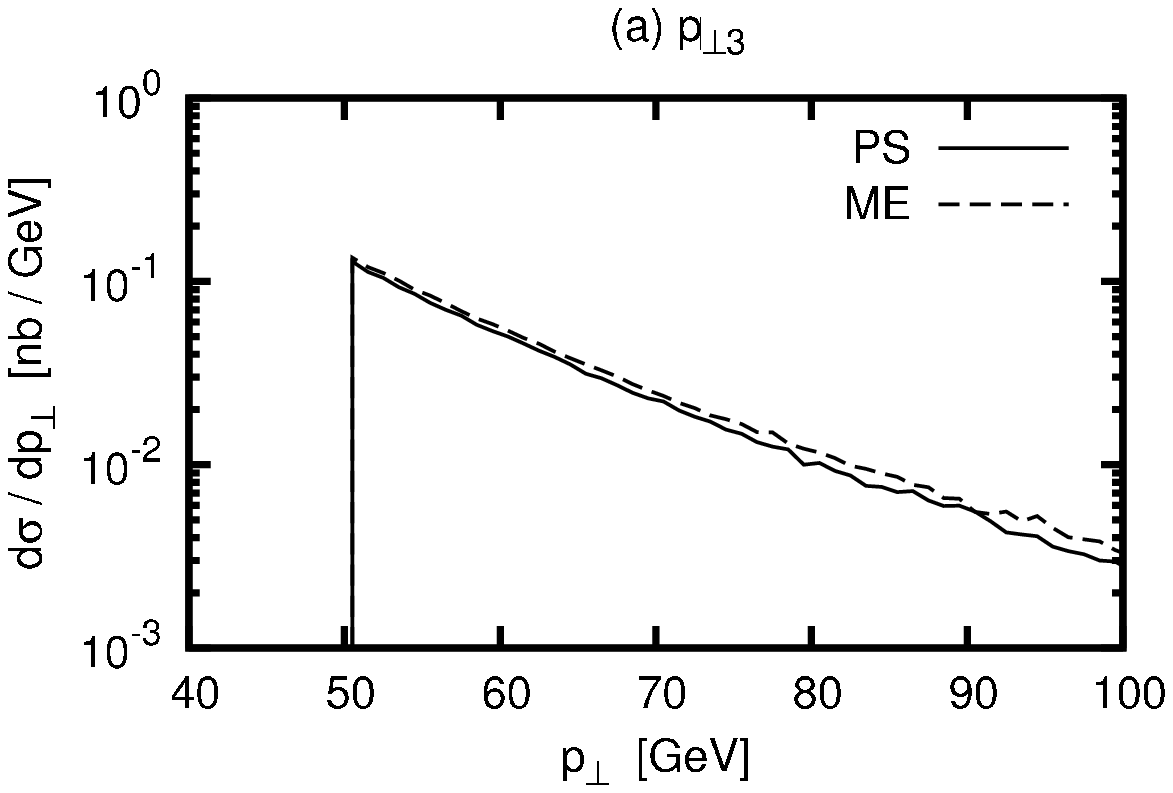}\hspace{-2mm}
\includegraphics[scale=0.43,angle=270]{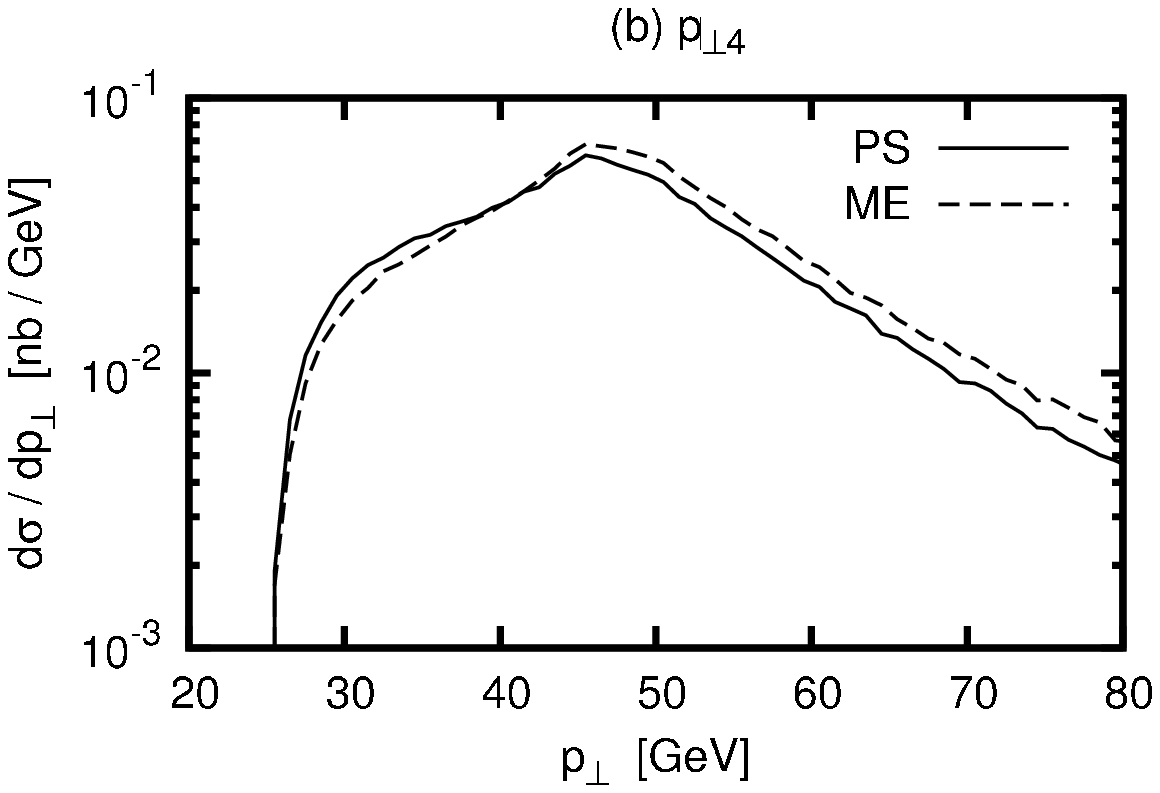}\hspace{-2mm}
\includegraphics[scale=0.43,angle=270]{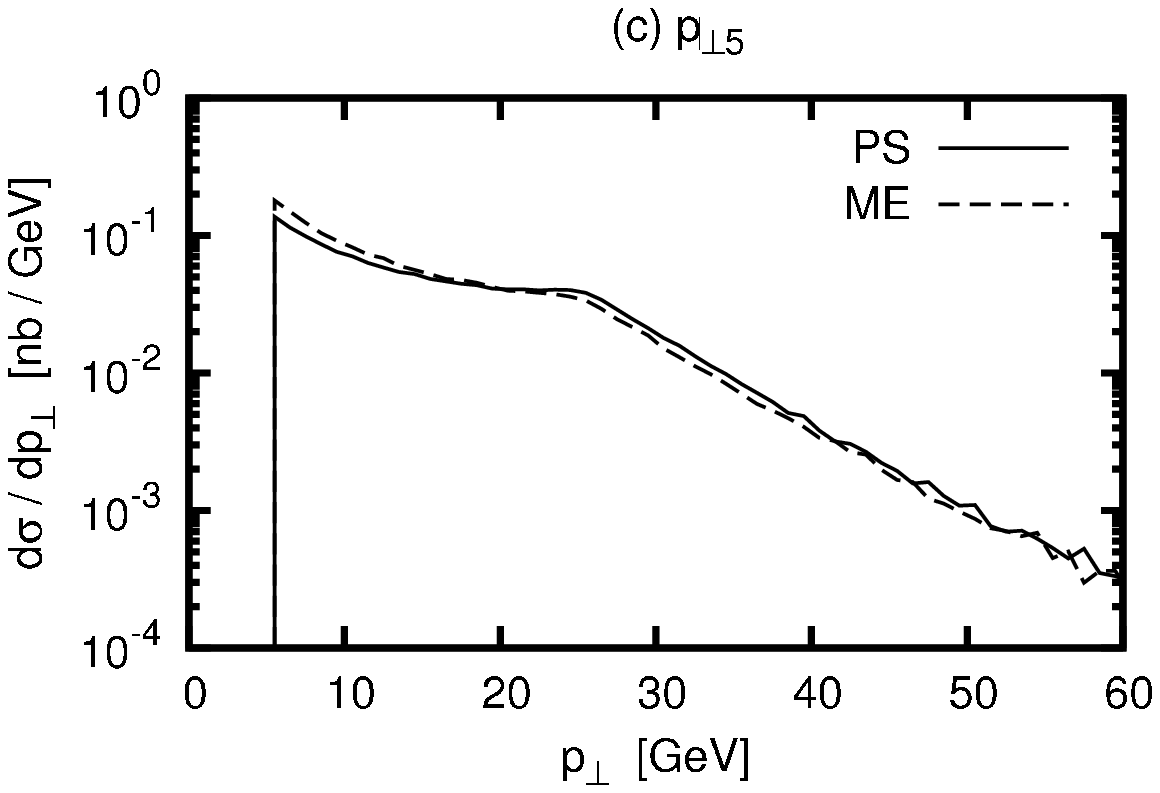}\\
\includegraphics[scale=0.43,angle=270]{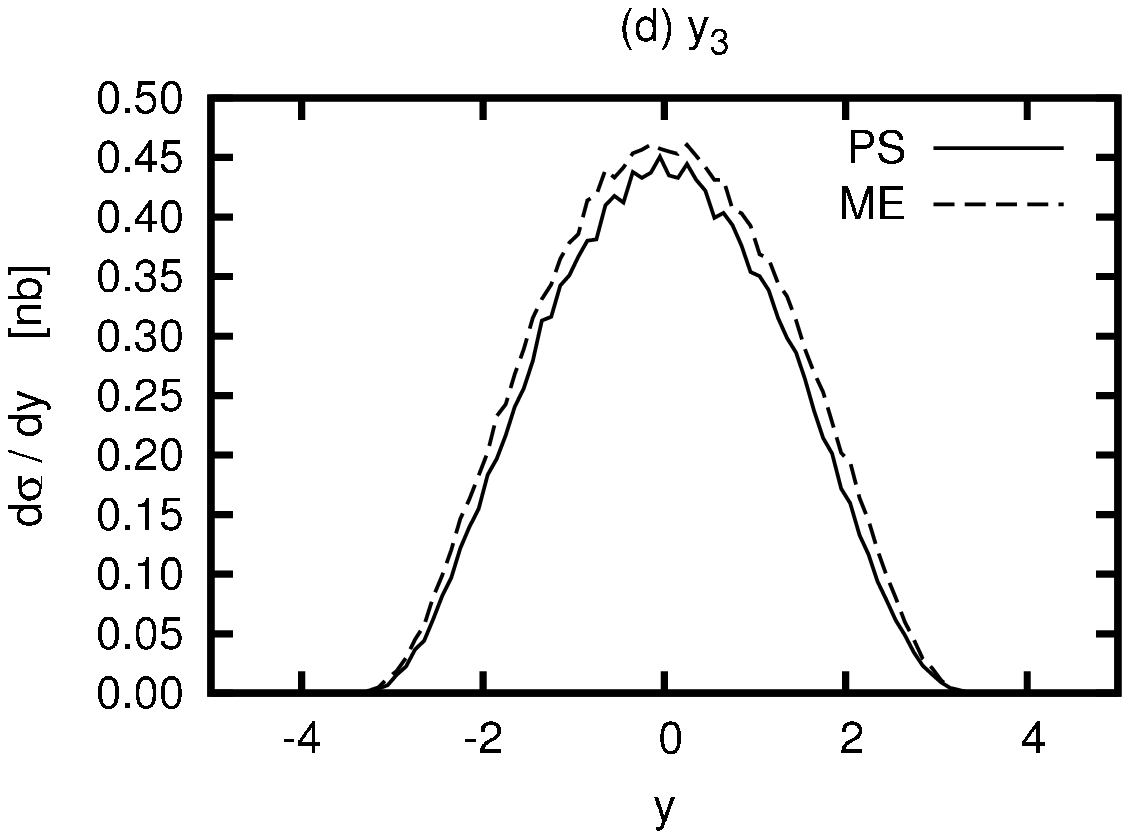}\hspace{-2mm}
\includegraphics[scale=0.43,angle=270]{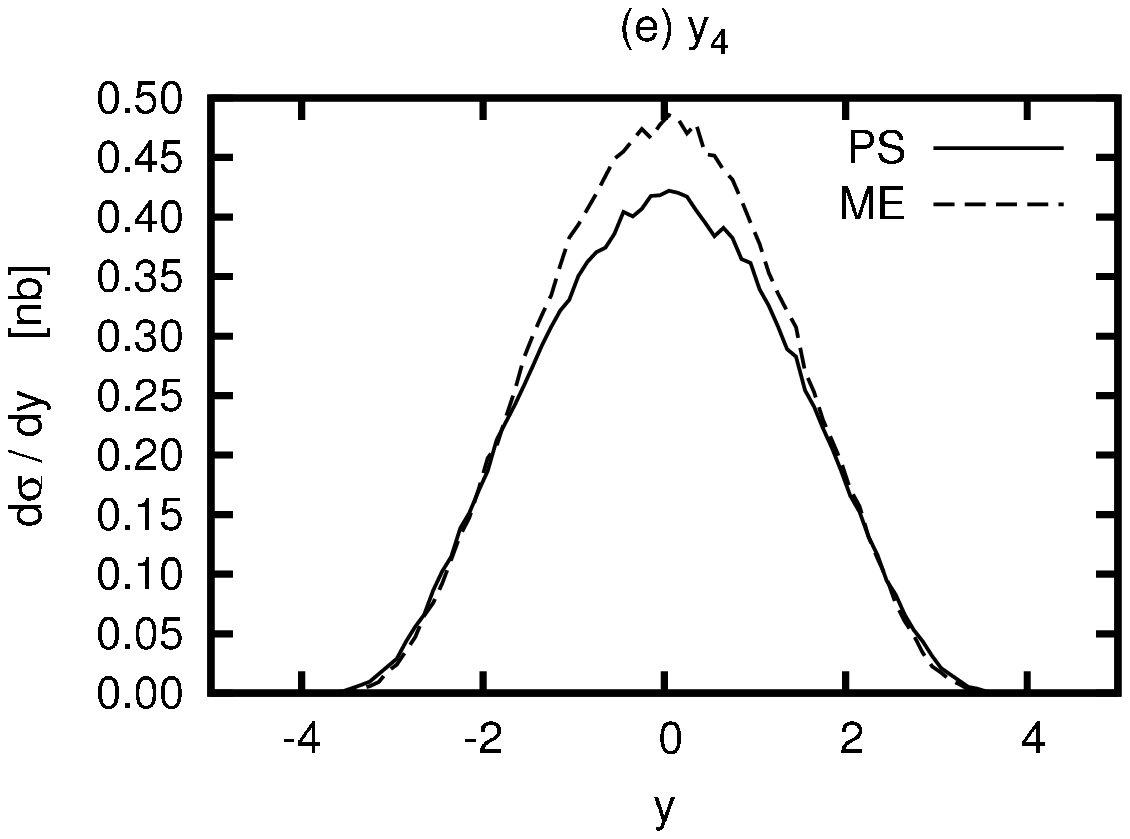}\hspace{-2mm}
\includegraphics[scale=0.43,angle=270]{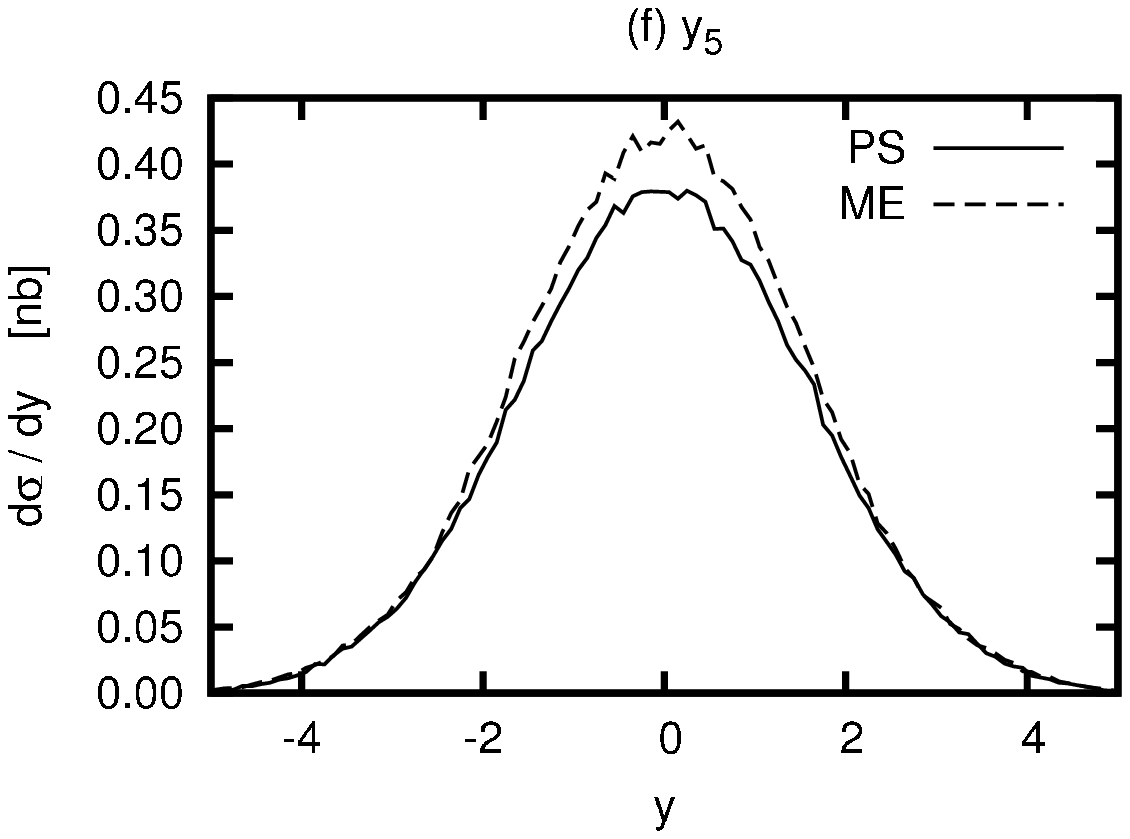}\\
\includegraphics[scale=0.43,angle=270]{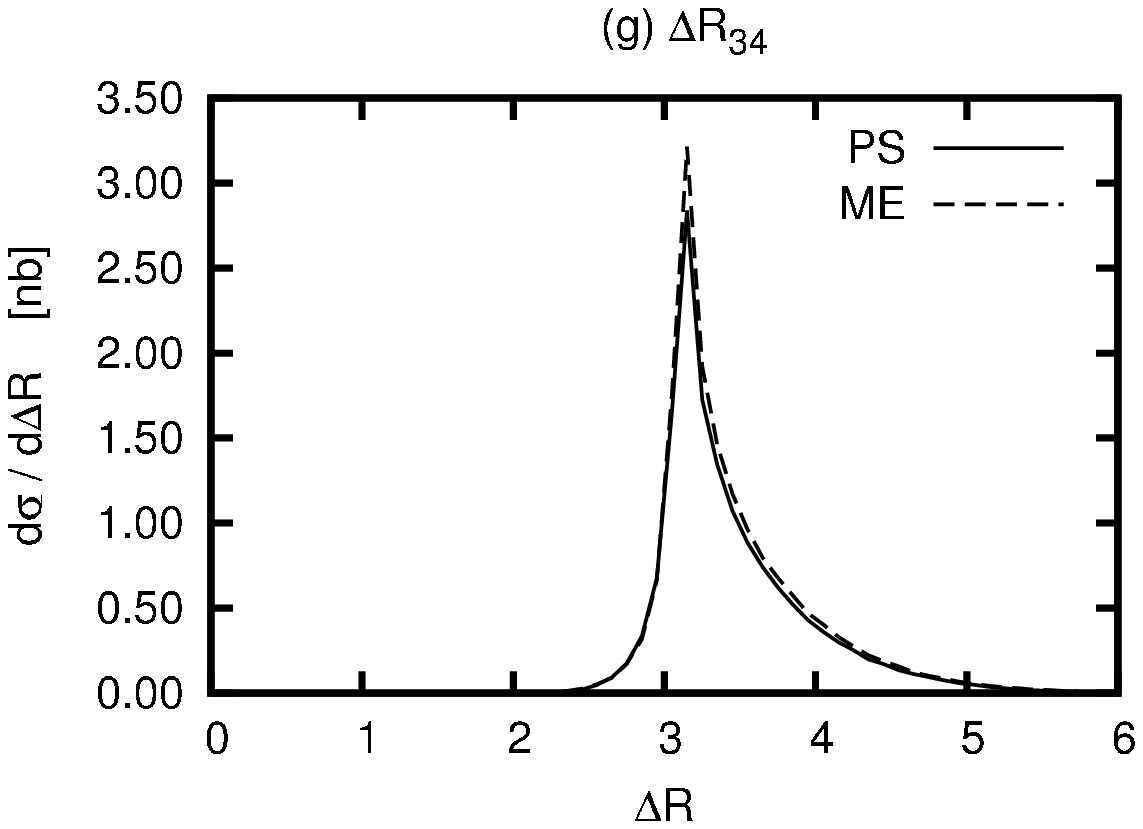}\hspace{-2mm}
\includegraphics[scale=0.43,angle=270]{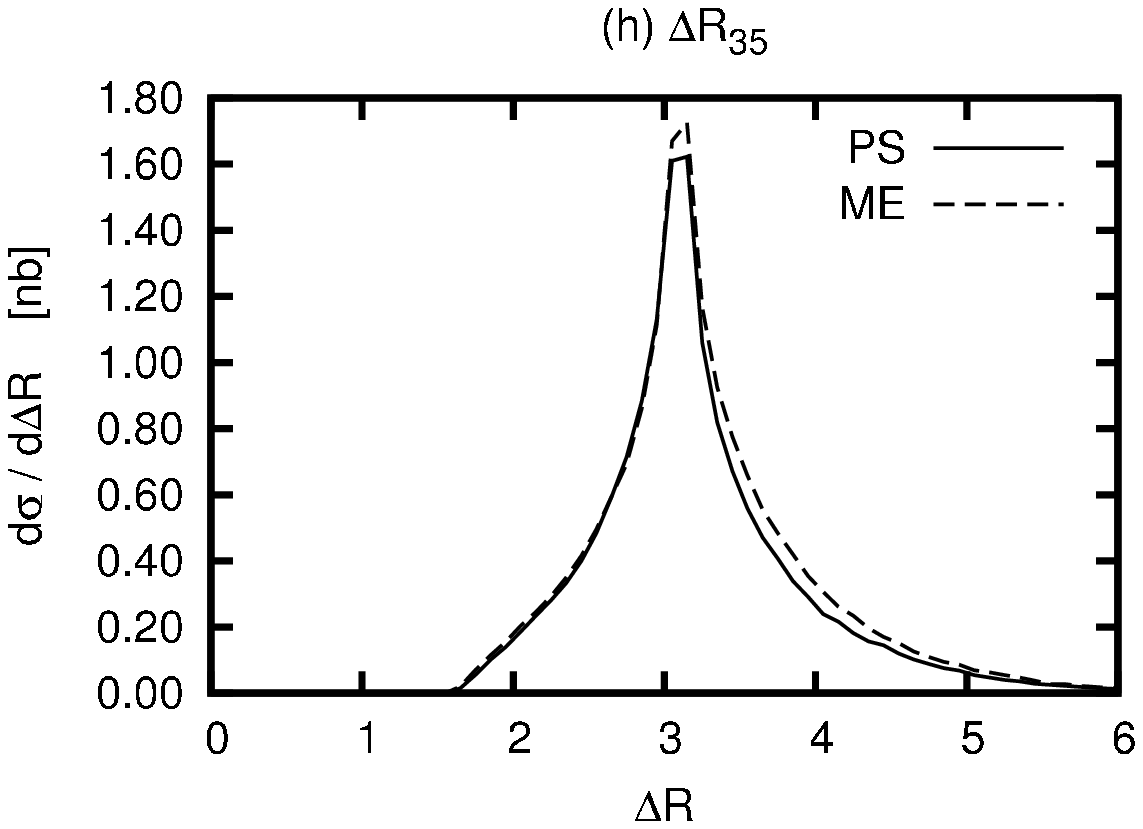}\hspace{-2mm}
\includegraphics[scale=0.43,angle=270]{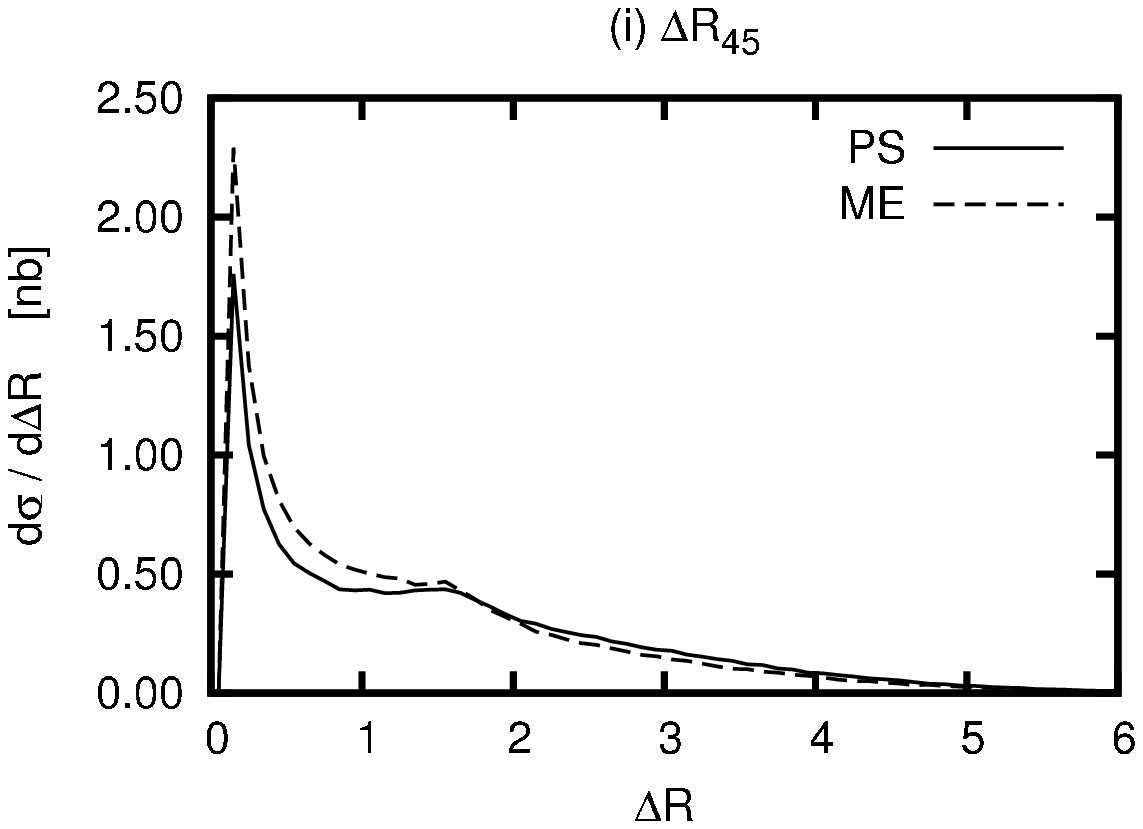}
\caption{Kinematic distributions for cut set (2) at Tevatron energies
($\p\pbar$, $\sqrt{s} = 1.96\TeV$): ${\pT}_3^\mmin = 50.0\GeV$,
${\pT}_5^\mmin = 5.0\GeV$, $R_{\mrm{sep}}^{\mmin} = 0.1$
\label{fig:2to3.cut2}}
\end{figure}

Results are shown using the default showers for the three sets of cuts
defined previously, for the transverse momenta, rapidities and $R$
separations of the three jets.  The first set covers as large an area of
phase space as possible, while the second isolates a region where the
parton shower is expected to perform well. The final set should isolate a
region where the three jets are all relatively hard, and well separated,
exactly the region where the parton shower is not expected to perform well.
In the plots that follow, we stress that the main aim is for a good
qualitative agreement with the matrix elements.

\begin{figure}
\centering
\includegraphics[scale=0.43,angle=270]{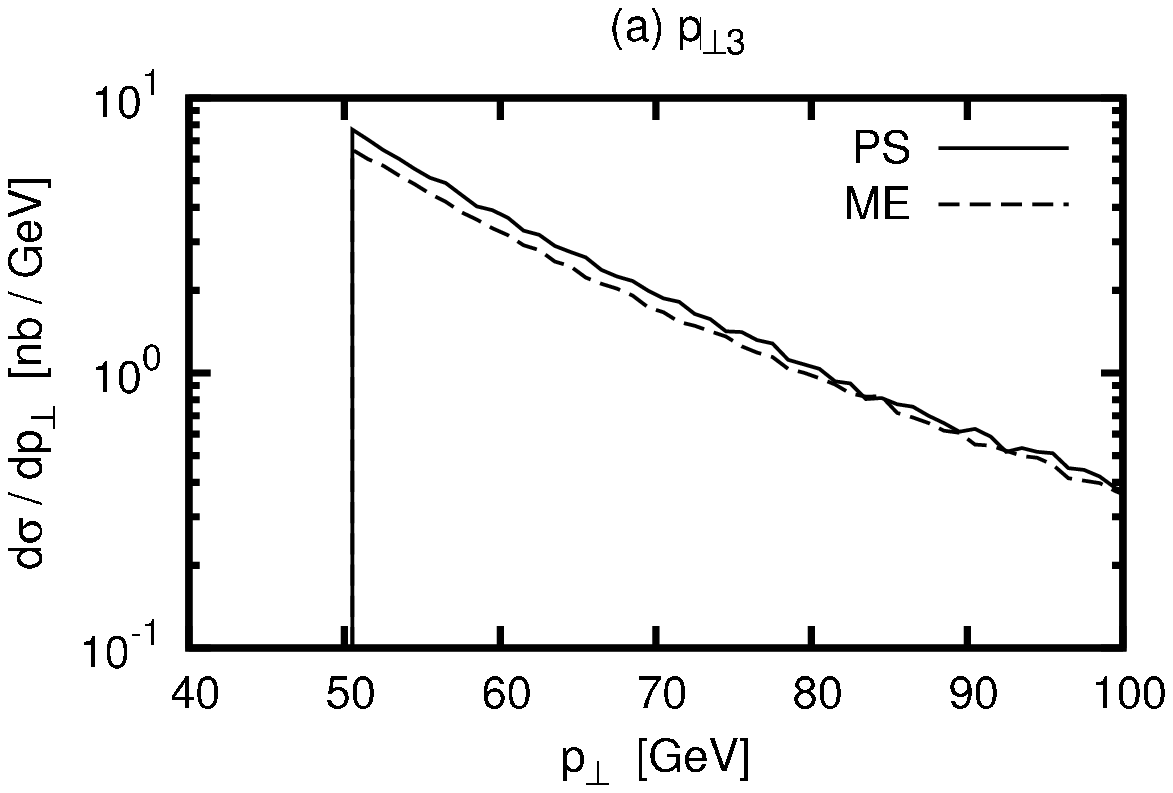}\hspace{-2mm}
\includegraphics[scale=0.43,angle=270]{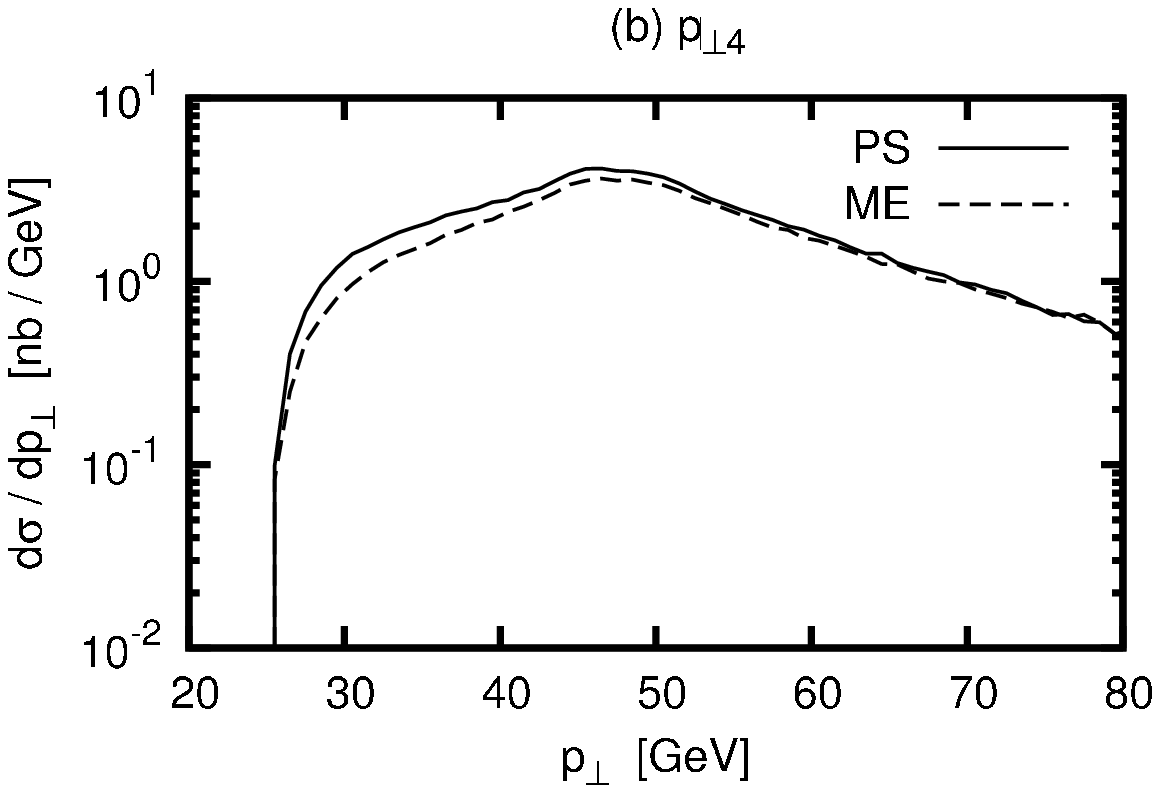}\hspace{-2mm}
\includegraphics[scale=0.43,angle=270]{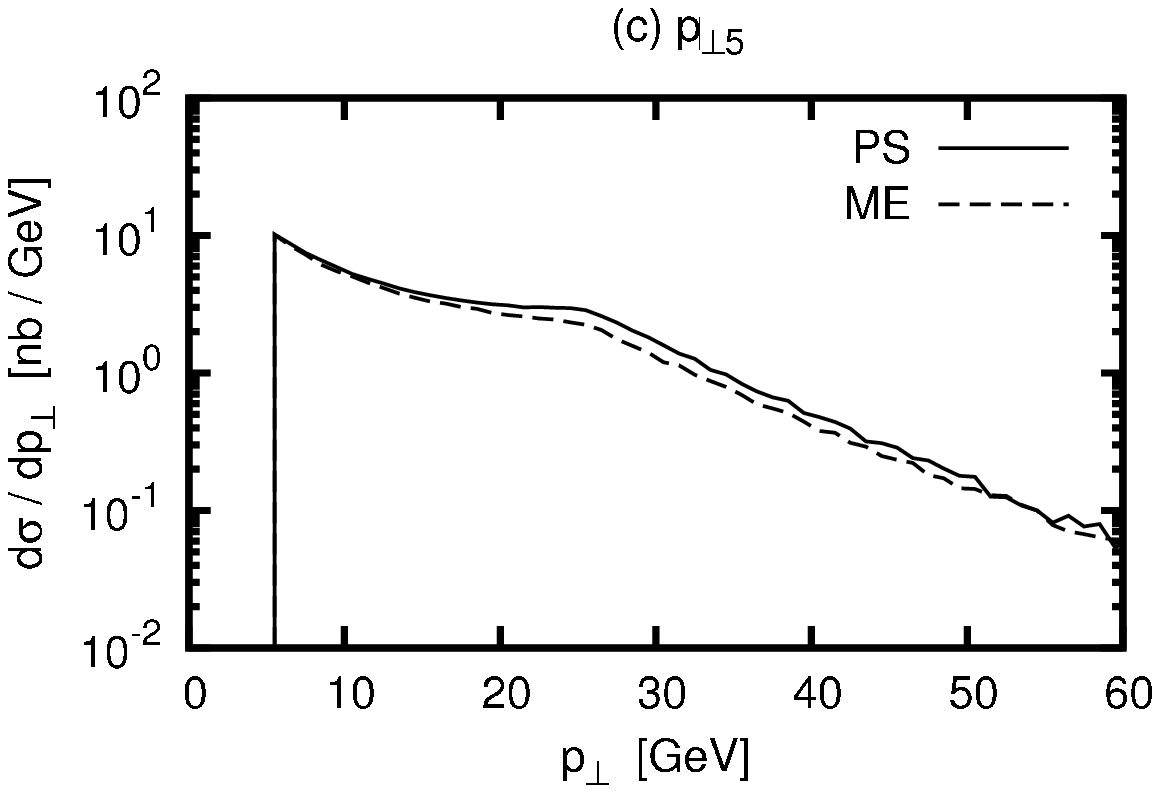}\\
\includegraphics[scale=0.43,angle=270]{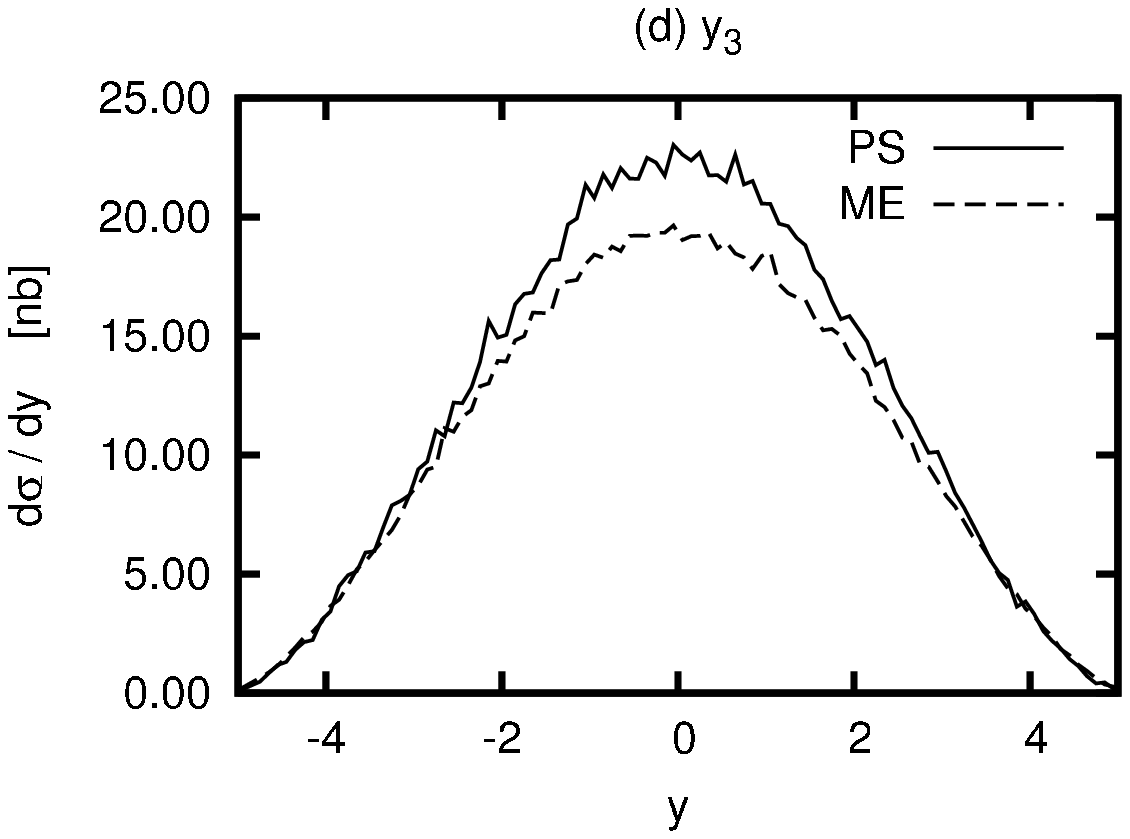}\hspace{-2mm}
\includegraphics[scale=0.43,angle=270]{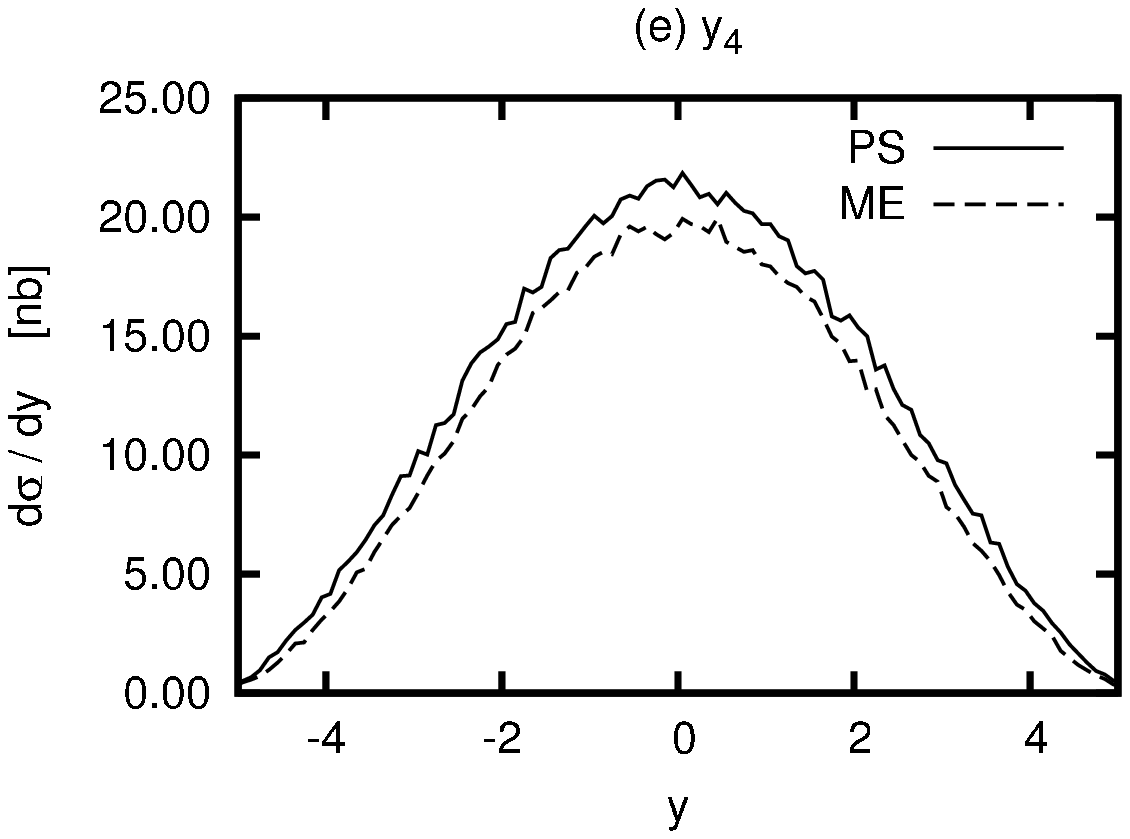}\hspace{-2mm}
\includegraphics[scale=0.43,angle=270]{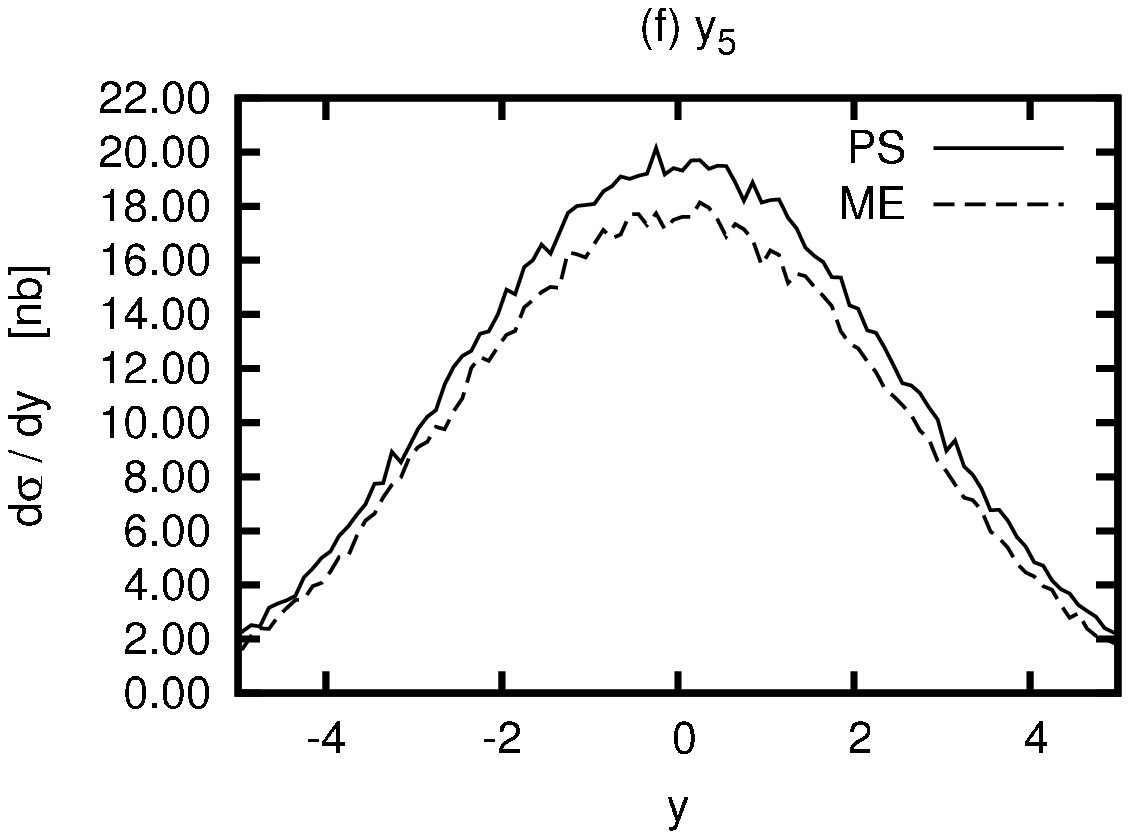}\\
\includegraphics[scale=0.43,angle=270]{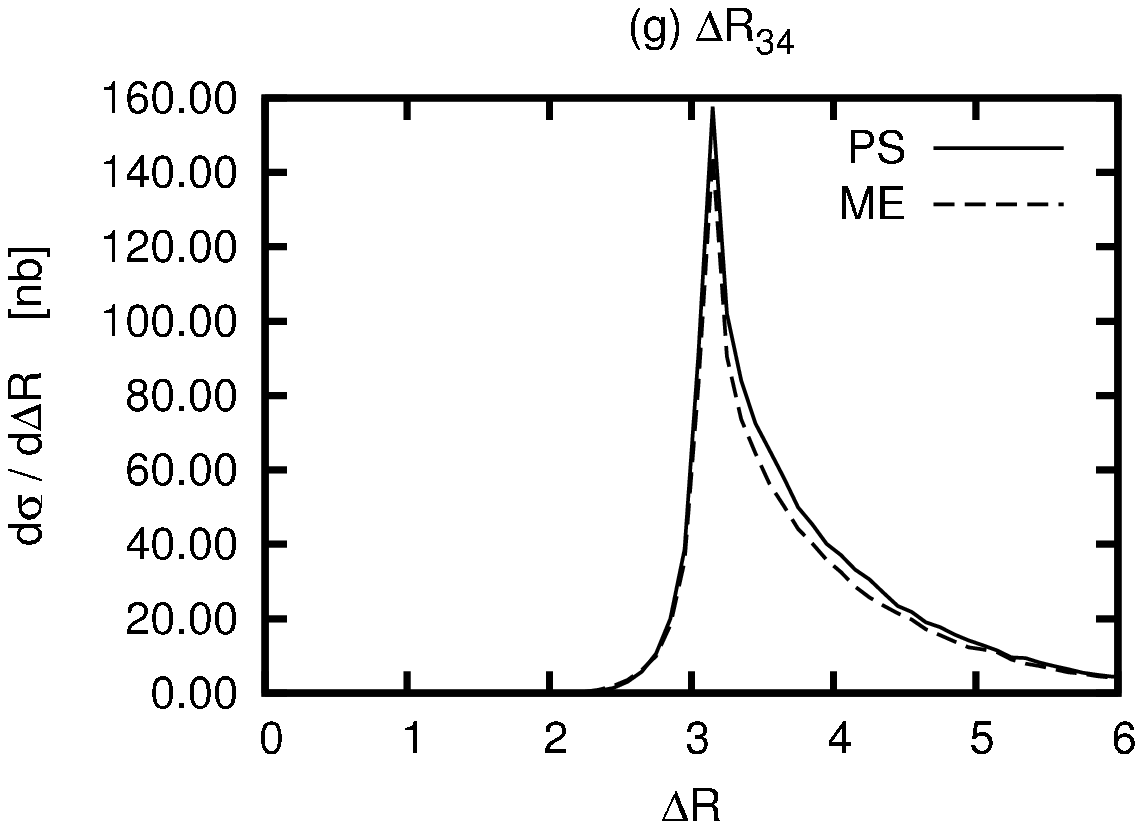}\hspace{-2mm}
\includegraphics[scale=0.43,angle=270]{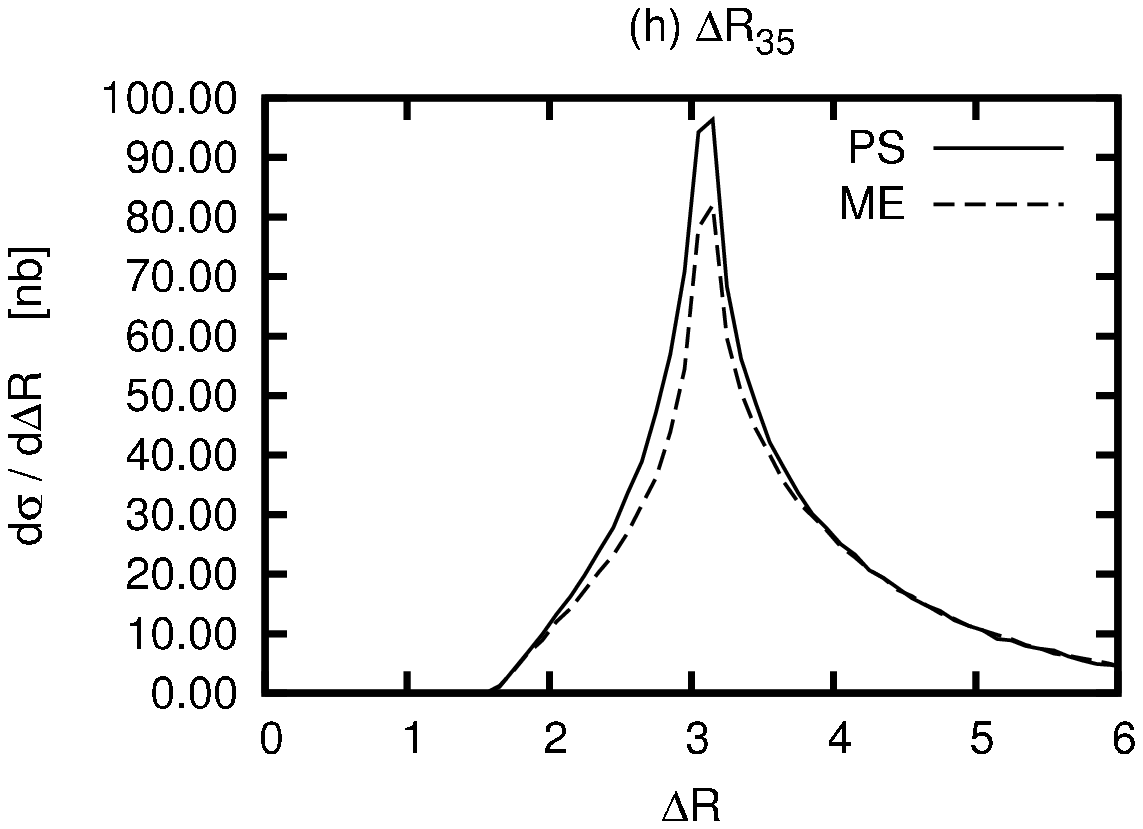}\hspace{-2mm}
\includegraphics[scale=0.43,angle=270]{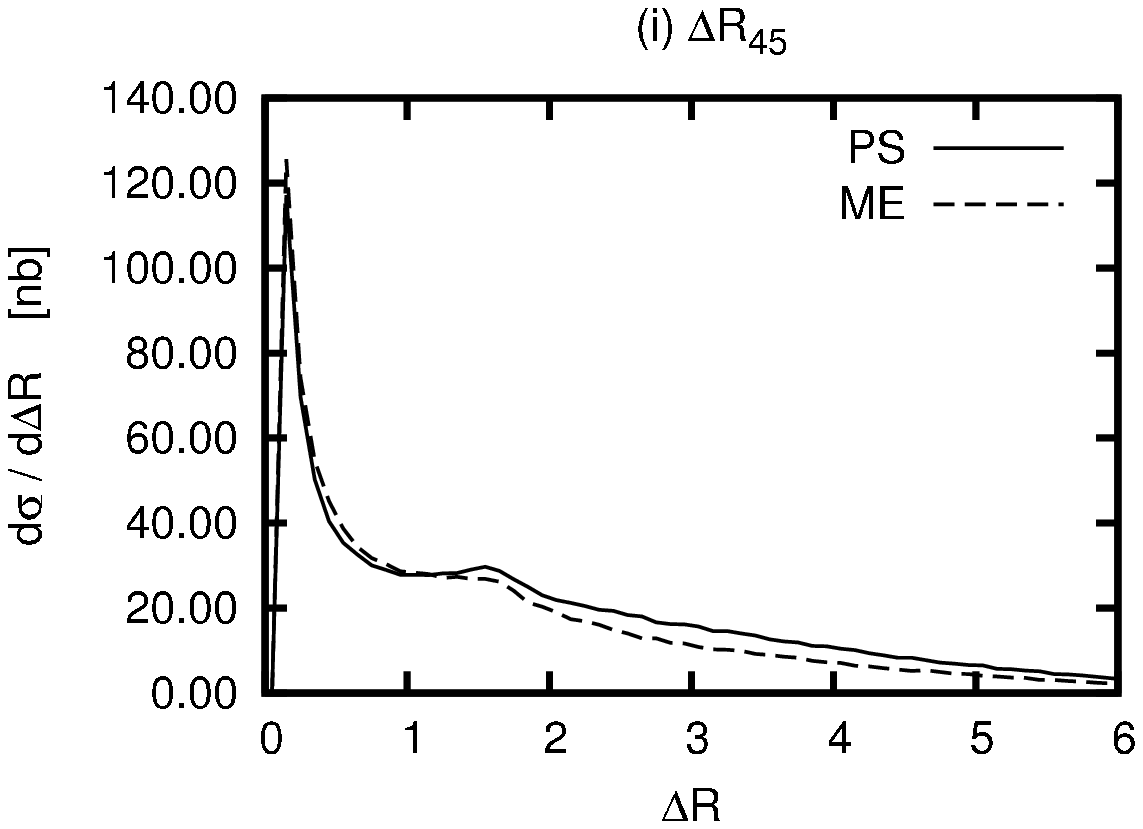}
\caption{Kinematic distributions for cut set (2) at LHC energies
($\p\p$, $\sqrt{s} = 14\TeV$): ${\pT}_3^\mmin = 50.0\GeV$,
${\pT}_5^\mmin = 5.0\GeV$, $R_{\mrm{sep}}^{\mmin} = 0.1$
\label{fig:2to3.cut2.lhc}}
\end{figure}

For the second set of cuts, results are shown in
Fig.~\ref{fig:2to3.cut2} for Tevatron energies ($\p\pbar$, $\sqrt{s} =
1.96\TeV$) and in Fig.~\ref{fig:2to3.cut2.lhc} for LHC energies
($\p\p$, $\sqrt{s} = 14\TeV$). While at the lower energy, the PS rate is
slightly below that of the ME over most of the phase space, and at the
higher, the ME is slightly above, all distributions are well
reproduced. The small excesses in events at low $p_{\perp 4}$, 
$p_{\perp 5} > 25\GeV$ and $R_{45} > \pi/2$ are consistent with a slightly
too high contribution from hard- and widely-separated jets, which we
examine further below.

\begin{figure}
\centering
\includegraphics[scale=0.43,angle=270]{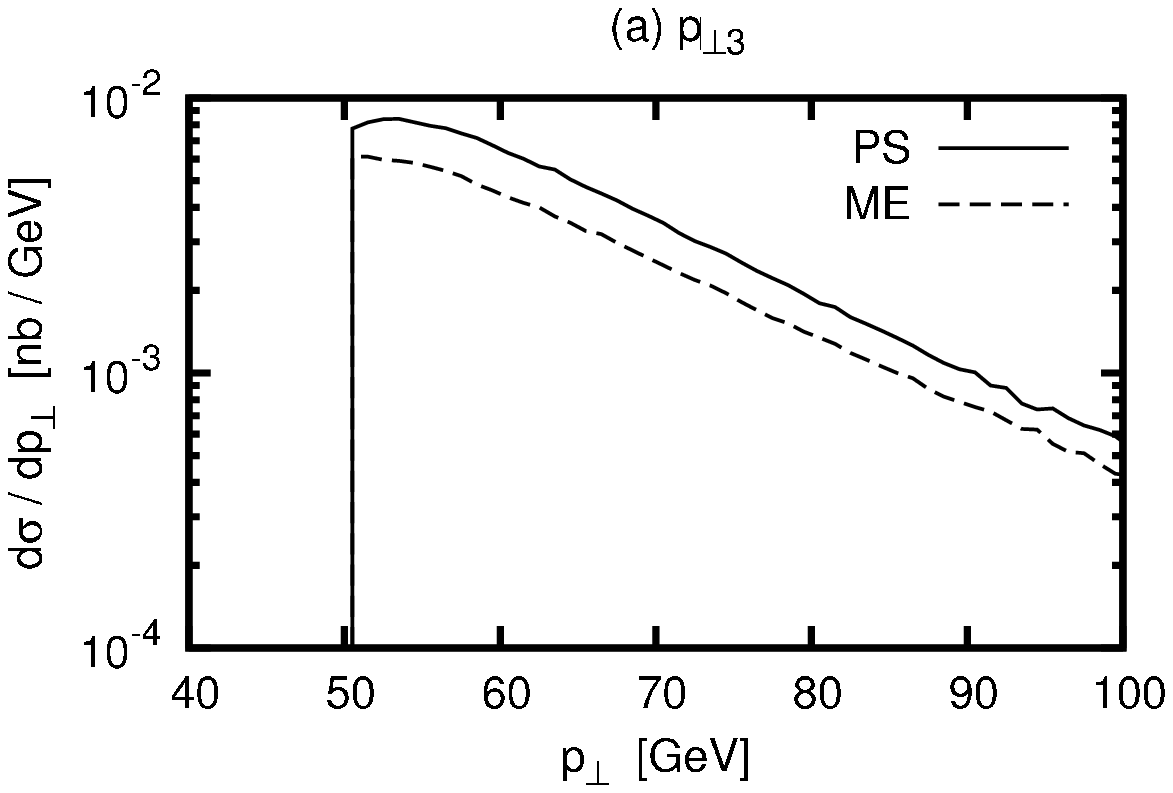}\hspace{-2mm}
\includegraphics[scale=0.43,angle=270]{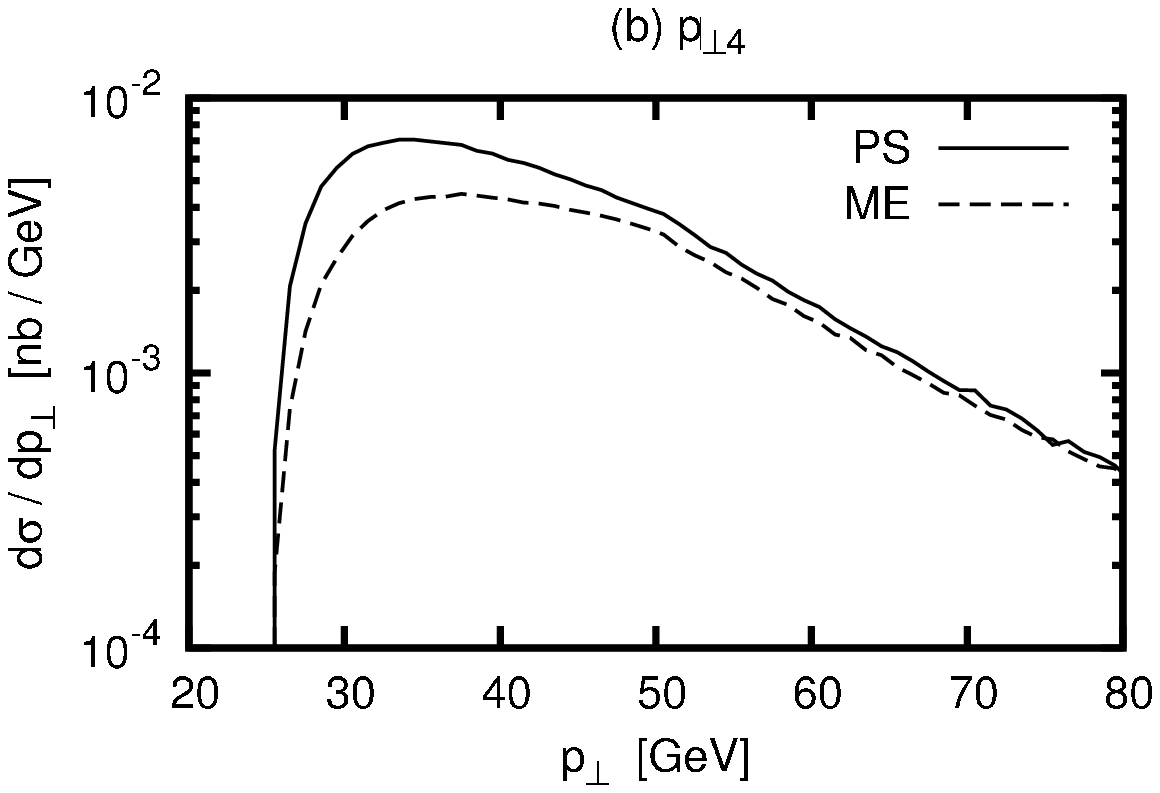}\hspace{-2mm}
\includegraphics[scale=0.43,angle=270]{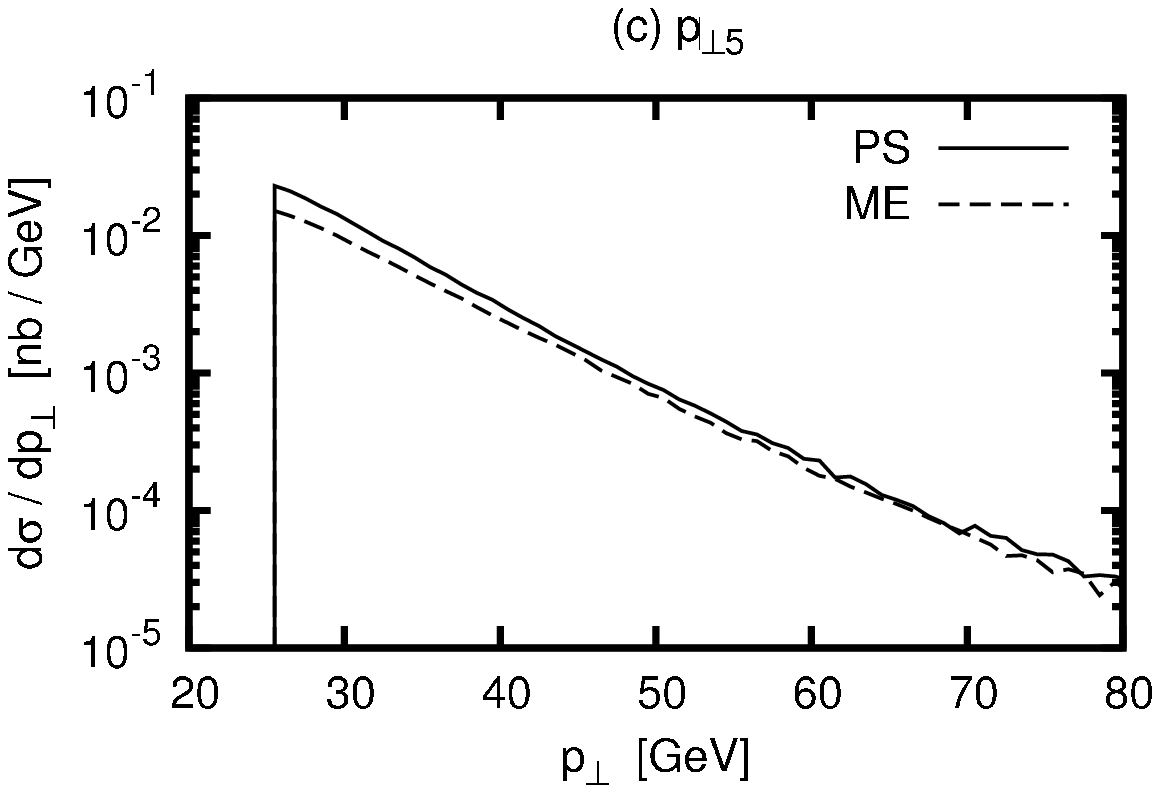}\\
\includegraphics[scale=0.43,angle=270]{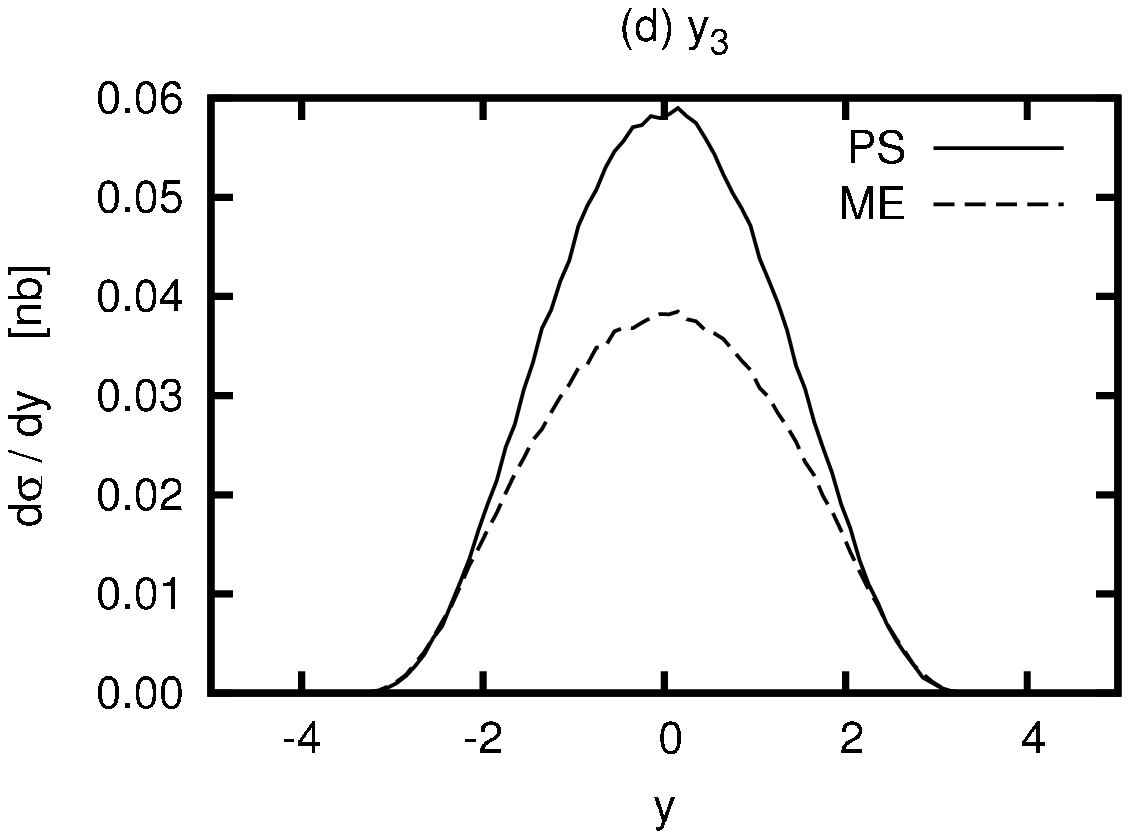}\hspace{-2mm}
\includegraphics[scale=0.43,angle=270]{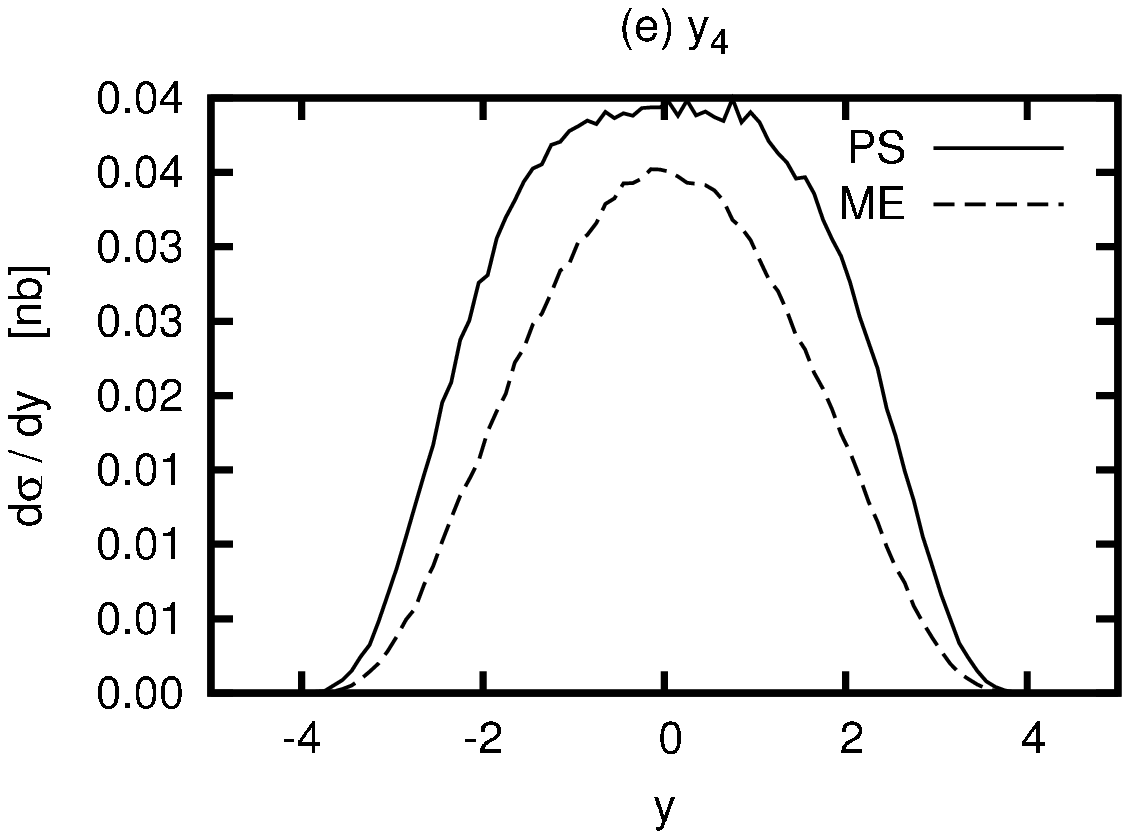}\hspace{-2mm}
\includegraphics[scale=0.43,angle=270]{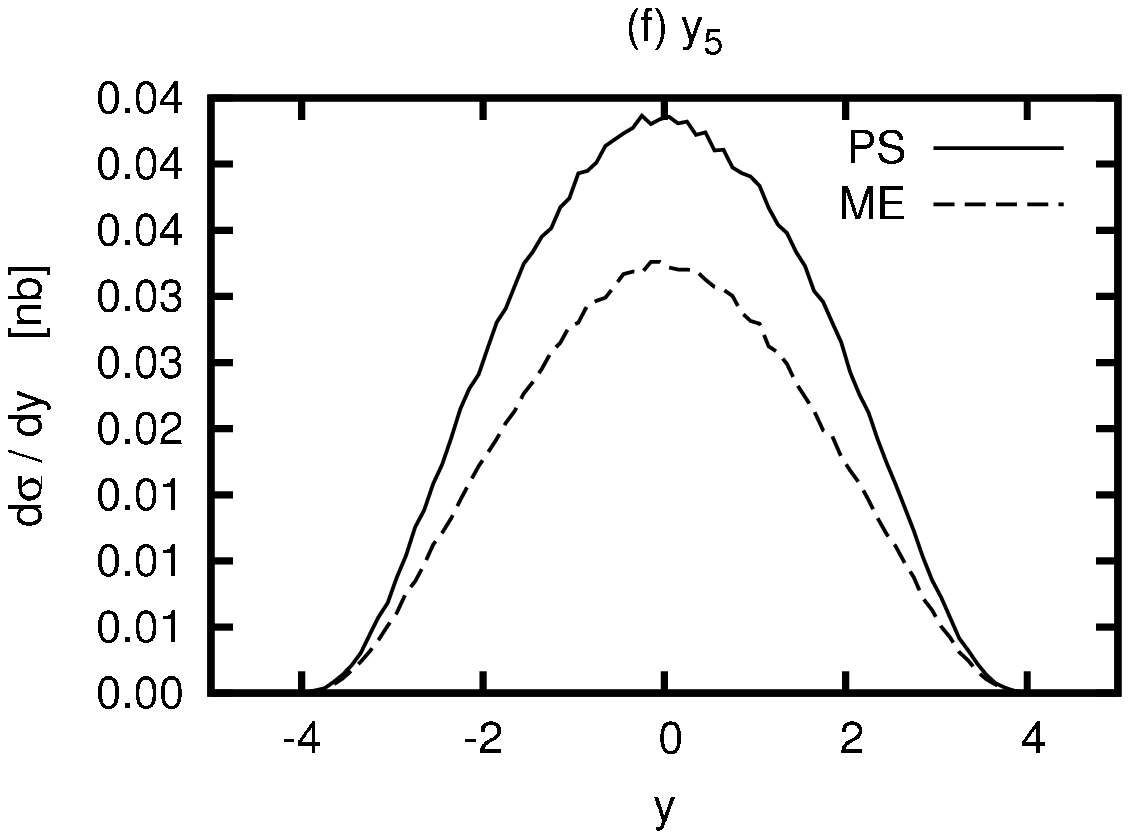}\\
\includegraphics[scale=0.43,angle=270]{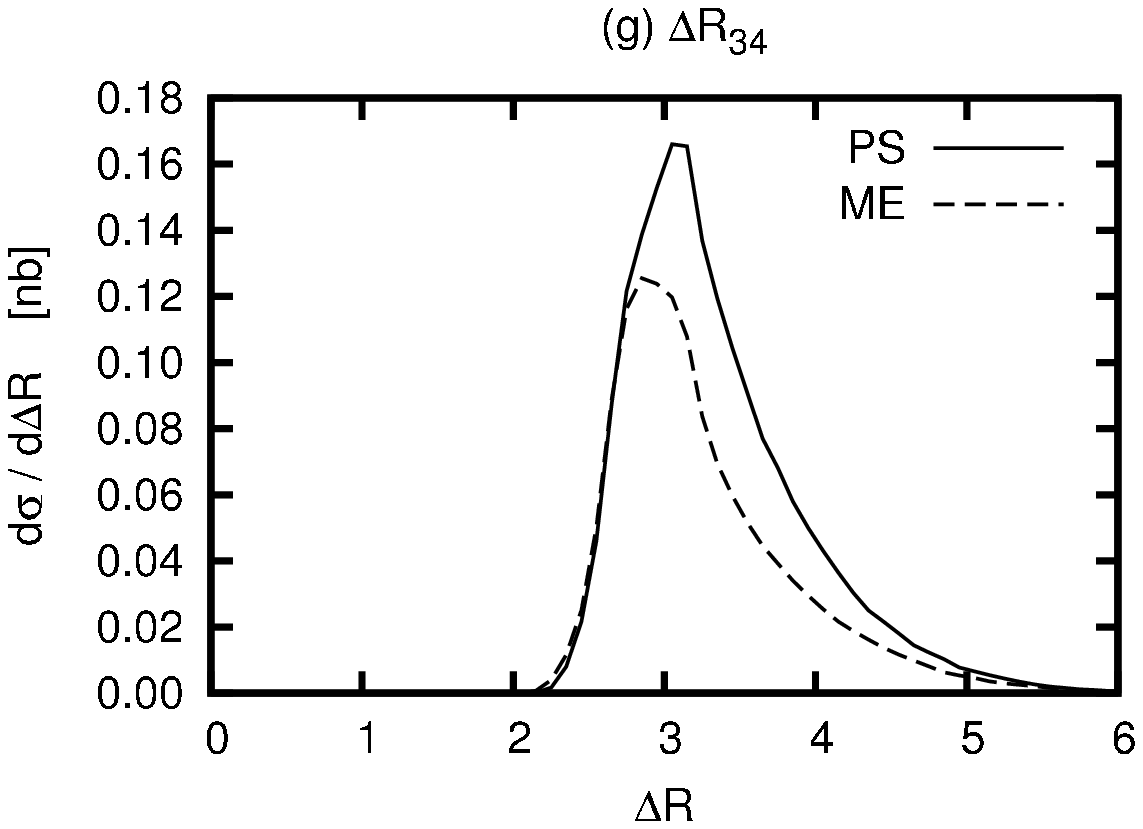}\hspace{-2mm}
\includegraphics[scale=0.43,angle=270]{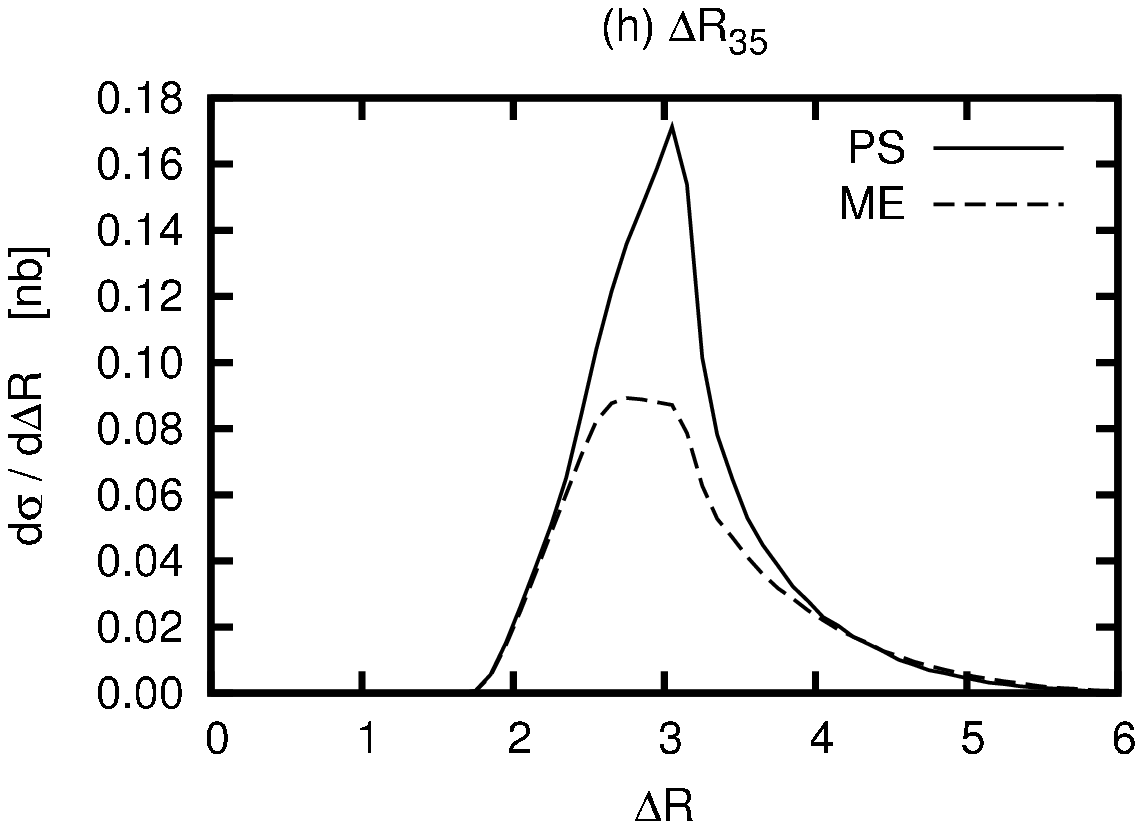}\hspace{-2mm}
\includegraphics[scale=0.43,angle=270]{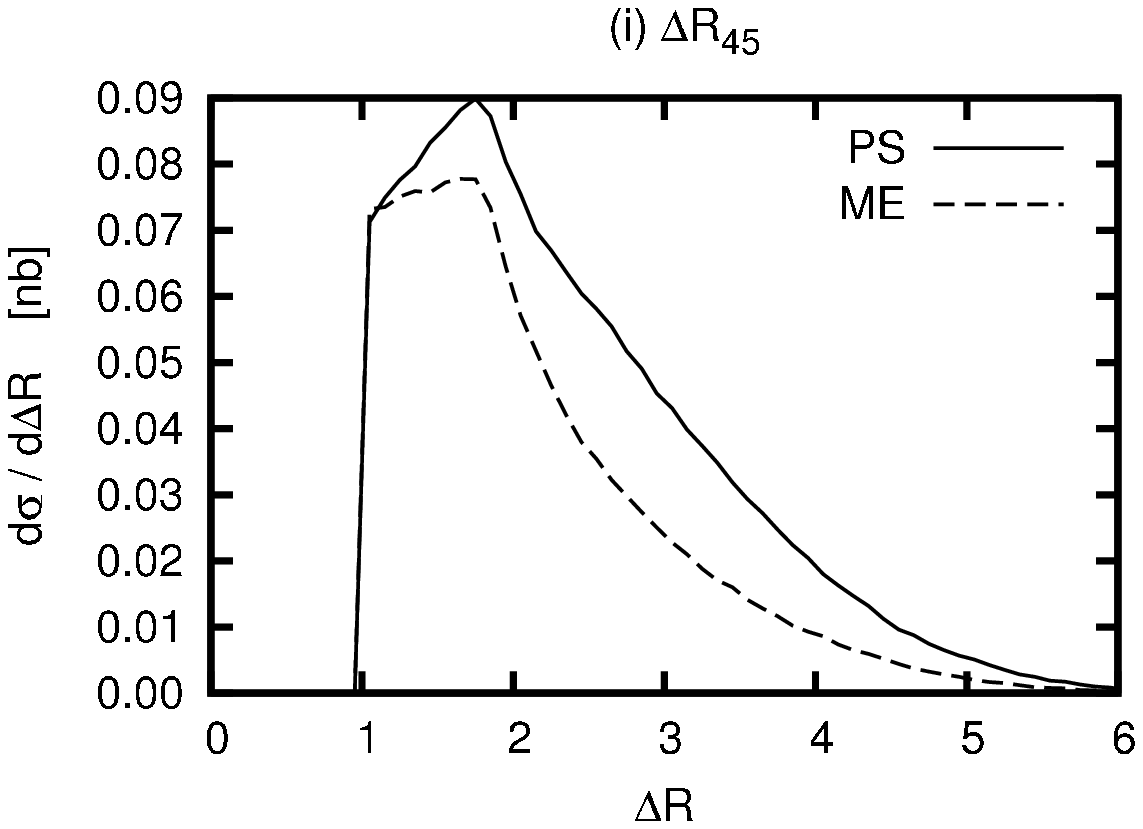}
\caption{Kinematic distributions for cut set (3) at Tevatron energies
($\p\pbar$, $\sqrt{s} = 1.96\TeV$): ${\pT}_3^\mmin = 50.0\GeV$,
${\pT}_5^\mmin = 25.0\GeV$, $R_{\mrm{sep}}^{\mmin} = 1.0$
\label{fig:2to3.cut3}}
\end{figure}

In Fig.~\ref{fig:2to3.cut3}, the results for the third set of
cuts at Tevatron energies are shown. In this region of hard, widely
separated jets, the PS is not expected to perform so well. Indeed, here the
PS rate is somewhat larger than the ME one. As previously, there is a
noticeable excess in events at low ${\pT}_4$, suggesting configurations
where a softer $2 \to 2$ process is shifted to meet
the jet requirements through an ISR emission and its recoil. The rapidity
distribution $y_4$ is also too broad. A simple check shows that there is no
large contribution from an ISR emission becoming harder than the original
two jets. That the shapes of ${\pT}_5$ and $y_5$ are relatively well
described suggests that there are effects arising from the
rotation and boosts employed in the recoil handling, also giving an excess
in the tail of $\Delta R_{45}$. We do not study this further at
this time. Although the distributions are by no means perfect, we content
ourselves with the fact that these events are in a disfavoured region of
phase space and that the description gives the broad features of the shapes.

\begin{figure}
\centering
\includegraphics[scale=0.43,angle=270]{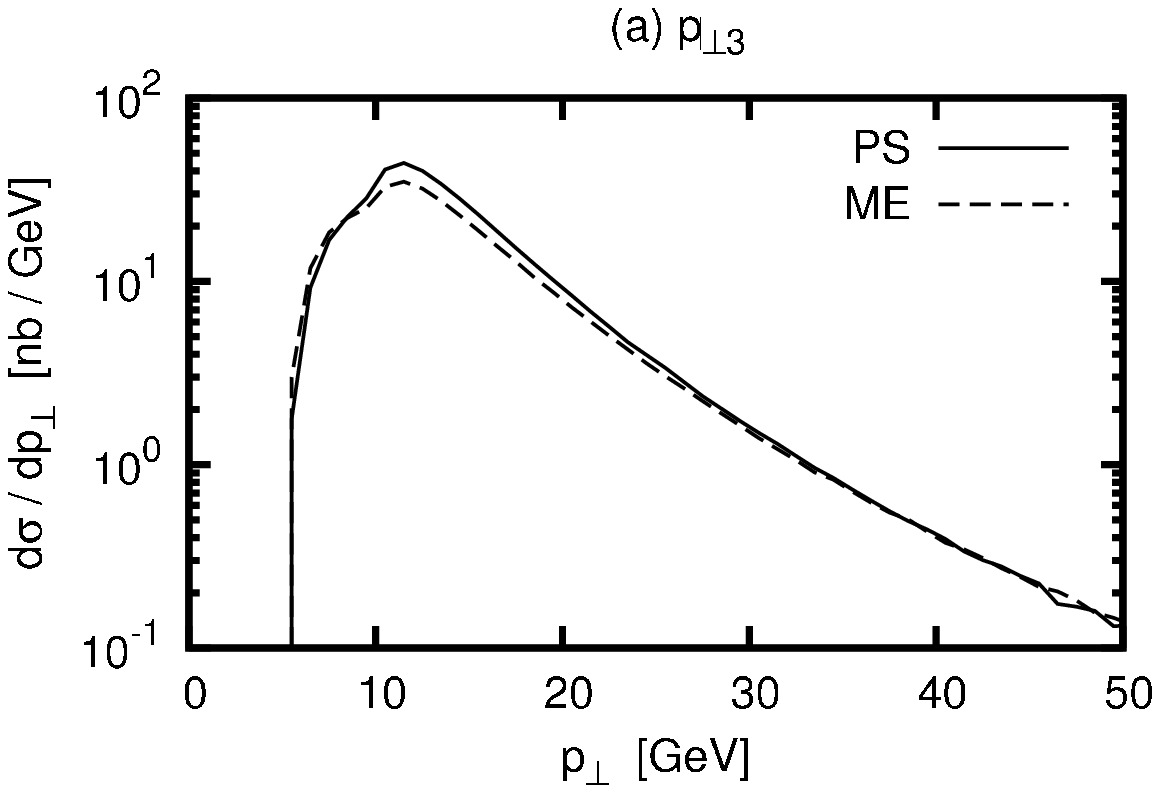}\hspace{-2mm}
\includegraphics[scale=0.43,angle=270]{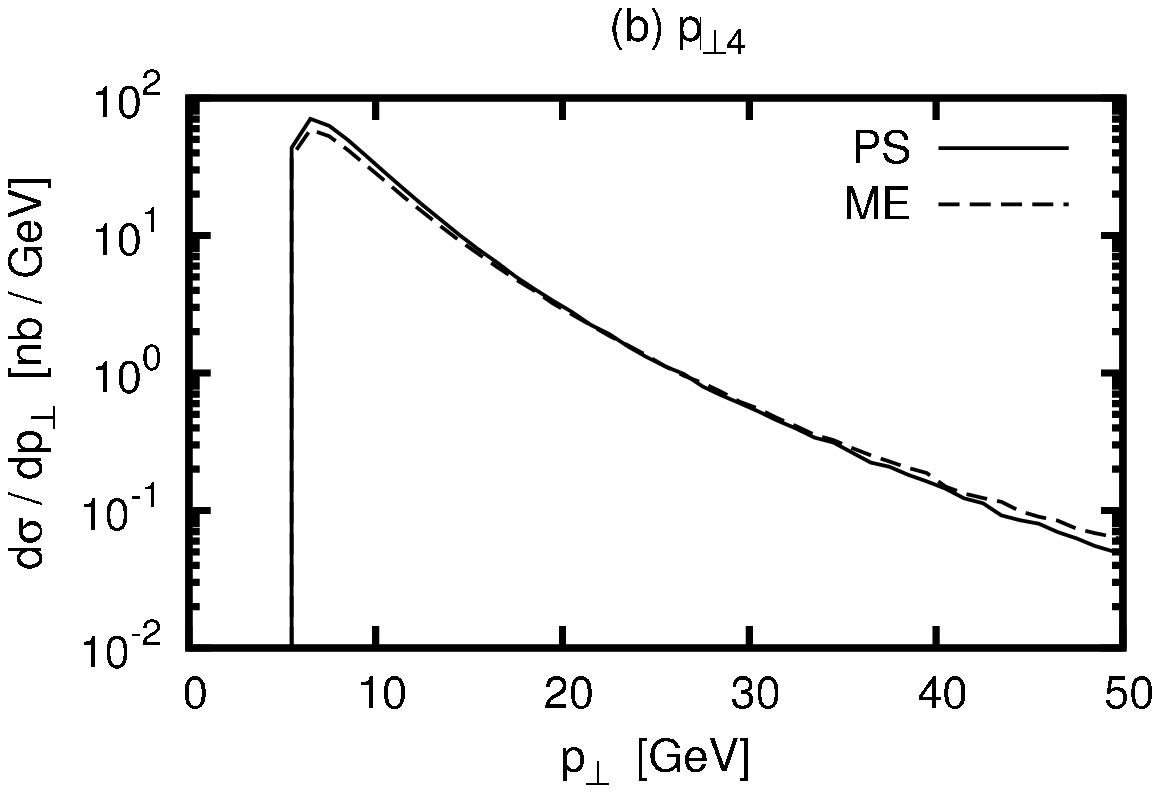}\hspace{-2mm}
\includegraphics[scale=0.43,angle=270]{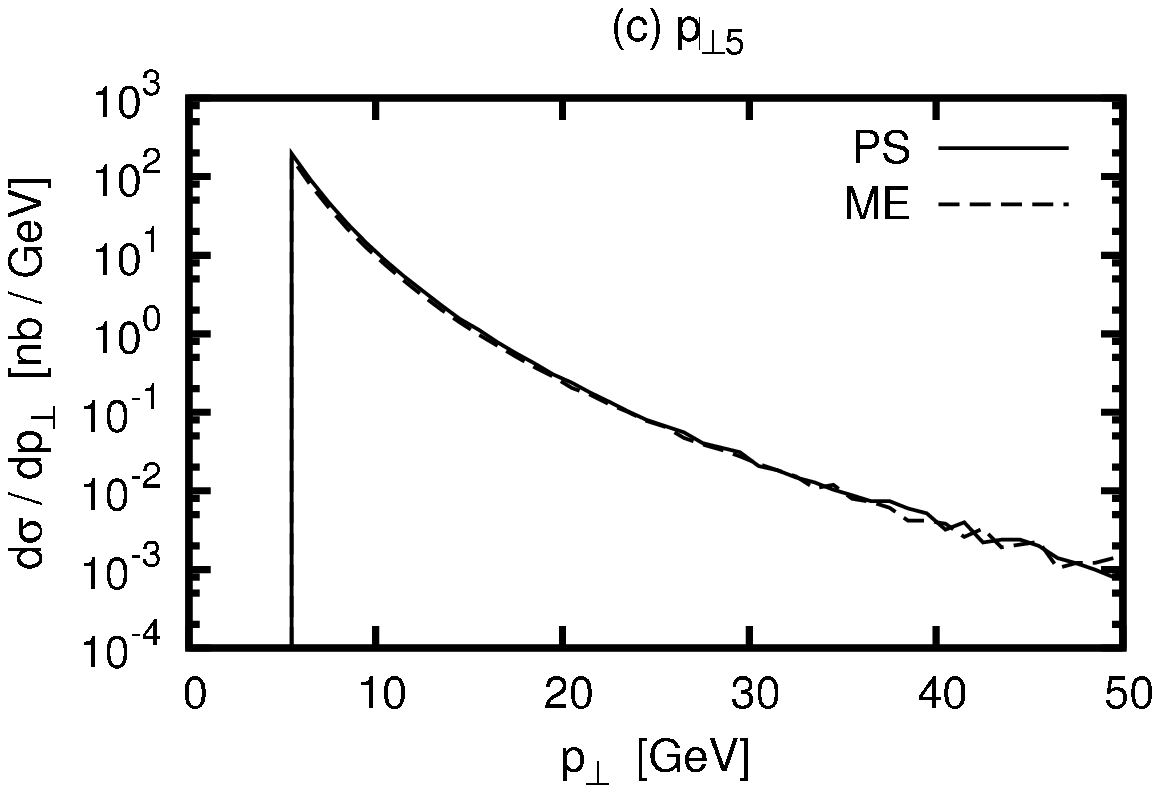}\\
\includegraphics[scale=0.43,angle=270]{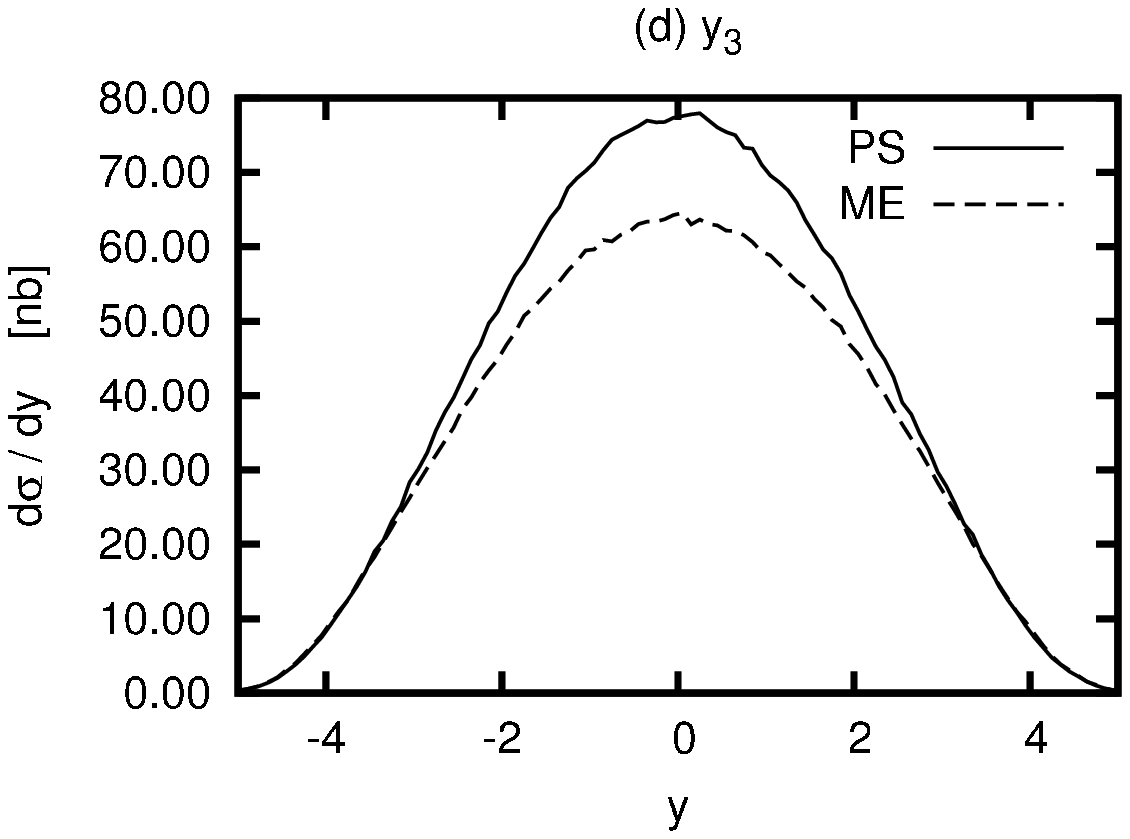}\hspace{-2mm}
\includegraphics[scale=0.43,angle=270]{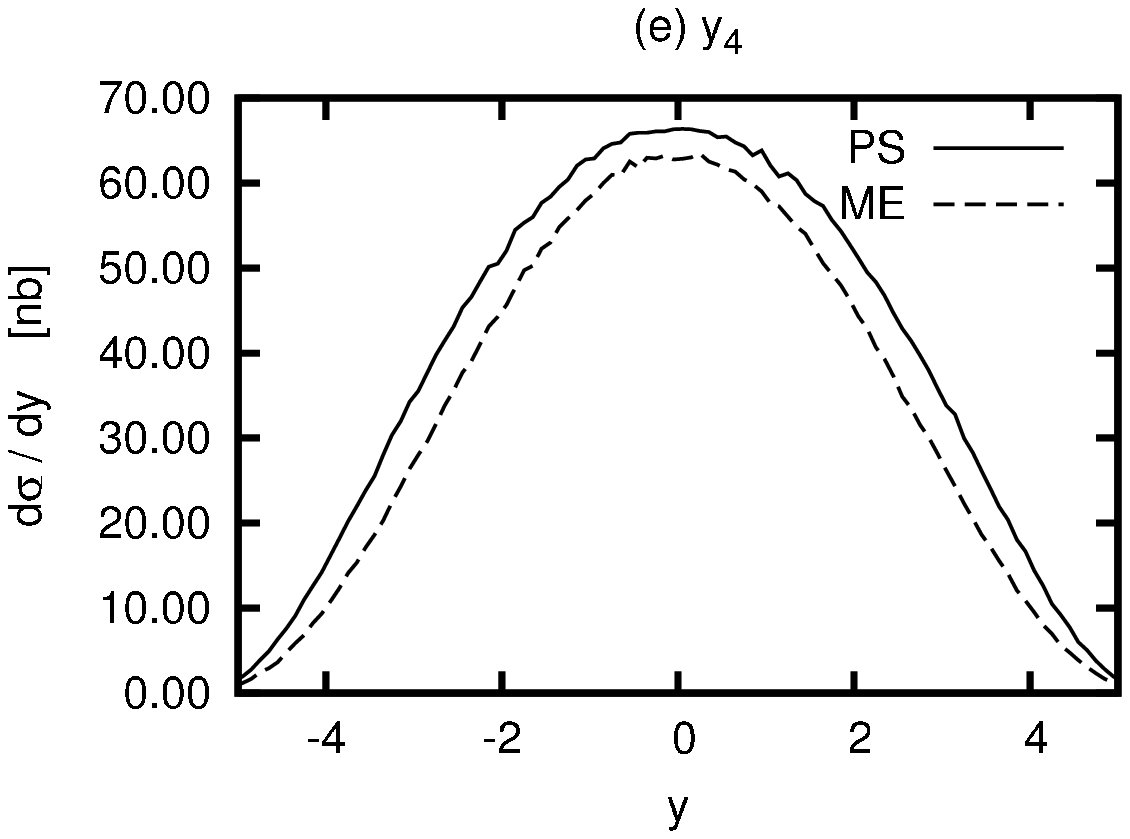}\hspace{-2mm}
\includegraphics[scale=0.43,angle=270]{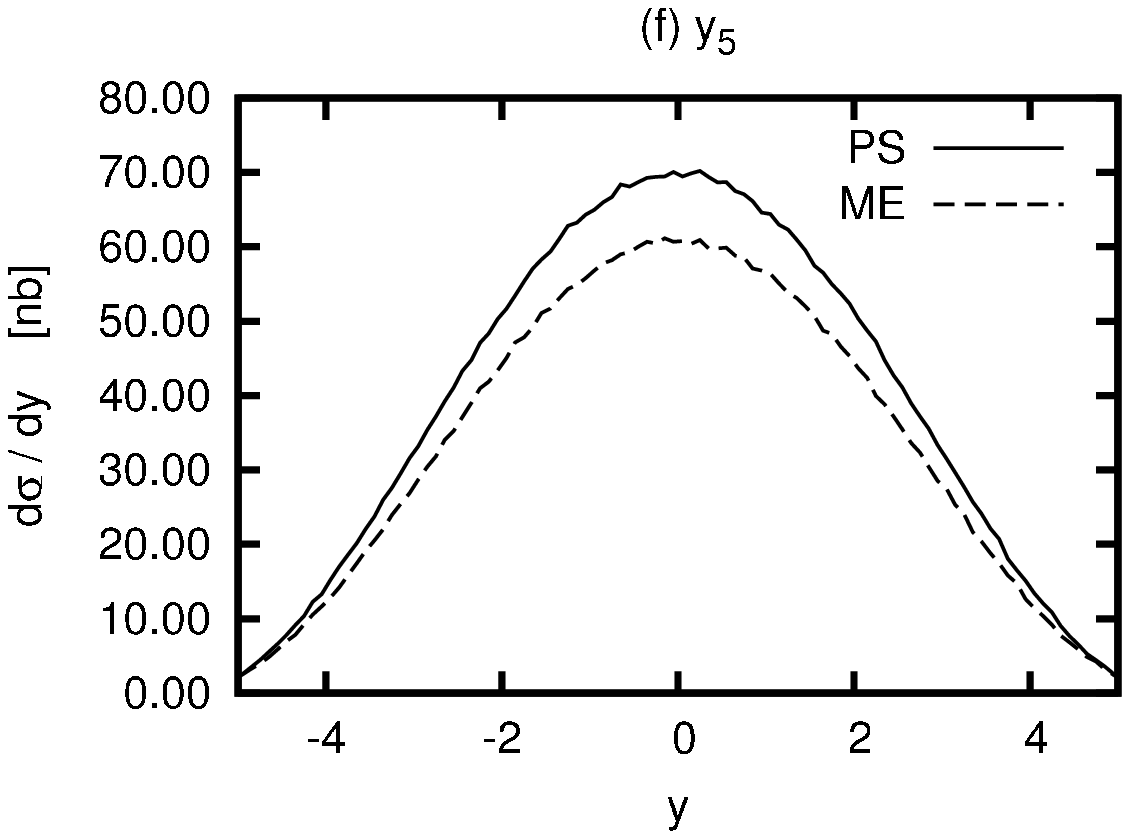}\\
\includegraphics[scale=0.43,angle=270]{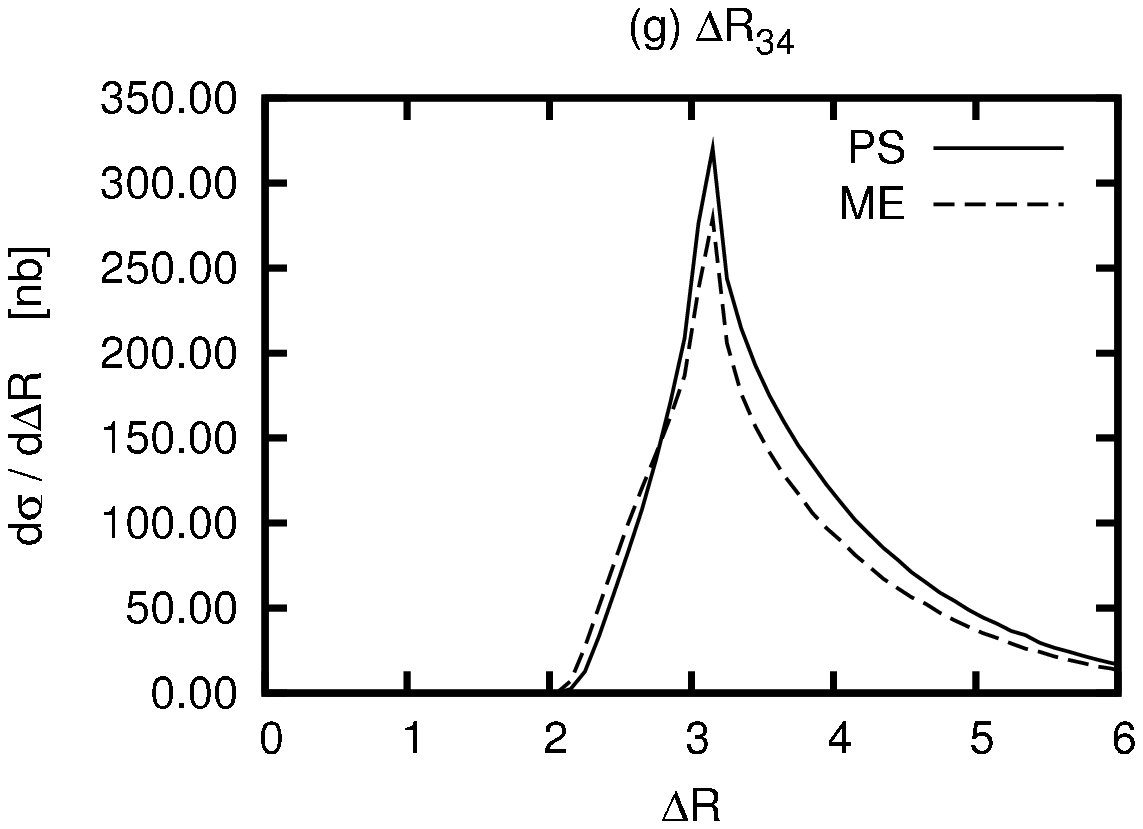}\hspace{-2mm}
\includegraphics[scale=0.43,angle=270]{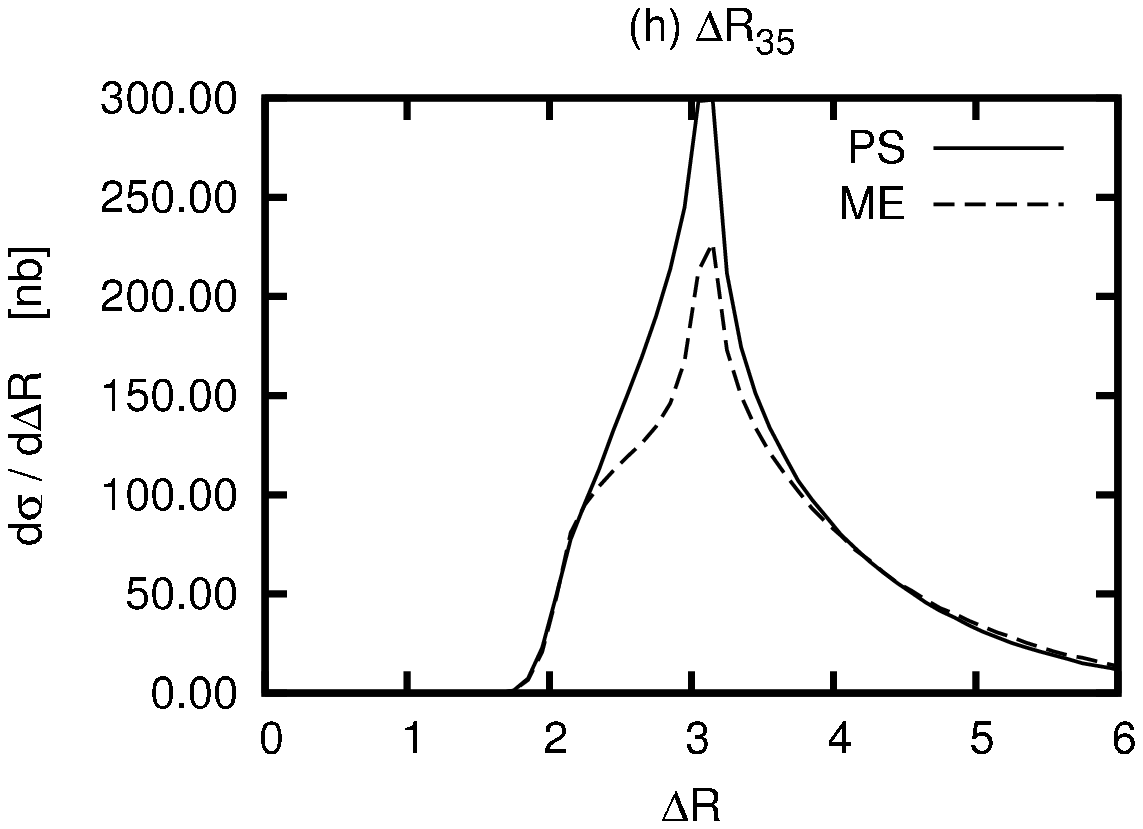}\hspace{-2mm}
\includegraphics[scale=0.43,angle=270]{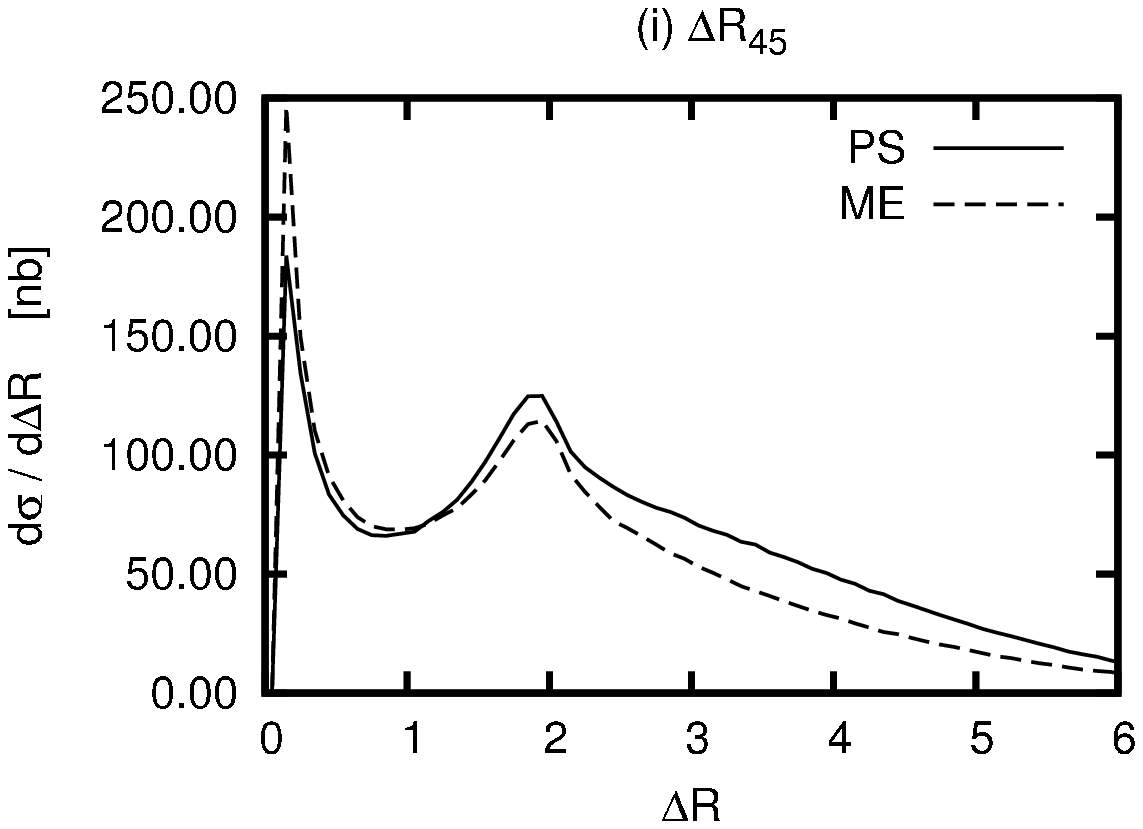}
\caption{Kinematic distributions for cut set (1) at Tevatron energies
($\p\pbar$, $\sqrt{s} = 1.96\TeV$): ${\pT}_3^\mmin = 5.0\GeV$,
${\pT}_5^\mmin = 5.0\GeV$, $R_{\mrm{sep}}^{\mmin} = 0.1$
\label{fig:2to3.cut1}}
\end{figure}

In Fig.~\ref{fig:2to3.cut1}, the results for the first set of cuts
are shown at Tevatron energies. This cut set is designed to be as inclusive
as possible, representative of the MPI that are important for MB and UE
physics. The distributions show a mixture of the features of the
second and third cut sets as shown previously. In general, both the rates
and shapes are well reproduced, a sign of the dominance of the
soft and collinear regions of phase space.

\begin{figure}
\centering
\includegraphics[scale=0.43,angle=270]{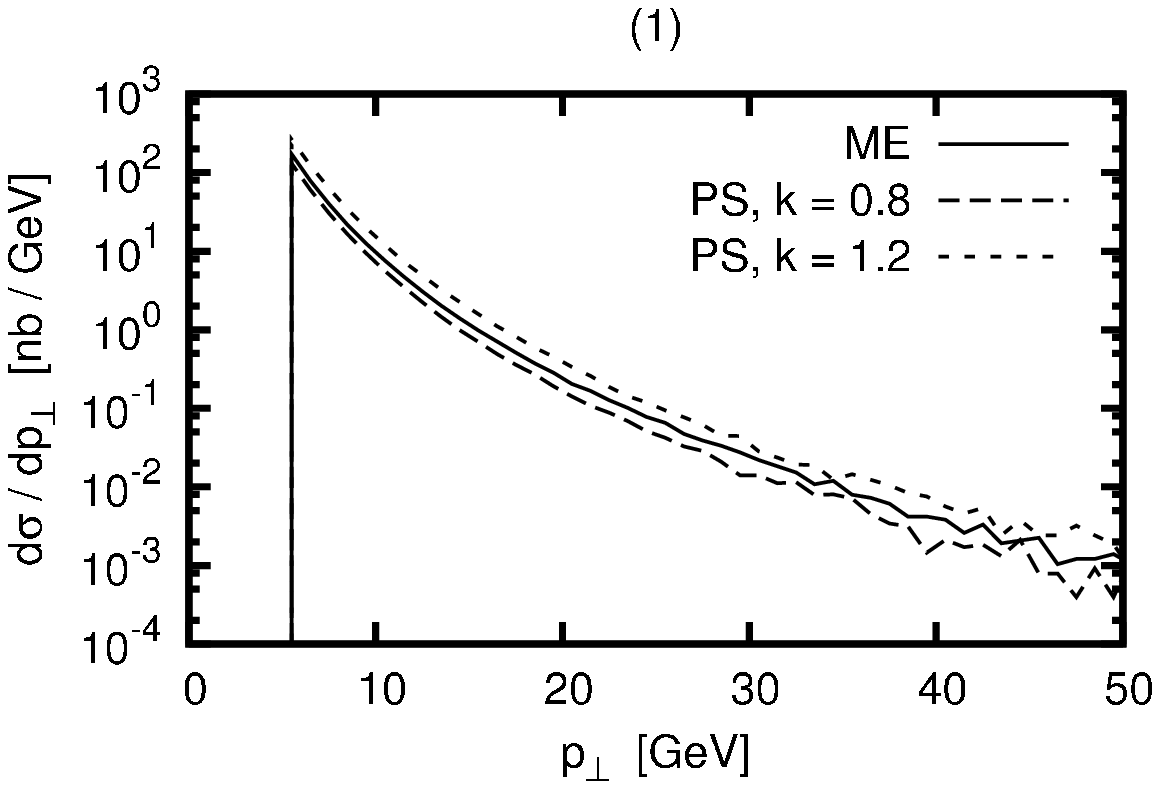}\hspace{-2mm}
\includegraphics[scale=0.43,angle=270]{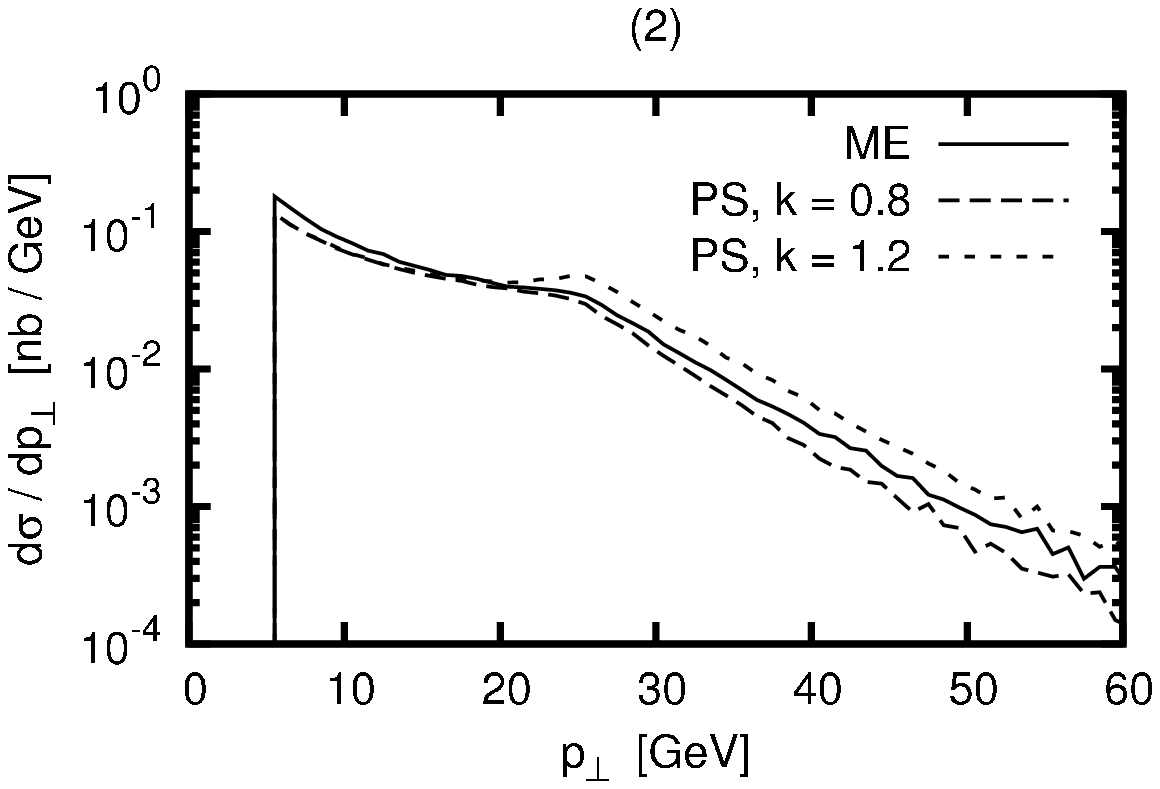}\hspace{-2mm}
\includegraphics[scale=0.43,angle=270]{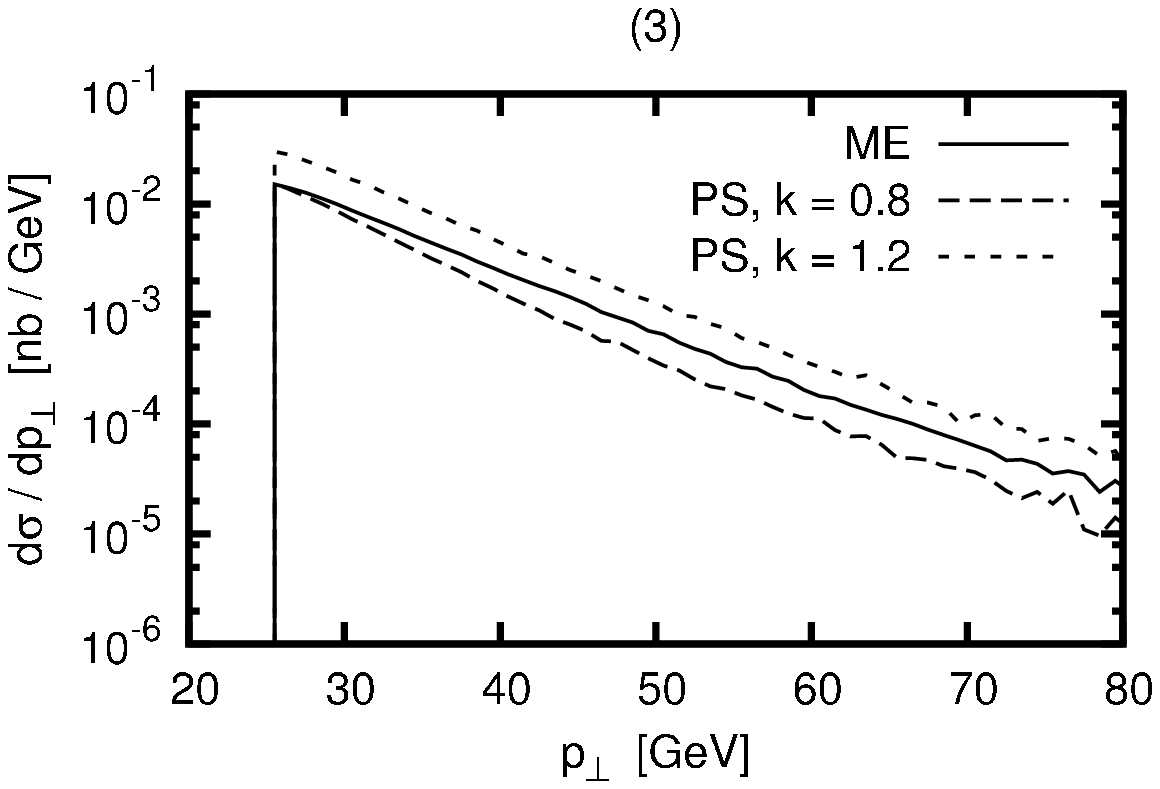}
\caption{${\pT}_5$ for each set of cuts at Tevatron energies ($\p\pbar$,
$\sqrt{s} = 1.96\TeV$) with $k = 0.8$ and $k = 1.2$
\label{fig:2to3.scale}}
\end{figure}

Some freedom to probe the ambiguities of the shower starting scale 
is offered by the possibility of introducing a constant factor, $k$, such that
$\pTmax = k~*~Q_{\mrm{fac}}$. This constant may be set separately for
the initial- and final-state showers, and in all results shown so far has
been set to $k = 1$. In Fig.~\ref{fig:2to3.scale}, the $\pT$ of the softest
jet is shown for each set of cuts at Tevatron energies when $k = 0.8$ and
$k = 1.2$, both for the initial- and final-state shower. Changes in $k$
give variations both in rate and shape, and the agreement with the ME
result is dependent on both the exact cuts used and the energy.
Although scale variations away from $k = 1$ may give better agreement in
certain regions of phase space and at certain energies, there is no
indication that a systematic shift, higher or lower, would improve the
overall PS description.

\begin{figure}
\centering
\includegraphics[scale=0.75]{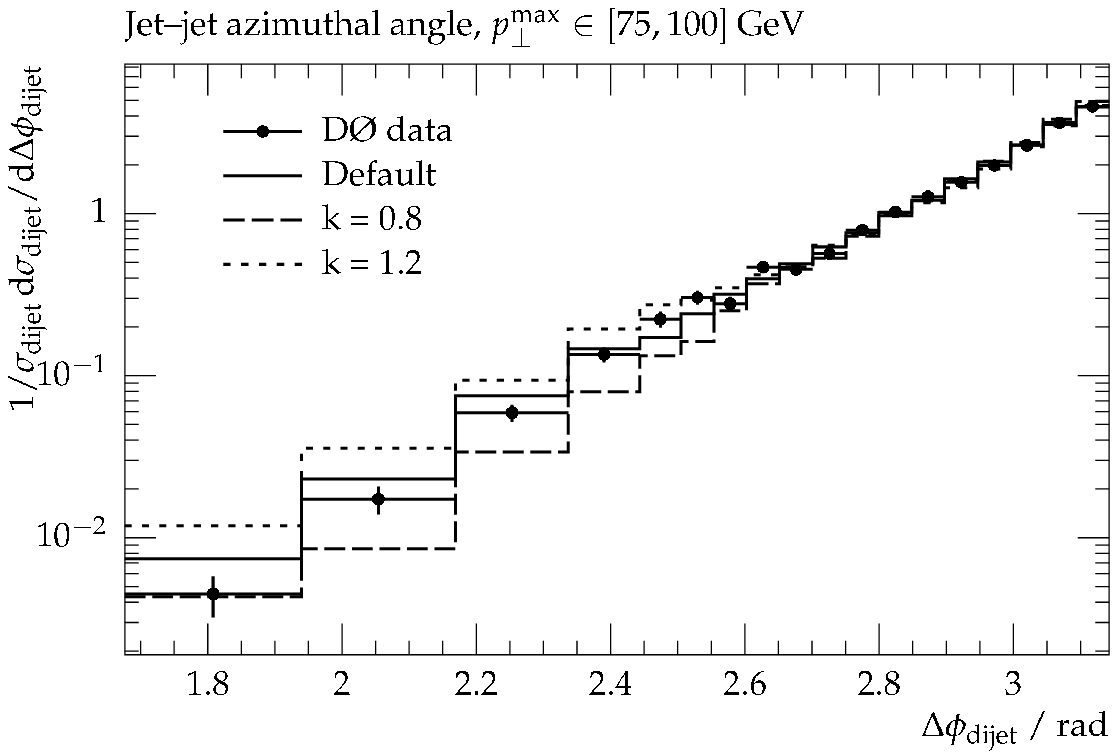}
\caption{Jet-jet azimuthal-angle distribution showing the changes when the
shower starting scale is adjusted by a factor $k = 0.8$ and $k = 1.2$
\label{fig:azi-scale}}
\end{figure}

Finally, as an example of the impact of these variations on final
observables, the jet-jet azimuthal-angle distribution, as studied by D0, is
shown in Fig.~\ref{fig:azi-scale}. Other than the different factors of $k$,
all other settings remain at their default value. As with many other
observables, many different parameters in the overall framework will play a
role, and these must then be adjusted to give the best description of
the data. The shower $k$ factors could be varied in the context of a
more complete tuning to data, at the cost of an increase in the size of
the parameter space, but based on the evidence above we choose to keep $k =
1$ in all that follows.
 
\section{Tunes to Tevatron data}
\label{sec:tuning}

In this section, having ``validated'' the now-modified PS framework in the
previous section, we study the tuning prospects of \tsc{Pythia 8} to
Tevatron data.  All changes to the framework outlined previously apply
solely to the case of hadron collisions, so the existing LEP tune of the
final-state shower and hadronisation can remain unchanged.  However, even
the subset of parameters relating only to hadron collisions is large, so
the strategy is to pick a minimal set of key parameters, and try to find a
region in this parameter space that describes key pieces of Tevatron data.
One key goal is to show that a combined MB/UE tuning of Tevatron data is
within reach for \textsc{Pythia 8}, as is possible in previous versions. It
should be stressed that this is not designed to be a ``complete'' tuning of
the generator. The search of this limited parameter space is guided by the
principles we outline below, but there is still no guarantee that other
regions of parameter space do not give a similar or better descriptions of
the data. We hope that what we present here offers a good starting point
for a more global Professor-type tuning, encompassing a wider set of data
and parameters, as well as offering a general guide to those getting
involved with such efforts.

We also take this opportunity to move away from the CTEQ5L PDF
set. As explained in Sec.~\ref{sec:pdfs}, a selection of newer sets is
now available directly in \textsc{Pythia 8}, and we pick one standard LO
set (CTEQ6L1) and one modified set (MRST LO**) with which to begin the
tuning attempts.

The following Rivet analyses have been used in the tuning process:
\begin{description}
  \item[CDF\_2000\_S4155203: Z pT measurement in Run I $\Z \to e^+e^-$
events \cite{Affolder:1999jh}] \hfill \\
$\pT$ and cross section measurement of $e^+e^-$ pairs in the 
region of $66 < m_{ee} < 116\GeV$.

  \item[CDF\_2001\_S4751469: Field \& Stuart Run I underlying event
analysis \cite{Affolder:2001xt}] \hfill \\
The direction of the leading charged particle jet in an event is used to
define three regions in $\eta-\varphi$ space. With $\Delta \varphi =
\varphi - \varphi_{\mrm{LeadingJet}}$, the toward region is defined by
$|\Delta\varphi| < 60^{\circ}$ (centered on the leading jet), the away
region by $|\Delta\varphi| > 120^{\circ}$ and the transverse region by
$60^{\circ} < |\Delta\varphi| < 120^{\circ}$. All regions are constrained
such that $|\eta| < 1$. Two datasets are used; one using a minimum bias
trigger, the other using a ``JET20'' trigger, where one calorimeter tower
cluster must have $E_{T} > 20\GeV$. Results are given as a function of the
transverse momentum of the leading jet, $\p_{\perp}^{\mrm{lead}}$.

  \item[CDF\_2002\_S4796047: Run I charged multiplicity measurement
\cite{Acosta:2001rm}] \hfill \\
Charged multiplicity measurements at $\sqrt{s} = 630~\&~1800\GeV$. The
$\langle \pT \rangle (N_{\mrm{ch}})$ measurements have largely been
superseded by CDF\_2009\_S8233977, below.

  \item[D0\_2004\_S5992206: Run II jet azimuthal decorrelation analysis
\cite{Abazov:2004hm}] \hfill \\
Distributions of $\Delta \varphi = \varphi_1 - \varphi_2$, where
$\varphi_1$ is the azimuthal angle of the hardest jet and $\varphi_2$ that of the
second hardest jet. Results are given in 4 different bins of ${\pT}_1$.

  \item[CDF\_2008\_LEADINGJETS: Run II underlying event in leading jet
events \cite{CDFLEADINGJETS}]
Regions are defined as in CDF\_2001\_S4751469, but with
the two transverse regions additionally separated into ``transMIN'' and
``transMAX'' regions, based on which contains the lowest/highest
number/$\sum \pT$ density, event-by-event. A ``transDIF'' region is defined
by the difference between the transMAX and transMIN regions.

  \item[CDF\_2008\_NOTE\_9351: Run II underlying event in Drell-Yan
\cite{CDFNOTE9351}] \hfill \\
As CDF\_2008\_LEADINGJETS, but using a leading $\Z^0$ reconstructed from
a Drell-Yan lepton pair.

  \item[CDF\_2009\_S8233977: Run II min bias cross-section analysis
\cite{Aaltonen:2009ne}] \hfill \\
Measurements of track $\pT$, $\langle \pT \rangle (N_{\mrm{ch}})$ and $\sum
E_{\mrm{T}}$ for $\sqrt{s} = 1.96\TeV$.
\end{description}

Our limited parameter space consists of the following:
\begin{description}
  \item[\begin{minipage}{\linewidth}\raggedright
\texttt{SigmaProcess:alphaSvalue}
\end{minipage}] \hfill \\
$\alphas(M_\Z)$ for the hard process. Directly affects the single-particle
$\pT$ spectrum in the low-$\pT$ region and jet cross sections at high
$\pT$. For MRST LO**, a decrease of $\sim 15\%$ is expected due to
similarly-sized increase in the PDF content (the QCD $2 \to 2$ cross
section is quadratic both in PDFs and in $\alphas$).

  \item[\begin{minipage}{\linewidth}\raggedright
\texttt{SpaceShower:alphaSvalue}
\end{minipage}] \hfill \\
$\alphas(M_\Z)$ for the initial-state shower. Together with the
regularisation parameters (see below), gives control over the amount of ISR
activity. Although convenient to be set equal to $\alphas$ of the hard
process, above, this is not a requirement.

  \item[\begin{minipage}{\linewidth}\raggedright
\texttt{SpaceShower:pT0Ref}, \texttt{SpaceShower:ecmRef},
\texttt{SpaceShower:ecmPow}, \texttt{SpaceShower:samePTasMI},
\texttt{SpaceShower:pTmin}
\end{minipage}] \hfill \\
Parameters used to regularise the $\pT \to 0$ divergence in the
initial-state shower (Sec.~\ref{sec:spaceshower}).
While the smooth dampening of eq.~(\ref{eq:turnoffISR}) is physically
preferable to a step-function regularisation, as provided by
\texttt{SpaceShower:pTmin}, there is currently no real constraint on these
options. For now, we choose to fix the ${\pT}_0^{\mrm{ref}}$ to an
energy independent value, rather than use the same running as used in the
MI framework. Together with $\alphas(M_\Z)$ above, these values are tuned
mainly based on the $\pT(\Z^0)$ spectrum (although this also contains large
contributions from primordial $\kT$) and the D0 jet azimuthal decorrelation
analysis. While a change of $\pTo$ will give more activity only at small
$\pT$'s, a change to $\alphas$ will affect the whole spectrum, although
still not evenly due to its running.

  \item[\begin{minipage}{\linewidth}\raggedright
\texttt{SpaceShower:rapidityOrder}
\end{minipage}] \hfill \\
If switched on, rapidity-unordered ISR emissions, subsequent to the first
emission off each dipole, are vetoed (Sec.~\ref{sec:spaceshower}).
Studies suggest that having this option activated helps to dampen the rise
of the underlying event, leading to a better agreement with data.

  \item[\begin{minipage}{\linewidth}\raggedright
\texttt{MultipleInteractions:alphaSvalue}
\end{minipage}] \hfill \\
$\alphas(M_\Z)$ for the MPI framework. This can be used to help control the
amount of MPI activity, in conjunction with the regularisation parameters
(see below). In the context of the MPI model there is no
strong reason to vary this independently of
\texttt{SigmaProcess:alphaSvalue}, and we choose to restrict them to be
equal or very close to each other. As with ISR, a change to $\alphas$ will
affect the whole spectrum of MPI as opposed to just the soft region. This
balance can have effects on e.g. $\langle \pT \rangle (N_{\mrm{ch}})$
distributions, which in turn depends heavily on colour reconnection
effects.

  \item[\begin{minipage}{\linewidth}\raggedright
\texttt{MultipleInteractions:pT0Ref},
\texttt{MultipleInteractions:ecmRef},
\texttt{MultipleInteractions:ecmPow}
\end{minipage}] \hfill \\
Parameters for the regularisation of the MPI framework
(Sec.~\ref{sec:MI}). $\pTo$ is a key ingredient in getting both the charged
multiplicity and the underlying-event distributions correct. As we only
consider Tevatron data here, the energy dependence is only constrained by
the CDF charged multiplicity at $\sqrt{s} = 630\GeV$. The reference energy
is left unchanged at $1800\GeV$.

  \item[\begin{minipage}{\linewidth}\raggedright
\texttt{MultipleInteractions:bProfile}, \texttt{MultipleInteractions:expPow}
\end{minipage}] \hfill \\
Parameters used to select the matter profile used in the MPI framework 
(Sec.~\ref{sec:MI}). As noted previously, the introduction of showers off
all scattering subsystems reduces the need for a double Gaussian profile;
instead we choose the overlap function, eq.~(\ref{eq:exppow}), giving one
degree of freedom, $E^{\mrm{pow}}_{\exp}$. This parameter is
important in the same distributions as $\pTo$ above; for charged
multiplicities it affects high-multiplicity tail, while it also controls
how quickly the underlying-event activity grows as a function of the
transverse momentum of the leading jet in MB events.
A value $E^{\mrm{pow}}_{\exp} = 1$ gives results
somewhat similar to the default double-Gaussian matter profile, giving more
fluctuations out in the high-multiplicity tail and a faster rise of the
underlying event, while $E^{\mrm{pow}}_{\exp} = 2$ reduces back to the
single Gaussian case. We examine values between these two limits
to try to find the best balance.

  \item[\begin{minipage}{\linewidth}\raggedright
\texttt{BeamRemnants:reconnectRange}
\end{minipage}] \hfill \\
The $R$ parameter of eq.~(\ref{eq:crec}). This is primarily tuned to the
latest measurement of $\langle \pT \rangle (N_{\mrm{ch}})$ from Run II
Tevatron data, but should also give reasonable results for the lower energy
runs. It should be noted that this parameter is non-linear;
at $R \sim 5$, the amount of reconnection has almost completely saturated.
\end{description}

Not included in the above list are those parameters related
to primordial $\kT$ (Sec.~\ref{sec:pkt}). For the hard component, early
studies indicated that the $\pT(\Z^0)$ peak was well described with its default
setting ($\sigma_{\mrm{hard}} = 2\GeV$), while for the soft component,
variations showed little sensitivity in the mean $\pT$ per parton. In
Tab.~\ref{tab:tevtune}, the parameters for two tunes, 2C (CTEQ6L1) and 2M
(MRST LO**) are given. Any parameter not shown is set to the default value
of \tsc{Pythia 8.142}.

\renewcommand{\arraystretch}{1.15}
\begin{table}
\begin{center}
\begin{tabular}{|l|c|c|c|}\hline
\textbf{Parameter} & \textbf{Tune 2C} &
\textbf{Tune 2M}   & \textbf{Tune 4C} \\ \hline
\texttt{SigmaProcess:alphaSvalue}         & 0.135 & 0.1265 & 0.135 \\
\texttt{SpaceShower:rapidityOrder}        & on    & on     & on    \\
\texttt{SpaceShower:alphaSvalue}          & 0.137 & 0.130  & 0.137 \\
\texttt{SpaceShower:pT0Ref}               & 2.0   & 2.0    & 2.0   \\
\texttt{MultipleInteractions:alphaSvalue} & 0.135 & 0.127  & 0.135 \\
\texttt{MultipleInteractions:pT0Ref}      & 2.320 & 2.455  & 2.085 \\
\texttt{MultipleInteractions:ecmPow}      & 0.21  & 0.26   & 0.19  \\
\texttt{MultipleInteractions:bProfile}    & 3     & 3      & 3     \\
\texttt{MultipleInteractions:expPow}      & 1.60  & 1.15   & 2.00  \\
\texttt{BeamRemnants:reconnectRange}      & 3.0   & 3.0    & 1.5   \\
\texttt{SigmaDiffractive:dampen}          & off   & off    & on    \\
\texttt{SigmaDiffractive:maxXB}           & N/A   & N/A    & 65    \\
\texttt{SigmaDiffractive:maxAX}           & N/A   & N/A    & 65    \\
\texttt{SigmaDiffractive:maxXX}           & N/A   & N/A    & 65    \\ \hline
\end{tabular}
\end{center}
\caption{Parameters for Tunes 2C, 2M and 4C (described in
Sec.~\ref{sec:LHC}). Any parameter not shown is set to the default value of
\tsc{Pythia 8.142}
\label{tab:tevtune}}
\end{table}

\begin{figure}
\begin{minipage}{\textwidth}
\centerline{
\includegraphics[scale=0.6]{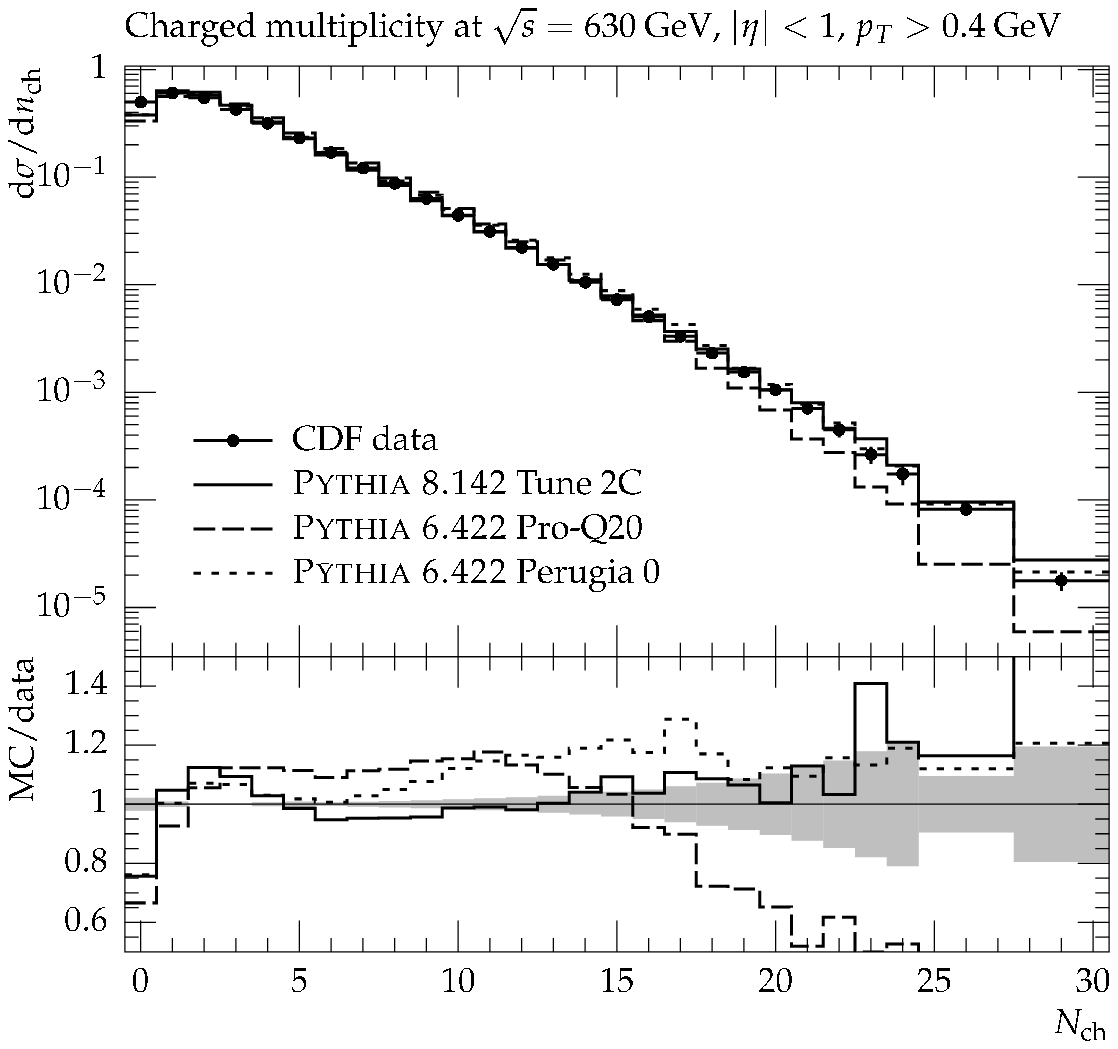}\hspace{7.5mm}
\includegraphics[scale=0.6]{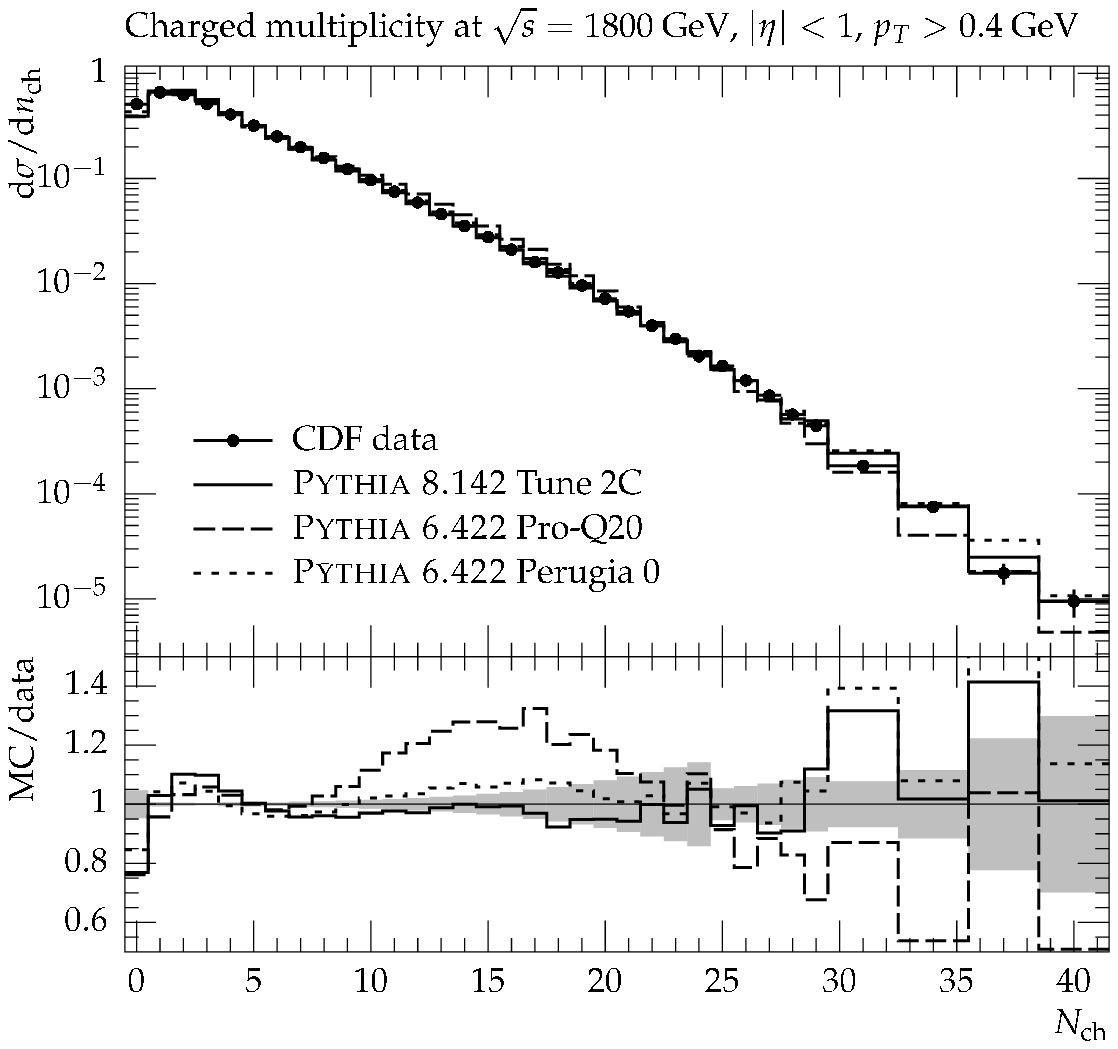}
}
\vspace{5mm}
\end{minipage}
\begin{minipage}{\textwidth}
\centerline{
\includegraphics[scale=0.6]{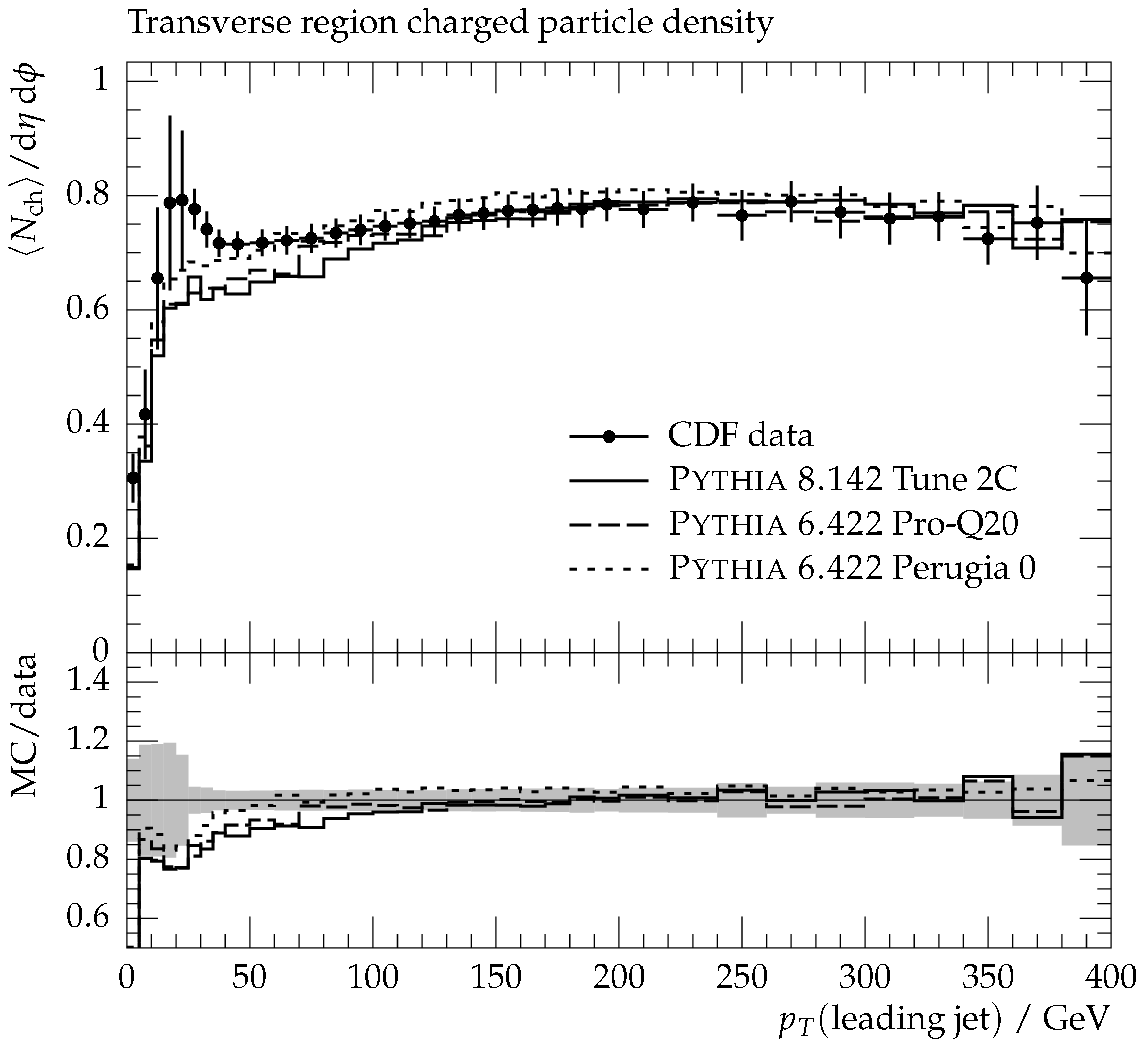}\hspace{7.5mm}
\includegraphics[scale=0.6]{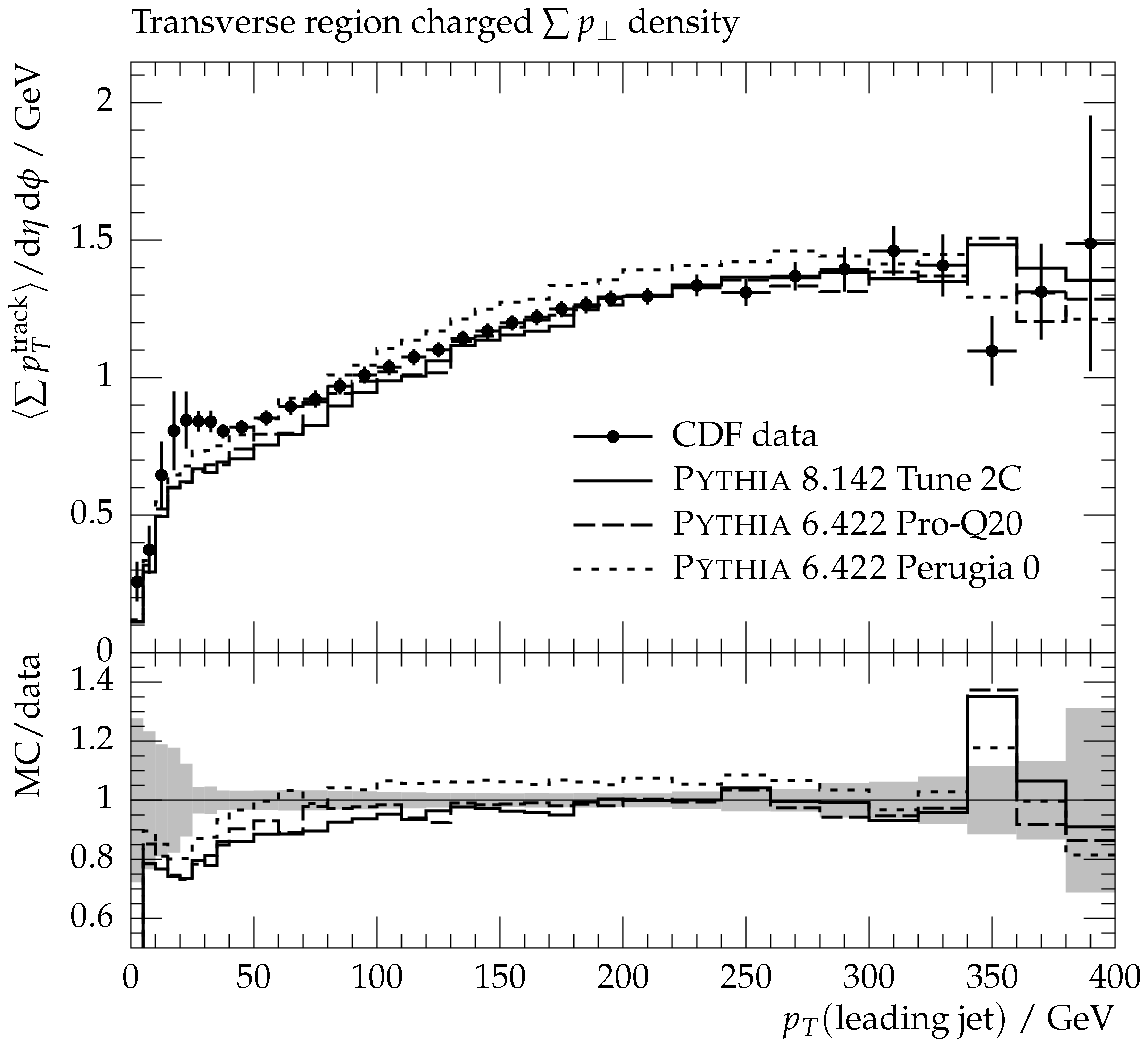}
}
\vspace{5mm}
\end{minipage}
\begin{minipage}{\textwidth}
\centerline{
\includegraphics[scale=0.6]{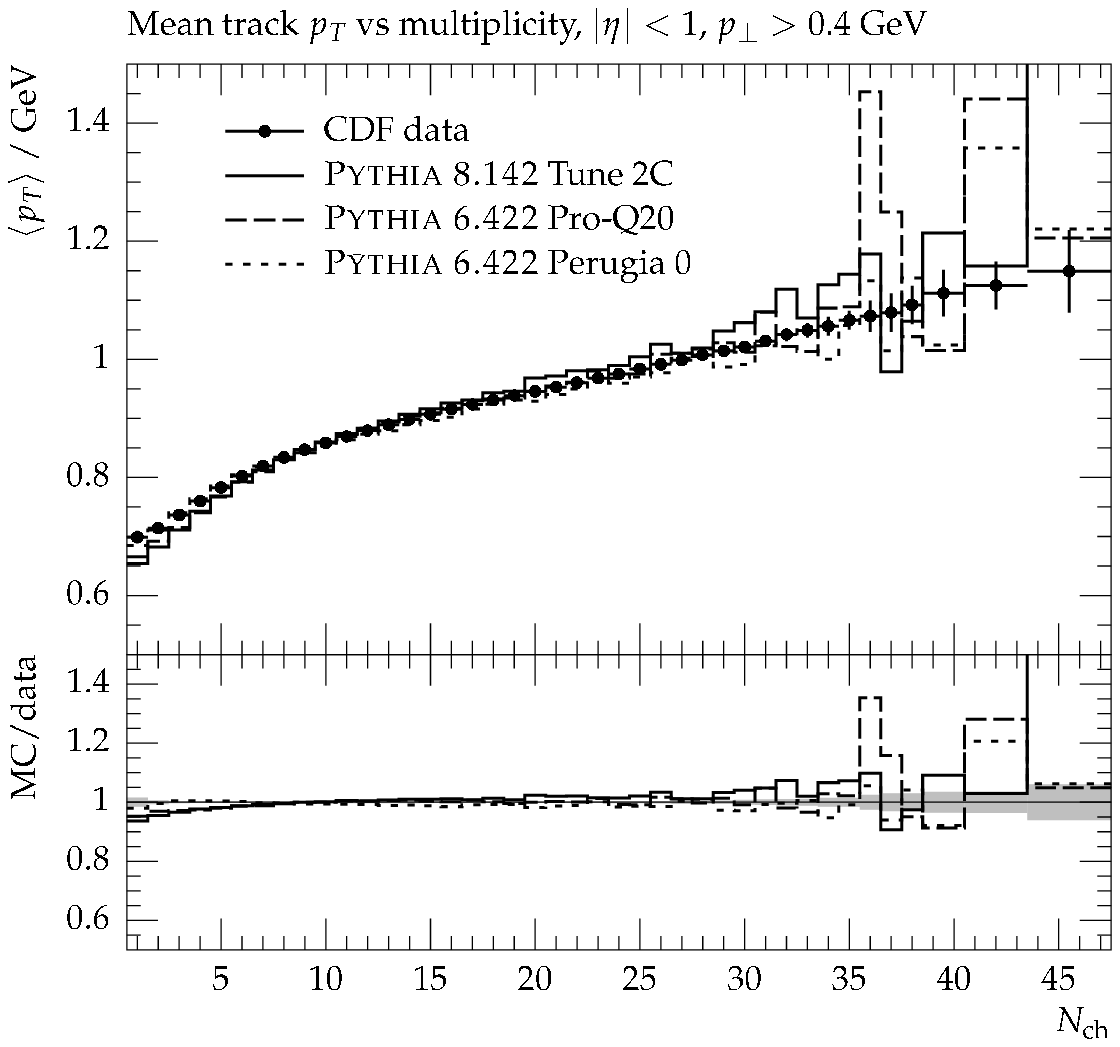}\hspace{7.5mm}
\includegraphics[scale=0.6]{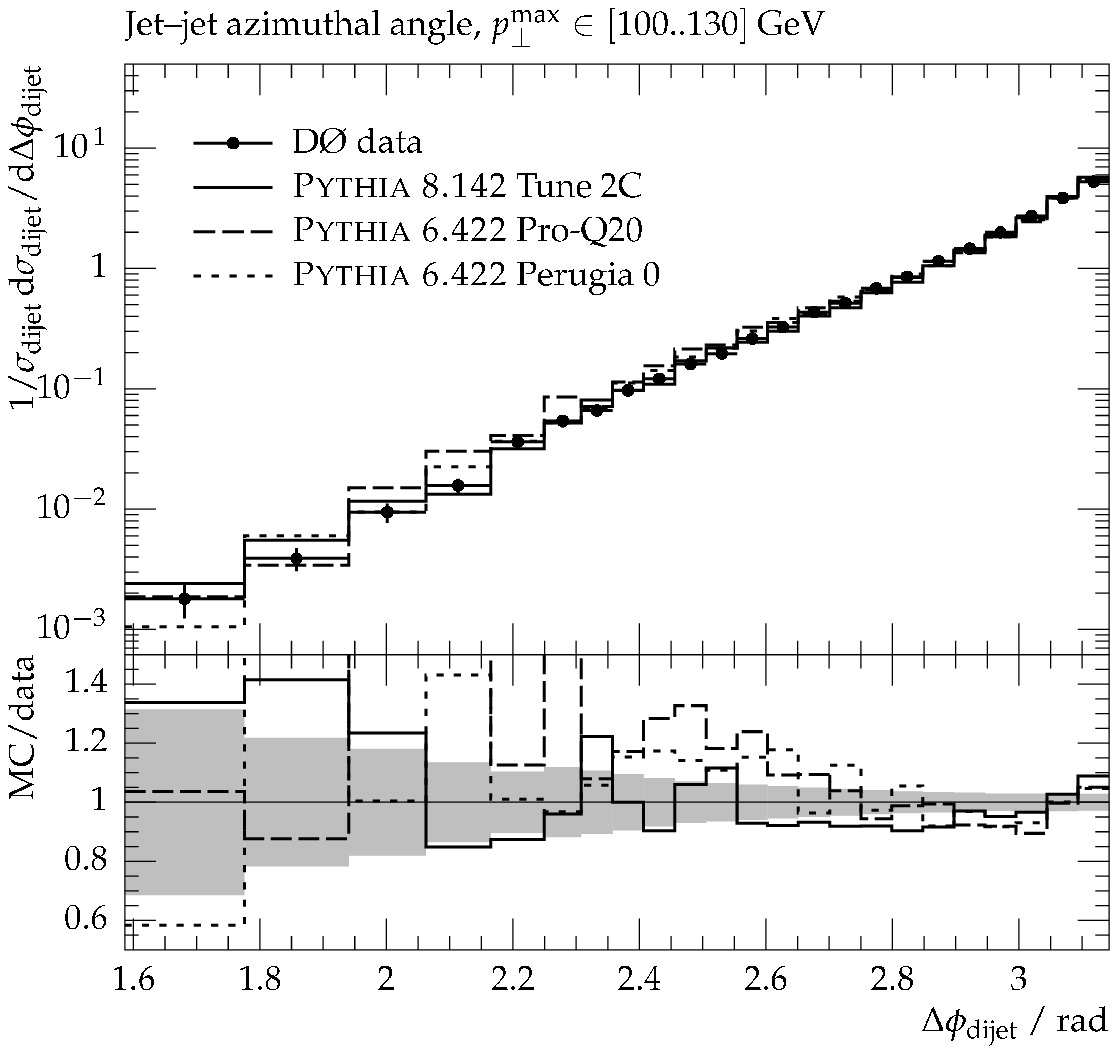}
}
\end{minipage}
\caption{Selected Rivet plots for Tune 2C, compared against data, and tunes
Pro-Q20 and Perugia 0
\label{fig:Tune2C}}
\end{figure}

\begin{figure}
\begin{minipage}{\textwidth}
\centerline{
\includegraphics[scale=0.6]{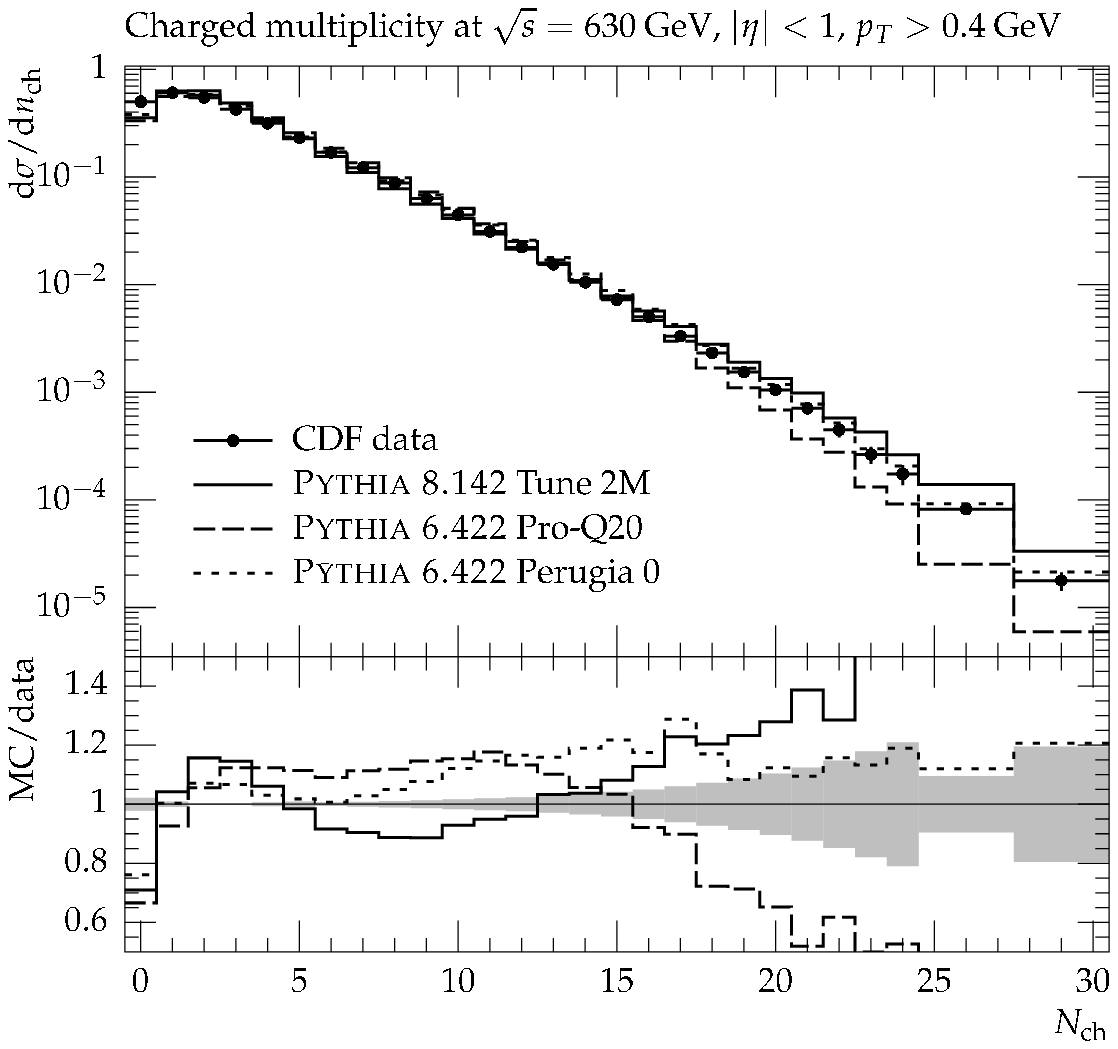}\hspace{7.5mm}
\includegraphics[scale=0.6]{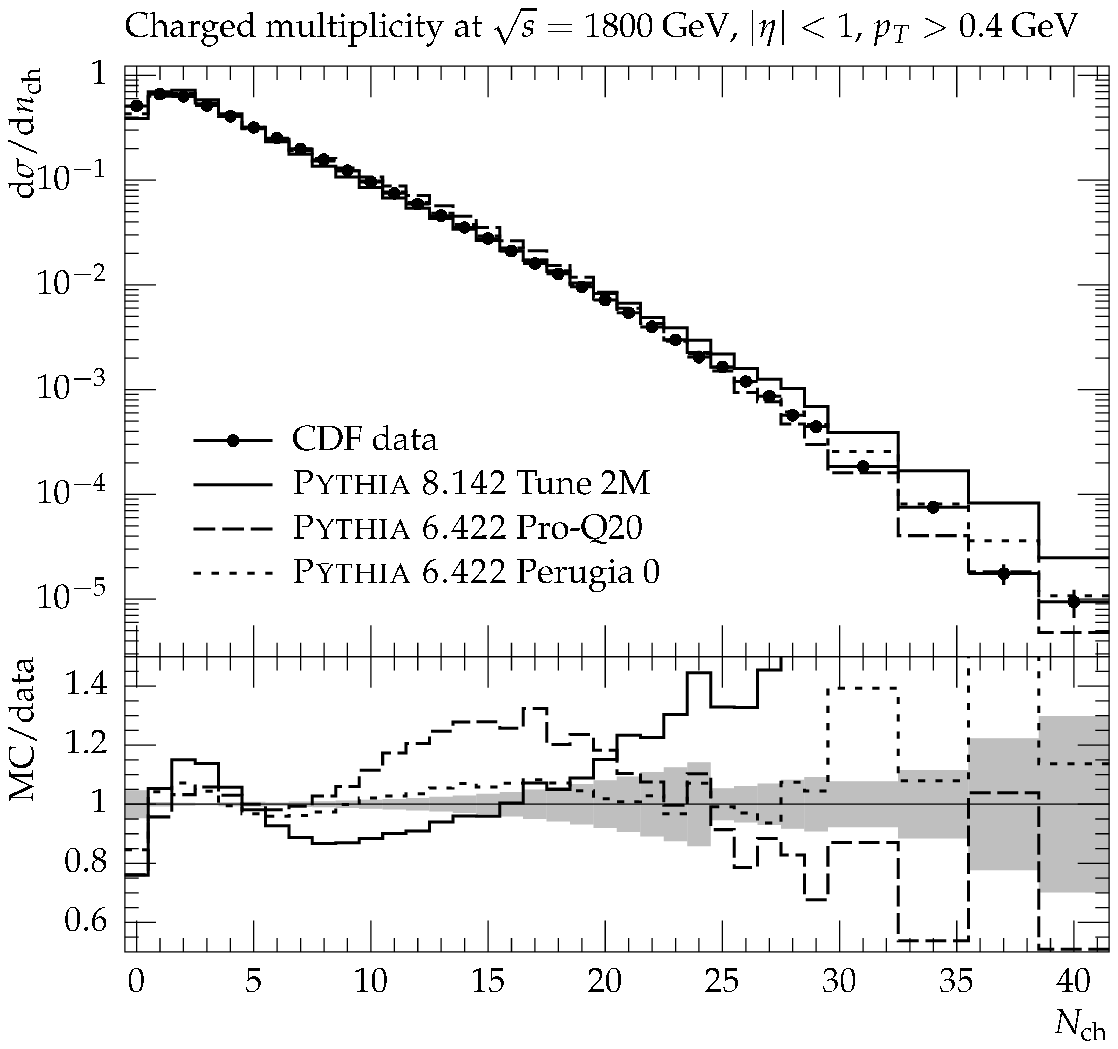}
}
\vspace{5mm}
\end{minipage}
\begin{minipage}{\textwidth}
\centerline{
\includegraphics[scale=0.6]{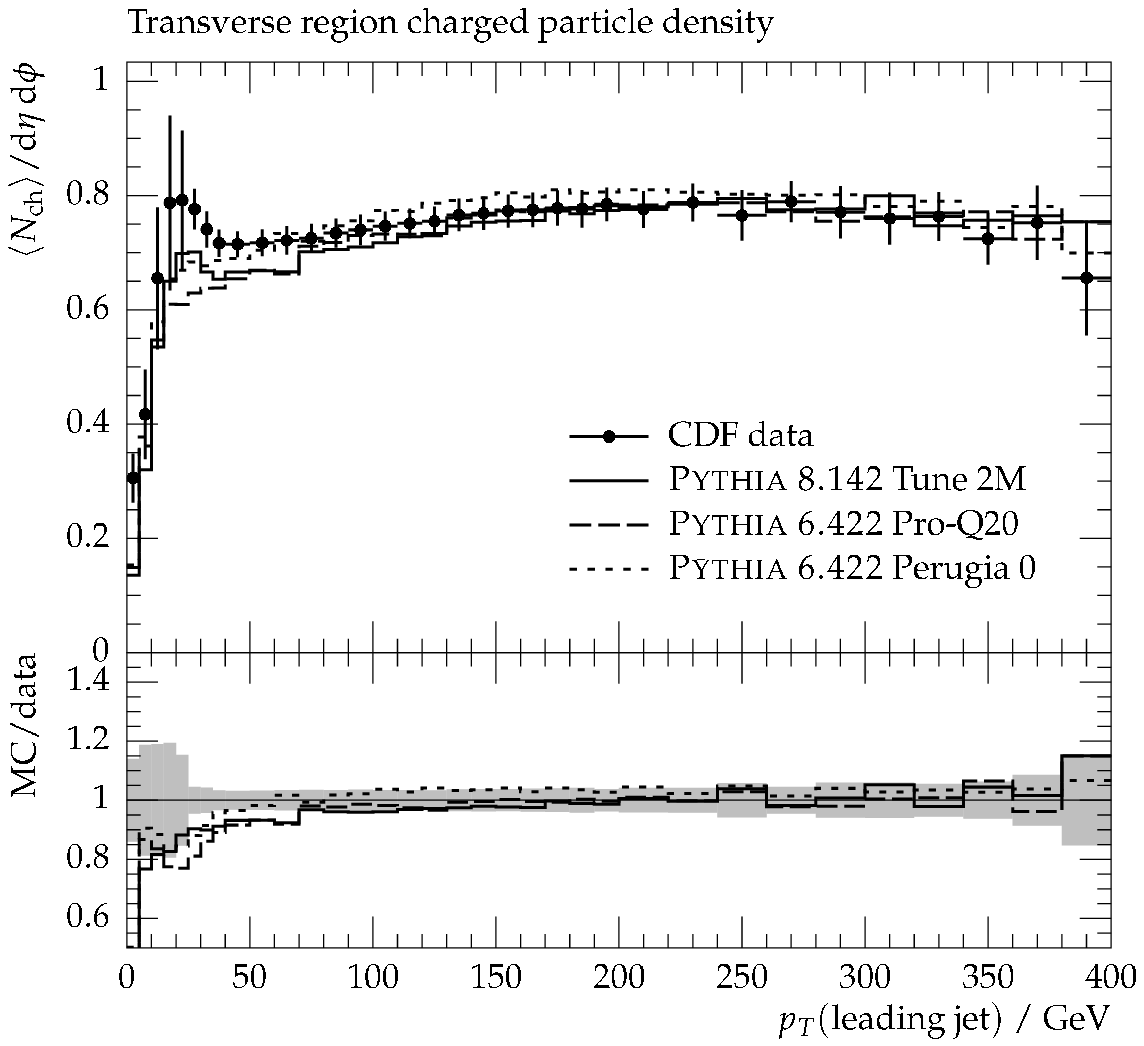}\hspace{7.5mm}
\includegraphics[scale=0.6]{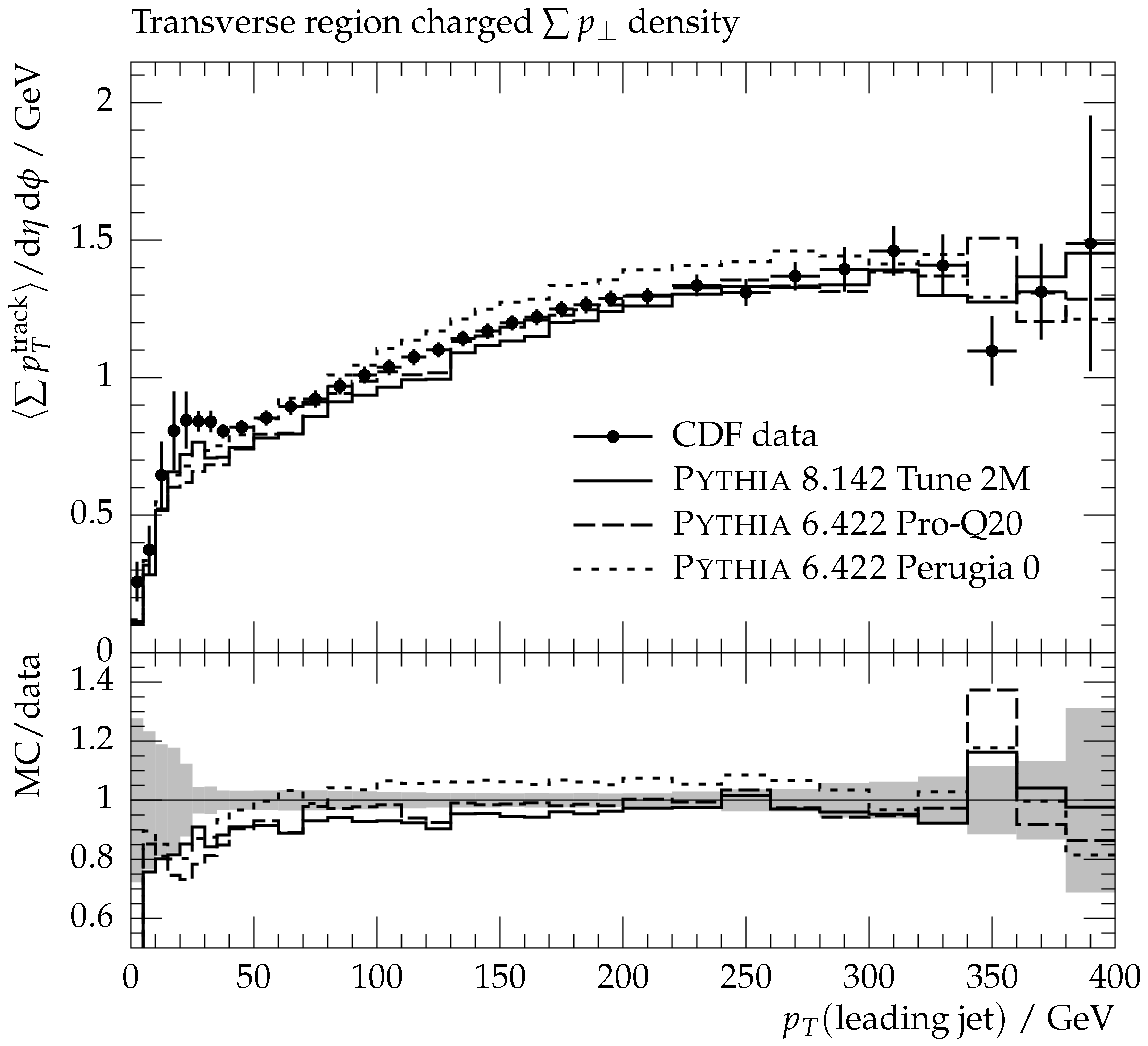}
}
\vspace{5mm}
\end{minipage}
\begin{minipage}{\textwidth}
\centerline{
\includegraphics[scale=0.6]{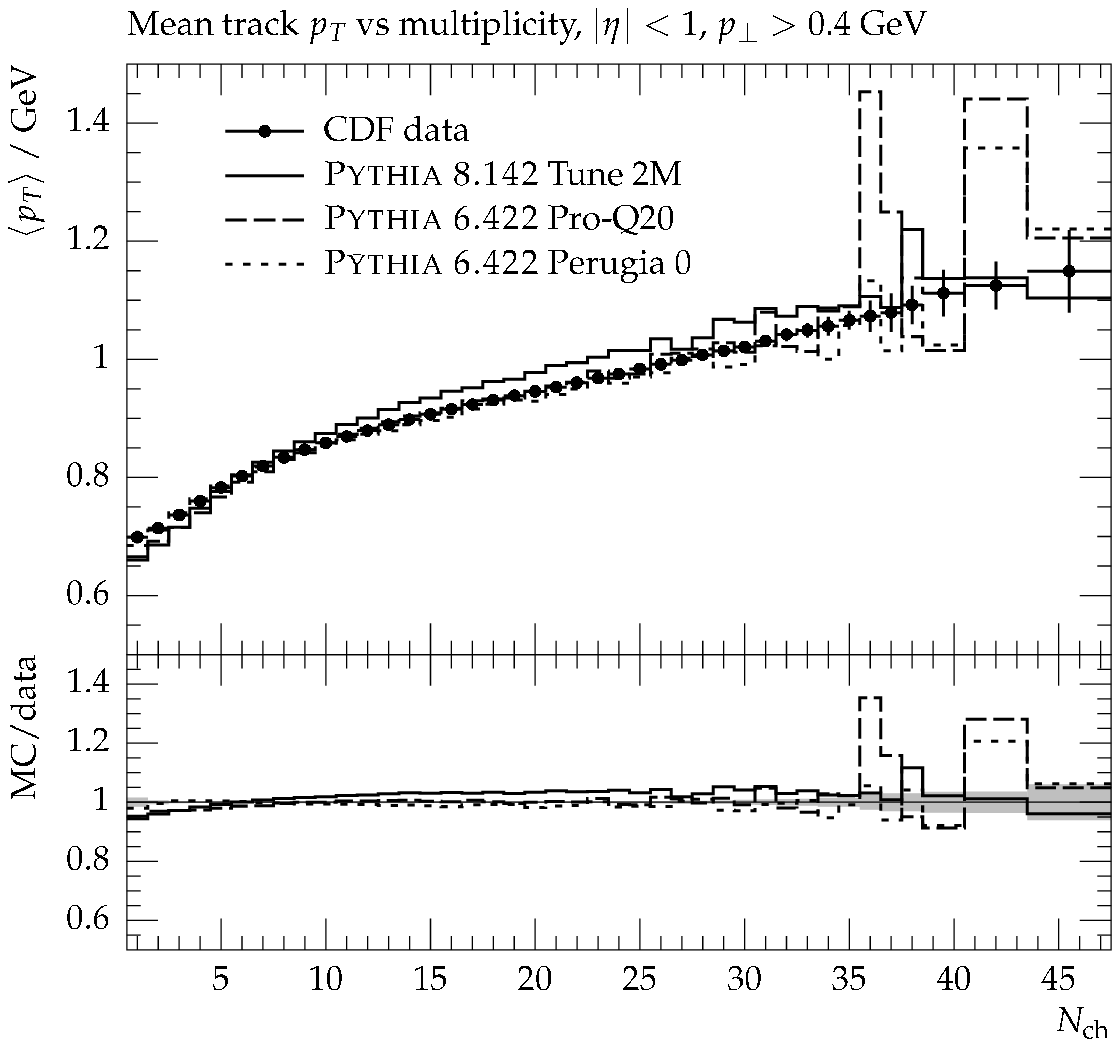}\hspace{7.5mm}
\includegraphics[scale=0.6]{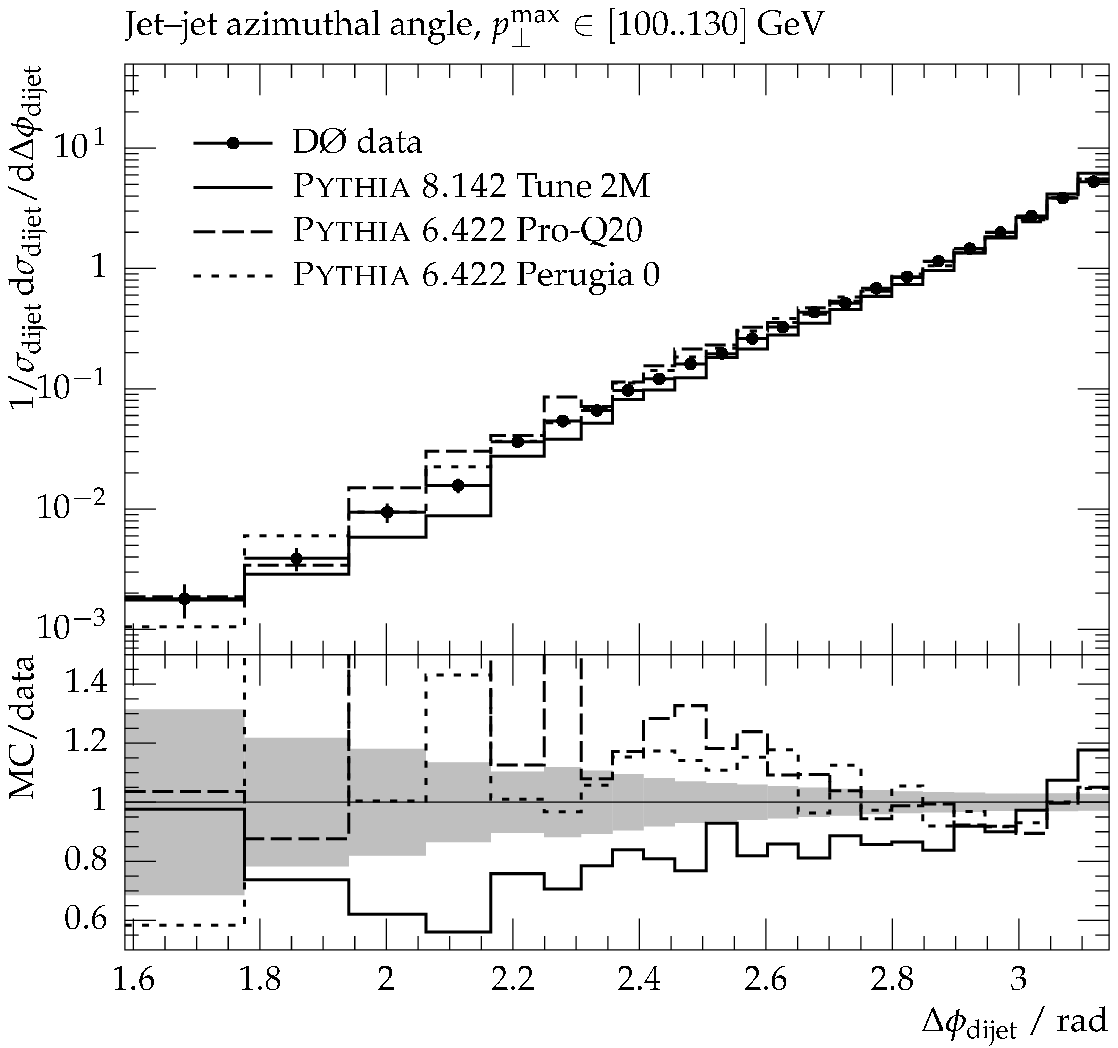}
}
\end{minipage}
\caption{Selected Rivet plots for Tune 2M, compared against data, and tunes
Pro-Q20 and Perugia 0
\label{fig:Tune2M}}
\end{figure}

Selected plots are shown in Figs.~\ref{fig:Tune2C}~\&~\ref{fig:Tune2M} for
tunes 2C and 2M respectively. The tunes are compared against data and
\tsc{Pythia 6} tunes Pro-Q20 and Perugia 0 for (left-to-right,
top-to-bottom):
\begin{Itemize}
\item[1)] CDF\_2002\_S4796047: Charged multiplicity at $\sqrt{s} = 630\GeV$
\item[2)] CDF\_2002\_S4796047: Charged multiplicity at $\sqrt{s} = 1800\GeV$
\item[3)] CDF\_2008\_LEADINGJETS: Transverse charged particle density 
\item[4)] CDF\_2008\_LEADINGJETS: Transverse charged $\sum \pT$ density 
\item[5)] CDF\_2009\_S8233977: $\langle \pT \rangle (N_{\mrm{ch}})$
\item[6)] D0\_2004\_S5992206: Jet-jet azimuthal angle for
          $100 < \pT^{\mmax} < 130\GeV$
\end{Itemize}
It is notable that less colour reconnection is needed to match the $\langle
\pT \rangle (N_{\mrm{ch}})$ data than in the old default. One contribution
to this is the increased $\alphas$ in the MPI framework, giving more
activity at all $\pT$ values rather than just extra low-$\pT$ activity from
a change in $\pTo$. The good agreement in the D0 jet-jet azimuthal
distribution gives a good sign that the balance between contributions from
ISR and MPI is well described, supplemented by $\pT(\Z^0)$ measurements,
not shown.

There is a large difference in the $E^{\mrm{pow}}_{\exp}$ parameter between
the two tunes. In fact, in Tune 2M, the multiplicities at both
$630~\&~1800\GeV$ do overshoot in the tails; the low value of
$E^{\mrm{pow}}_{\exp}$ is required to correctly describe the underlying
event in the region ${\pT}_{\mrm{lead}} < 200\GeV$. One key reason is that
the the MRST PDF set combined with the retuned $\alphas$ differs in both
shape and normalisation relative to CTEQ6L1, and when probing the amount of
activity as a function of the leading jet, these differences become more
apparent. A value $E^{\mrm{pow}}_{\exp} \sim 1.4$ would give a better
description of the overall multiplicity distributions, but with this
setting, the average activity in this ${\pT}_{\mrm{lead}}$ region is lower
than for CTEQ6L1. In Tune 2M, we have compensated for this by allowing more
impact-parameter fluctuations out to the high multiplicity tails, but
expect regions in parameter space to exist which better address this
balance.

For both tunes, the transverse activity in the underlying event study no
longer shows the large rise noted in Fig.~\ref{fig:intro-rivet}. Both tunes
also show success in a combined MB/UE description at the Tevatron.
Over all datasets, Tunes 2C and 2M never do significantly worse
than Pro-Q20 and Perugia 0, and it is hoped that a full tuning, considering
a larger parameter set, could improve agreement further.

\section{First LHC data}
\label{sec:LHC}

\subsection{Diffractive cross sections}

For the LHC data, we begin by studying the diffractive cross section and
the simple scheme to dampen its growth outlined in Sec.~\ref{sec:sigmatot}.
On the experimental side, we turn to a recent ATLAS study designed to
enhance the diffractive content of minimum bias events \cite{AtlasDiff}.
The study relies on the Minimum Bias Trigger Scintillators (MBTS), covering
two rings in pseudorapidity, $2.09 < |\eta| < 2.82$ and $2.82 < |\eta| <
3.84$. By selecting events with a hit on only one side of the MBTS,
diffractive events are preferentially chosen. While only total cross
sections are examined here, the ATLAS study also shows that track distributions
are better described with the new high-mass diffractive framework of
Sec.~\ref{sec:diffraction}. It is noted that this study
is not corrected for detector/reconstruction effects, but that such
corrections are not expected to change the conclusions of the study.

The particular quantity of interest is
$R_{\mrm{ss}} = N_{\mrm{ss}} / N_{\mrm{any}}$, where $N_{\mrm{ss}}$ is the
number of events with a hit on exactly one side of the MBTS and
$N_{\mrm{any}}$ the number with a hit on either side. We follow the
approach of the ATLAS study, in using fixed acceptance values for the
different event classes while varying the contributions of the different
diffractive modes.

\renewcommand{\arraystretch}{1.15}
\begin{table}
\begin{center}
\begin{tabular}{|l|c|c|c|c|c|c|}
\cline{2-7}
\multicolumn{1}{c|}{} & \bf $\sqrt{s}$ (TeV) &
\bf $\mathbf{\sigma_{\mrm{ND}}}$ (mb) &
\bf $\mathbf{\sigma_{\mrm{SD}}}$ (mb) &
\bf $\mathbf{\sigma_{\mrm{DD}}}$ (mb) &
\bf $\mathbf{R_{\mrm{ss}}}$ (\%) &
\bf $\mathbf{\sigma_{\mrm{diff}} / \sigma_{\mrm{inel}}}$ (\%)
\\ \hline
\multirow{3}{*}{\bf Default} &
  0.90 ($\p\p$)    & 34.4 & 11.7 & 6.4 & 5.5 & 34.5 \\
& 1.96 ($\p\pbar$) & 39.0 & 12.5 & 7.5 & 5.4 & 33.8 \\
& 7.00 ($\p\p$)    & 48.5 & 13.7 & 9.3 & 5.1 & 32.1 \\ \hline
\multirow{3}{*}{\bf Damped} &
  0.90 ($\p\p$)    & 36.0 & 10.7 & 5.8 & 5.0 & 31.5 \\
& 1.96 ($\p\pbar$) & 40.9 & 11.4 & 6.7 & 4.8 & 30.1 \\
& 7.00 ($\p\p$)    & 50.9 & 12.4 & 8.1 & 4.5 & 28.7 \\ \hline
\end{tabular}
\end{center}
\caption{Non-, single- and double-diffractive cross sections at different
energies for default and damped ($\sigma_{\mrm{SD}}^{\mrm{max}} = 
\sigma_{\mrm{DD}}^{\mrm{max}} = 65~\mrm{mb}$) settings. The percentage of
same-side events (using the same acceptance values as the $7\TeV$ study)
and ratio of diffractive to inelastic cross sections are also given
\label{tab:sigmadamp}}
\end{table}

\begin{figure}
\centering
\includegraphics[scale=0.60,angle=270]{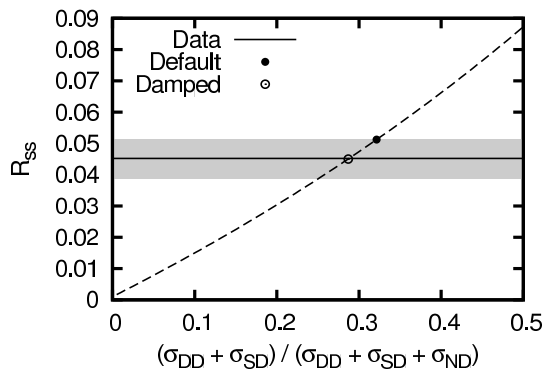}
\caption{$R_{\mrm{ss}}$ as a function of diffractive contribution to the
total inelastic cross section. The dashed curve gives the generator
prediction when the ratio is fixed at the default value, $\sigma_{\mrm{SD}}
/ \sigma_{\mrm{DD}} = 1.53$
\label{fig:diff1}}
\end{figure}

From data, a value
$R_{\mrm{ss}} = [4.52 \pm 0.02 (\mrm{stat}) \pm 0.61 (\mrm{syst})]\%$ is
preferred, while the default settings from \tsc{Pythia 8} gives a value
$R_{\mrm{ss}} = 5.11\%$, lying just inside the error range. Assuming that
the single- and double-diffractive components saturate at the same value, a
damping $\sigma_{\mrm{SD}}^{\mrm{max}} = \sigma_{\mrm{DD}}^{\mrm{max}} =
 65~\mrm{mb}$ brings $R_{\mrm{ss}}$ into closer agreement with the data.
With this value, in this simpleminded model, saturation would then still be
far away. Tab.~\ref{tab:sigmadamp} gives the non-, single- and
double-diffractive cross sections for different energies in both the
default and damped scenarios. $R_{\mrm{ss}}$ values are also provided, but
calculated with the same acceptance values as the $7\TeV$ ATLAS study, and
the diffractive to inelastic cross section ratio. In Fig.~\ref{fig:diff1},
$R_{\mrm{ss}}$ is shown as a function of the diffractive contribution to
the total inelastic cross section. In the damped scenario, the generator
$R_{\mrm{ss}}$ now sits upon the central data value.

\begin{figure}
\centering
\includegraphics[scale=0.60,angle=270]{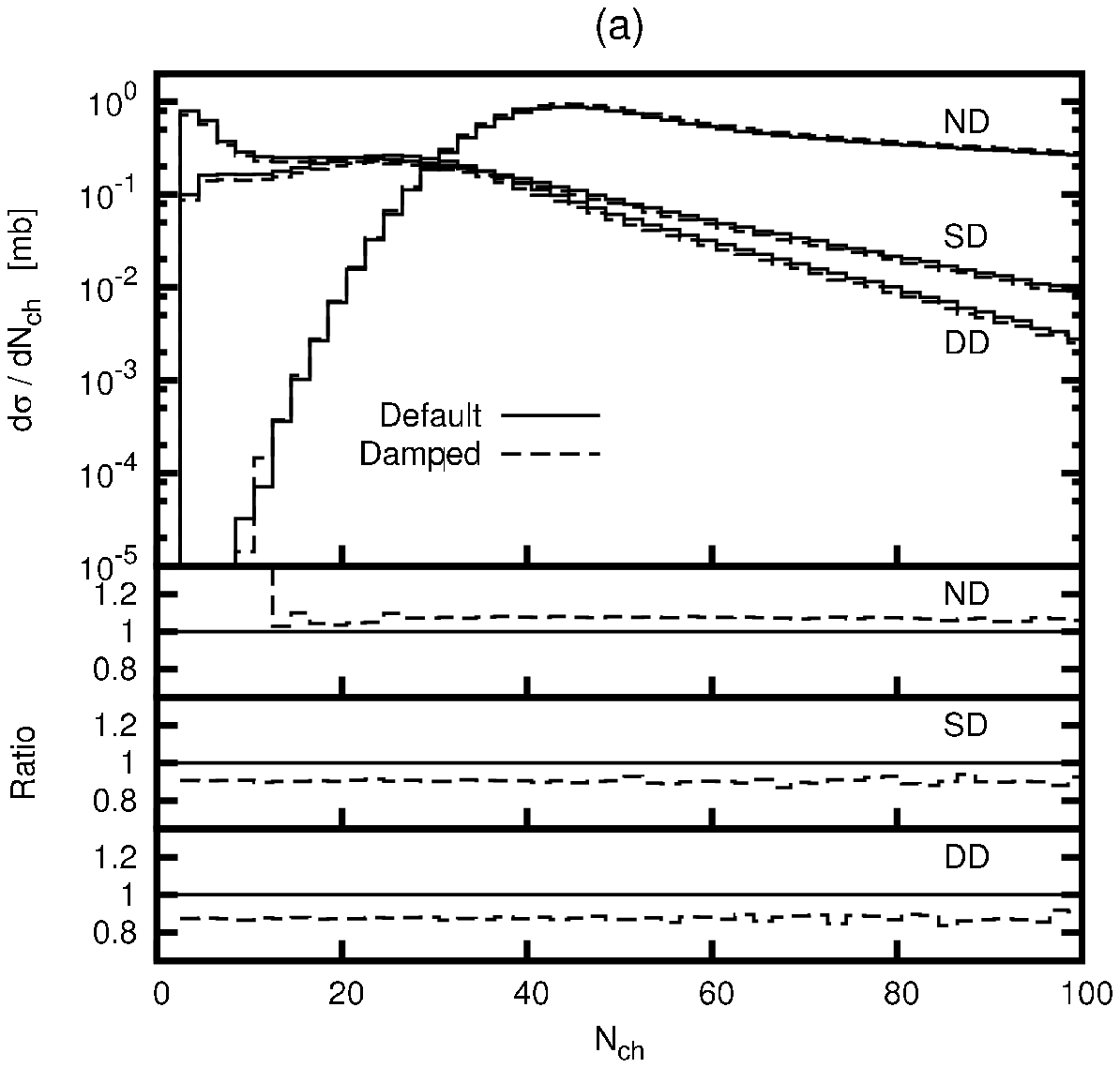}
\includegraphics[scale=0.60,angle=270]{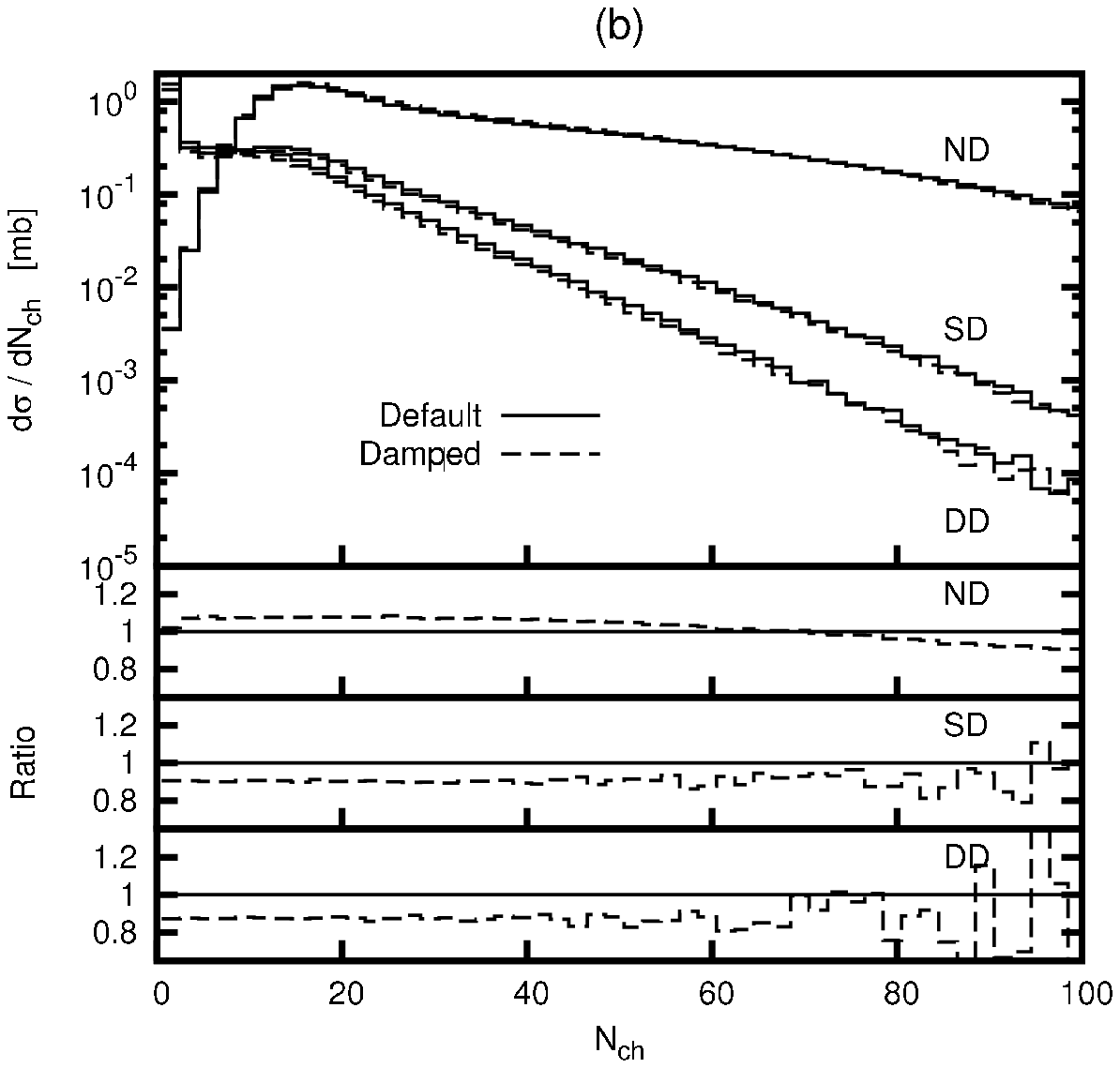}
\caption{Charged multiplicity distributions at $\sqrt{s} = 7\TeV$ showing
the breakdown of non-, single- and double-diffractive events, (a) over the
whole rapidity range and (b) for $|y| < 2.5$
\label{fig:diff2}}
\end{figure}

In Fig.~\ref{fig:diff2}, charged multiplicity distributions at $\sqrt{s} =
7\TeV$ are shown, giving the breakdown of non-, single- and
double-diffractive events, (a) over the whole rapidity range and (b) for
$|y| < 2.5$. When the entire phase space is taken into account, the single-
and double-diffractive events dominate in the lower bins, with the
non-diffractive component reaching the same level only at $N_{\mrm{ch}} \sim
30$. In (b), just the rapidity cut is enough to allow the non-diffractive
component to rise above the diffractive one down to much lower multiplicities.

It should be remembered that the non-diffractive cross section enters in the
MPI framework, eq.~(\ref{eq:MPIevol}). This is most clearly seen in the
ratio plots; while the diffractive components show a more or less constant
drop across the multiplicity range, in (b), the non-diffractive component
begins to drop in the high-multiplicity tail. This drop will also
occur in (a), but at higher multiplicities than shown. Any retuning to
compensate for the effects of the damping must then also take into account
the change of slope of the multiplicity distribution.

\subsection{Tuning prospects}

We now try a more complete tune to early LHC MB/UE data. Where
possible, data has been taken from the online HEPDATA database, but certain
key datasets are not presently available. For these, instead, data has been
read off from the relevant plots available in ATLAS publications. In
particular, from HEPDATA, the following datasets have been taken
\begin{itemize}
\item[1)] ALICE ($|\eta| < 1$) charged multiplicity and rapidity
          distributions at $\sqrt{s} = 0.90,~2.36~\&~7.00\TeV$
          \cite{Aamodt:2010ft,Aamodt:2010pp}. The charged multiplicity
          distributions at $\sqrt{s} = 0.90~\&~2.36\TeV$ are taken from
          the INEL dataset, but are shown with the zeroth bin removed. All
          others are taken from the INEL$>$0 dataset, where one track in
          the acceptance region is required to trigger.
\item[2)] ATLAS ($|\eta| < 2.5$, $\pT > 500\MeV$) charged multiplicity,
          track $\pT$, mean $\pT$ as a function of charged multiplicity
          and rapidity distributions at $\sqrt{s} = 0.90\TeV$ using the
          INEL$>$0 dataset \cite{Aad:2010rd}.
\end{itemize}
The following data has been taken from publications:
\begin{itemize}
\item[1)] As (2) above, but at $\sqrt{s} = 7.00\TeV$. Errors for the
          rapidity distribution are taken, but this was not possible
          for the remaining observables \cite{AtlasMB7}.
\item[2)] ATLAS ($|\eta| < 2.5$, $\pT > 500\MeV$) charged track based
          underlying event \cite{AtlasUE}. Charged particle number
          and sum-$\pT$ density distributions are taken with errors for
          the toward, away and transverse regions. A charged track of
          $\pT > 1\GeV$ in the $\eta$ acceptance is required to trigger
          an event.
\end{itemize}

\begin{figure}
\centering
\includegraphics[scale=0.42,angle=270]{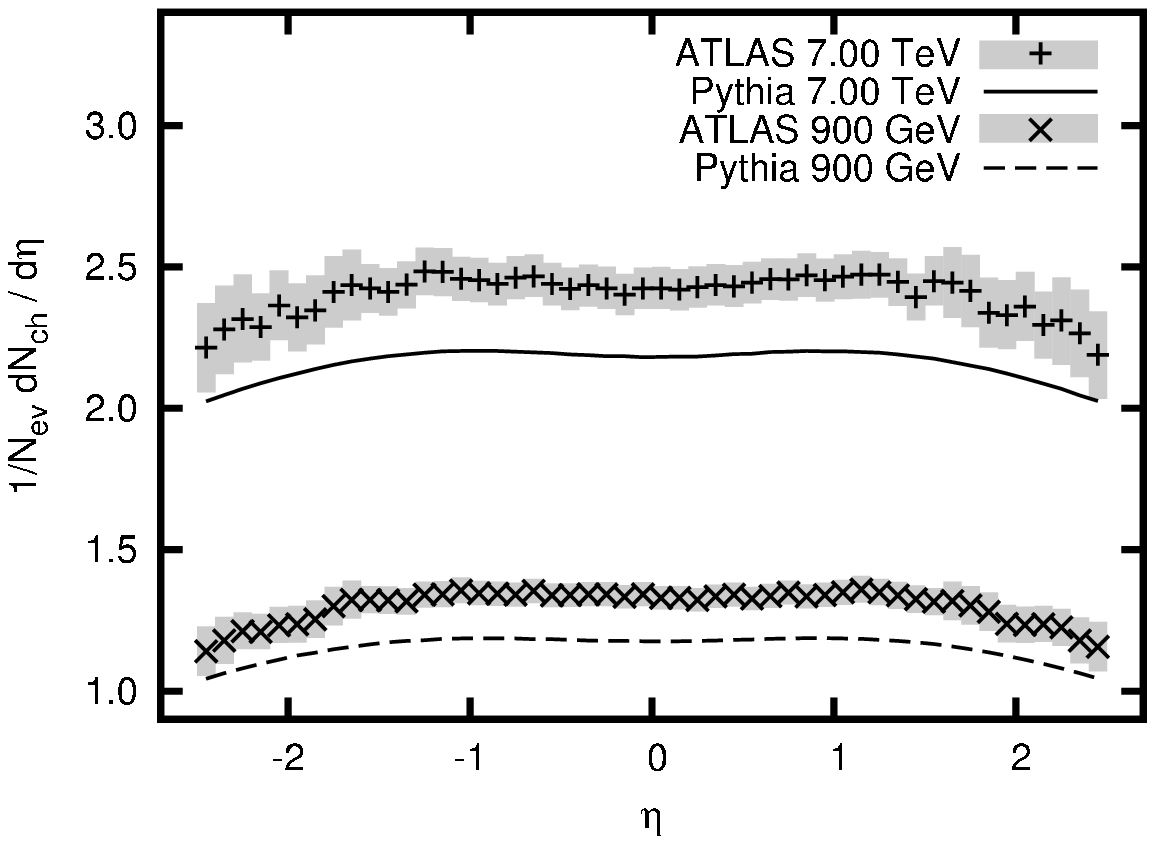}
\includegraphics[scale=0.42,angle=270]{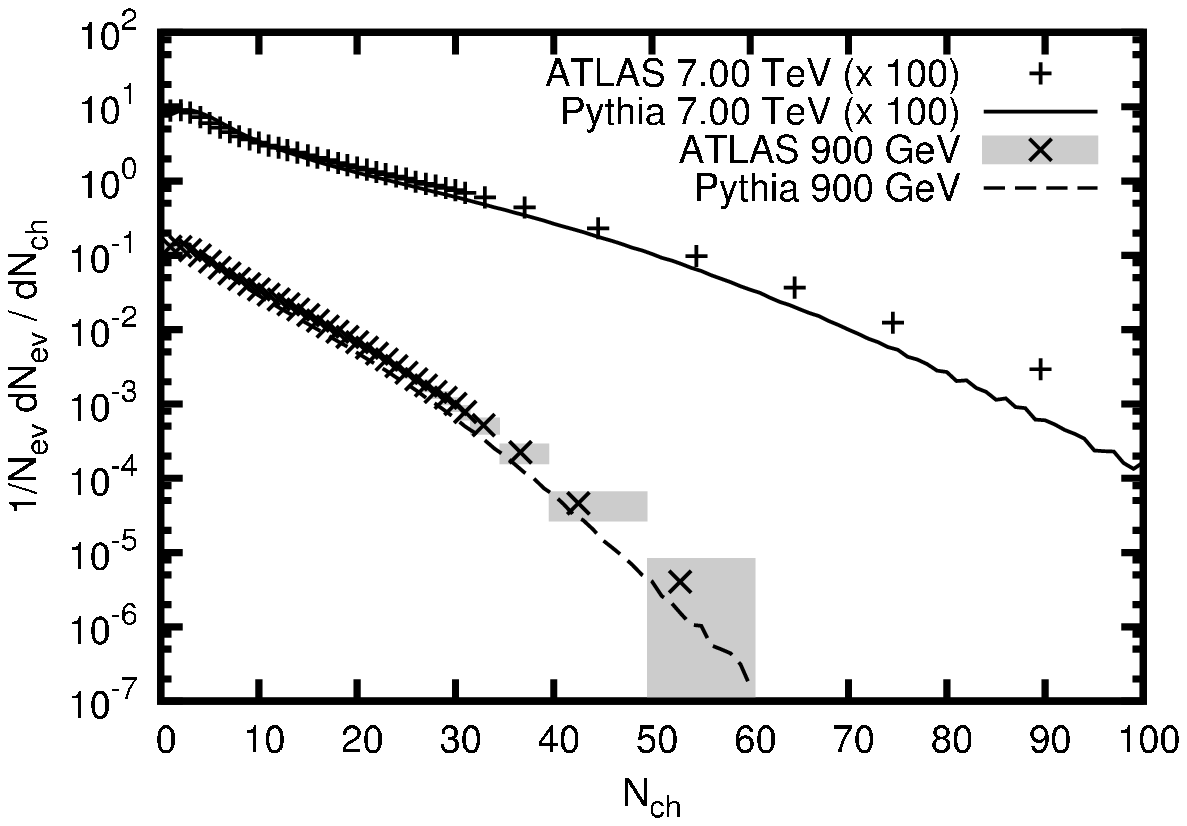}
\includegraphics[scale=0.42,angle=270]{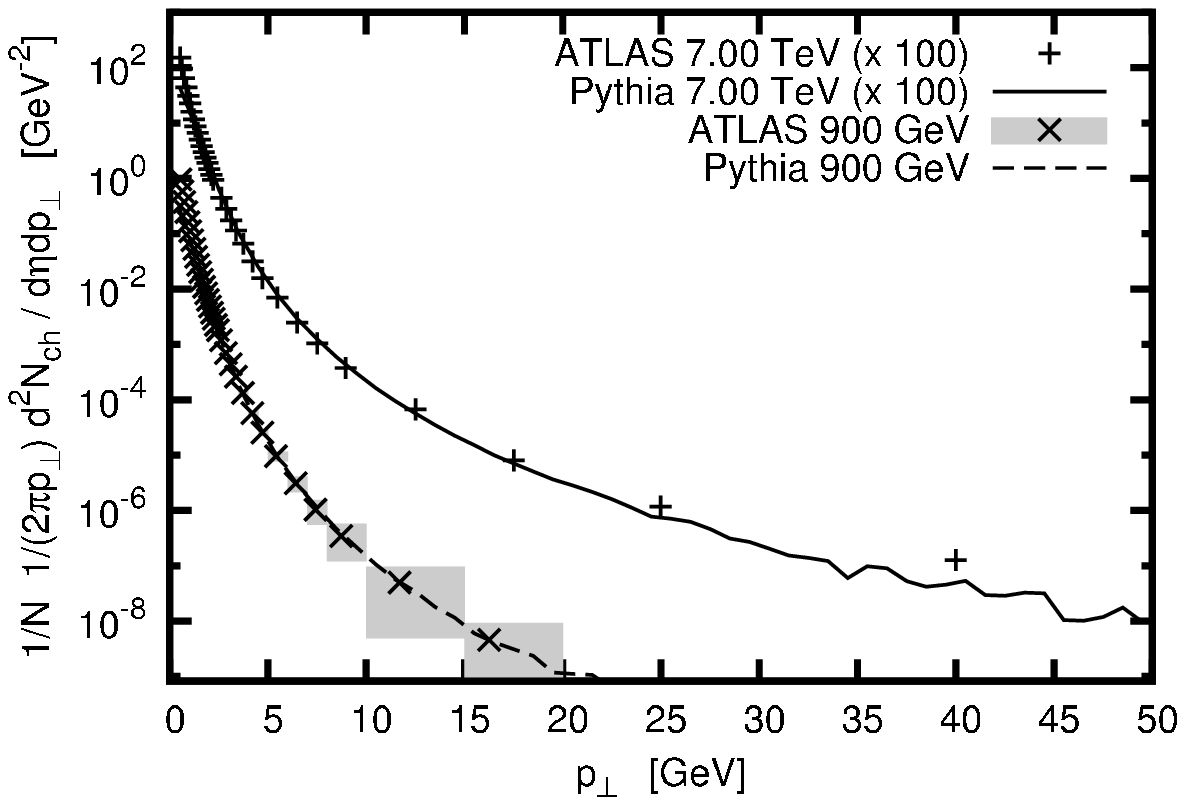}
\includegraphics[scale=0.42,angle=270]{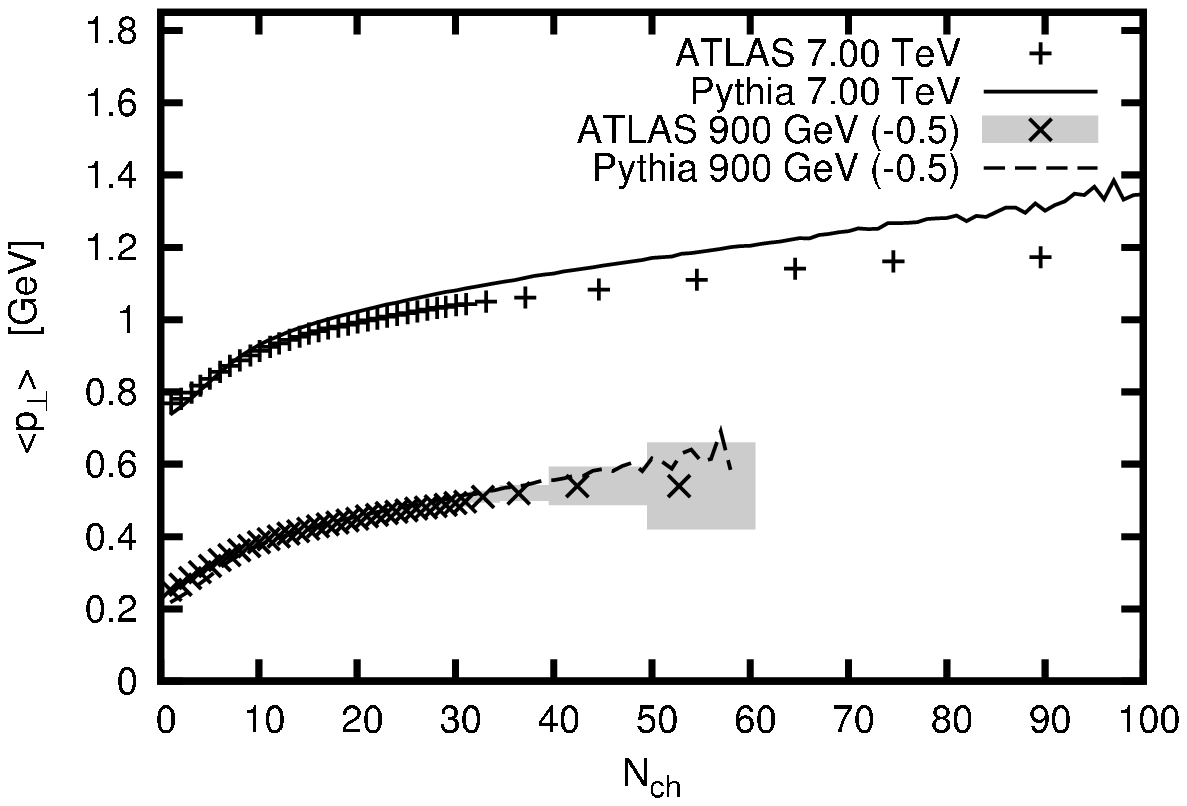}
\includegraphics[scale=0.42,angle=270]{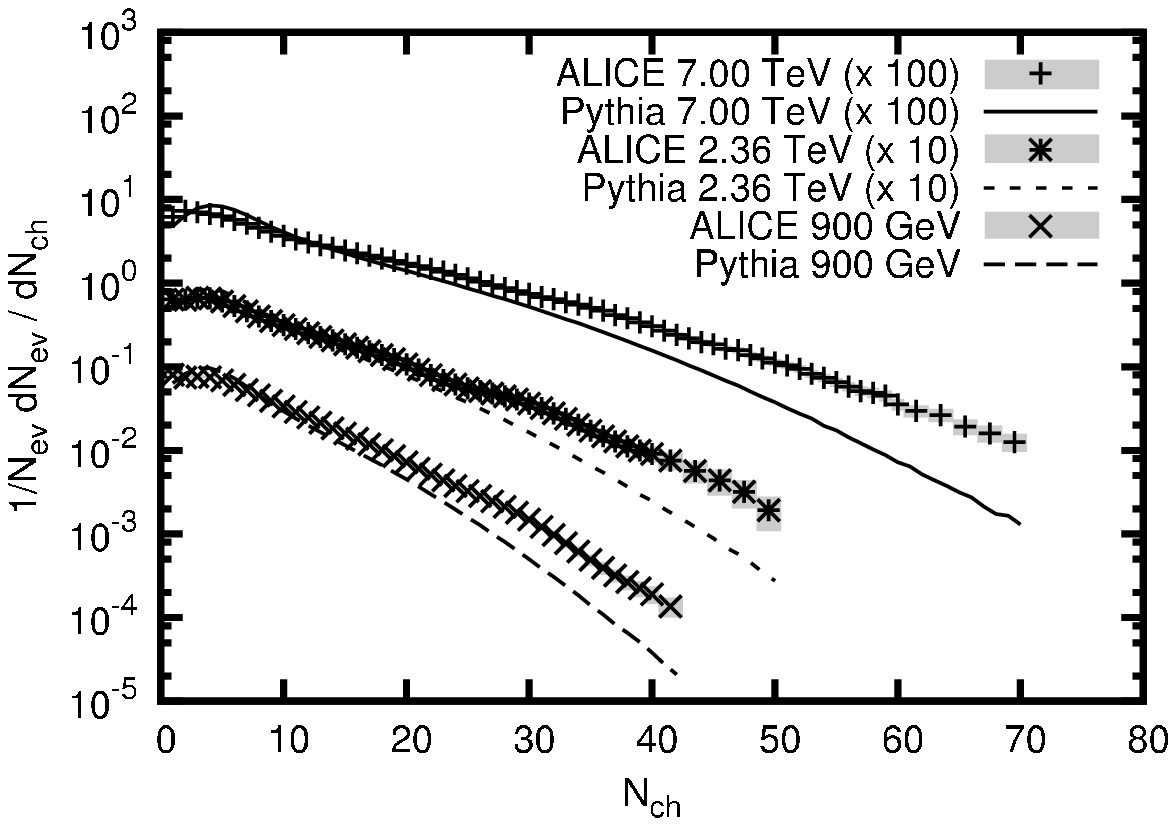}
\includegraphics[scale=0.42,angle=270]{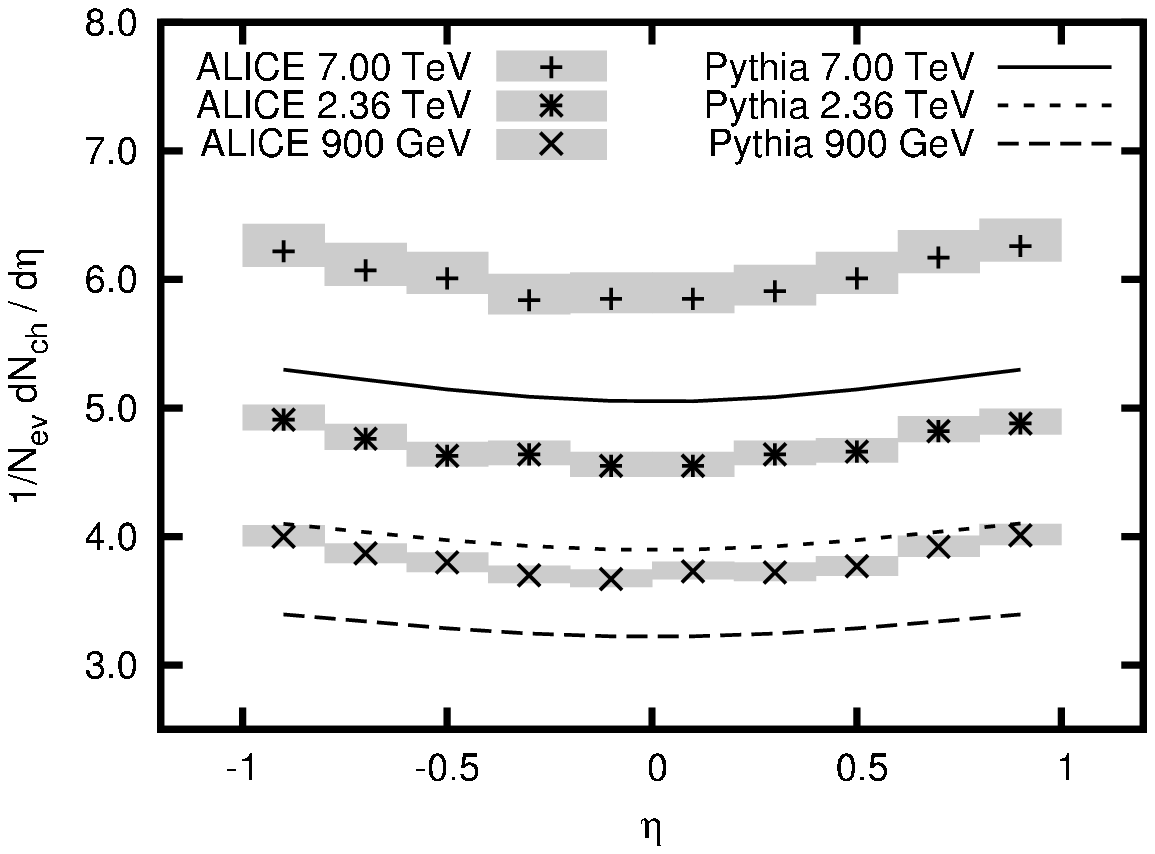}
\includegraphics[scale=0.42,angle=270]{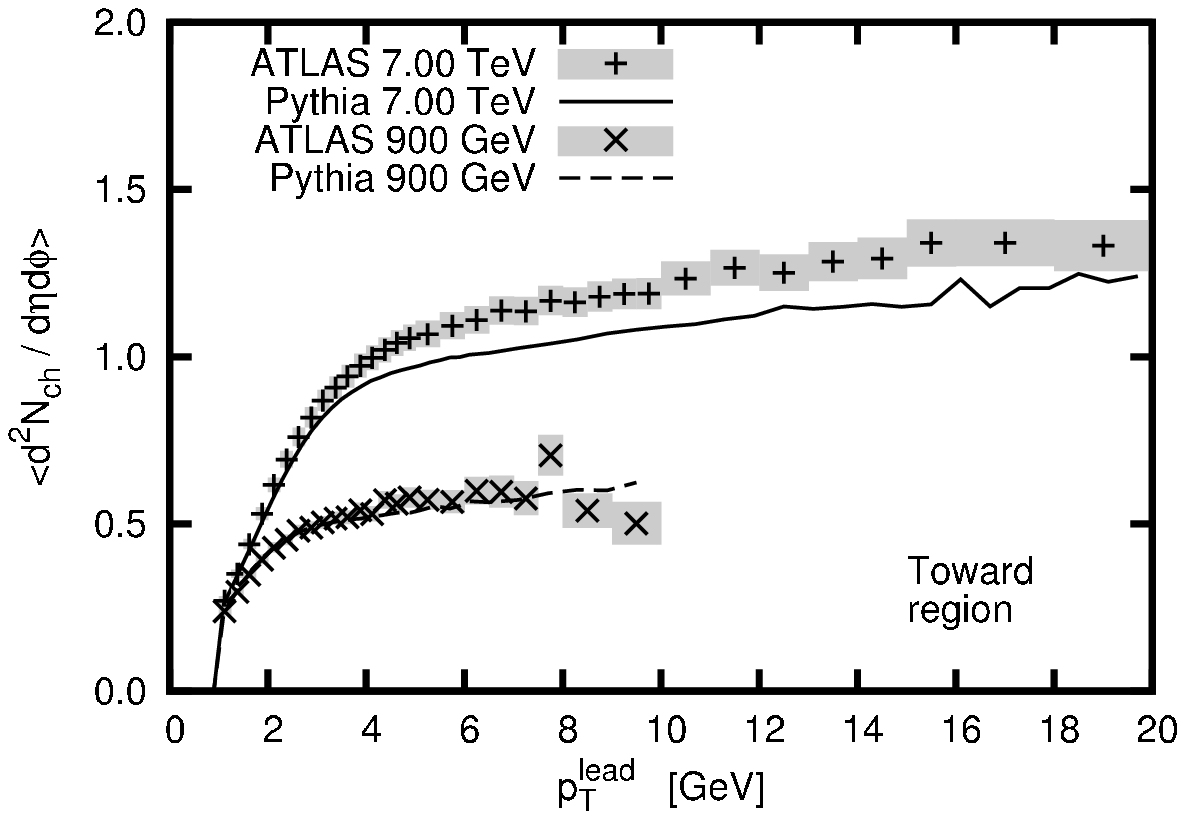}
\includegraphics[scale=0.42,angle=270]{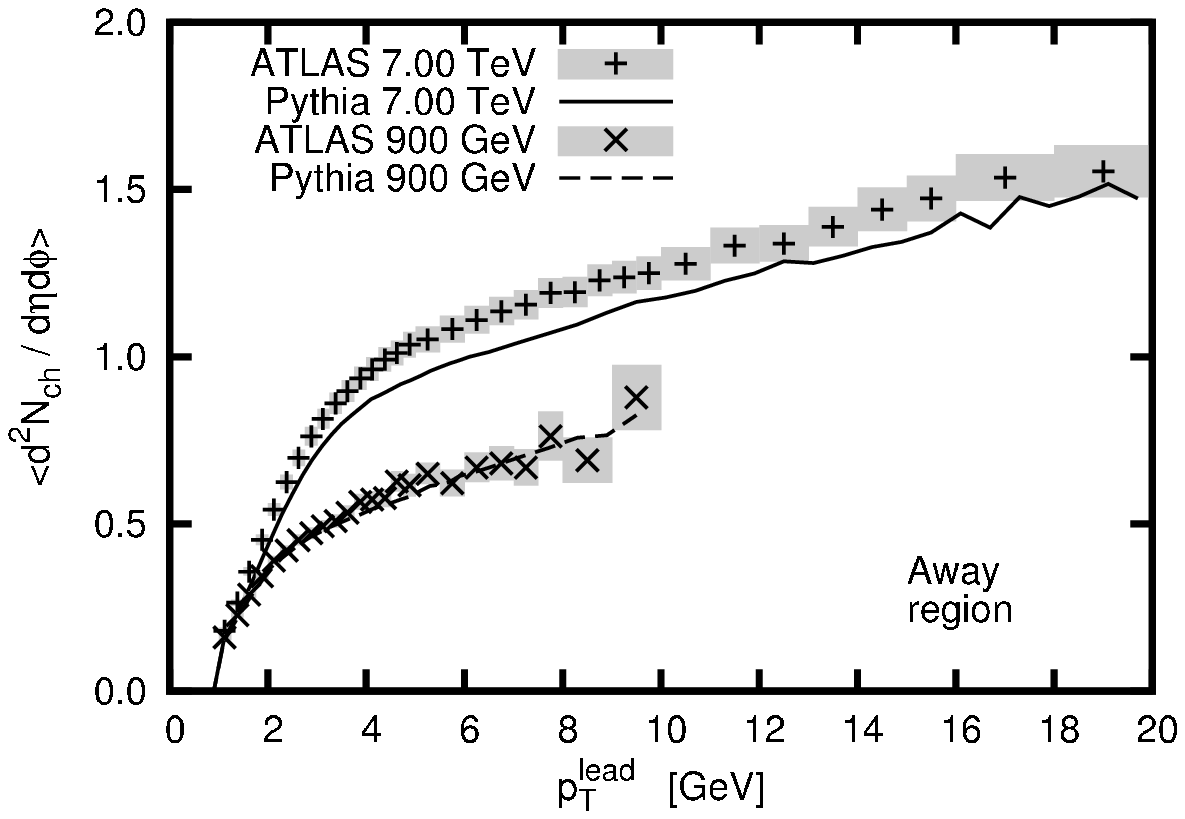}
\includegraphics[scale=0.42,angle=270]{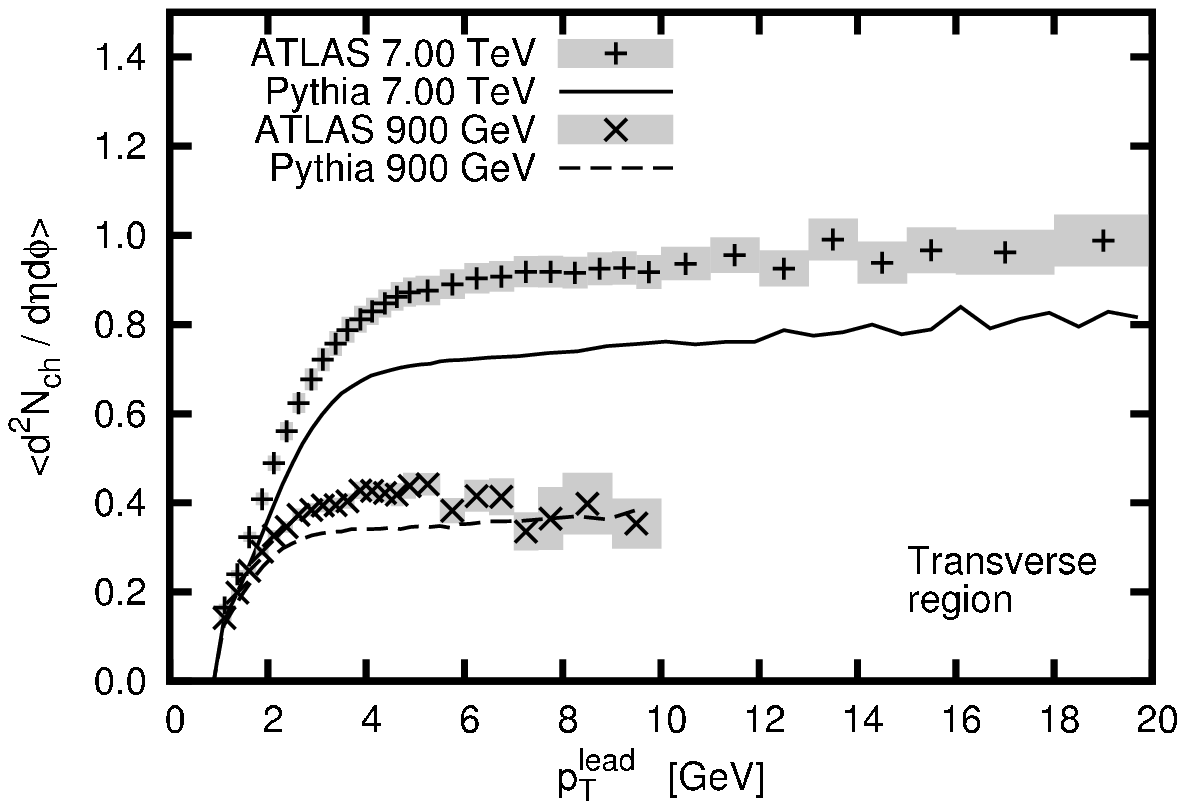}
\includegraphics[scale=0.42,angle=270]{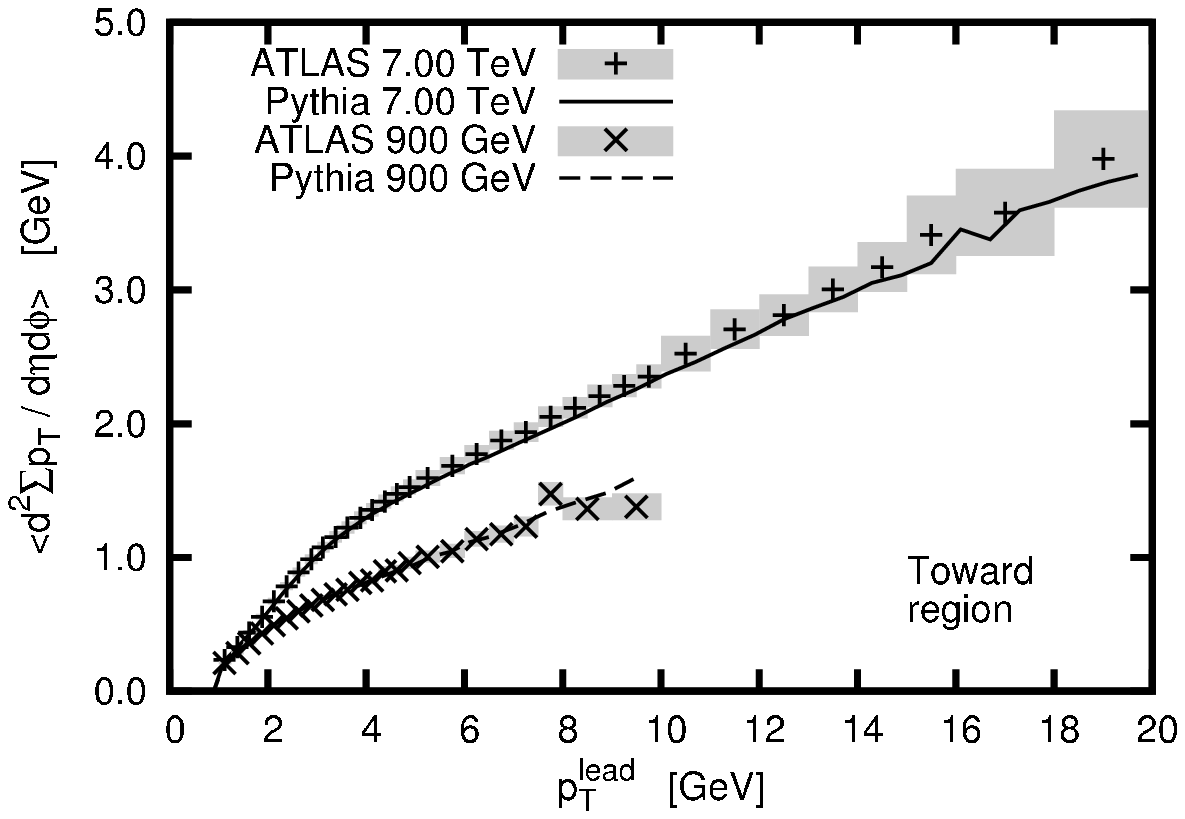}
\includegraphics[scale=0.42,angle=270]{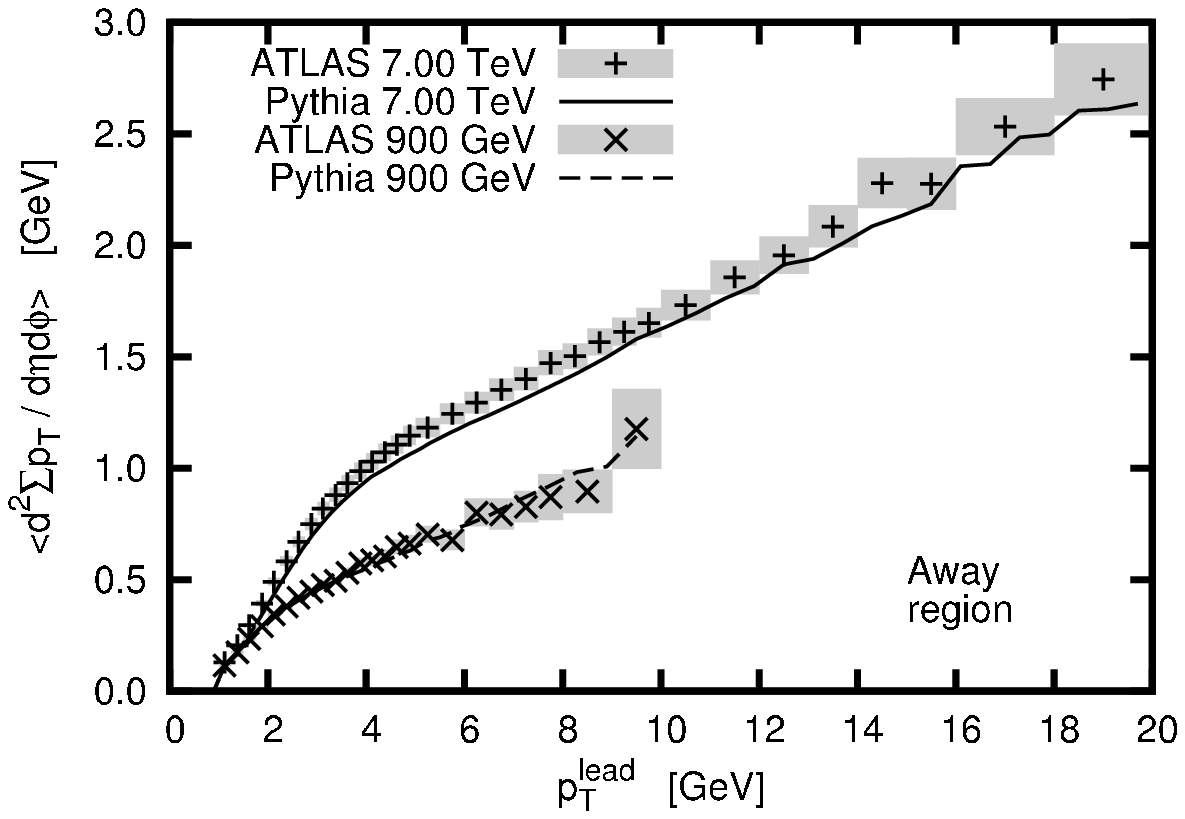}
\includegraphics[scale=0.42,angle=270]{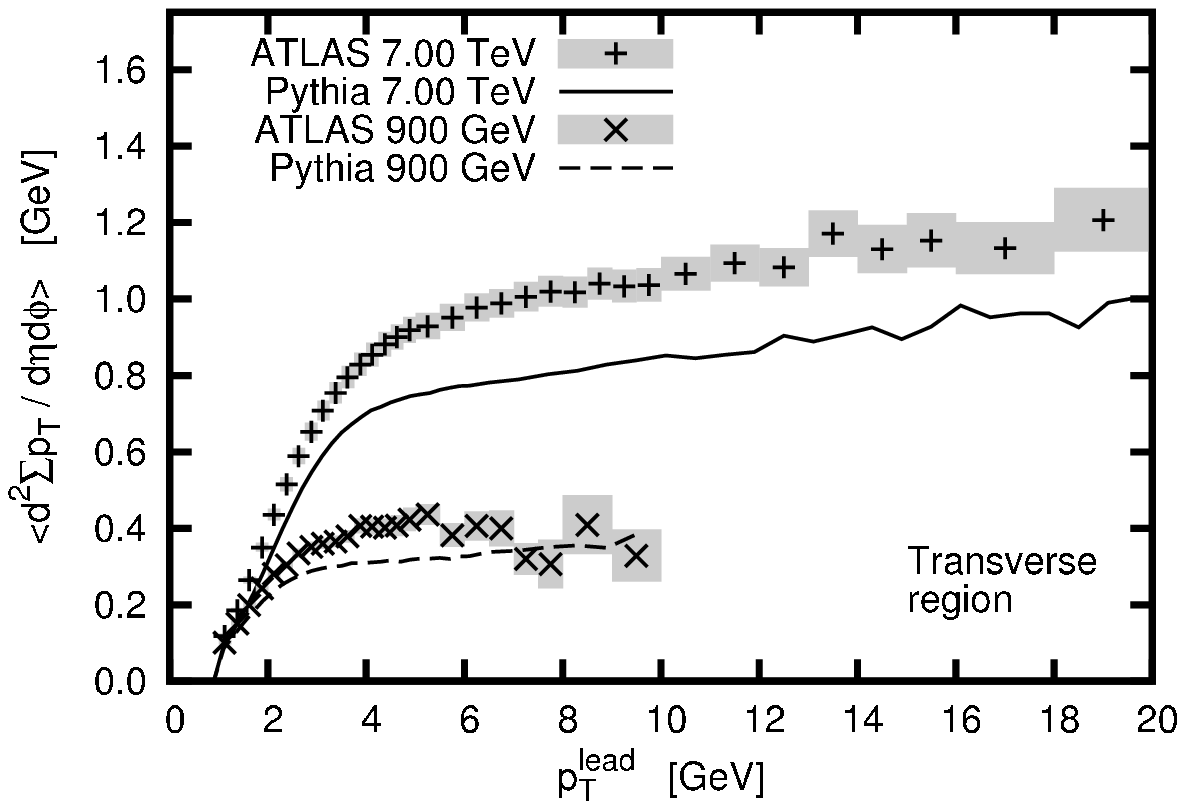}
\caption{Tune 2C compared against early LHC data (see text). Where
available, systematic and statistical errors are shown summed in quadrature
as the grey band
\label{fig:LHC-2C}}
\end{figure}

The results for Tune 2C, introduced in Sec.~\ref{sec:tuning}, are shown in
Fig.~\ref{fig:LHC-2C}, compared against these datasets. Where available,
systematic and statistical errors are shown summed in quadrature. At all
energies, the multiplicities lie below the data. In the underlying event,
the charged number density at $7\TeV$ undershoots in the toward and away
regions while at both energies there is too little activity in the
transverse region. The same plots for Tune 2M show the same general
features.

In attempting to improve agreement with these datasets, we stay with the
CTEQ6L1 PDF set, using Tune 2C as a starting point, while turning
on the diffractive damping outlined previously. Additionally,
we only vary those parameters relating to the MPI
framework and colour reconnection (specifically the $R$ parameter of
Sec.~\ref{sec:cr}). By doing this, we ignore the possibility
of shifting contributions between ISR and MPI; this balance is perhaps
better determined with observables such as $\pT(\Z^0)$ and jet-jet azimuthal
correlations. While this procedure is unlikely to give us
the best tune possible, it should give a good indication of whether the
generator is capable of describing the broad features of the data.

\begin{figure}
\centering
\includegraphics[scale=0.42,angle=270]{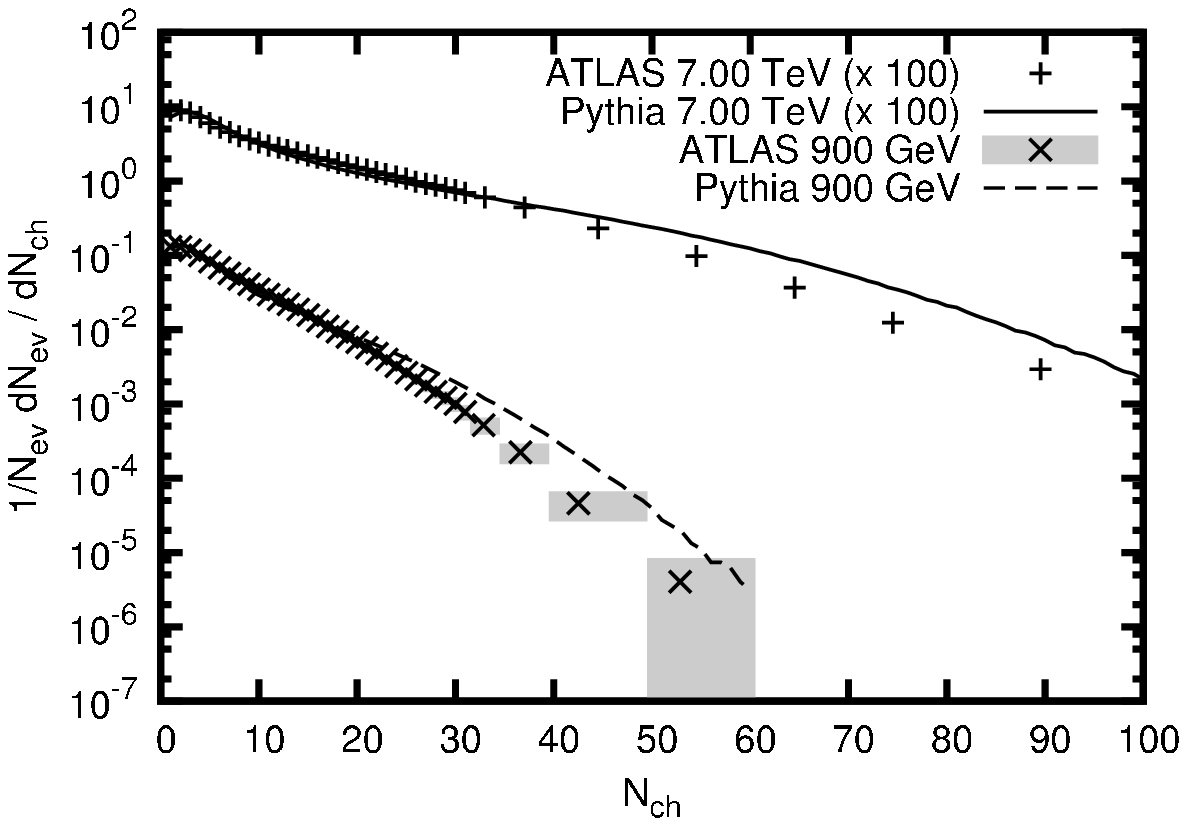}
\includegraphics[scale=0.42,angle=270]{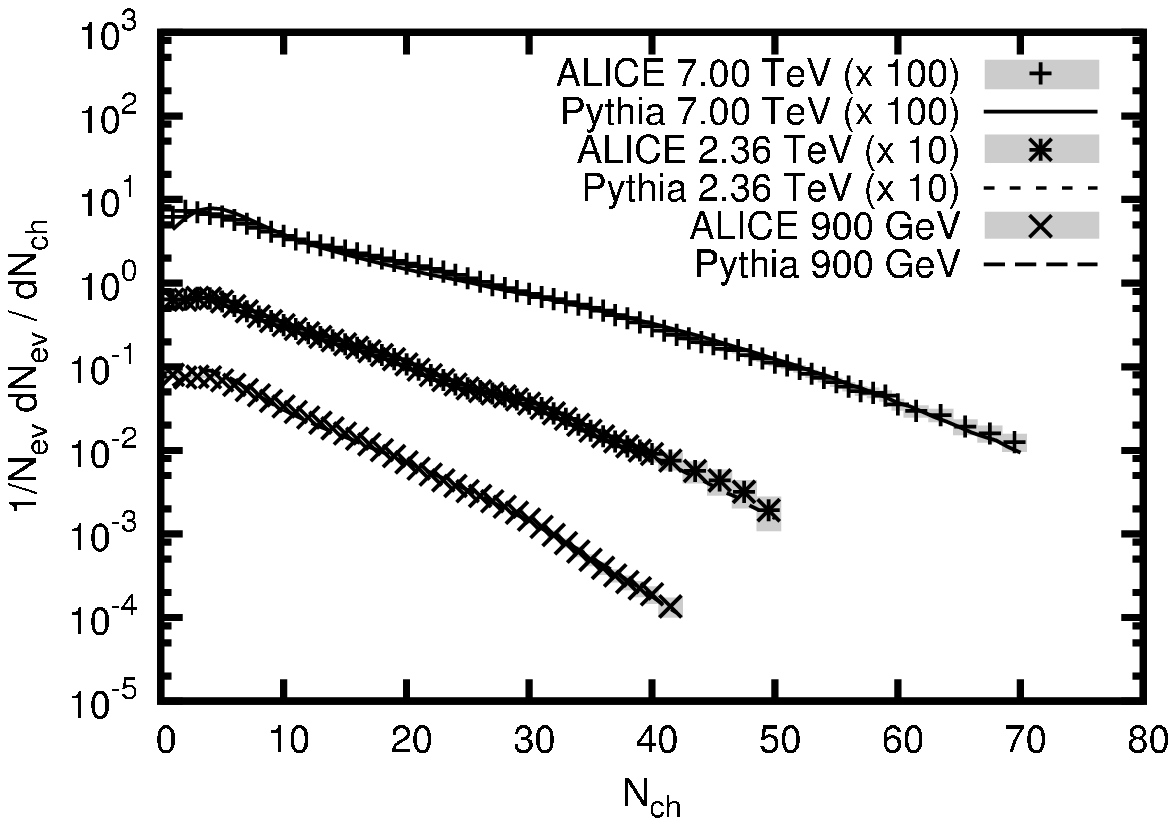}
\includegraphics[scale=0.42,angle=270]{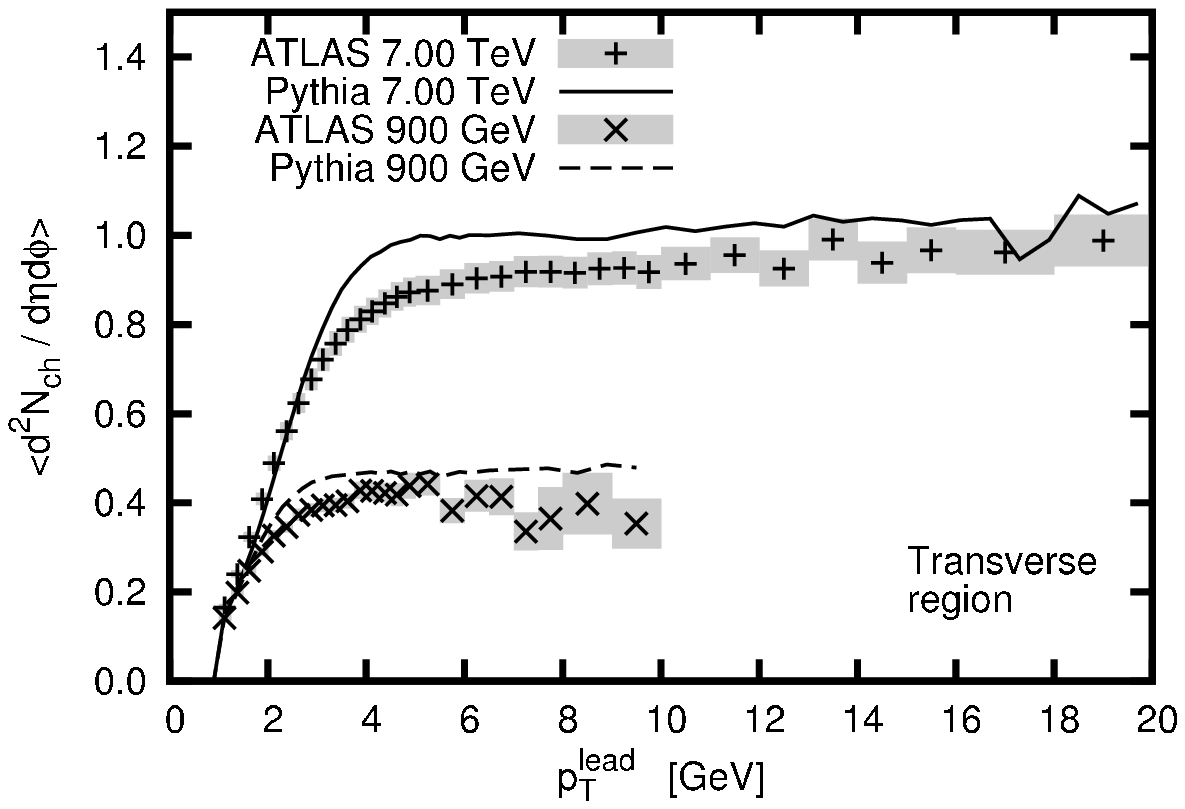}
\caption{ATLAS and ALICE multiplicities and the transverse region charged
particle density when tuned only to the ALICE data (see text)
\label{fig:LHC-ALICE-fit}}
\end{figure}

The primary tools available, then, are the $p_{\perp0}^{\mathrm{ref}}$ and
$E^{\mrm{pow}}_{\exp}$ parameters of the MPI framework, supplemented by
$E_{\mathrm{CM}}^{\mathrm{pow}}$ to control the energy scaling (keeping
$E_{\mathrm{CM}}^{\mathrm{ref}}$ fixed at $1800\GeV$). As an
example of the balance that needs to be found between them,
Fig.~\ref{fig:LHC-ALICE-fit} shows the ATLAS and ALICE charged multiplicity
distributions and the transverse region charged particle density, when
solely tuned to match the ALICE multiplicity data
($p_{\perp0}^{\mathrm{ref}} = 2.047$, $E^{\mrm{pow}}_{\exp} = 1.5$ and
$E_{\mathrm{CM}}^{\mathrm{pow}} = 0.19$). In order to match the ALICE
multiplicities, especially all the way out in the tails (made very slightly
more difficult by the diffractive damping, as shown previously), a lower
$E^{\mrm{pow}}_{\exp}$ is needed, but this leads to an overshoot in the
ATLAS multiplicity tails and an underlying event that rises too quickly.

\begin{figure}
\centering
\includegraphics[scale=0.42,angle=270]{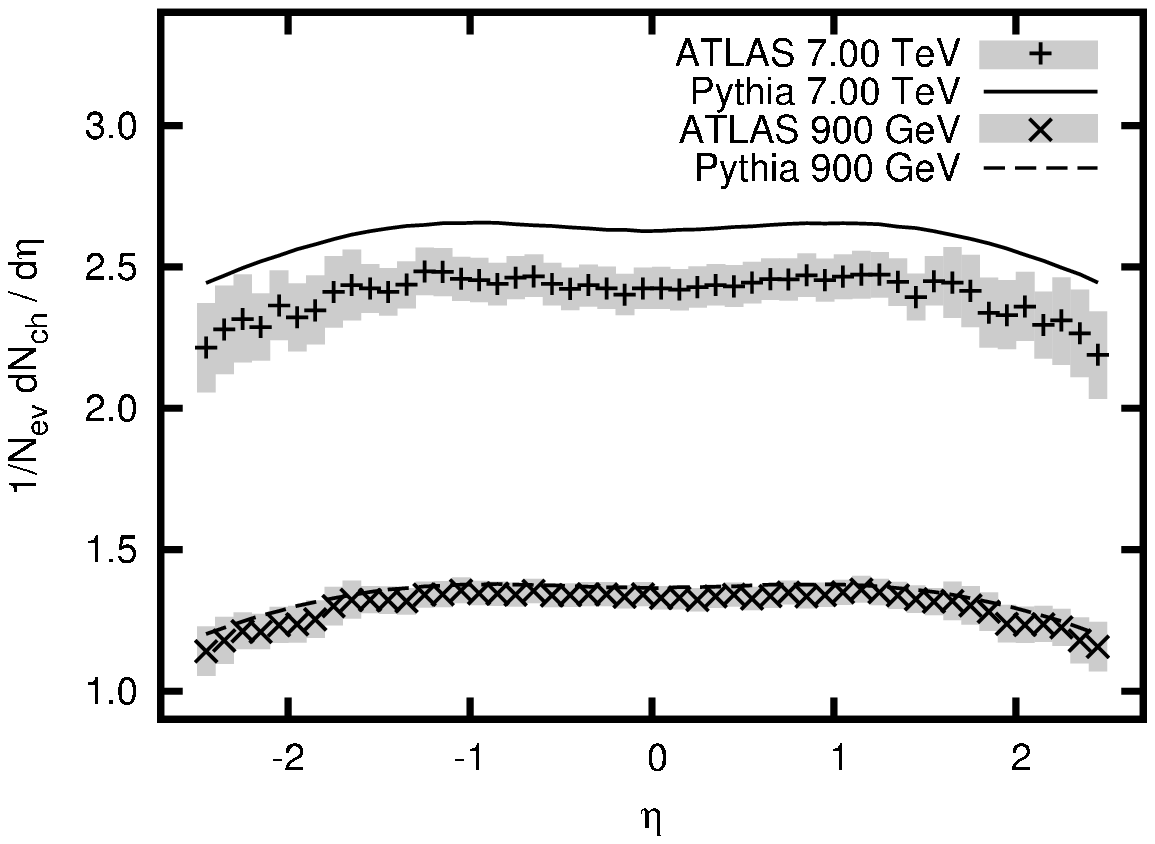}
\includegraphics[scale=0.42,angle=270]{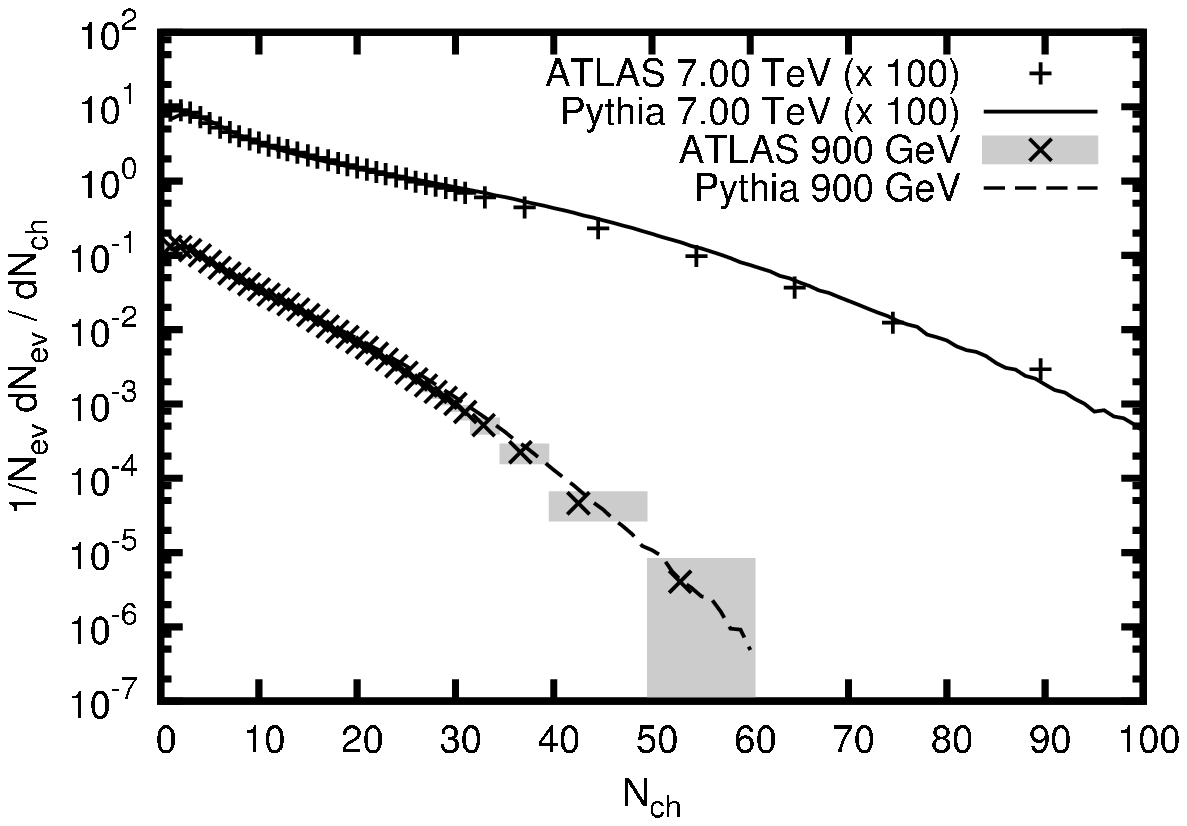}
\includegraphics[scale=0.42,angle=270]{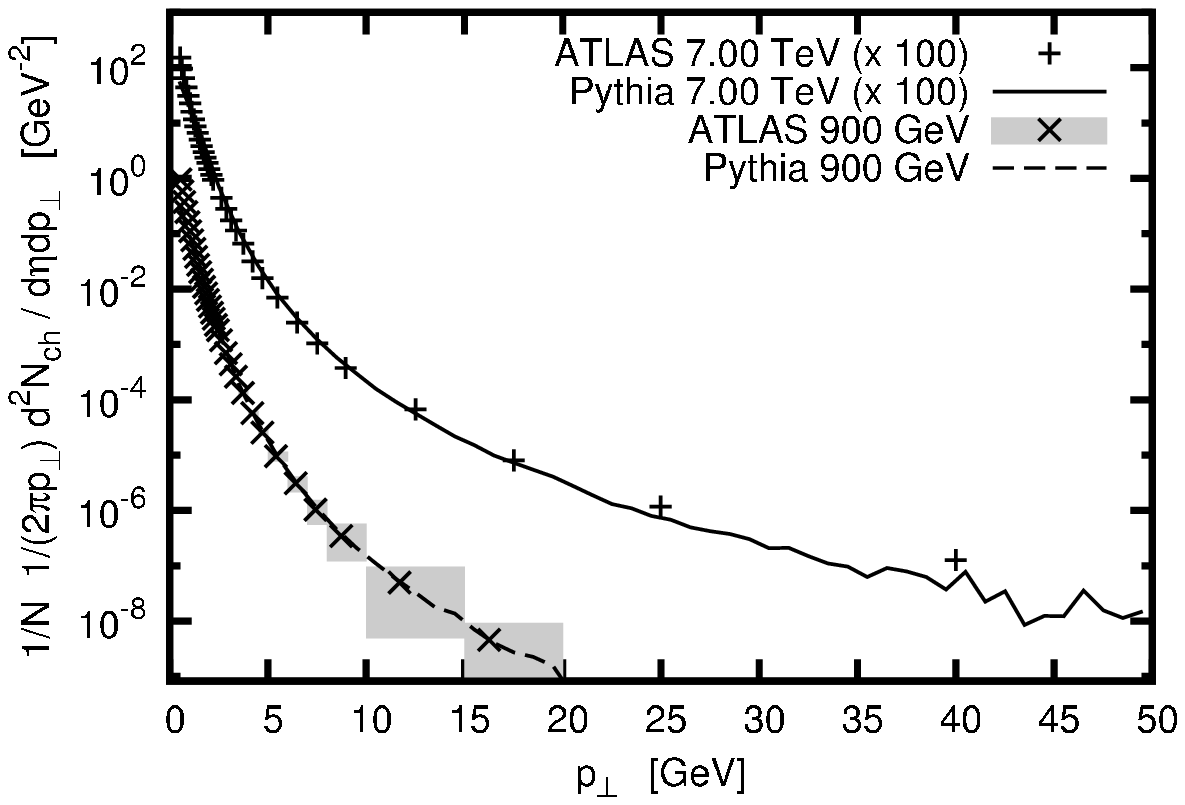}
\includegraphics[scale=0.42,angle=270]{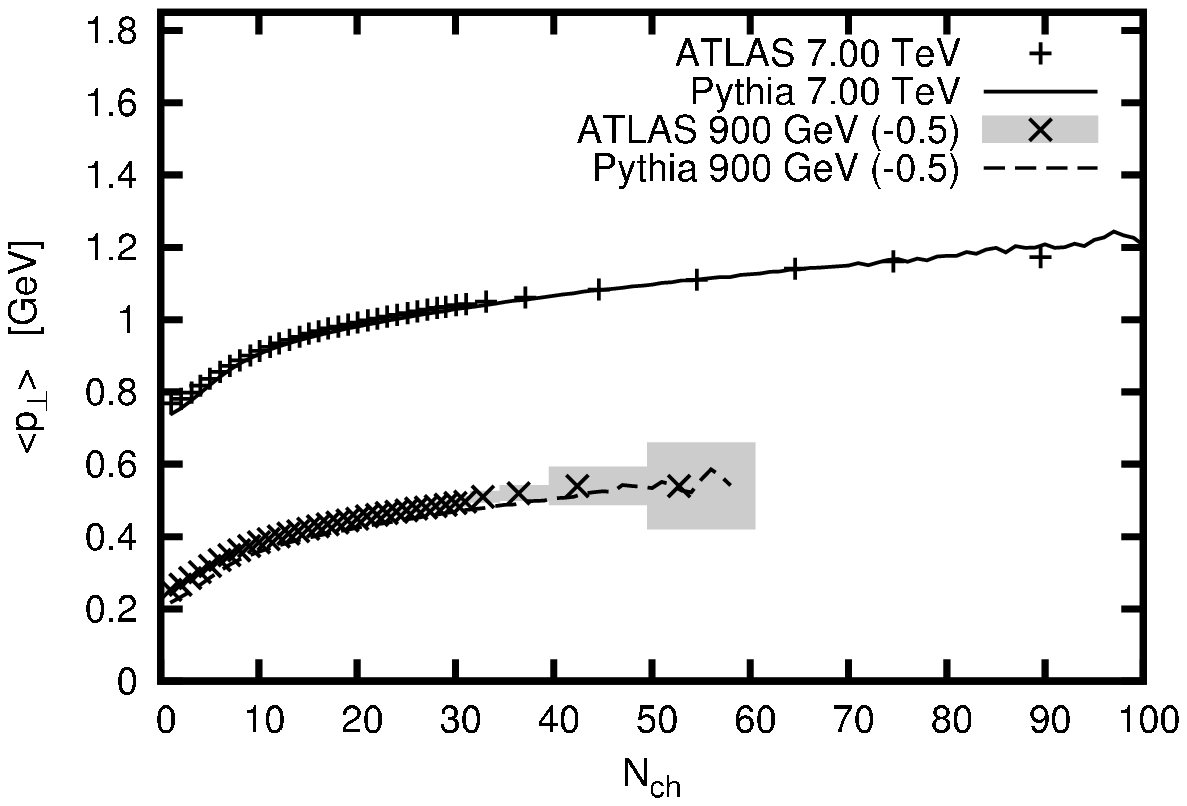}
\includegraphics[scale=0.42,angle=270]{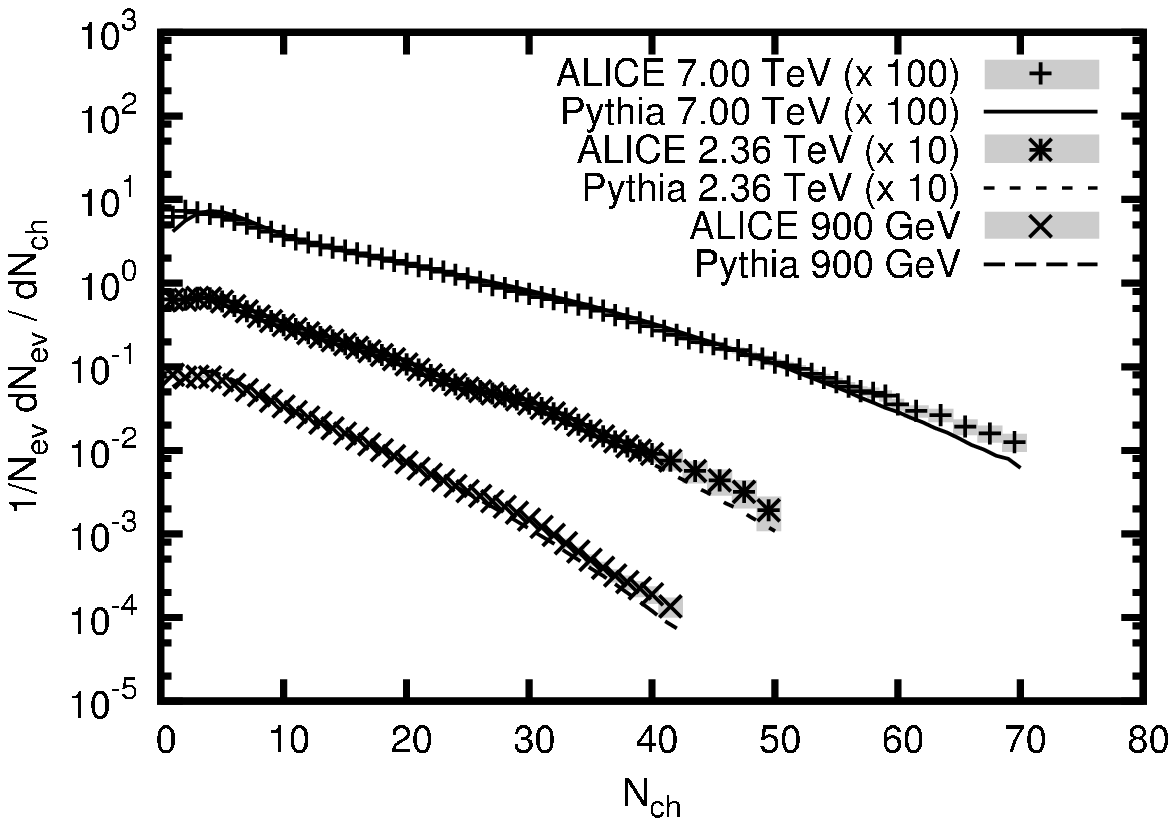}
\includegraphics[scale=0.42,angle=270]{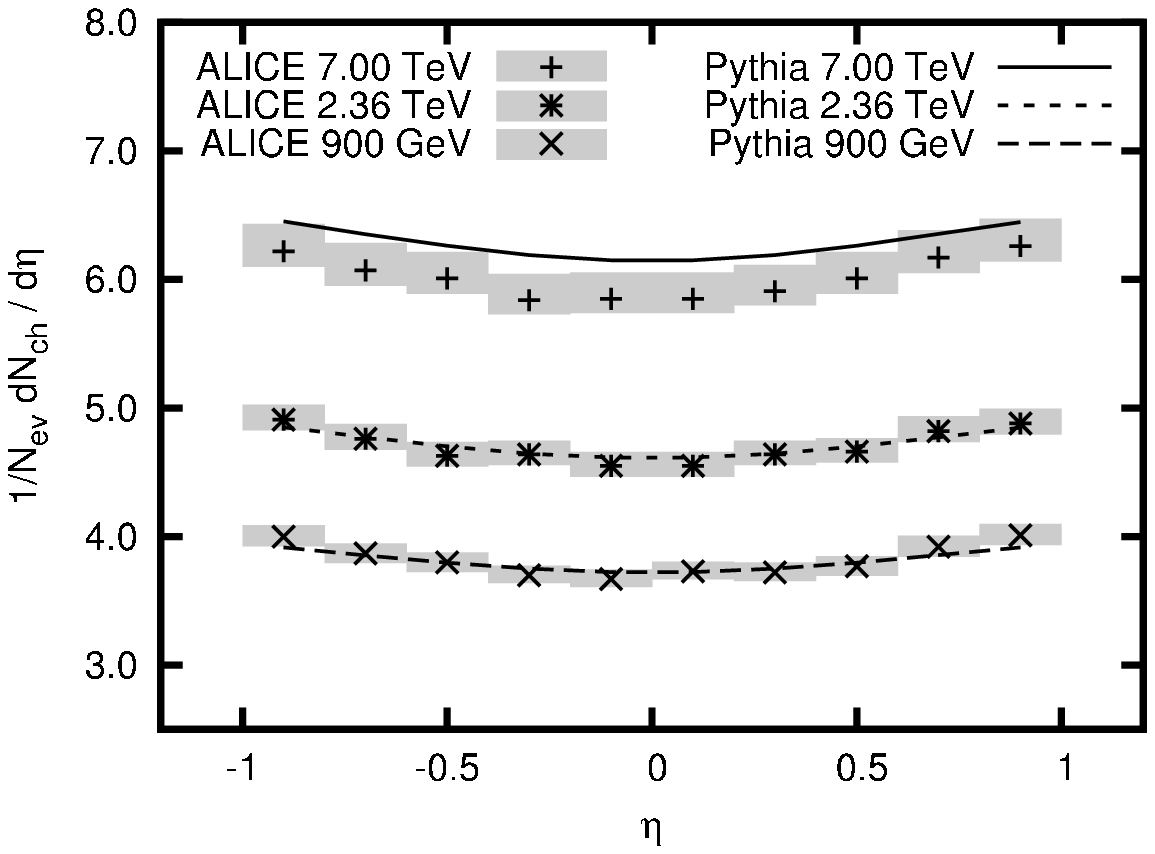}
\includegraphics[scale=0.42,angle=270]{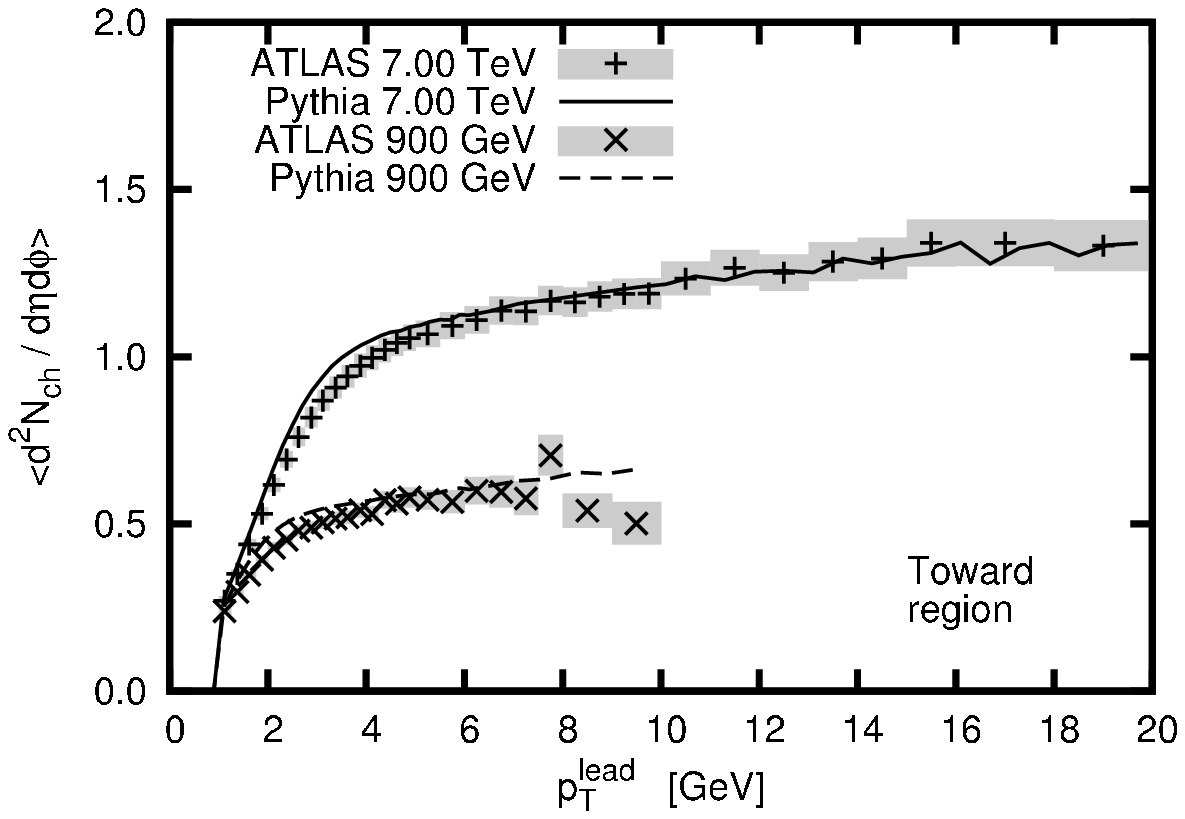}
\includegraphics[scale=0.42,angle=270]{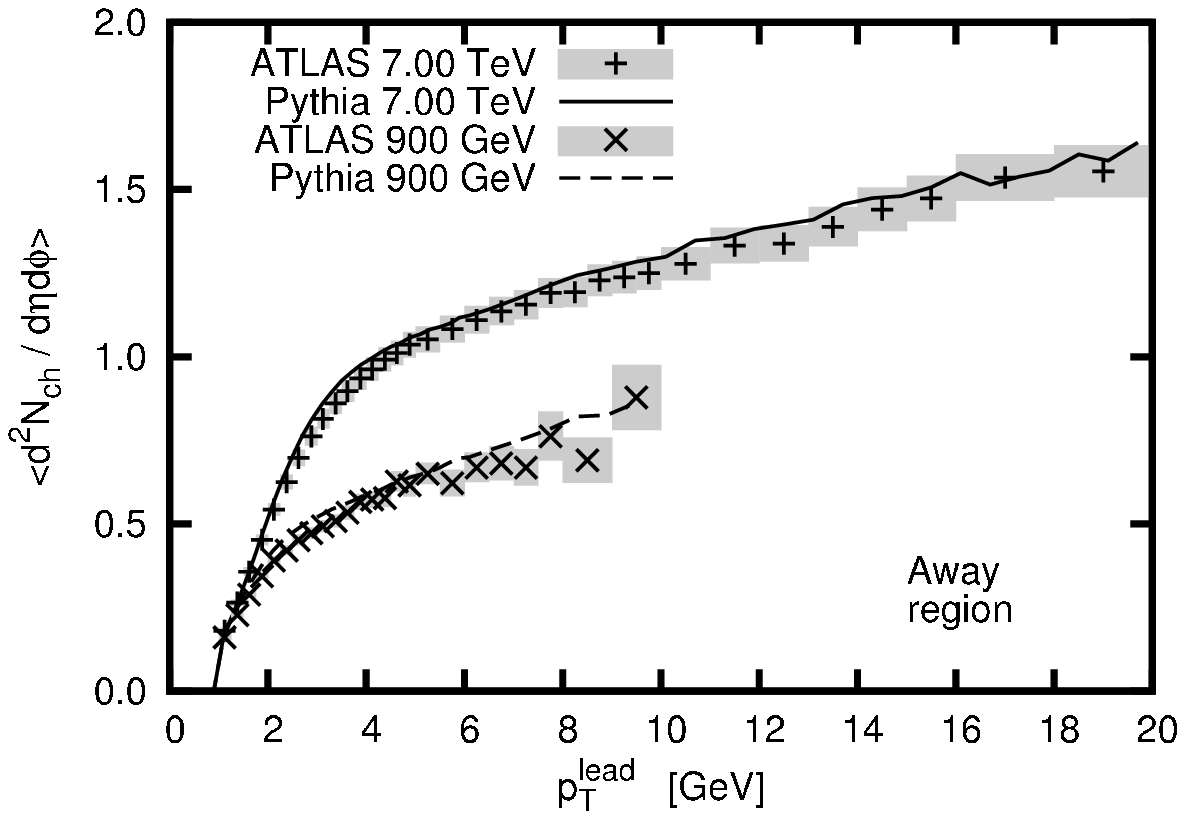}
\includegraphics[scale=0.42,angle=270]{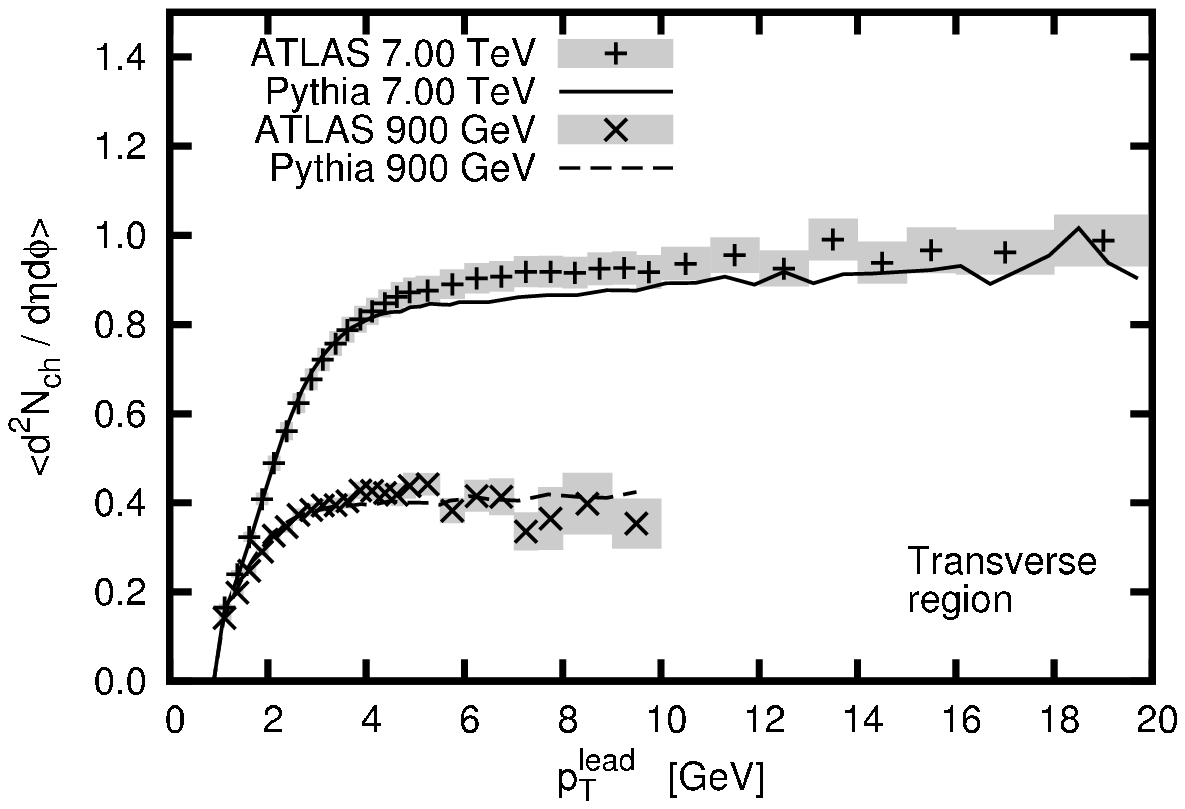}
\includegraphics[scale=0.42,angle=270]{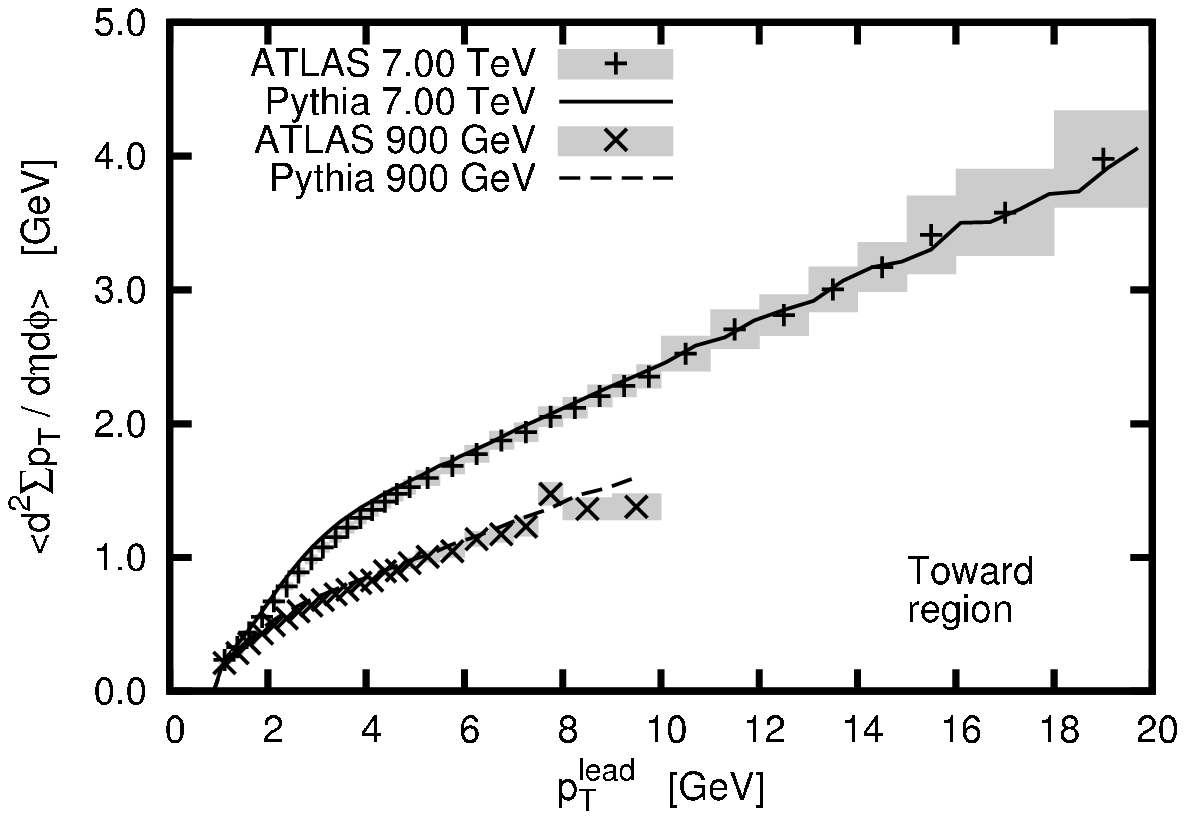}
\includegraphics[scale=0.42,angle=270]{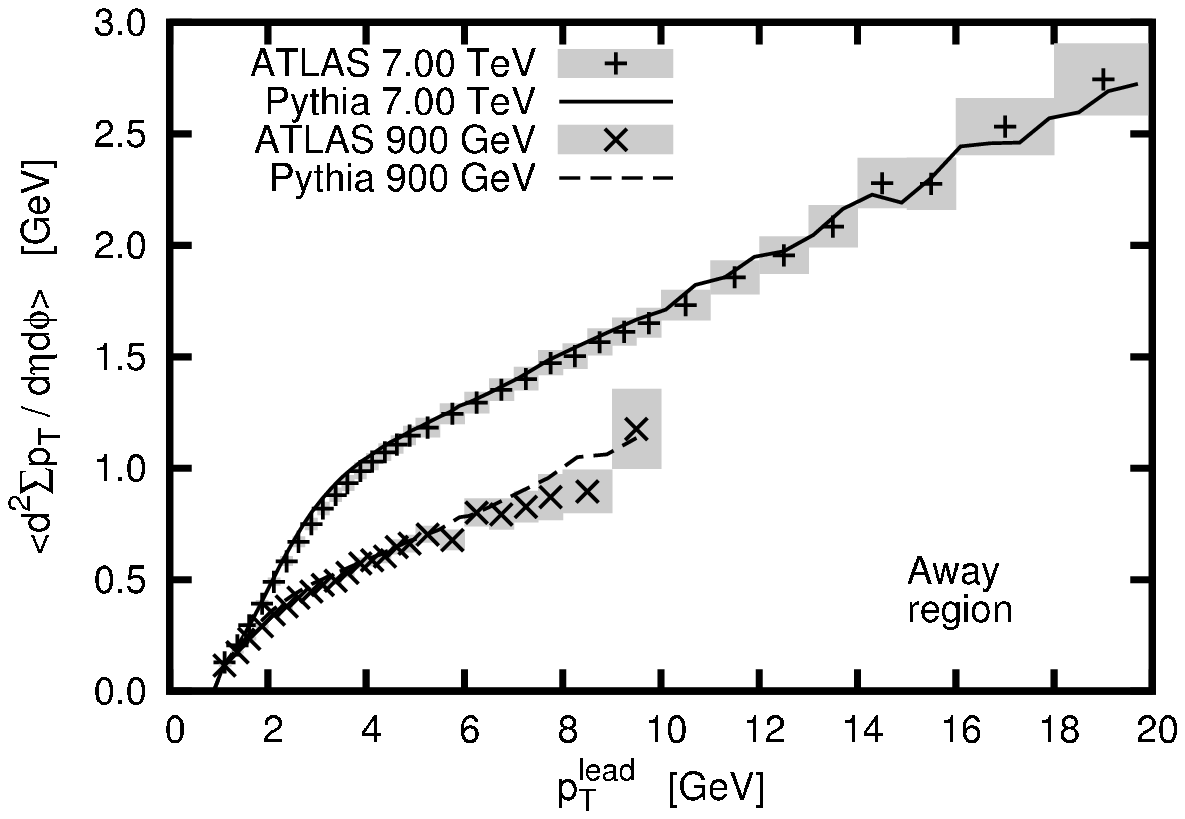}
\includegraphics[scale=0.42,angle=270]{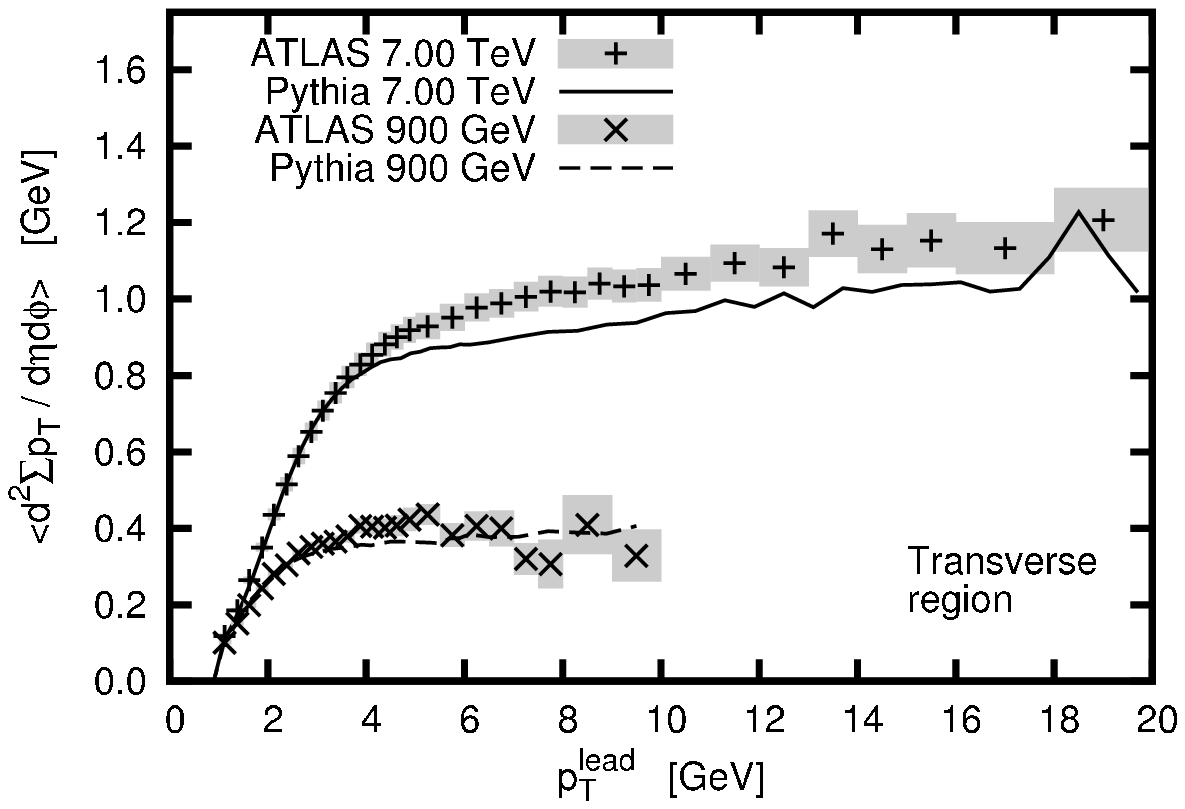}
\caption{Tune 4C compared to LHC data
\label{fig:LHC-4C}}
\end{figure}

In Fig.~\ref{fig:LHC-4C}, results are shown for a set of
parameters which give a reasonable description to all sets of data, Tune 4C
in Tab.~\ref{tab:tevtune}. In this set, the colour reconnection parameter,
$R$ is adjusted to match the ATLAS $\langle \pT \rangle(N_{\mrm{ch}})$
data. This decrease, although preferable in terms of describing the data
with less non-perturbative corrections, has direct effects on the other
MB/UE data; it does help to improve the description of the
high-multiplicity tails, but the extra multiplicity leads to a rise in the
transverse charged number density, while there is not a similar rise in the
$\sum \pT$ density. The $E^{\mrm{pow}}_{\exp}$ parameter is pushed out such
that it in fact reduces to the single-Gaussian case. With this setting, the
rise in the toward and away regions is still slightly too high. While
overall, at $7\TeV$, the multiplicities are slightly on the high side, the
level of activity in the transverse region in the underlying event is
slightly too low. As shown in the previous example, getting out to the very
high-multiplicity tails of the ALICE data is a hard task.

\begin{figure}
\centerline{
\includegraphics[scale=0.6]{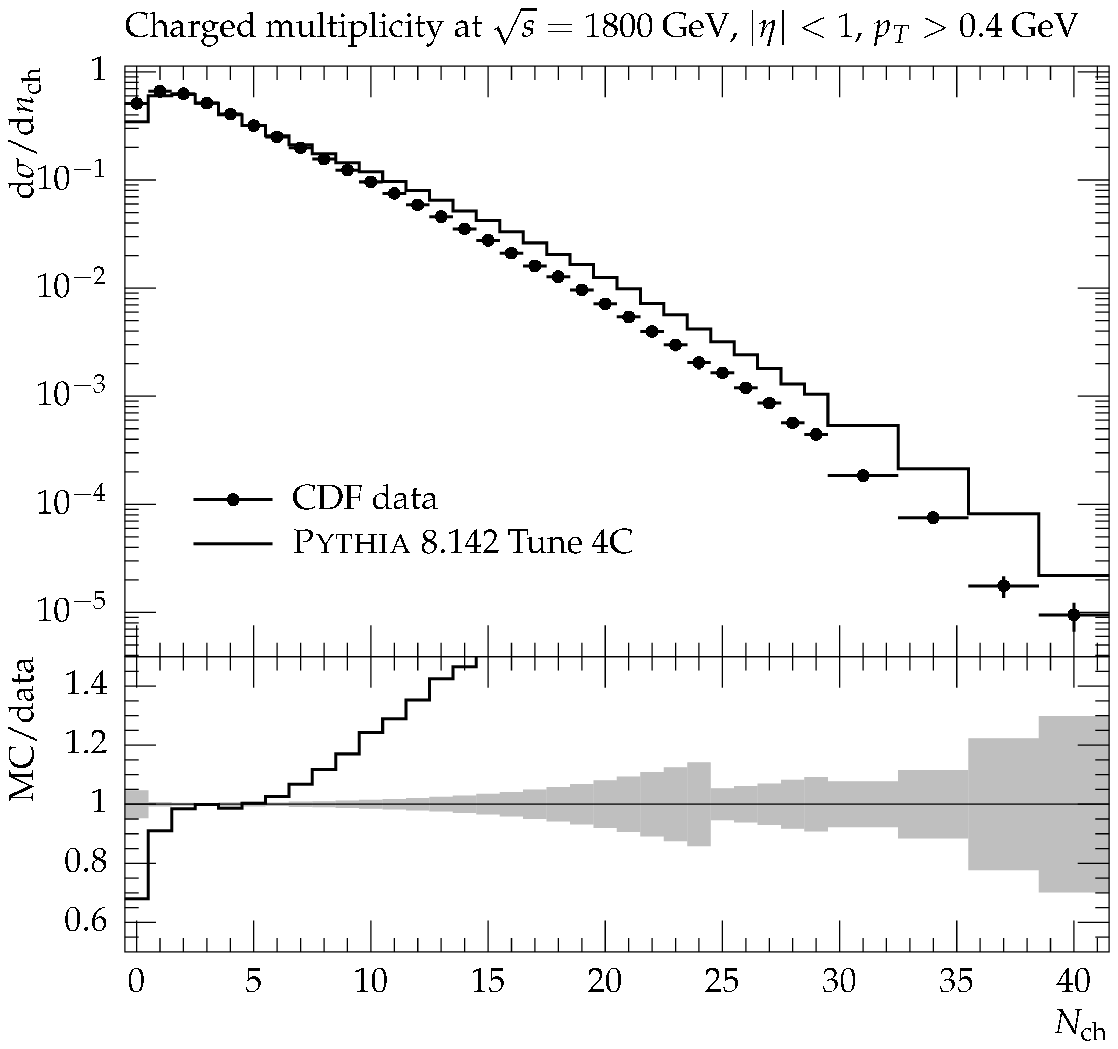}\hspace{7.5mm}
\includegraphics[scale=0.6]{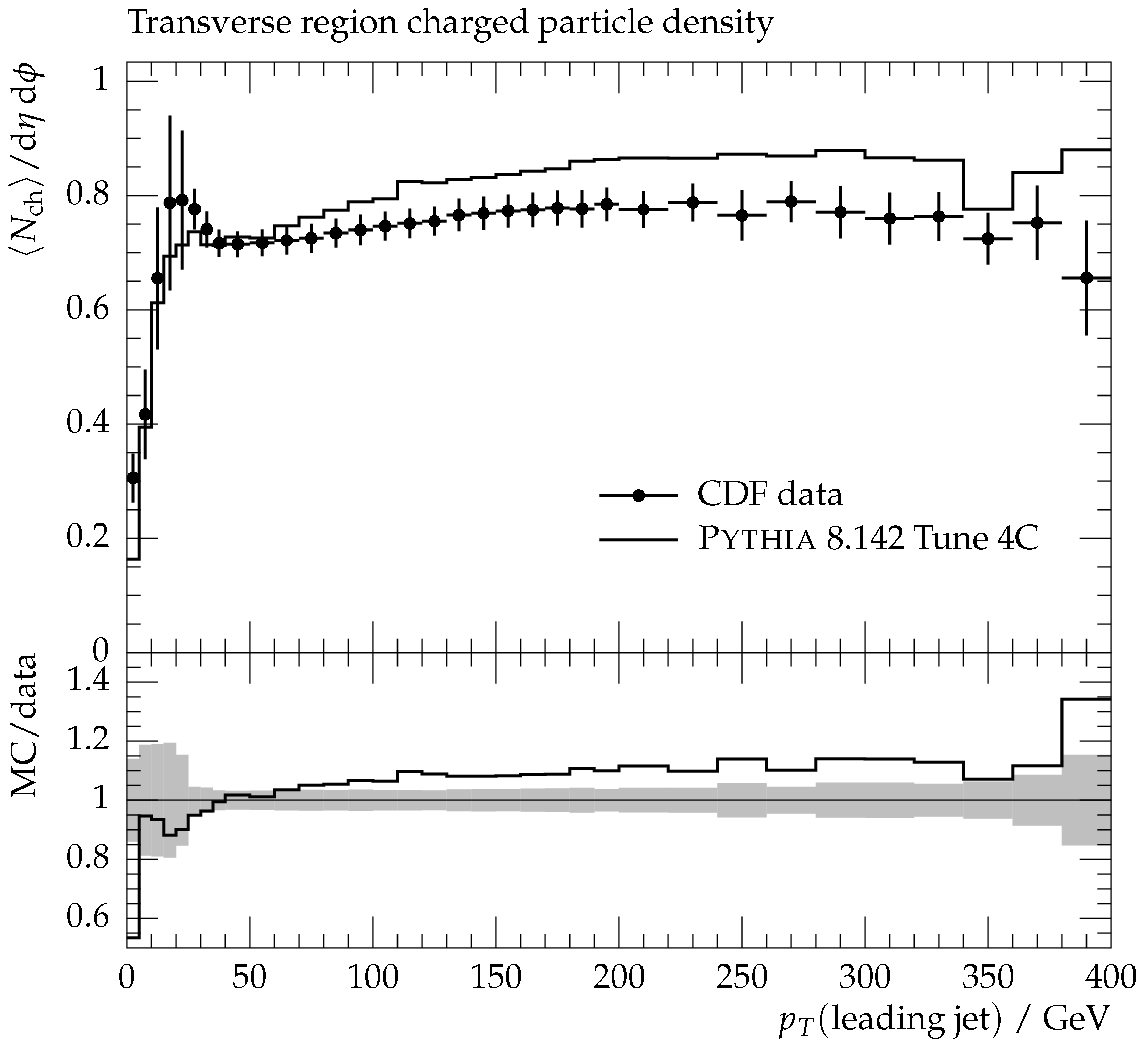}
}\caption{Charged particle multiplicity at $\sqrt{s} = 1800\GeV$ and
transverse region charged particle density at the Tevatron for Tune 4C
\label{fig:4CTeV}}
\end{figure}

Fig.~\ref{fig:4CTeV} shows the charged particle multiplicity at
$\sqrt{s} = 1800\GeV$ and the transverse region charged particle density
when Tune 4C is run against the Tevatron analyses of the previous section.
In general Tune 4C gives too much activity, also for the multiplicity at
$\sqrt{s} = 630\GeV$. In both the transverse region (where the $\sum \pT$
density is in fact well described) and overall, the average $\pT$ per
particle sits below the data. This suggests a tension between Tevatron and
LHC data, or at the very least a new kind of energy interpolation between
the $900\GeV$ and $7\TeV$ data, as suggested by R.~Field, based on his
studies with the \tsc{Pythia 6} generator \cite{Field:2009zz}.

\section{Summary and outlook}

In this article, modest changes to the parton-shower framework have been
introduced, including an improved description of dipoles stretching from
the final to the initial state in the final-state shower and additional
azimuthal correlations in the initial-state shower. To accompany
these changes, studies of the first shower emission in QCD events have been
made against $2 \to 3$ matrix elements, and simple tunes to Tevatron and
early LHC data have been produced.

The azimuthal weighting is primarily based on the expected angular
distribution for the emission of soft gluons, but it is known that away
from this limit it obtains corrections; hard emissions are free to ``write
their own rules''. These effects have been studied using the $\alpha$
angle, both in comparison to $2 \to 3$ matrix elements and Tevatron data.
In the comparison against data, the width of the pseudorapidity
distribution for the softest jet is well described, even without the
additional weighting, although an improvement in the central ``dip'' is
seen. For the $\alpha$ angle, good agreement with $2 \to 3$ ME's is noted,
but while the agreement to data is improved, it is still not as good as
hoped for. Further sources of asymmetry thus have to be found.

While an eventual goal would be to have a matching of the first shower
emission in QCD events to matrix elements, we have begun here with some
simple kinematic comparisons. Even with such a matching, subsequent
emissions would still be handled by the default shower, so it is important
to understand its behaviour in as much detail as possible. In regions where 
the parton shower is expected to perform well, the description is good both
in terms of rate and shape, across a wide range of energies. Away from the
soft and collinear regions, there are non-trivial kinematic effects, but
still a reasonable overall agreement, generally not worse than could be
expected from nontrivial higher orders.

As shown in Sec.~\ref{sec:intro}, the inclusion of FSR in the interleaved PS
and MPI framework has not been without some issues. In particular,
the simultaneous description of MB and UE data at the Tevatron has been
problematic. The changed dipole handling in FSR has turned out to be a key
factor, and we have performed a simple hand tuning to the Tevatron data to
show that a combined MB/UE description is possible. This has been done for
both the CTEQ6L1 and MRST LO** PDF sets. Across all datasets
considered, these tunes are never significantly worse than the Pro-Q20 and
Perugia 0 tunes of \tsc{Pythia 6.4}. These tunes are in no way intended to
be the final answer; to simplify the task, we have limited ourselves to a
few key parameters and datasets. It is hoped that what is presented here
will serve both as a starting point for a more complete global tune
to data, and as a general guide to some of the principles involved.

Finally, the first batches of LHC data have been released, including many
new MB and UE studies. Starting with the CTEQ6L1 Tevatron tune, and
limiting ourselves to changes in MPI and colour reconnection
parameters, supplemented by a slight change in diffractive cross
sections, a tune that broadly describes the features of much of the data
has been produced. It is encouraging that this has been possible given the
limited scope of the tuning. It is hoped that, with a more
complete tuning effort, \tsc{Pythia} will be able to describe this data
more accurately.

What is shown in this article is a very early look at this new data, and we
look forward to more data from the LHC experiments. New studies to help
separate diffractive contributions and their modeling will be welcome.
$\pT(\Z^0)$ and jet-jet azimuthal distributions will help to separate the
contributions from ISR and MPI, while other MB/UE data may help to resolve
some of the tensions seen so far. It remains an open question as to whether
the current models will be able to simultaneously describe both Tevatron
and LHC data. There may well be a region in the overall parameter space
that will be able to do this to some extent, but we do not rule out
differences due to experimental effects or deficiencies in the models, e.g.
related to the energy dependence of different parameters.

\subsection*{Acknowledgments}
This work was supported by the Marie Curie Early Stage Training program
``HEP-EST'' (contract number MEST-CT-2005-019626), the Marie
Curie research training network ``MCnet'' (contract number
MRTN-CT-2006-035606), and the Swedish Research Council (contract numbers
621-2008-4252 and 621-2007-4157). The authors wish to thank Deepak Kar for
his help in understanding the ATLAS UE analysis and Andy Buckley and
Hendrick Hoeth for their help with the Rivet framework and analyses. For
reading data from plots, the EasyNData tool was used
(\url{http://puwer.web.cern.ch/puwer/EasyNData/}).

\bibliography{2to3}{}
\bibliographystyle{utcaps}

\end{document}